# Reactivity-Informed Machine Learning for Performance Prediction and Design Space Exploration of Alkali-Activated Slag

Qiyao He[a,b,c], Zhanzhao Li[a,b,c], Kai Gong[a,b,c,*]

[a] Department of Civil and Environmental Engineering, Rice University, Houston, Texas 77005, United States.

[b] Rice Advanced Materials Institute, Rice University, Houston, Texas 77005, United States

[c] Ken Kennedy Institute, Rice University, Houston, Texas 77005, United States

[*] Corresponding author. E-mail: kg51@rice.edu

# ABSTRACT

Establishing quantitative relationships among mix design, raw material properties, curing conditions, and performance remains a long-standing challenge in cementitious materials, particularly for alkali-activated materials with variable precursor and activator chemistry. Here, we curated the largest literature-derived alkali-activated slag (AAS) dataset to date, comprising 3053 compressive strength records, 155 chemically distinct ground granulated blast-furnace slags (GGBSs), and 24 attributes that explicitly incorporate precursor chemistry, fineness, and reactivity. Based on this dataset, multiple machine learning (ML) algorithms were benchmarked across progressively enriched feature scenarios, demonstrating that integrating GGBS compositions, fineness, curing conditions, and specimen geometry substantially improves predictive performance. In particular, the average metal oxide dissociation energy (AMODE), a physically interpretable representation of precursor reactivity, provides a compact alternative descriptor to explicit oxide compositions while enabling comparable predictive performance. Model interpretation revealed physically consistent trends from the heterogeneous literature-scale data, including non-monotonic effects of $Na_2O$ dosage and silicate modulus, reduced predicted strength at higher water content and larger specimen size, and coupled oxide-level effects that are more coherently represented by AMODE than by individual oxide contents. Statistically constrained design space exploration further reveals reactivity-dependent trade-offs among strength, embodied $CO_2$ emissions, and cost. In particular, the design maps identify high-strength regions with substantially lower $CO_2$ emissions than ordinary Portland cement-based references at similar cost. Overall, this work demonstrates how reactivity-informed ML can extract physically meaningful trends from heterogeneous AAS data and guide source-dependent binder design. The curated dataset is publicly accessible to support broader advances in cement and concrete research.

**Keywords:** Alkali-activated slag; Precursor reactivity; Machine learning; Design space exploration; Strength–$CO_2$–cost trade-offs.

# 1 INTRODUCTION

Concrete is the most consumed material worldwide after water, and its demand continues to expand with global infrastructure development [1,2]. Despite its central role, concrete mix design in practice remains largely empirical, often relying on iterative trial batching and adjustment that are time-consuming and resource-intensive, and often yield suboptimal outcomes [3,4]. A fundamental reason for this limitation is the absence of robust quantitative relationships linking raw material attributes, mixture proportions, curing conditions, and engineering performance. Such relationships are difficult to establish because concrete performance emerges from a complex interplay among reaction kinetics, transport processes, and microstructural evolution across multiple spatial and temporal scales.

This challenge is further amplified as cementitious systems become increasingly complex in response to decarbonization goals and rising performance requirements. Ordinary Portland cement (OPC) production accounts for approximately 8–9% of global anthropogenic $CO_2$ emissions [1], driving increased incorporation of diverse supplementary cementitious materials [5,6], and the development of alternative binder systems [7,8]. Among these alternatives, alkali-activated materials (AAMs) have attracted considerable attention due to their lower embodied $CO_2$, ability to valorize various industrial byproducts and wastes, and potential for high mechanical, chemical, and thermal performance [9–11]. However, the wide compositional variability of precursors and diversity of activator chemistries introduce substantial chemical and microstructural heterogeneity, enlarging the design space and complicating the establishment of composition–structure–processing–property relationships required for engineering applications.

Recent advances in artificial intelligence and machine learning (ML) provide powerful tools to address these challenges. ML approaches can effectively capture nonlinear interactions among mixture proportions, processing variables, and performance objectives [3]. When integrated with inverse optimization, they enable systematic exploration of complex design spaces under coupled performance, economic, and environmental constraints [3,4,12,13]. Applications of ML in cement and concrete materials have expanded rapidly [14–18], drawing primarily on curated laboratory datasets [18–20] and, to a lesser extent, industry data [4,21,22]. These studies span a broad spectrum of engineering tasks, including prediction of mechanical properties [21–24], fresh properties [25–27], and durability performance [28–30]; phase and hydration modeling [31–33]; and image-based microstructural characterization [34].

Similarly, interest in applying ML to predict AAM properties has grown steadily since 2017 (**Figure 1a**). Existing ML studies typically encompass diverse precursor systems, including fly ash, ground granulated blast-furnace slag (GGBS), metakaolin, and silica fume (**Figure 1b**). These trends were identified using a large language model-powered text-mining pipeline [35,36], supplemented by manual verification. Despite this growing interest, most ML studies rely on relatively small and fragmented datasets and remain constrained by limited feature integration. Although several studies have considered chemical, physical, or processing variables [37–44], few have systematically integrated precursor reactivity with physical properties, curing conditions, and testing geometry.

These shortcomings are particularly pronounced in studies of alkali-activated slag (AAS), a major subclass of AAMs in which GGBS serves as the precursor. Unless otherwise specified, "slag" in this work denotes GGBS. A systematic review of the ML-for-AAS literature published through the end of 2025 [45–61] reveals several recurring gaps (**Figure 1c, d**). First, dataset sizes remain small, with the largest reported dataset containing 570 experimental data points [49]. Given the high dimensionality and chemical heterogeneity of AAS systems, such datasets, while valuable, provide sparse coverage of the compositional–processing space, which can limit reliable learning of nonlinear interactions. Second, precursor properties are frequently incomplete. Only a small fraction of prior studies (12%) incorporate slag chemical composition, despite extensive evidence of slag chemical variability [62,63] and its strong influence on reaction kinetics, gel formation, setting behavior, and mechanical performance [64–71]. Similarly, precursor fineness, which also affects reaction kinetics and strength development [72], is often omitted. More fundamentally, to the authors' knowledge, no prior ML study of AAS has incorporated physically grounded structural descriptors of precursor reactivity, despite reactivity being a primary driver of reaction kinetics and structural evolution. Third, practical variables known to influence cementitious material performance, particularly curing conditions [73,74] and specimen geometry [75,76], are commonly neglected. Collectively, sparse datasets and the omission of influential variables increase the risk of overfitting, introduce omitted-variable bias, and limit model robustness and applicability across different compositional domains.

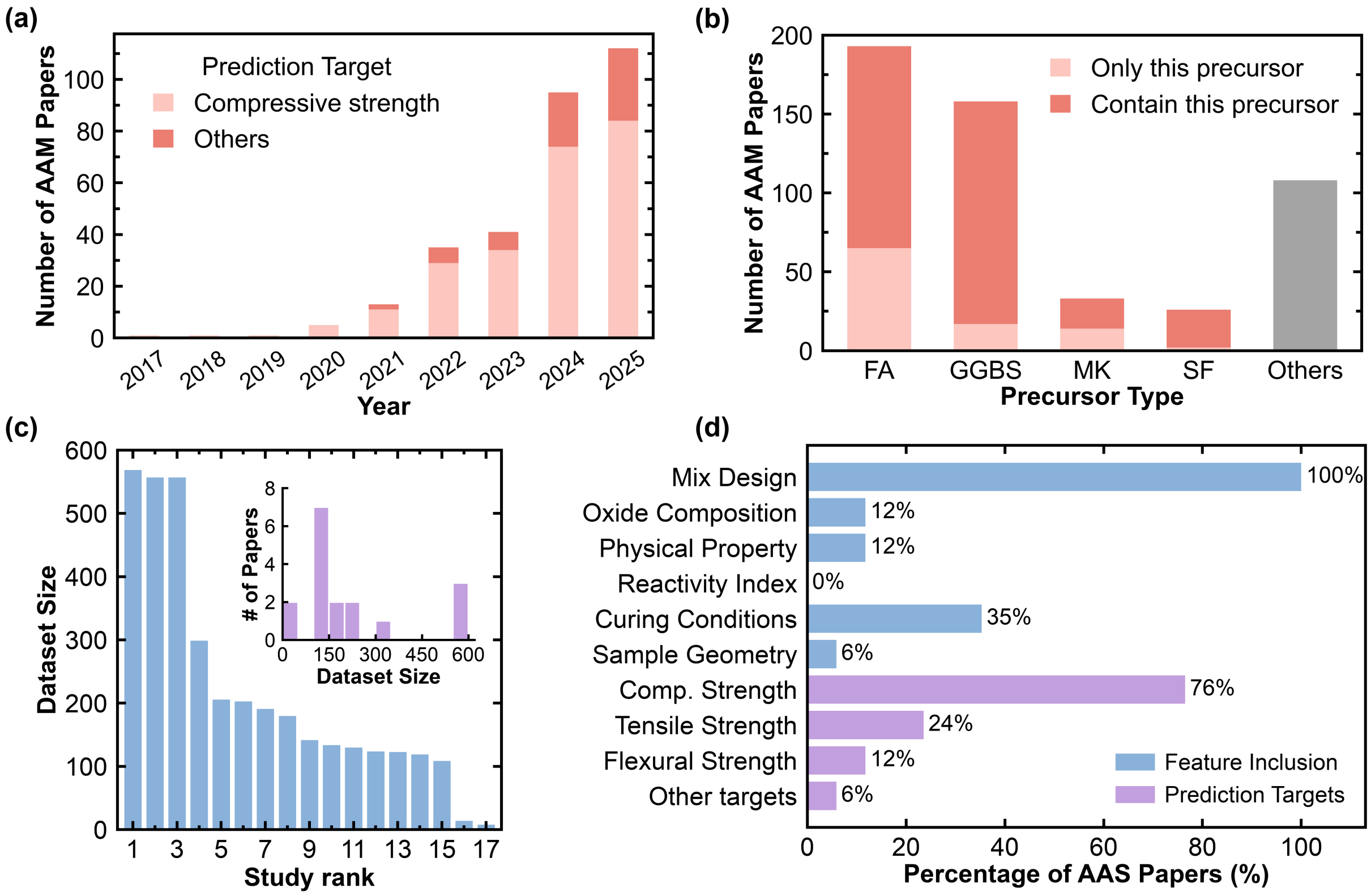

**Figure 1.** Trends in publications on machine learning (ML) applications in alkali-activated materials (AAMs) since 2017, based on studies published through the end of 2025: (a) number of ML-for-AAMs papers by publication year; (b) distribution of ML-for-AAMs papers by precursor type; (c) dataset sizes used in individual ML-for-alkali-activated slag (AAS) studies, ordered from largest to smallest, with an inset showing their distribution; and (d) input feature categories and prediction targets across ML-for-AAS papers, where curing conditions exclude age. Precursor abbreviations are defined as follows: FA, fly ash; GGBS, ground granulated blast-furnace slag; MK, metakaolin; and SF, silica fume.

To address these gaps, this study establishes a comprehensive data-driven framework for strength prediction and design space exploration in AAS systems. A large AAS dataset was curated, comprising over 3000 compressive strength records and 155 chemically distinct GGBSs collected from 136 studies. It integrates mixture proportions, precursor chemical and physical attributes, curing conditions, and specimen geometry within a unified feature space.

A central innovation is the incorporation of a physically grounded reactivity descriptor, the average metal oxide dissociation energy (AMODE) [77–80], which was originally developed based on atomistic simulation results to quantify intrinsic GGBS reactivity. This enables a large-scale analysis of how slag reactivity interacts with activator dosage, silicate modulus, and water content to influence compressive strength development. Based on the dataset, multiple ML models were developed for strength prediction and benchmarked across progressively enriched feature scenarios. SHapley Additive exPlanations (SHAP) and accumulated local effects (ALE) analyses were employed to evaluate feature influence and nonlinear response trends.

Previous studies have developed performance-based mixture design approaches for slag-containing AAMs [81,82]. In this study, the trained models were deployed as surrogate models for statistically constrained design space exploration, revealing reactivity-dependent favorable parameter ranges and coupled effects. In parallel, screening-level estimates of embodied $CO_2$ emissions and cost were used to map performance–cost–emissions trade-offs relative to OPC reference systems. Accordingly, the models are used primarily as reactivity-informed synthesis and exploration tools within the literature-supported AAS domain, rather than as universal predictors for arbitrary future mixtures. The openly accessible dataset further provides a reusable data resource for future data-driven advances in AAMs.

# 2 METHODOLOGY

## 2.1 Dataset Curation and Feature Engineering

### *2.1.1 Data Sources and Extraction*

Data curation focused on AAS mixtures reported in peer-reviewed journal articles indexed in the Web of Science Core Collection, ensuring consistent documentation standards. A small number of additional publicly available literature sources, including conference papers and theses, were also included when they provided sufficient and traceable information. Tabulated data were extracted using Tabula [83], and graphical data were digitized using WebPlotDigitizer [84]. All extracted values were manually verified for completeness, internal consistency, and numerical accuracy. Source identifiers were retained for each record to preserve traceability and support future analysis of study-level variability.

#### *2.1.1.1 Inclusion and Exclusion Criteria*

Mixtures were included only if they provided all the information needed to construct the full predefined feature set. This feature set was defined a *priori* based on domain knowledge of factors influencing AAS strength development and on the availability of cross-study data to support robust model training and comparability. Specifically, studies were required to report: (i) GGBS oxide composition; (ii) sufficient activator chemistry information to enable calculation of activator-supplied $Na_2O$ content, silicate modulus (Ms = $SiO_2/Na_2O$ molar ratio), and water/GGBS ratio; (iii) detailed mixture proportions; and (iv) compressive strength. Certain mixtures were excluded due to limited data availability or added system complexity. Specifically, we excluded mixtures employing (i) activators other than $Na_2SiO_3$ and NaOH (e.g., $K_2SiO_3$, KOH, $Na_2CO_3$, $K_2CO_3$); (ii) non-standard constituents such as recycled aggregates, fibers, or admixtures other than water reducers, or (iii) non-standard curing protocols (e.g., long-term unsealed water immersion used to study leaching behavior). Together, these criteria balance data completeness and dataset size, ensuring a coherent dataset suitable for mechanistically informed ML models.

#### *2.1.1.2 Feature Engineering*

The curated dataset enabled the construction of 24 features representing the chemical, physical, mixture, curing, and geometrical factors that influence strength development in AAS systems. These features were grouped into six categories based on their physical significance and mechanistic role:

1) GGBS Chemistry: Mass fractions of major oxides ($SiO_2$, $Al_2O_3$, CaO, MgO, $Fe_2O_3$, $Na_2O$, $K_2O$, $Mn_xO_y$, and $TiO_2$) were included as compositional descriptors of the GGBS, given their documented influence on GGBS reactivity and AAS strength development [65,67,68,70,85–90].

2) GGBS Physical Property: Blaine fineness ($cm^2/g$), an air-permeability-based fineness index representing an equivalent specific surface area. Blaine fineness was selected over BET surface area and laser diffraction-based particle size metrics because it is reported more consistently across studies.
3) GGBS Reactivity Index: The modified AMODE [79], a physics-informed reactivity descriptor, was calculated as a composition-derived proxy for GGBS reactivity (see **Section 2.3** for details).
4) Mix Design Parameters: $Na_2O$/GGBS (wt.%), Ms, water/GGBS ratio, fine aggregate/GGBS ratio, coarse aggregate/GGBS ratio, and superplasticizer/GGBS (wt.%).
5) Curing Conditions:
   - $\ln(\text{Age} + 1)$: The natural logarithm of curing age (days) was used to reduce the extreme skewness in reported ages (ranging from 2 hours to 1278 days) and to improve model performance, given that most data were concentrated within 28 days.
   - Base temperature (°C): Nominal curing temperature, typically around 20 °C.
   - Additional temperature (°C): Deviation from the base temperature, representing elevated or reduced temperature curing applied during the curing process.
   - $\ln(\text{Energy} + 1)$: A signed logarithmic thermal-exposure proxy was calculated as sign(Additional temperature) × ln(Energy + 1), where Energy = |Additional temperature| × Additional time, and Additional time denotes the duration of additional-temperature curing. The temperature–time product serves as a simple cumulative thermal-exposure descriptor rather than a measure of actual energy consumption. It combines the magnitude and direction of the temperature deviation from the base curing condition with its duration, while the logarithmic transformation reduces skewness. Sensitivity analysis showed that the predictive performance and main model-inferred trends were largely unchanged when temperature and time were entered separately **(Section S1.2** of the **Supplementary Material)**.
   - Relative humidity (RH, %) during curing was included when reported or inferable from curing descriptions.
6) Specimen Geometry:

- Aspect ratio: Ratio of specimen height (along loading direction) to its original cross-sectional dimension (for prismatic specimens, the mean side length).
- Equivalent length (mm): $L_{eq} = \sqrt[3]{\text{sample volume (mm}^3\text{)}}$ was used to represent specimen size across cubes, prisms, and cylinders and to account for size effects on compressive strength [76,91].

Compared with prior ML-based studies of AAS (**Figure 1d**), the present feature set comprehensively integrates precursor chemical composition, physical properties, reactivity information, curing conditions, mix design, and specimen geometry within a unified modeling framework. Because the dataset was compiled from heterogeneous literature sources, it is intended for trend extraction, comparative modeling, and constrained exploration within the data-supported domain. Model-derived feature effects are therefore interpreted as conditional trends.

## 2.2 Data Preprocessing

Application of the above criteria yielded about 3400 data points from more than 145 articles. Although relatively large, the dataset is compiled entirely from literature sources and is therefore still subject to inherent uncertainty arising from variability across laboratory conditions, testing protocols (e.g., loading rate), measurement error, reporting inconsistencies, and unreported raw material attributes (e.g., glass content, cooling rate during GGBS production, and particle size distribution beyond Blaine fineness). While such sources of uncertainty cannot be eliminated, the preprocessing workflow was designed to harmonize reported variables and improve traceability while retaining the broad variability inherent to literature-derived AAS data. Three preprocessing steps were applied: (1) *data imputation* to address missing values, (2) *data cleaning* to correct errors and exclude ineligible records, and (3) *data standardization* to rescale features and avoid magnitude-related bias during model training [92].

### *2.2.1 Missing Data Imputation*

Blaine fineness was missing for 38.1% of the records and was imputed using a constant value of 4500 $cm^2/g$. Sensitivity analysis comparing constant imputation, multivariable imputation, and feature omission indicated limited changes in predictive performance and the main feature–response trends (see **Section S1.1** for details). This imputed value was used as a practical placeholder and the learned effect of Blaine fineness was interpreted cautiously. The imputed values were used for modeling but were not included in the feature distributions shown in **Figure 2**.

RH, a crucial curing parameter, was also unreported in many studies. Missing RH values were imputed based on common curing descriptions used in AAS research: (1) RH = 100% for water or bath curing (water-saturated environment); (2) RH = 95% for moist-room or fog-room curing (near-saturated vapor conditions); and (3) RH = 90% for sealed or covered curing, which limits moisture loss without imposing external saturation. Studies lacking sufficient information to infer curing humidity were removed. These assigned RH values are approximate representations of the reported curing environments. For base curing temperature, values reported only as “room temperature” or “ambient temperature” were assumed to be 20 °C. Sensitivity tests examined adding an indicator for inferred RH, excluding RH, varying the inferred RH value for sealed curing to 80%, 85%, and 95%, and varying the assumed room temperature to 18, 22, and 24 °C. These tests indicated limited changes in predictive performance and the main feature–response trends (**Section S1.2**). The inferred RH values and base curing temperatures were included in **Figure 2** to enable a consistent numerical representation of curing conditions across the dataset.

Unreported minor oxide contents were assigned zero as a modeling assumption. After preprocessing, all feature scenarios used the same 3053 records, differing only in the features included. Zero aggregate ratios represent the physical absence of the corresponding aggregate fraction, rather than imputed missing values.

#### *2.2.2 Data Cleaning*

Data cleaning was performed using a hybrid model-assisted, expert-reviewed workflow. First, duplicate records were identified through feature comparison and source verification and removed. Next, extreme values were manually inspected to identify potential extraction or reporting errors. Four regression algorithms (Gaussian Process Regression, Artificial Neural Network, Random Forest, and Extreme Gradient Boosting) were then trained using the full feature set and compressive strength as the target for residual-based anomaly screening. Algorithmic formulations and hyperparameter tuning procedures are described in **Section 2.4** and **Section S1.3**. Model training and evaluation were repeated across 10 random states (10, 20, …, 100). In each run, potential outliers were flagged using the z-scores of absolute prediction errors:

$$z_i = \frac{\epsilon_i - \mu_{b(i)}}{\sigma_{b(i)}} \tag{1}$$

where $\epsilon_i = \left| y_i^{\text{true}} - y_i^{\text{prediction}} \right|$ denotes the absolute prediction error for sample $i$; $b(i)$ is the strength bin containing sample $i$, and $\mu_{b(i)}$ and $\sigma_{b(i)}$ are the mean and standard deviation of absolute errors within bin $b(i)$, respectively. Strength binning accounts for differences in prediction-error magnitude across strength ranges. Samples with $z_i$ >3 were flagged following Aggarwal [93].

For each data point, the frequency of being flagged across models and random states was recorded. Data points with high flagging frequencies were manually examined against the original source for extraction errors, duplicates, and consistency with the dataset's inclusion criteria. Prediction errors or unusual strength trends alone were not considered sufficient grounds for exclusion. Records were excluded because they fell outside the activator scope or had limited quantitative precision, with specific justifications documented in **Table S7**. Sensitivity analysis indicated that excluding these records had little effect on predictive performance and the main findings (**Section S1.4**). Final dataset statistics are provided in **Section 3.1**.

### *2.2.3 Data Standardization*

During extraction, all reported units were harmonized to a consistent basis for each feature. Prior to model training, all input features and the target variable were standardized using z-score normalization to achieve zero mean and unit variance. This step was necessary because feature values span several orders of magnitude; without scaling, large-magnitude features can dominate distance-based computations and bias model optimization [94]. Standardization was performed using StandardScaler from Scikit-learn [95]. To avoid data leakage, scaling parameters were fit exclusively on the training set and subsequently applied to the test set. All reported root-mean-square error (RMSE) values and predicted strengths were transformed back to MPa for interpretation and visualization.

### 2.3 Reactivity Index: Modified AMODE

To support mechanistic interpretation of compositional effects, we incorporated AMODE as a physically grounded, composition-derived proxy for GGBS reactivity. AMODE was originally developed by Gong and White based on atomistic simulations [80] to evaluate the relative reactivity of glassy structures in the compositional space of $CaO$–$MgO$–$Al_2O_3$–$SiO_2$, and later extended to volcanic glasses [78]. AMODE represents the weighted average metal–oxygen bond strength within the glass network, with higher AMODE values reflecting stronger overall network structure and hence lower chemical reactivity. The original parameter is defined as:

$$\text{AMODE} = \frac{\sum N_M \times CN_M \times BS_{M\text{-}O}}{\sum N_M} \tag{2}$$

where $N_M$ is the number of metal cations of type $M$ (e.g., Ca, Mg, Al, Si) in the system, and $CN_M$ and $BS_{M\text{-}O}$ denote the average coordination number ($CN$) and metal–oxygen bond strength ($BS$) of cation $M$, respectively. $CN$ values can be obtained from molecular dynamics simulations, while $BS$ values can be obtained from the literature [96].

In this study, we employed the simplified form, the modified AMODE [79], which was developed by assuming a constant $CN$ for each cation type (independent of glass compositions). This approximation allows AMODE to be computed directly from bulk chemical compositions without requiring structural characterization or atomistic modeling, making it straightforward to apply. Previous studies have shown that the modified AMODE performs comparably to the original formulation, with $R^2$ values of 0.86–0.97 for linear relationships with experimental reactivity indicators, including calorimetric peak time, bound water content, and degree of reaction [79]. Across these datasets, it generally outperforms 15 conventional compositional descriptors, including the basicity coefficient and extent of depolymerization (NBO/T) [79]. The modified AMODE parameter has recently been expanded to cover the compositional space of $CaO$–$MgO$–$Al_2O_3$–$SiO_2$–$Fe_2O_3$–$Na_2O$–$K_2O$–$MnO$–$TiO_2$ [77], as given by:

$$\text{Modified AMODE} = \frac{209N_{\mathrm{Ca}}+174N_{\mathrm{Mg}}+368N_{\mathrm{Al}}+424N_{\mathrm{Si}}+331N_{\mathrm{Fe}}+157N_{\mathrm{Na}}+133N_{\mathrm{K}}+198N_{\mathrm{Mn}}+353N_{\mathrm{Ti}}}{\left(N_{\mathrm{Ca}}+N_{\mathrm{Mg}}+N_{\mathrm{Al}}+N_{\mathrm{Si}}+N_{\mathrm{Fe}}+N_{\mathrm{Na}}+N_{\mathrm{K}}+N_{\mathrm{Mn}}+N_{\mathrm{Ti}}\right)} \quad (3)$$

where $N_i$ denotes the molar amount of cation $i$ in the glass. This updated formulation incorporates contributions from minor oxides (e.g., Fe, Na, K, Mn, Ti), which have also been reported to influence GGBS reactivity [87,88]. Accordingly, **Equation (3)** was used here to calculate the reactivity index for GGBSs used in the dataset based on their bulk chemical compositions. A strong negative correlation between the modified AMODE and Ca/Si ratio was observed (**Figure S14)**, consistent with the lower average bond strength and greater network depolymerization associated with Ca-rich slags.

Because both AMODE and modified AMODE were developed for amorphous materials, their application to bulk chemical composition here assumes that GGBSs are fully amorphous. Although detailed mineralogical information is rarely reported in the compiled studies, this approximation is reasonable, as GGBS is generally known to be predominantly amorphous [63,90]. Variations in crystalline fraction, glass structure, and cooling history are therefore not explicitly resolved; modified AMODE provides a composition-derived estimate of slag reactivity. The associated limitations are discussed further in **Section 4.2**. For simplicity, the term "AMODE" will be used throughout this paper to refer to the modified AMODE. The final cleaned dataset contains 155 distinct GGBSs with a wide range of AMODE values (**Figure 2a**).

## 2.4 Machine Learning Models

To predict the compressive strength of AAS systems from the input features defined in **Section 2.1.1.2**, four ML algorithms were evaluated: Gaussian Process Regression (GPR), Artificial Neural Network (ANN), Random Forest (RF), and Extreme Gradient Boosting (XGB). Linear Regression (LR) was included to serve as a baseline. These models represent kernel-based, neural-network, and tree-based methods, allowing comparison of predictive performance and model behavior across distinct modeling paradigms.

GPR is a probabilistic, kernel-based regression model that places a prior distribution over functions and conditions this prior on the observed data to obtain predictive means and uncertainties [97].

ANN is a nonlinear function approximator composed of interconnected layers of neurons, with model parameters (weights and biases) optimized through gradient-based minimization of a predefined loss function [98]. RF is an ensemble of decision trees trained on bootstrapped subsets of the data, with the final prediction obtained by averaging the outputs of individual trees [99]. This bagging strategy reduces variance, improves robustness, and mitigates overfitting relative to a single decision tree. XGB is also tree-based but relies on a sequential learning process known as gradient boosting [100], in which each tree is trained to correct the residual errors of the previous ensemble. It incorporates regularization, shrinkage, and column subsampling to control model complexity and enhance generalization, rendering XGB highly effective for structured tabular datasets [101]. LR was employed as a classical baseline, assuming a linear relationship between the input features and the target, with coefficients estimated via ordinary least-squares minimization [102].

### *2.4.1 Input Feature Configurations*

To systematically evaluate the influence of chemical information on model performance, four input feature scenarios were defined and tested for each model (**Table 1**):

- **Scenario 1 – Baseline Mix Design and Age (7 features).** This baseline scenario includes only mix design parameters and curing age, expressed as $\ln(\text{Age}+1)$. No precursor chemistry, physical properties, curing conditions beyond age, or specimen geometry are included. This configuration reflects the minimal feature set commonly adopted in existing ML studies on AAS systems.
- **Scenario 2 – Physical Property, Curing, and Geometry (14 features).** This scenario extends the baseline by incorporating GGBS Blaine fineness, along with curing conditions and specimen geometry, while still excluding explicit GGBS chemical information.
- **Scenario 3 – Explicit Chemical Composition (23 features).** In addition to the Scenario 2 features, this scenario incorporates the nine oxide contents of GGBS, providing an explicit composition-based description of precursor chemistry.

- **Scenario 4 – Physics-Informed Reactivity (15 features).** This scenario substitutes the nine individual oxide contents in Scenario 3 with a single calculated feature, AMODE, which condenses compositional data into a physically grounded reactivity parameter.

**Table 1.** Summary of the feature configurations across the four scenarios investigated. GGBS denotes ground granulated blast-furnace slag.

| Feature category | # of Features | Scenario 1 | Scenario 2 | Scenario 3 | Scenario 4 |
|---|---|---|---|---|---|
| GGBS chemical compositions | 9 | | | ✓ | |
| GGBS physical property | 1 | | ✓ | ✓ | ✓ |
| GGBS reactivity | 1 | | | | ✓ |
| Mix design parameters | 6 | ✓ | ✓ | ✓ | ✓ |
| Curing conditions | 5 | ✓[a] | ✓ | ✓ | ✓ |
| Sample geometry | 2 | | ✓ | ✓ | ✓ |

a. For Scenario 1, curing conditions include only ln(Age + 1).

In feature-limited scenarios (Scenarios 1 and 2), samples with identical input variables but different underlying precursor properties or curing/testing conditions may become indistinguishable in the reduced feature space. These samples were retained to ensure a consistent dataset size and enable direct performance comparison across scenarios. Duplicate-input groups and the associated empirical RMSE floors were quantified for Scenarios 1 and 2 (details on the calculation methods and results are given in **Section S3.1**). These two scenarios serve as baseline references; Scenario 2 is also used for comparison in design space exploration.

### *2.4.2 Model Training and Hyperparameter Optimization*

All models were trained using a consistent workflow to ensure fair comparison and reliability, similar to our previous work [103,104]. Briefly, the cleaned dataset was randomly split into an 80:20 train–test partition. This evaluation compares models and feature scenarios within the curated literature domain; it does not establish cross-study or cross-source transferability. For ANN, XGB, and RF, hyperparameter tuning was performed on the training set using random search combined with shuffled, stratified 10-fold cross-validation. Stratification was applied based on compressive strength to preserve its distribution across folds. Random search was selected over exhaustive grid search due to its computational efficiency in high-dimensional hyperparameter spaces and its demonstrated ability to identify well-performing regions of the search space with substantially fewer evaluations [105]. The hyperparameter search ranges and tuning strategies are summarized in **Table S6**.

For GPR, a composite kernel of the form $K$ = *Constant* × *Rational Quadratic* + *White Noise* was adopted to balance flexibility and generalization. The *Constant* kernel sets the overall signal magnitude, while the *Rational Quadratic* kernel captures multi-scale smoothness by accommodating a distribution of length scales. This scale-mixture property enables the capture of multi-scale trends, which can improve generalization near the boundaries of the training domain [106]. Finally, the *White Noise* kernel accounts for inherent noise and ensures numerical stability. All kernel hyperparameters were optimized from fixed initial values by maximizing the marginal likelihood [107], without random restarts.

The workflow was repeated using 10 random seeds (10, 20, …, 100). Each seed controlled the train–test split and, for ANN, XGB, and RF, the shuffled cross-validation folds and randomized hyperparameter search.

### *2.4.3 Model Evaluation*

Model performance was evaluated using the coefficient of determination ($R^2$) and RMSE, with overfitting assessed by the relative differences between training and test metrics ($\Delta R^2$ and ΔRMSE). The calculation details are provided in **Section S1.6**. Results are reported as the mean metric values across the 10 runs, with standard deviations shown as error bars. Post-hoc model interpretation was performed using SHAP [108] and ALE [109], as described in **Section 3.4**. All model construction, training, testing, and interpretability analyses were performed in Python using standard ML libraries, including Scikit-learn [95], XGBoost [101], SHAP [108], and PyALE [109].

## 2.5 Design Space Exploration

To explore conditional mix design trends within the data-supported domain, the trained surrogate models were used to predict compressive strength across a synthetic design space spanning key mix design parameters. For each model type, predictions were generated across 10 random states, as described in **Section 2.4**. Their mean and standard deviation were used to quantify the central trend and inter-run variability. The synthetic design space was constructed in three steps: (i) generating a full-factorial grid within the selected feature ranges, (ii) screening formulations for non-negative ingredient masses and mass-balance consistency, and (iii) applying Mahalanobis filtering to limit multivariate extrapolation beyond the experimental data distribution.

### *2.5.1 Generation of Artificial Design Space*

An artificial design space was generated by systematically varying five key features, selected based on domain knowledge and feature importance rankings from SHAP analysis, within their ranges represented in the dataset: (i) Age (days): 1, 3, 7, 14, 28, 56, 90, 180; (ii) $Na_2O$/GGBS (wt.%): 1–20, sampled at 1 wt.% increments from 1–8 wt.% and 2 wt.% increments from 10–20 wt.%; (iii) Ms: 0–5 in 0.5 increments; (iv) Water/GGBS ratio: 0.2–0.8 in 0.05 increments; and (v) GGBS AMODE (kcal/mol): 287–311 in 3 kcal/mol increments. All remaining input features (e.g., curing conditions and specimen geometry) were fixed at representative values, which are summarized in **Table S9**. AMODE (Scenario 4) was used in place of the nine individual oxide compositions to reduce input dimensionality. Scenario 2 was explored as a reference configuration using the same procedure. The resulting full-factorial grid comprised 144,144 synthetic design points for Scenario 4.

### *2.5.2 Mixture Conversion and Mass-Balance Screening*

To check mass-balance consistency and support the $CO_2$ emission and cost calculations in **Section 2.6**, $Na_2O$/GGBS, Ms, and water/GGBS were converted to ingredient masses on a 100 g GGBS basis. A 50 wt.% NaOH solution and a sodium silicate solution with 37 wt.% solids and a $SiO_2/Na_2O$ mass ratio of 3.3 were initially used. When their combined water contribution exceeded the target total water content, the sodium silicate solution was replaced with sodium metasilicate pentahydrate ($Na_2SiO_3{\cdot}5H2O$), and the ingredient masses were recalculated. Water from both activator sources, including crystallization water, was included in the total water balance. Only formulations with non-negative ingredient masses and closed $Na_2O$-equivalent, $SiO_2$, and water balances proceeded to Mahalanobis filtering. For Scenario 4, this screening removed 4464 of the 144,144 generated points, leaving 139,680 points. These checks establish mass-balance consistency for the assumed ingredient set but do not guarantee solution stability, workability, or practical feasibility, which require further information on materials and processing. Conversion equations and worked examples are provided in **Section S1.8**.

### *2.5.3 Multivariate Filtering via Mahalanobis Distance*

Although each feature in the artificial grid was constrained within its observed univariate range, many synthetic combinations represent substantial multivariate extrapolation beyond the joint feature distribution of the experimental data. Because ML models are less reliable under such extrapolation [110,111], a multivariate filtering step was applied to remove combinations lying far from the experimental distribution. Filtering was performed using the Mahalanobis Distance ($D_M$) [112], a widely used multivariate distance metric that accounts for feature correlations through the covariance matrix [113]. Prior to $D_M$ calculation, the experimental and artificial features were standardized using the mean and standard deviation of the empirical dataset. The $D_M$ for each artificial data point $x$ was computed as:

$$D_M(x) = \sqrt{(x-\mu)^T \Sigma^{-1} (x-\mu)} \tag{4}$$

where $\mu$ is the mean vector and $\Sigma$ is the covariance matrix of the experimental features. This metric quantifies the distance of a point from the multivariate centroid, while accounting for feature variances and correlations, thereby reflecting the joint structure of the experimental dataset distribution [114].

To determine a cutoff $D_M$ for filtering, we compared the squared $D_M$ values to a chi-square distribution with degrees of freedom ($df$) equal to the number of input variables ($df$ = 15 for Scenario 4). The 97.5th percentile of the chi-square distribution was adopted as the cutoff for $D_M^2$. For $df$ = 15, this yields [115]:

$$\chi^2_{0.975,\,15} = 27.488 \;\rightarrow\; D_{M,\mathrm{cutoff}} = \sqrt{27.488} \approx 5.243 \tag{5}$$

Although the chi-square relationship assumes multivariate normality, the present dataset combines mixture types and includes discrete curing and geometry settings. Mahalanobis distance is therefore used here as a heuristic multivariate screening criterion to reduce unsupported extrapolation, rather than as a formal probabilistic applicability boundary. The filter does not guarantee adequate local data support for every retained combination. For Scenario 4, this process removed 85,025 of the 139,680 points retained after mass-balance screening, leaving 54,655 for subsequent analysis. The filtered design space was used to visualize predicted strength trends and evaluate interactions among key variables, including curing age, mix design parameters, and GGBS reactivity.

## 2.6 Environmental and Economic Impact Calculations

In addition to mechanical performance, environmental and economic impacts are essential considerations for practical implementation of AAS mixtures. Accordingly, we quantified the embodied $CO_2$ emissions and estimated cost for all mixtures in the curated experimental dataset and the filtered synthetic design space. For both calculations, a screening-level boundary was used, including material-related contributions and additional elevated-curing energy contributions where applicable, with totals normalized by the total mixture mass. Ingredient masses were obtained using the conversion procedure in **Section S1.8**. Emission and cost factors are summarized in **Table 2**, with additional assumptions applied where necessary.

The material-related contributions for each mixture were calculated as:

$$\mathrm{CO_2}_{\mathrm{material}} = \sum_i m_i f_i^{\mathrm{CO_2}} \tag{6}$$

$$\mathrm{Cost}_{\mathrm{material}} = \sum_i m_i f_i^{\mathrm{Cost}} \tag{7}$$

For mixtures subjected to curing above their base temperature, additional energy consumption and its associated $CO_2$ and cost contributions were included:

$$E = f_{\mathrm{E}} \Delta T t \tag{8}$$

$$\mathrm{CO_2}_{\mathrm{energy}} = E f_{\mathrm{E}}^{\mathrm{CO_2}} \tag{9}$$

$$\mathrm{Cost}_{\mathrm{energy}} = E f_{\mathrm{E}}^{\mathrm{Cost}} \tag{10}$$

where $m_i$ is the mass fraction of constituent $i$ in the mixture; $f_i^{\mathrm{CO_2}}$ and $f_i^{\mathrm{Cost}}$ are the raw material $CO_2$ emission factor and cost factor for constituent $i$ (**Table 2**), respectively; $E$ is the additional curing energy consumption; $f_{\mathrm{E}}$ is the curing-energy factor; $\Delta T$ is the applied curing temperature above ambient, corresponding to the "Additional temperature" in the dataset; $t$ is the duration of elevated-temperature curing; and $f_{\mathrm{E}}^{\mathrm{CO_2}}$ and $f_{\mathrm{E}}^{\mathrm{Cost}}$ are the $CO_2$ emission factor and cost factor for energy consumption, respectively.

The curing-energy factor $f_{\mathrm{E}}$ was taken as $2.81\times10^{-6}$ kWh·°C$^{-1}$·h$^{-1}$·kg$^{-1}$, calculated from the steam-curing case study conducted by Zhu et al. [116]. The $CO_2$ emission factor $f_{\mathrm{E}}^{\mathrm{CO_2}}$ and cost factor $f_{\mathrm{E}}^{\mathrm{Cost}}$ for energy consumption were adopted as 0.367 kg·kWh$^{-1}$ and 0.1268 USD·kWh$^{-1}$, respectively, based on data from the U.S. Energy Information Administration [117]. Only elevated-temperature curing ($\Delta T > 0$) was considered; no additional energy was assigned to low-temperature curing ($\Delta T < 0$).

Finally, total $CO_2$ emissions and cost were obtained by summing the material-related and energy-related contributions:

$$\mathrm{CO_2}_{\mathrm{total}} = \mathrm{CO_2}_{\mathrm{material}} + \mathrm{CO_2}_{\mathrm{energy}} \tag{11}$$

$$\mathrm{Cost}_{\mathrm{total}} = \mathrm{Cost}_{\mathrm{material}} + \mathrm{Cost}_{\mathrm{energy}} \tag{12}$$

Sensitivity to emission and cost factors was evaluated using 20,000 Monte Carlo trials for both AAS mixtures and OPC references. Nonzero factors were independently sampled from uniform distributions spanning ±10% of the baseline values specified above and in **Table 2**; zero-valued factors remained unchanged. Mixture proportions and curing-energy quantities were held fixed, and the same sampled factors were applied to all mixtures within each trial. The 2.5th–97.5th percentile ranges quantify sensitivity to the prescribed factor variations.

The resulting environmental and economic distributions for both the curated experimental dataset and the filtered synthetic design space are presented in **Sections 3.2** and **3.5**. OPC paste and mortar references were derived by removing the relevant aggregate fractions from the industrial concrete composition in ref. [4] and renormalizing the remaining constituents.

**Table 2.** Summary of the $CO_2$ emission and cost factors for raw material production used in the screening-level calculation of environmental and economic impacts.

| Mixture constituent | $CO_2$ emission factor, $f_i^{CO_2}$ (kg $CO_2$-eq/kg) | Cost factor, $f_i^{Cost}$ (USD/kg) |
|---|---|---|
| GGBS[a] | 0.147 [118] | 0.14 [119] |
| NaOH solution[b] | 0.410 [120] | 0.204 [121] |
| $Na_2SiO_3$ solution[c] | 0.34571 [120] | 0.572 [122] |
| $Na_2SiO_3 \cdot 5H_2O$[d] | 1.24 [123] | 0.891 [124] |
| Extra $SiO_2$[e] | 0.00397 [4] | 0.397 [4] |
| Water | 0.00212 [125] | 0 [4] |
| Fine aggregate | 0.00503 [125] | 0.00595 [4] |
| Coarse aggregate | 0.00503 [125] | 0.00992 [4] |
| Superplasticizer | 1.53 [126] | 2.982 [4] |
| Cement[f] | 0.922 [127] | 0.16 [119] |

a. The emission factor follows the recovered-material approach in ref. [118], with no allocation of ironmaking burdens to slag.
b. Assumed to contain 50 wt.% NaOH. The emission factor is converted from the pure-NaOH basis to the solution-mass basis. The price assumes a pure-NaOH basis and is scaled by 0.50 for the solution.
c. Assumed to contain 37 wt.% solids with $SiO_2/Na_2O$ mass ratio = 3.3. The emission factor is converted from the dry-solids basis to the solution-mass basis. The price is assumed to apply to the solution mass.
d. Crystallization water is included in the total water balance; both factors are per kilogram of pentahydrate product.
e. Additional $SiO_2$, where required to achieve the target $SiO_2/Na_2O$, is assumed to be supplied by silica fume.
f. Used for the OPC reference systems.

# 3 RESULTS & DISCUSSION

## 3.1 Characteristics of the Curated Dataset

After the imputation and cleaning steps described in **Section 2.2**, the final curated dataset comprises 3053 data points extracted from 136 independent studies conducted worldwide. With 24 candidate input features, it provides a nominal sample-to-feature ratio above 100. Although records from the same study, material source, or experimental series may be correlated, the dataset offers substantially broader statistical coverage than prior ML datasets for AAS systems (**Figure 1c, d**). It also spans a broad range of mixture types, including pastes, mortars, and concretes in proportions of 1151:1165:737, respectively, providing reasonably balanced representation across binder-only and aggregate-bearing systems.

The distributions of all unscaled input features except GGBS chemistry are presented in **Figure 2a**. Most variables span wide and continuous ranges, covering much of the design domain reported in the AAS literature. A small number of observations lie near the boundaries of the feature space, such as fine aggregate/GGBS ratios of 8–10, base temperature near 30 °C, and GGBS Blaine fineness exceeding 7000 $cm^2/g$. The broad additional temperature range (–50 to 80 °C) reflects uncommon curing conditions adopted in selected studies. Expert inspection confirmed that these values are experimentally plausible rather than artifacts of data extraction or reporting errors. Retaining them preserves the breadth of literature-reported design space and reduces artificial truncation. Despite generally wide coverage, several features exhibit sparsely populated subdomains, including fine aggregate/GGBS ratios of 4–8 and GGBS Blaine fineness between 6000 and 8000 $cm^2/g$. These gaps reveal composition and processing regimes that remain underexplored experimentally, suggesting opportunities for targeted data collection to improve modeling fidelity.

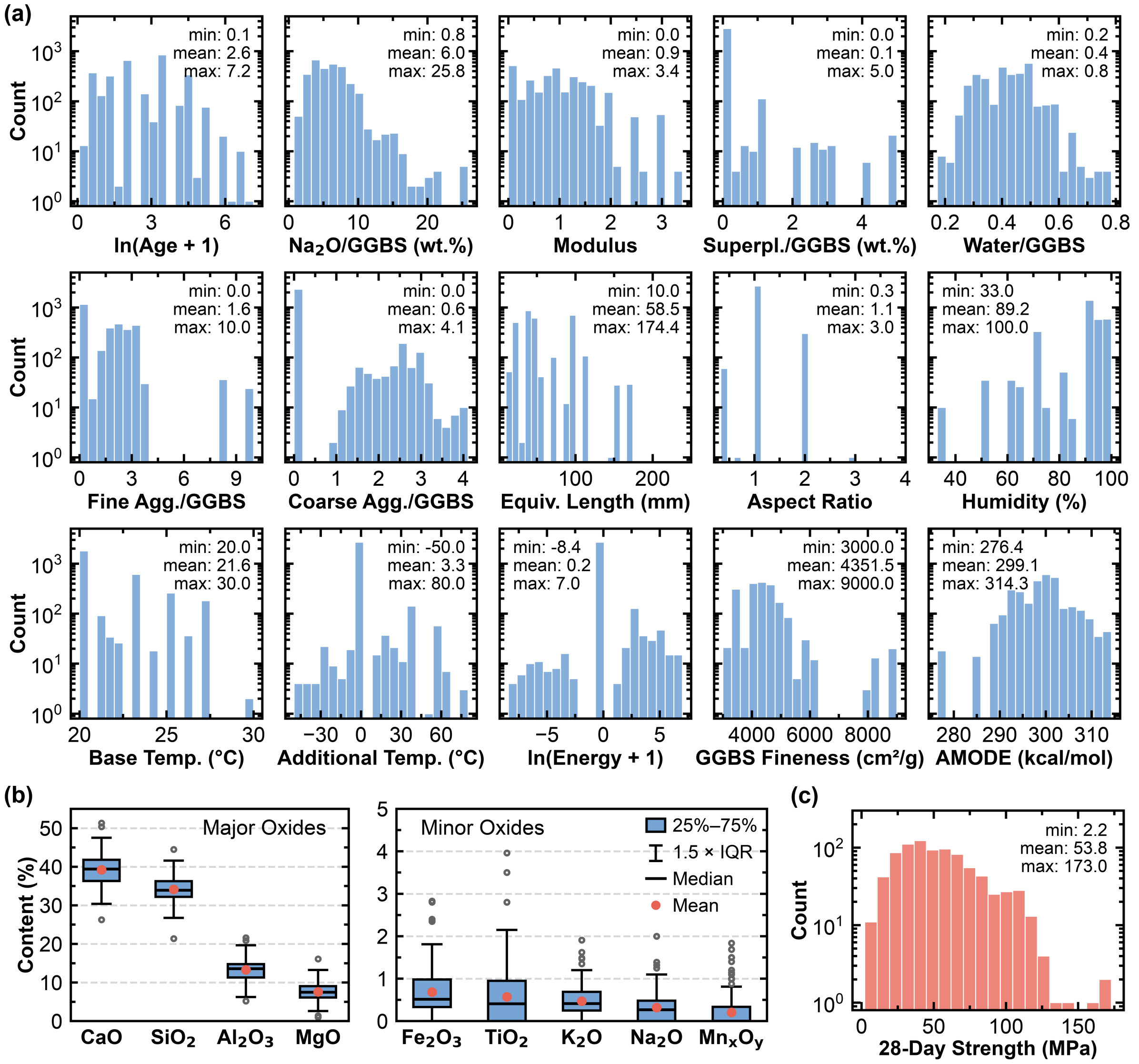

**Figure 2.** Distributions of variables in the curated AAS dataset. (a) Statistical distributions of mixture proportions, curing conditions, specimen geometry, and GGBS properties (Blaine fineness and the average metal oxide dissociation energy, AMODE). (b) Distributions of oxide contents across 155 distinct GGBS compositions. (c) Distribution of 28-day compressive strength data in the dataset, including pastes, mortars, and concretes. Histograms use logarithmic count axes and show minimum, mean, and maximum values. Boxplots show quartiles, medians, means, and 1.5 × interquartile range (IQR) whiskers, with outliers plotted individually.

**Figure 2b** summarizes the chemical compositions of 155 distinct GGBSs included in the dataset, showing that GGBSs are dominated by CaO (30–50 wt.%), $SiO_2$ (25–40 wt.%), $Al_2O_3$ (5–20 wt.%), and MgO (2–15 wt.%), along with minor oxides, including $Fe_2O_3$, $TiO_2$, $Na_2O$, $K_2O$, and manganese oxides (MnO and $Mn_2O_3$; typically < 2 wt.%). The substantial variability in major oxide contents highlights pronounced chemical heterogeneity of GGBS sources represented in the AAS literature. This compositional diversity translates into a broad range of AMODE values (~280–310 kcal/mol), indicating meaningful variation in network bond strength and slag reactivity. Consequently, optimal mix design is unlikely to be universal across slag sources; instead, mixture proportions and activator chemistry need to account for source-specific chemical and physical characteristics. This heterogeneity contributes to the wide distribution of 28-day compressive strengths observed in the curated dataset (**Figure 2c),** which ranges from near zero to around 170 MPa.

Scatter plots of 28-day strength against individual features (**Figure S15**) show weak correlations for most variables, with Pearson correlation coefficients $|r| < 0.2$. Even for the most correlated feature, water/GGBS ratio ($r = -0.41$), 28-day strength can vary by over 100 MPa at similar water/GGBS ratios. These correlations are generally weaker than those reported for blended cement systems based on large industrial datasets [4,21], underscoring the greater variability of AAS systems and the limited explanatory power of individual features.

Collectively, these observations motivated the use of ML frameworks to capture nonlinear, multivariate relationships among precursor characteristics, activator chemistry, water content, and curing/testing conditions.

### 3.2 Environmental and Economic Evaluation of AAS Mix Designs in the Curated Dataset

**Figure 3a** summarizes the average material-related $CO_2$ contributions from each mixture constituent in AAS pastes, mortars, and concretes in the curated dataset. Energy-related emissions are excluded from this material-level breakdown but included in the totals in **Figure 3b.** GGBS accounts for approximately 49–59% of emissions, followed by the activator at 39–42%, across all sample types. Other constituents contribute comparatively little. This trend differs somewhat from prior environmental assessments of AAMs [38,128], which often identify the activator as the dominant source of emissions. This difference partially reflects the lower activator dosage generally required for the GGBS-based systems examined here compared with those commonly used for lower-calcium precursors, such as fly ash and metakaolin.

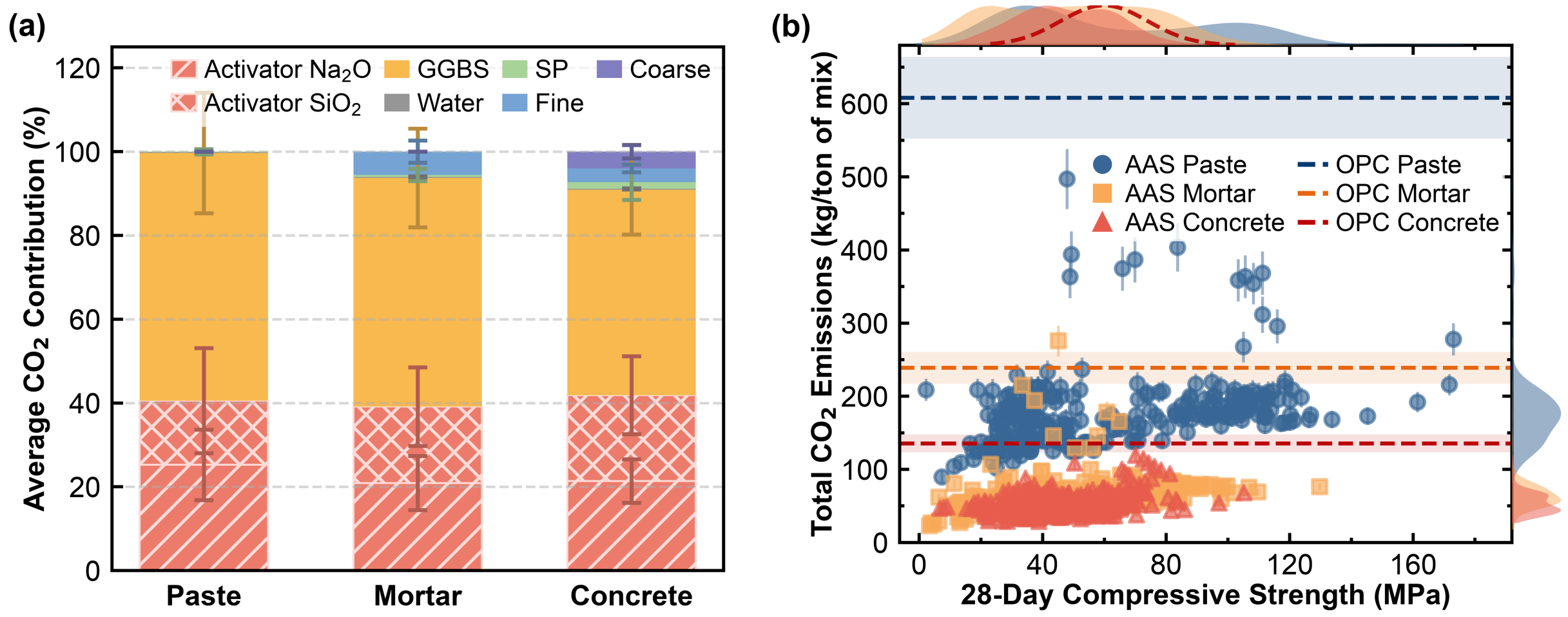

**Figure 3.** Environmental assessment of the curated AAS dataset using mixtures with reported 28-day strength. (a) Average material-related $CO_2$ contribution percentages. Activator contributions are allocated to $Na_2O$ and $SiO_2$ for reporting; water denotes separately added water. Error bars represent one standard deviation across records. SP denotes superplasticizer. (b) Scatter plot of estimated total $CO_2$ emissions versus 28-day compressive strength, including emissions from additional curing energy where applicable. Horizontal dashed lines indicate the ordinary Portland cement (OPC) reference benchmarks. Vertical bars for AAS mixtures and shaded bands around the OPC reference lines show the 2.5th–97.5th percentile ranges from factor sensitivity analysis. Marginal distributions are shown as shaded curves along the axes (right: $CO_2$ emissions; top: strength). The dashed Gaussian curve on the top axis shows the 28-day strength distribution of industrial OPC concrete [4] for context only; comparisons between AAS mixtures and OPC benchmarks are not strength-matched.

**Figure 3b** shows the estimated total $CO_2$ emissions, including both material- and elevated-curing-energy-related contributions, plotted against 28-day compressive strength. The horizontal dashed lines denote the corresponding $CO_2$ emissions of OPC reference systems, calculated using industrial average mix compositions of blended cement concrete reported by Pfeiffer et al. [4]. Because that reference reports only concrete compositions, paste and mortar baselines were estimated by removing the relevant aggregate fractions and renormalizing the remaining ingredient masses. These benchmarks provide contextual comparisons, and the industrial OPC concrete strength distribution shown above the plot does not establish corresponding strengths for the derived paste and mortar benchmarks.

Under the adopted assumptions, mean mass-normalized $CO_2$ emissions are approximately 71%, 72%, and 60% lower for AAS pastes, mortars, and concretes, respectively, than the corresponding OPC benchmarks. Reductions exceed 80% for some paste and mortar formulations. Pastes generally exhibit substantially higher $CO_2$ emissions per unit mass than the corresponding mortar and concrete systems, reflecting the relatively minor contribution of aggregates to total emissions in both AAS (**Figure 3a**) and OPC systems [128]. The pronounced vertical scatter in **Figure 3b** indicates that mixtures with similar strengths can differ widely in carbon footprint, especially among pastes. Differences in activator chemistry and curing conditions contribute to this variability. Conversely, mixtures with similar emissions can exhibit markedly different 28-day strengths. The large strength–emissions dispersion illustrates the exploratory and heterogeneous nature of literature-reported AAS mixtures and highlights the challenges of relying on empirical formulation strategies alone.

**Figure 4a** presents the corresponding cost analysis. GGBS and the activator account for approximately 43–59% and 40–44% of material cost, respectively. Together with **Figure 3a**, these results identify GGBS and the activator as the primary contributors to the environmental and economic profile of AAS mixtures.

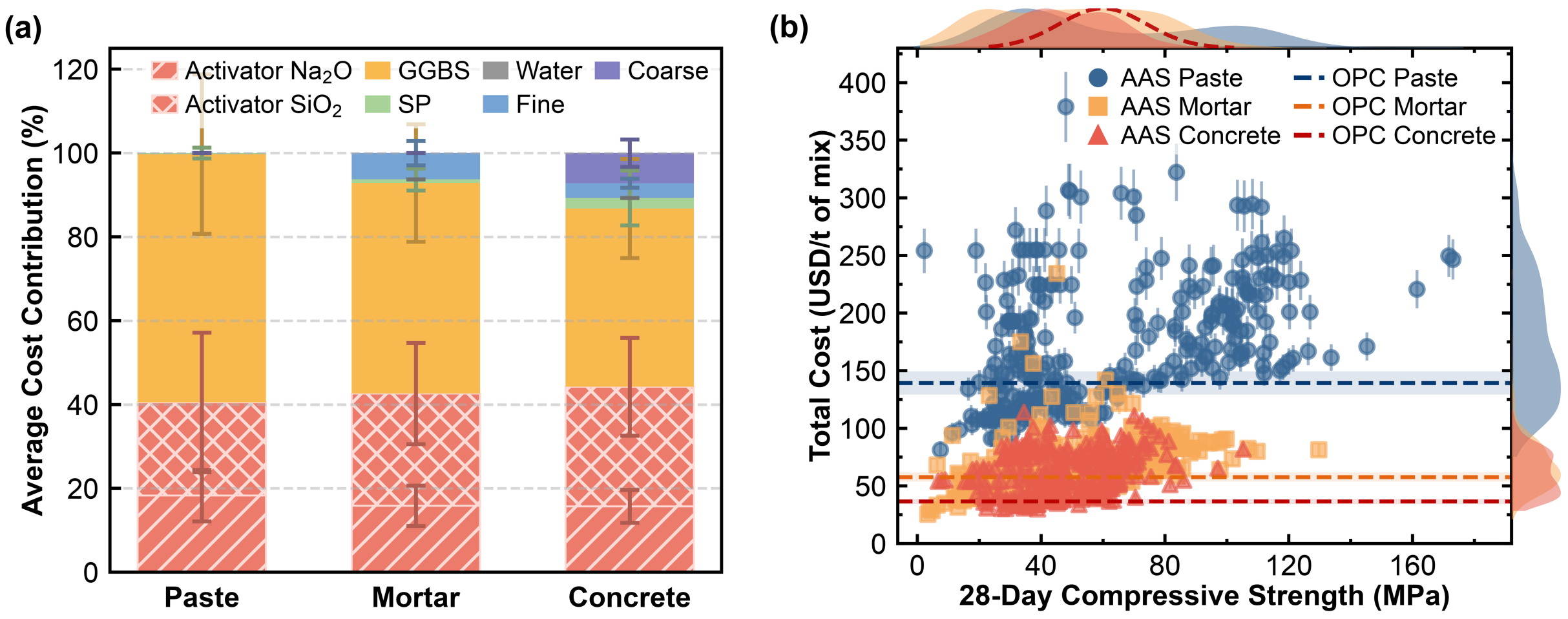

**Figure 4.** Economic assessment of the curated AAS dataset using mixtures with reported 28-day strength. (a) Average material-related cost contribution percentages. Activator contributions are allocated to $Na_2O$ and $SiO_2$ for reporting; water denotes separately added water. Error bars represent one standard deviation across records. (b) Scatter plot of estimated total cost versus 28-day compressive strength, including additional curing-energy costs where applicable. Horizontal dashed lines indicate costs for the same OPC reference compositions used in **Figure 3b**. Vertical bars for AAS mixtures and shaded bands around the OPC reference lines show the 2.5th–97.5th percentile ranges from factor sensitivity analysis. Marginal distributions are shown as shaded curves along the axes (right: cost; top: strength). The dashed Gaussian curve on the top axis shows the 28-day strength distribution of industrial OPC concrete [4] for context only; comparisons between AAS mixtures and OPC benchmarks are not strength-matched.

**Figure 4b** plots total cost against 28-day strength, including elevated-curing-energy costs where applicable. The horizontal dashed lines show costs for the same OPC reference compositions used in **Figure 3b**. As observed for emissions, paste systems exhibit substantially higher costs per unit mass than mortar and concrete systems. Furthermore, AAS mixtures display substantial cost variability at similar strengths, and conversely, strengths vary widely at similar cost. This cost variability likewise reflects differences in mixture compositions and curing conditions. Compared with OPC benchmarks, the cost of AAS mixtures can be either lower or higher, depending on the mix designs and performance targets. Thus, cost competitiveness is not an intrinsic property of AAS alone but depends strongly on formulation choices.

Overall, the combined strength–cost and strength–emissions distributions show that mechanical performance, embodied $CO_2$, and cost must be evaluated jointly. Mixtures with similar strength can have markedly different economic and environmental profiles, while low-cost or low-emission designs do not necessarily deliver low strength. These observations provide the basis for the multi-objective design space exploration developed in later sections.

### 3.3 Machine Learning Model Performances

The predictive performance of the five models across the four feature scenarios (**Table 1**) is presented in **Figure 5**. Model accuracy is evaluated using test $R^2$ and RMSE, while model robustness and potential overfitting are assessed using $\Delta R^2$ and ΔRMSE. A complete statistical summary across all random states is provided in **Table S13**.

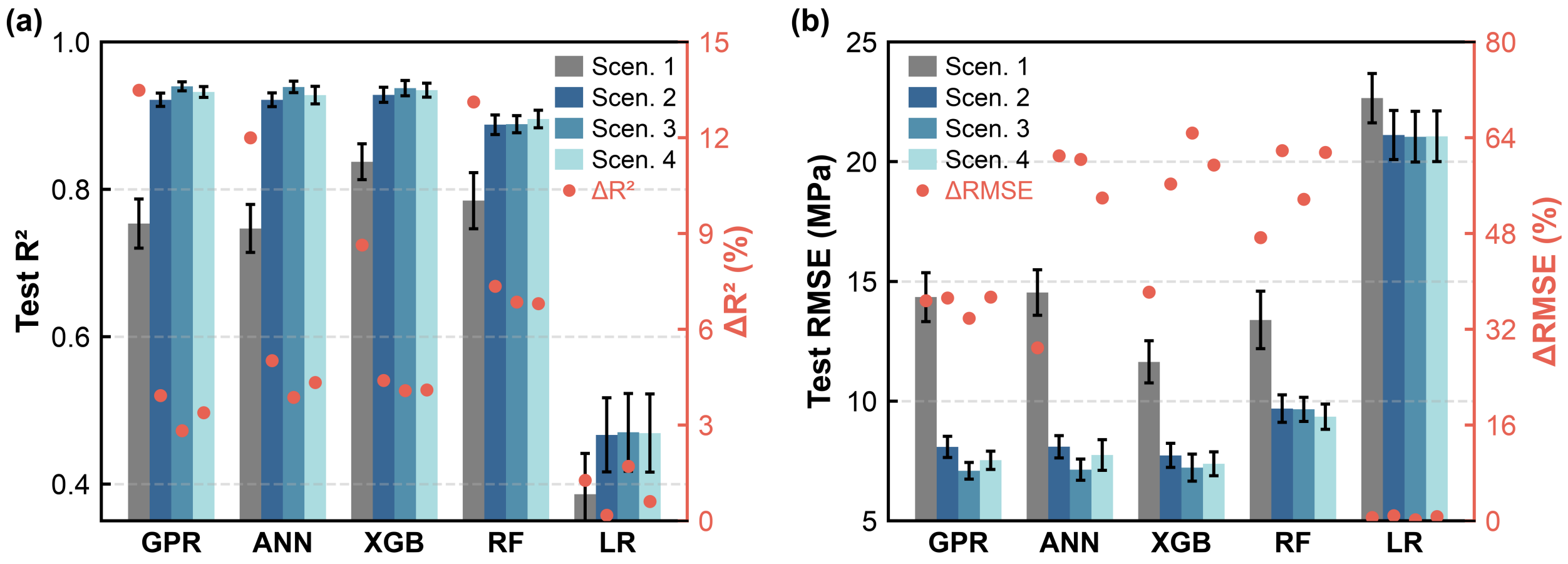

**Figure 5.** Mean predictive performances of the four nonlinear ML models (GPR, ANN, RF, and XGB) and the linear regression (LR) baseline under four feature scenarios, evaluated across 10 random states: (a) test $R^2$ and (b) test RMSE. Error bars denote one standard deviation across runs. Circular markers represent the train–test performance gap, expressed as percentage differences ($\Delta R^2$ and ΔRMSE).

Across all scenarios, the four nonlinear models, i.e., GPR, ANN, RF, and XGB, consistently outperform the LR baseline. These models achieve substantially higher test $R^2$ values (~0.75–0.94), lower test RMSE values (~7–15 MPa), and reduced inter-run variability in test $R^2$. In contrast, LR yields limited predictive capacity ($R^2$ values of ~0.39–0.47), indicating that linear formulations are insufficient to capture the complex relationships governing AAS compressive strength. Among the nonlinear ML models, GPR, ANN, and XGB exhibit comparable accuracy and outperform RF in Scenarios 2–4.

Regarding overfitting, LR shows the smallest train–test gap, consistent with its lower model flexibility and variance. Among the nonlinear models, GPR demonstrates the most favorable balance between predictive accuracy and overfitting control, with smaller $\Delta R^2$ and ΔRMSE values in Scenarios 2–4. Together with its stable performance across random states, this behavior makes GPR advantageous for subsequent surrogate modeling and design space exploration.

Model performance generally improves with feature richness. Using only mix design parameters and curing age (Scenario 1) results in lower predictive accuracy across the nonlinear models (test $R^2 \approx 0.75$–0.84; test RMSE ≈11.6–14.5 MPa). This is expected because the reduced feature representation cannot fully distinguish differences in precursor characteristics and experimental conditions across literature-reported mixtures. Incorporating GGBS fineness, curing conditions, and specimen geometry (Scenario 2) substantially improves performance (test $R^2 \approx 0.89$–0.93; test RMSE ≈ 7.7–9.7 MPa), highlighting the value of these additional variables for explaining AAS strength variability. Further inclusion of either the full set of GGBS oxide contents (Scenario 3) or the condensed reactivity descriptor (AMODE, Scenario 4) provides modest additional gains relative to Scenario 2.

**Figure 6** illustrates the predictive performance of a representative GPR model across the four feature scenarios. Scenario 1 shows pronounced scatter and systematically larger deviations from the equality line, whereas Scenarios 3 and 4 exhibit tighter clustering around parity, in agreement with their higher predictive accuracy. Scenario 2 displays intermediate performance closer to Scenarios 3 and 4, consistent with the quantitative trends in **Figure 5**. Notably, Scenario 4 achieves performance close to Scenario 3 while replacing nine oxide variables with a single physically interpretable reactivity descriptor.

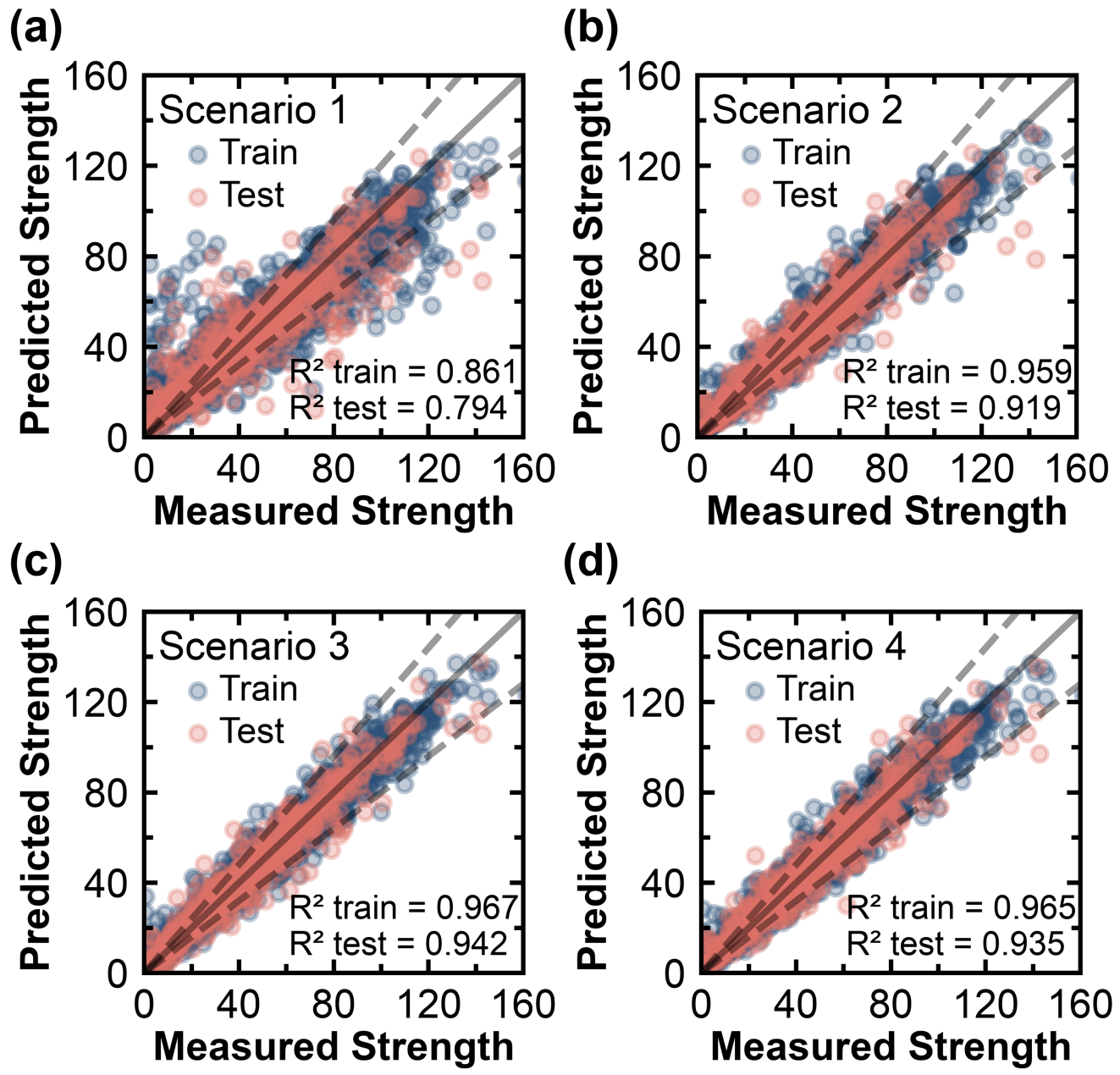

**Figure 6.** Measured versus GPR-predicted compressive strength (in MPa) under four feature scenarios: (a) Scenario 1, Baseline Mix Design and Age (7 features), (b) Scenario 2, Physical Property, Curing, and Geometry (14 features), (c) Scenario 3, Explicit Chemical Composition (23 features), and (d) Scenario 4, Physics-Informed Reactivity (15 features). For each scenario, results correspond to the representative random state whose predictive performance is closest to the ensemble average across all runs. The solid line in each figure denotes the line of equality, while dashed lines indicate ±20% deviation from equality. Detailed feature configurations are summarized in **Table 1**.

### 3.4 Model Interpretation and Feature Impact Exploration

To interpret model behavior and relate learned patterns to physically meaningful drivers, we employed two post-hoc interpretability techniques: SHAP [108] and ALE [109]. SHAP assigns feature attribution to individual predictions based on cooperative game theory, enabling both local interpretation and aggregated global importance ranking. ALE complements SHAP by estimating average local effects while reducing extrapolation bias that can arise in the presence of correlated inputs. For consistency, we focus on SHAP and ALE results for the GPR model in the main text, because GPR showed the most favorable balance of predictive accuracy, overfitting control, and smooth response behavior (**Figures 5** and **6**). Complete interpretability results for other models are provided in **Section S4**.

**Figure 7a–d** presents SHAP results for the GPR model under the four feature scenarios. The left panels show representative SHAP summary plots computed from the test subset, with features ordered in descending importance from top to bottom. Each point represents one sample, where the horizontal position (SHAP value) indicates the magnitude and direction of the feature contribution to predicted strength, and the color denotes the feature value. Positive SHAP values increase the predicted strength, while negative values decrease it. The right panels summarize mean absolute SHAP values averaged over the 10 model instances trained using different random states, with error bars denoting one standard deviation. Larger mean absolute SHAP values correspond to greater overall feature influence.

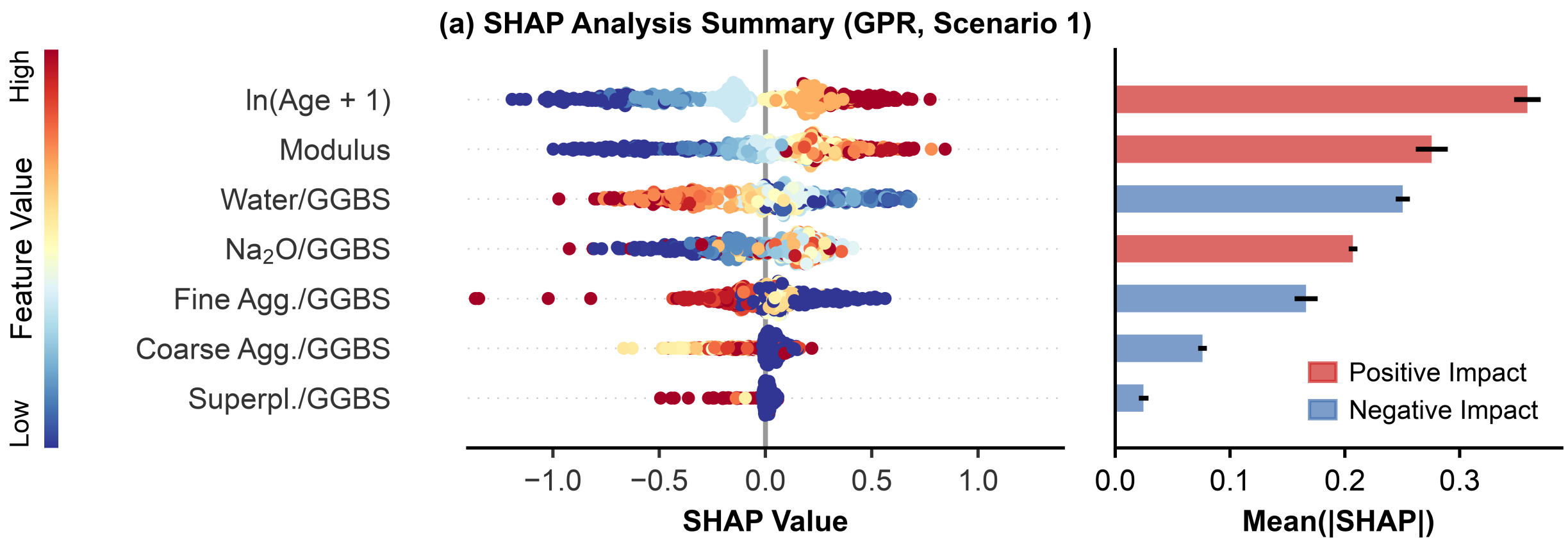

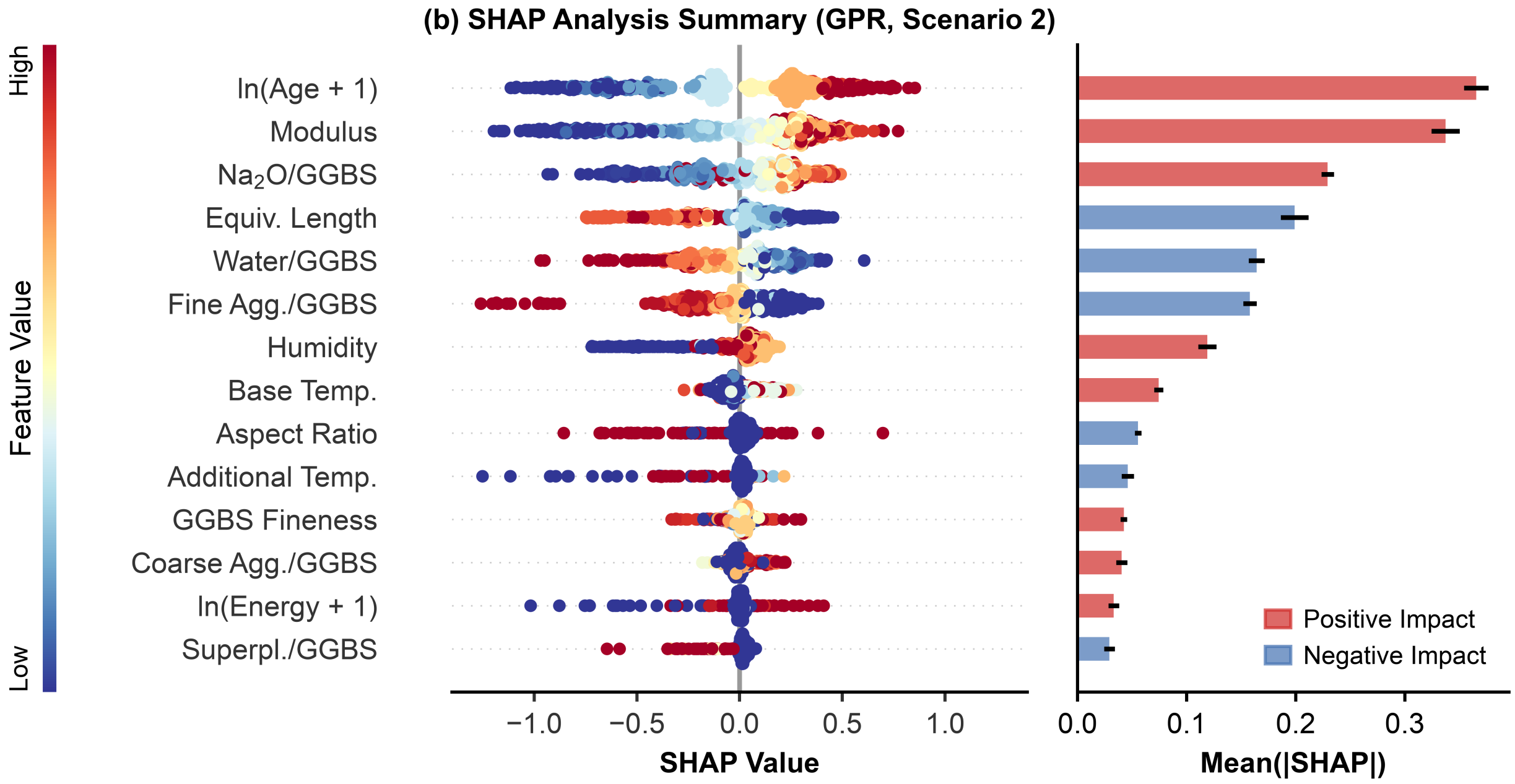

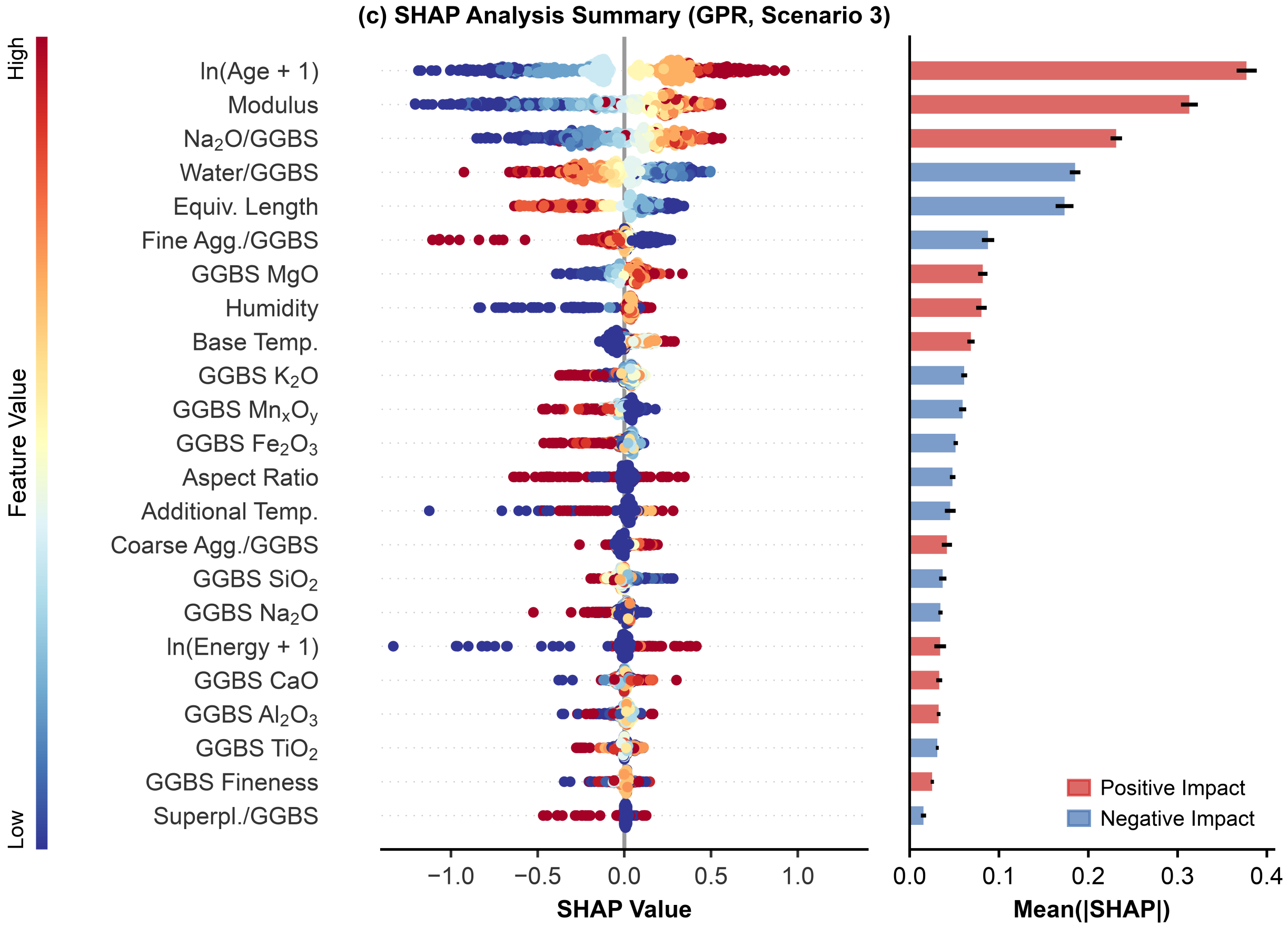

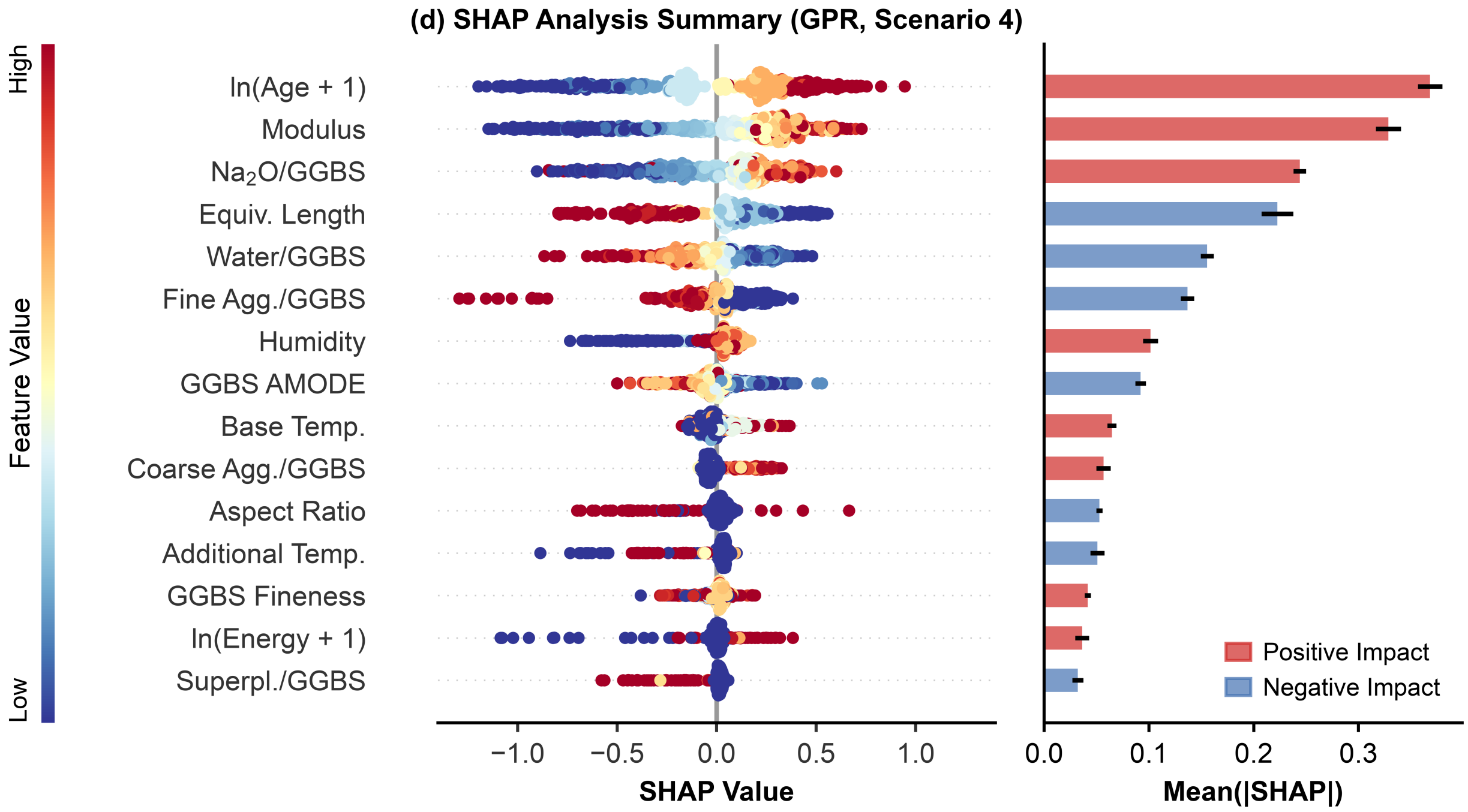

**Figure 7.** Summary of SHAP analysis for the GPR model under (a) Scenario 1, Baseline Mix Design and Age (7 features), (b) Scenario 2, Physical Property, Curing, and Geometry (14 features), (c) Scenario 3, Explicit Chemical Composition (23 features), and (d) Scenario 4, Physics-Informed Reactivity (15 features). All input features included in each model are shown. The left panels present representative SHAP summary plots computed from the test set, while the right panels display the corresponding mean absolute SHAP values averaged across 10 random states. Error bars denote one standard deviation across runs (see **Section 2.4** for details).

Because exact SHAP computation is expensive for kernel-based models such as GPR, KernelSHAP was computed using 200 randomly sampled background points to approximate conditional expectations, and SHAP values were evaluated for all samples in the 20% test set. A sensitivity analysis (**Figure S17**) confirms that this background sample size provides stable attributions. ALE was then used to examine the directionality, nonlinearity, and potential optima of the dominant features identified by SHAP. Given the large number of features and scenarios considered, full ALE plots are provided in **Section S4.3**, while key trends are discussed in the main text.

Across all scenarios, several features consistently emerge as major contributors to strength prediction. Curing age, expressed as ln(Age + 1), is the most influential feature and contributes positively to predicted strength. ALE reveals a rapid early-age increase followed by progressive saturation at later ages (**Figure S21**), a trend consistent with decelerating reaction kinetics and the transition towards reaction- or transport-limited regimes. Activator chemistry, represented by silicate modulus and $Na_2O$ dosage ($Na_2O$/GGBS), also consistently ranks among the most influential features, consistent with its central role in controlling reaction kinetics and gel development [62,129]. While SHAP indicates that higher Ms and $Na_2O$ dosage are associated with higher predicted strength (**Figure 7**, right panel), the color gradients in the SHAP summary plots are not strictly monotonic (**Figure 7**, left panel). ALE clarifies this behavior by revealing optimal ranges beyond which further increases in Ms or $Na_2O$ yield diminishing or adverse predicted effects (**Figures S21–S24**). This trend reflects the balance between enhanced dissolution and gel formation at moderate activator levels and kinetic or thermodynamic limitations at higher dosages, as discussed in **Sections 3.5.2.1** and **3.5.2.2**. Other mix design parameters, such as water/GGBS and fine aggregate/GGBS ratios, also show relatively high importance, with predominantly negative effects on predicted strength, in agreement with established domain knowledge [62,130,131].

Specimen geometry exerts a strong influence on predicted strength when included in the feature set (**Figure 7b–d**). Equivalent specimen length ranks among the most influential features and is negatively associated with predicted strength. This trend is consistent with classical size-effect behavior for quasibrittle materials [132,133]. Larger specimens often exhibit lower nominal strength because the fracture process zone becomes smaller relative to specimen dimensions, promoting fracture-energy-controlled failure, with additional contributions from statistical flaw sensitivity. Aspect ratio shows a weaker but generally negative contribution. This agrees with experimental evidence that lower-slenderness specimens can exhibit higher measured compressive strength due to friction-induced confinement at rigid steel loading platens, whereas more slender specimens experience less end restraint and lower apparent peak strength [134].

These geometry effects also help clarify the mixed behavior observed for the coarse aggregate/GGBS ratio. Although coarse aggregate content shows low overall importance, its SHAP contribution shifts from modestly negative in Scenario 1 to weakly positive in Scenarios 2–4. This apparent reversal likely arises from statistical confounding with specimen geometry: concrete specimens, characterized by higher coarse aggregate contents, are typically tested using larger specimen sizes and higher aspect ratios than pastes and mortars, and coarse aggregate/GGBS ratio is strongly correlated with equivalent specimen length ($r$ = 0.79; **Figure S16**). Because specimen size and aspect ratio contribute negatively to predicted strength, their correlation with coarse aggregate content redistributes attribution, producing the apparent sign reversal. Consistent with this interpretation, ALE analysis confirms a weak, non-monotonic response (**Figures S21–S24**).

Curing conditions also affect strength prediction, with base temperature, RH, and ln(Energy + 1) all showing positive SHAP contributions. The positive associations of temperature and supplementary curing energy are consistent with the general understanding that moderately elevated temperatures accelerate early-age reaction kinetics and strength gain [135,136]. Likewise, the positive contribution of RH aligns with experimental evidence that moisture-preserving curing supports continued reaction and matrix densification, whereas reduced RH can promote drying shrinkage and microstructural damage that impair mechanical performance [74,137,138]. Because some RH values were inferred from qualitative curing descriptions, this feature should be interpreted as a nominal moisture-curing descriptor rather than a precise humidity measurement for every record.

ALE analysis provides additional nuance regarding thermal-curing effects (**Figure S21**). Base curing temperature and ln(Energy + 1) both show weak positive effects over the sampled range, whereas additional temperature exhibits a non-monotonic trend: predicted strength increases as negative temperature deviations approach ambient conditions but decreases at higher positive temperature deviations. These trends are broadly consistent with prior studies showing that thermal curing effects in AAS are age-dependent: moderate temperature elevation can enhance early-age strength, whereas excessive or prolonged thermal curing may reduce later-age strength by promoting rapid formation of dense early products that hinder subsequent diffusion and reaction [135,139]. Thus, temperature-related variables should not be interpreted in isolation; they reflect different aspects of curing temperature and duration.

Several features exhibit consistently low importance. Superplasticizer dosage ranks among the least influential variables, reflecting its absence in most mixtures (~93%) and its primary role in modifying fresh properties rather than directly governing compressive strength in AAMs [140]. This contrasts with observations from an industrial blended-cement concrete dataset, where superplasticizer is widely used and its dosage exhibits a strong positive correlation with strength [4]. In such systems, high-strength mixtures often contain silica fume and have low water-to-cement ratios, necessitating higher superplasticizer dosages to maintain workability. In the present AAS dataset, sparse usage, incomplete reporting of admixture chemistry, and known compatibility issues between conventional superplasticizers and highly alkaline AAS systems [141] likely reduce the marginal predictive contribution of this feature. GGBS Blaine fineness exhibits a weak but consistently positive contribution. This trend aligns with the general expectation that increased powder fineness can enhance reaction kinetics and early-age strength development [72]. However, its influence is limited by the relatively narrow range of commonly reported Blaine values in the dataset (4000–6000 $cm^2/g$), and by the imperfect correspondence between Blaine fineness and true particle size characteristics (**Figure S25**) [62,129,142]. Detailed analysis of fineness effects is provided in **Section S4.4**.

The influence of GGBS chemical composition is examined explicitly in Scenario 3 and in aggregated form through the AMODE reactivity index in Scenario 4. When individual oxide mass fractions are used (Scenario 3), SHAP (**Figure 7c** and **Figures S18c–S20c**) and ALE analyses (**Figures S21c–S24c**) indicate that oxide effects are generally moderate, coupled, and nonlinear, with trends broadly consistent with physicochemical understanding. Network-modifying oxides such as CaO and MgO generally show positive associations with SHAP values, aligning with the expectation that increased modifier content promotes glass depolymerization and enhances slag reactivity [80,87,143]. In contrast, network-forming or stabilizing oxides, including $SiO_2$ and $TiO_2$, show negative contributions, reflecting their association with increased network polymerization and reduced intrinsic reactivity [87,88,143]. $Al_2O_3$ displays a weak and non-monotonic relationship with predicted strength in ALE analysis, consistent with experimental reports that its influence on strength is context-dependent and varies with activator chemistry and curing age [67,70,90,129]. However, the oxide-level trends should be interpreted in the context of correlations among compositional variables. For example, $Na_2O$ is positively correlated with $K_2O$ ($r$ = 0.41;

**Figure S16**) and inversely correlated with CaO ($r = -0.52$; **Figure S16**), while MgO and CaO are also inversely correlated ($r = -0.58$; **Figure S16**). Under such multicollinearity, marginal feature attributions may partially redistribute among correlated oxides and therefore should not be interpreted as direct causal rankings. More detailed oxide-level analyses and discussion are provided in **Section S4.5**.

Attributing slag reactivity to a single oxide mass fraction is inherently limited when compositional variables are intercorrelated. This limitation motivated the development of the AMODE descriptor, which aggregates contributions from all constituent oxides while incorporating structural energetics [80]. In Scenario 4, AMODE exhibits a consistent and physically coherent contribution in the SHAP analysis (**Figure 7d**). Both SHAP and ALE (**Figure S21d**) indicate that lower AMODE values, which correspond to weaker average metal–oxygen bond strengths and thus higher slag reactivity, are associated with higher predicted strength. This monotonic trend is consistent with the theoretical basis of AMODE and with prior experimental and computational studies linking lower dissociation energies to accelerated glass dissolution, greater extent of reaction, and enhanced strength development [77–80]. Although its contribution is moderate relative to activator chemistry and curing age, its stable and physically interpretable trend supports its use as a physics-informed descriptor for modeling and design space exploration.

SHAP and ALE analyses were also performed for ANN, XGB, and RF, with full results provided in **Section S4**. Across all models, key predictors, including curing age, activator chemistry, key mix design ratios, specimen size, and GGBS reactivity, are consistently identified among the higher-ranking features. This agreement supports the robustness of the main feature-importance trends. Subtle differences in feature ranking and attribution magnitude are observed, particularly for RF. Overall, the smoother attribution patterns obtained from GPR and ANN align more closely with established understanding of AAS systems (see **Section S4** for further discussion). At the oxide level, these cross-model comparisons further underscore the intrinsic chemical complexity of GGBS and the limitations of interpreting individual oxide effects in isolation.

## 3.5 Design Space Exploration

Building on SHAP and ALE analyses, a design space exploration was conducted to examine how coupled variations in key mixture parameters impact predicted compressive strength. The analysis focused on curing age, $Na_2O$/GGBS, Ms, water/GGBS ratio, and GGBS reactivity (represented by AMODE); however, the framework can be readily extended to other input spaces as data coverage permits. By leveraging the trained surrogate models, the high-dimensional design space spanned by the compiled dataset can be sampled continuously, enabling conditional exploration beyond discrete experimental observations in the literature. The main text focuses on GPR under Scenario 4 because this configuration combines strong predictive performance, stable response behavior, and a physically informed representation of slag reactivity. For clarity, only representative cases relevant to practical AAS design are presented here, focusing on variables identified by SHAP as highly influential. The discussion centers primarily on pastes because (i) the binder phase governs reaction-driven microstructure development and matrix strength, and (ii) pastes are the most $CO_2$- and cost-intensive among the three sample types, making them an important basis for examining performance-driven and sustainability-oriented mix design trends.

### *3.5.1 Predicted Strength Development with Age*

**Figure 8** presents the GPR-predicted strength evolution with curing age for AAS pastes activated using sodium silicate ($Na_2SiO_3$, $Na_2O$/GGBS = 6%, Ms = 1.5) and sodium hydroxide (NaOH, $Na_2O$/GGBS = 6%) under sealed curing (RH = 90%) at 20°C, at varying levels of slag reactivity (as quantified by AMODE). These parameter values, together with the remaining fixed inputs summarized in **Table S9**, fall within commonly reported ranges in the literature, as shown by the feature distributions in **Section 3.1**. For both activators, predicted compressive strength increases rapidly at early ages and then gradually approaches a plateau after ~90 days. This trend is consistent with experimentally observed strength trajectories for both AAS and OPC systems [67,144–146].

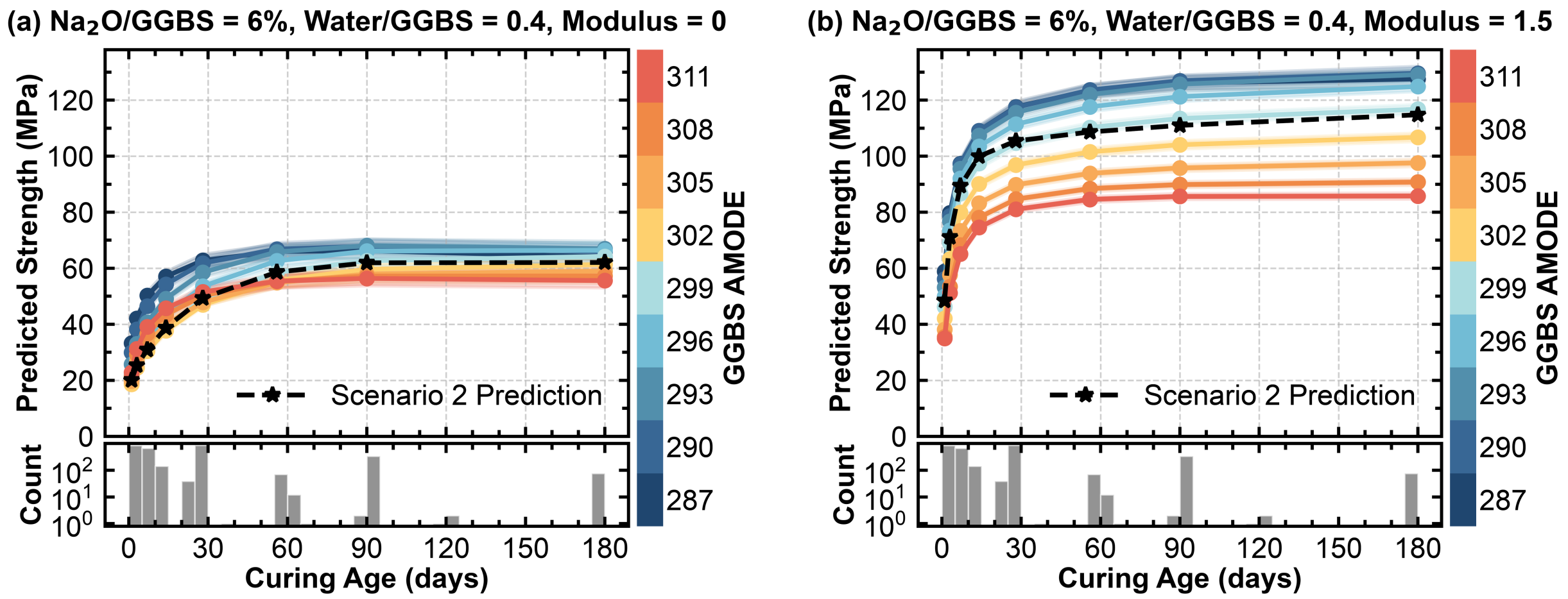

**Figure 8.** GPR-predicted strength development with increasing curing age for (a) NaOH-activated and (b) $Na_2SiO_3$-activated AAS pastes, based on design space exploration. The lower subplot shows the histogram of curing age in the experimental dataset used for model training and validation. The solid curve represents the mean prediction across 10 random states, while the shaded band indicates the corresponding standard deviation. Other fixed input values are listed in **Table S9**.

At equivalent alkali dosage and water/GGBS ratio, $Na_2SiO_3$-activated pastes are predicted to achieve substantially higher compressive strength than their NaOH-activated counterparts, in agreement with extensive experimental evidence [69,129,135,147,148]. This difference reflects the distinct reaction pathways and pore-structure development associated with the two activators. The additional silicate species supplied by $Na_2SiO_3$ solution contribute to the formation of a sodium-containing aluminum-substituted calcium–silicate–hydrate (C–(N)–A–S–H) gel, the primary strength-giving phase in AAS systems [149]. Although the high initial pH of NaOH solution can accelerate early slag dissolution and product formation [150], the more gradual and controlled reaction kinetics under $Na_2SiO_3$ activation promote more uniform precipitation of reaction products within the interstitial spaces between GGBS particles, reducing capillary porosity and enhancing strength [69]. As a result, alkali silicates are considered the most effective activators for most AAMs [62].

The importance of accounting for GGBS reactivity is highlighted by comparing Scenario 2 predictions (black dashed curve; generated without chemical input) with Scenario 4 predictions stratified by AMODE. While Scenario 2 collapses all mixtures onto a single predicted strength trajectory, Scenario 4 yields distinct, reactivity-dependent strength trajectories. This separation is more pronounced under $Na_2SiO_3$ activation, where higher-reactivity GGBSs (lower AMODE values) produce higher predicted strengths than low-reactivity GGBSs (higher AMODE values), especially within the AMODE range of 295–310. In contrast, NaOH-activated systems exhibit smaller predicted strength variations across slags of varying reactivity. These trends are consistent with experimental observations reporting greater sensitivity of silicate-activated systems to GGBS chemistry [68,129]. Mechanistically, this higher sensitivity can be attributed to the coupled dissolution-precipitation pathway during $Na_2SiO_3$ activation, in which both externally supplied silicate and slag-dissolved species contribute to gel formation. Because $Na_2SiO_3$ systems generally exhibit lower effective alkalinity than NaOH systems at equivalent $Na_2O$ dosage [148], GGBS dissolution becomes more sensitive to its intrinsic bond energetics, thereby amplifying the influence of slag reactivity on strength development. In contrast, the higher effective alkalinity of NaOH solution promotes rapid dissolution across diverse GGBSs, reducing the relative sensitivity of strength development to slag reactivity [149].

Similar strength trajectories are also obtained using other ML models, as shown in **Section S5.1**. While quantitative differences in prediction magnitude and reactivity sensitivity are observed, the primary trends remain consistent across models.

### *3.5.2 Individual Impacts of Mix Design Parameters*

#### *3.5.2.1 $Na_2O$/GGBS: Alkali Dosage*

**Figure 9** shows the GPR-predicted 28-day compressive strengths of AAS pastes activated with sodium hydroxide (Ms = 0) and sodium silicate (Ms = 1.5) as a function of activator dosages, along with the associated $CO_2$ emissions and cost. For both activators, predicted strength exhibits a clear non-monotonic dependence on $Na_2O$ dosage, increasing at low to moderate dosages and declining beyond an optimal range of 5–9%. Silicate activation yields substantially higher strength than hydroxide activation. Both $CO_2$ emissions and cost increase with $Na_2O$/GGBS. At comparable $Na_2O$ dosages, silicate activation is more costly and has higher $CO_2$ emissions, with the emissions difference increasing at higher dosages. The sharp increase in $CO_2$ at 10–12% $Na_2O$/GGBS (Ms =

1.5) reflects the assumed switch from $Na_2SiO_3$ solution to $Na_2SiO_3 \cdot 5H_2O$ to satisfy the water constraint. Together, these trends suggest that $Na_2O$ dosages exceeding ~9% offer limited additional strength benefit for the AAS paste systems while incurring economic and environmental penalties.

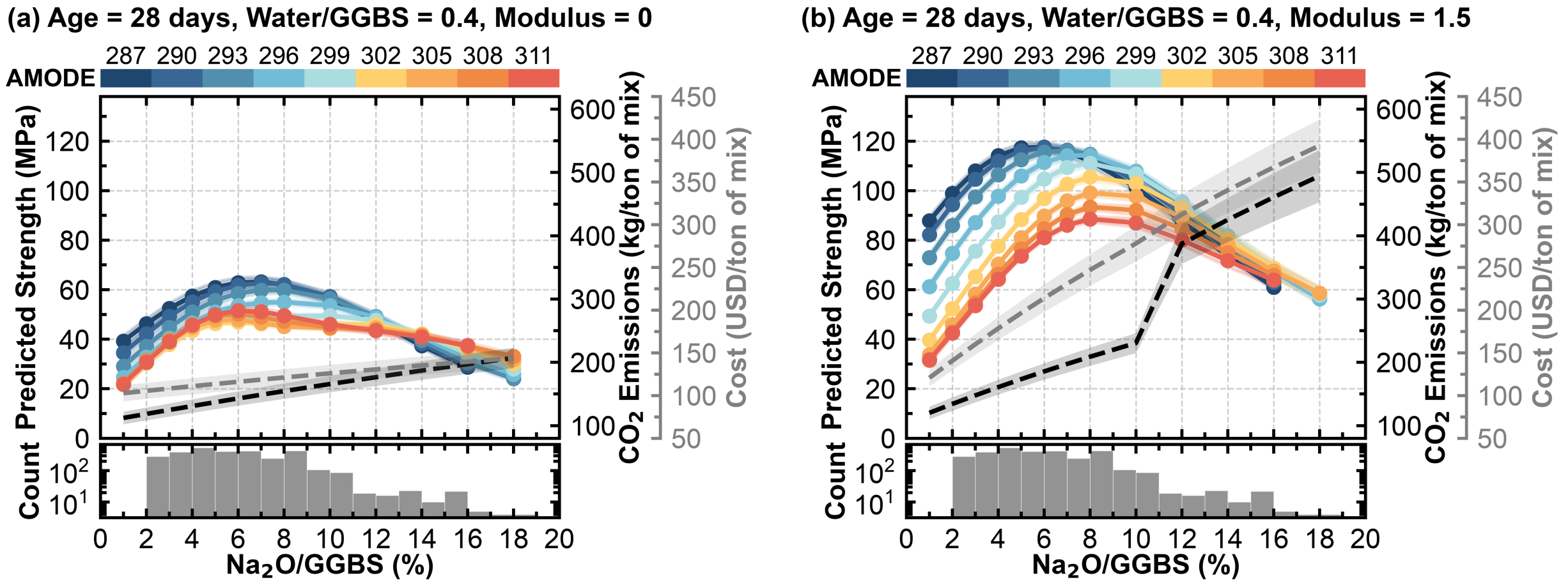

**Figure 9.** GPR-predicted 28-day strength as a function of $Na_2O$/GGBS for (a) NaOH-activated and (b) $Na_2SiO_3$-activated AAS pastes, based on design space exploration. The lower subplot shows the histogram of $Na_2O$/GGBS in the curated dataset. Solid curves and colored bands show mean predicted strengths and standard deviations across 10 random states. Black and grey dashed lines represent the associated $CO_2$ emissions and cost, respectively. Shading around the dashed lines indicates the 2.5th–97.5th percentile ranges from $CO_2$ and cost factor sensitivity analysis. Other fixed input values are listed in **Table S9**.

This non-monotonic dependence of strength on $Na_2O$/GGBS agrees with experimental data [62] and parallels reported porosity trends in AAS systems, where porosity initially decreases and then increases with rising alkali concentration [151]. At low $Na_2O$ contents, activation is insufficient to sustain slag dissolution and subsequent gel formation, resulting in limited strength development. Increasing $Na_2O$ dosage raises system alkalinity, an effect also experimentally observed for $Na_2SiO_3$-activated systems at fixed Ms [148,152], thereby promoting slag dissolution and gel formation and leading to increased compressive strength at intermediate dosages [148]. Beyond the optimal range, however, excessive alkalinity becomes detrimental. Elevated $OH^-$ activity reduces $Ca^{2+}$ solubility and can favor $Ca(OH)_2$ precipitation at the expense of C–(N)–A–S–H gel formation [153], potentially increasing susceptibility to carbonation and associated microcracking [154,155]. Concurrently, rapid early dissolution under high alkalinity can promote the formation of dense surface reaction layers on slag particles [62,81], which restrict ion transport and limit continued inner-particle reaction [151,156]. These mechanisms are consistent with the strength deterioration frequently reported at later ages for AAS samples activated at $Na_2O$/GGBS above ~10% [154,157].

Slag reactivity exerts a systematic influence on the predicted strength and optimal range of $Na_2O$/GGBS in silicate-activated systems, whereas this dependence is substantially weaker in hydroxide-activated systems. For $Na_2SiO_3$ activation, the optimal dosage range is approximately 5–9%, similar to the recommended ranges reported in the literature [81,146,151], but shifts with AMODE. For low- to medium-reactivity GGBSs (AMODE ≈ 302–311), higher dosages (7–9% $Na_2O$) are required to achieve peak strength, whereas highly reactive slags (AMODE < 293) reach optimal performance at slightly lower dosages (5–8% $Na_2O$). In contrast, hydroxide-activated systems exhibit a narrower and largely reactivity-insensitive optimal range of approximately 5–7% $Na_2O$, in agreement with previous studies [158,159]. This difference reflects the higher effective alkalinity and more strongly dissolution-driven activation mechanism of NaOH systems. These major trends are broadly observed across other ML models, as shown in **Section S5.2**.

*3.5.2.2 Silicate Modulus: Ms = $SiO_2/Na_2O$*

**Figure 10** presents the GPR-predicted 28-day compressive strength of AAS pastes as a function of Ms and GGBS reactivity (represented by AMODE), at a fixed $Na_2O$ dosage of 6%. The model predicts a non-monotonic dependence on Ms, showing a clear optimal range that depends on slag reactivity. Increasing Ms supplies more soluble silicate that promotes C–(N)–A–S–H gel formation, thereby enhancing strength. However, increasing Ms at a fixed $Na_2O$ dosage simultaneously reduces effective alkalinity, slowing slag dissolution and ultimately limiting gel formation and strength development at high Ms values [58]. This balance between enhanced silicate availability and reduced dissolution kinetics explains the observed non-monotonic behavior, in agreement with experimental findings [129,147]. Related studies have shown that increasing Ms at a constant $Na_2O$ dosage can reduce total porosity [148], but may also increase drying shrinkage [147,160].

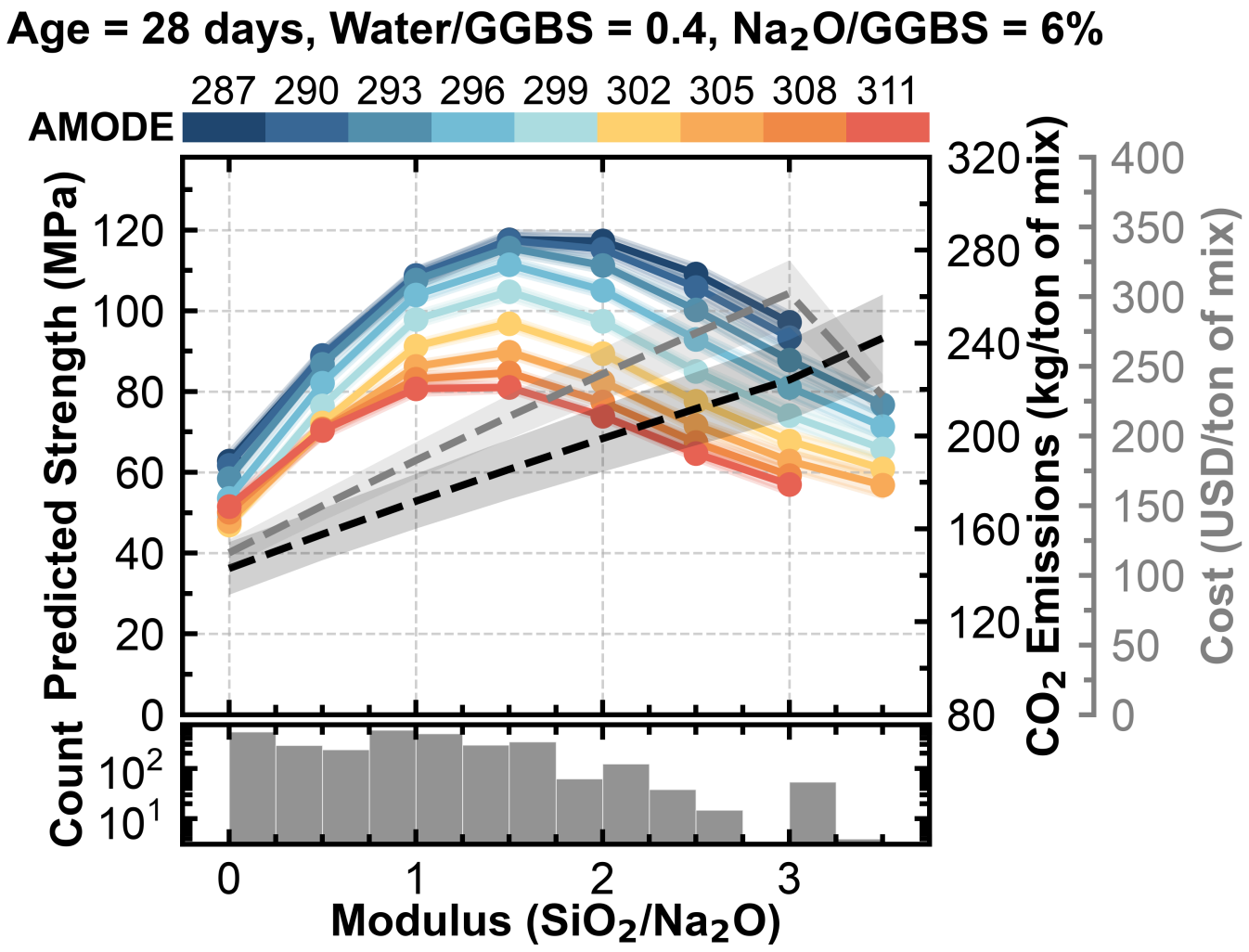

**Figure 10.** GPR-predicted 28-day strength as a function of silicate modulus (Ms) for $Na_2SiO_3$-activated AAS pastes based on design space exploration. The lower subplot shows the histogram of Ms in the curated dataset. Solid curves and colored bands show mean predicted strengths and standard deviations across 10 random states. Black and grey dashed lines represent the associated $CO_2$ emissions and cost, respectively. Shading around the dashed lines indicates the 2.5th–97.5th percentile ranges from $CO_2$ and cost factor sensitivity analysis. Other fixed input values are listed in **Table S9**.

Slag reactivity further modulates this behavior. Lower AMODE values (corresponding to higher slag reactivity) are generally associated with higher predicted strength. Although the predicted peaks occur near Ms = 1.5, higher-reactivity slags retain near-peak strength over a broader range extending towards Ms = 2.0. This behavior is consistent with experimental observations that more basic, reactive slags can favor higher optimal Ms than more acidic, less reactive slags [129], with respective AMODE values of approximately 300 and 320 calculated from the reported slag compositions. At fixed $Na_2O$ dosage, this trend can be rationalized by the reduced effective alkalinity at higher Ms, which remains sufficient to activate high-reactivity slags but may limit activation of low-reactivity slags.

Finally, at the fixed $Na_2O$ dosage considered here, $CO_2$ emissions increase with Ms, whereas cost also depends on the selected silicate source and is not monotonic. The cost decrease at higher Ms reflects the switch from $Na_2SiO_3$ solution to $Na_2SiO_3 \cdot 5H_2O$ with additional $SiO_2$ to satisfy the water constraint. Under the assumed unit prices (**Table 2**), this alternative ingredient combination is less costly. Increasing Ms beyond 1.5–2.0 nevertheless provides limited additional strength benefit while imposing greater environmental burdens. Similar strength trends are largely reproduced by the ANN, XGB, and RF models, although the tree-based models show less clearly defined optimal Ms ranges. Detailed discussion of cross-model differences is provided in **Section S5.3**.

*3.5.2.3 Water/GGBS Ratio*

**Figure 11** shows the GPR-predicted 28-day compressive strengths of AAS pastes as a function of water/GGBS ratio and GGBS reactivity (AMODE) at a fixed $Na_2O$ dosage of 6% and for both NaOH (Ms = 0) and $Na_2SiO_3$ (Ms = 1.5) activation. For both activators, compressive strength generally decreases with increasing water/GGBS ratio and with decreasing slag reactivity, aligning with the mechanistic understanding of AAS systems. Higher water contents dilute the alkaline activator, reducing pore solution alkalinity and increasing capillary porosity, leading to a less dense microstructure and reduced strength [62,81]. However, strength does not increase monotonically with decreasing water/GGBS ratio. Below a certain threshold, further reduction in water content yields progressively smaller strength gains, or even slightly reduced strength. Notably, this threshold varies with slag reactivity: highly reactive slags tolerate higher water/GGBS ratios than low-reactivity slags. At fixed $Na_2O$/GGBS (6%), reducing water/GGBS increases pore solution alkalinity and accelerates slag dissolution. For highly reactive slags, the combination of elevated alkalinity and intrinsic reactivity promotes rapid early gel formation, which, together with reduced water availability, can impair workability and increase air entrapment. Rapid precipitation may also lead to spatially heterogeneous gel formation and localized densification [156]. These effects offset the strength gains typically associated with lower water content. In contrast, less reactive slags exhibit slower dissolution kinetics and therefore tolerate lower water/GGBS ratios before similar limitations arise.

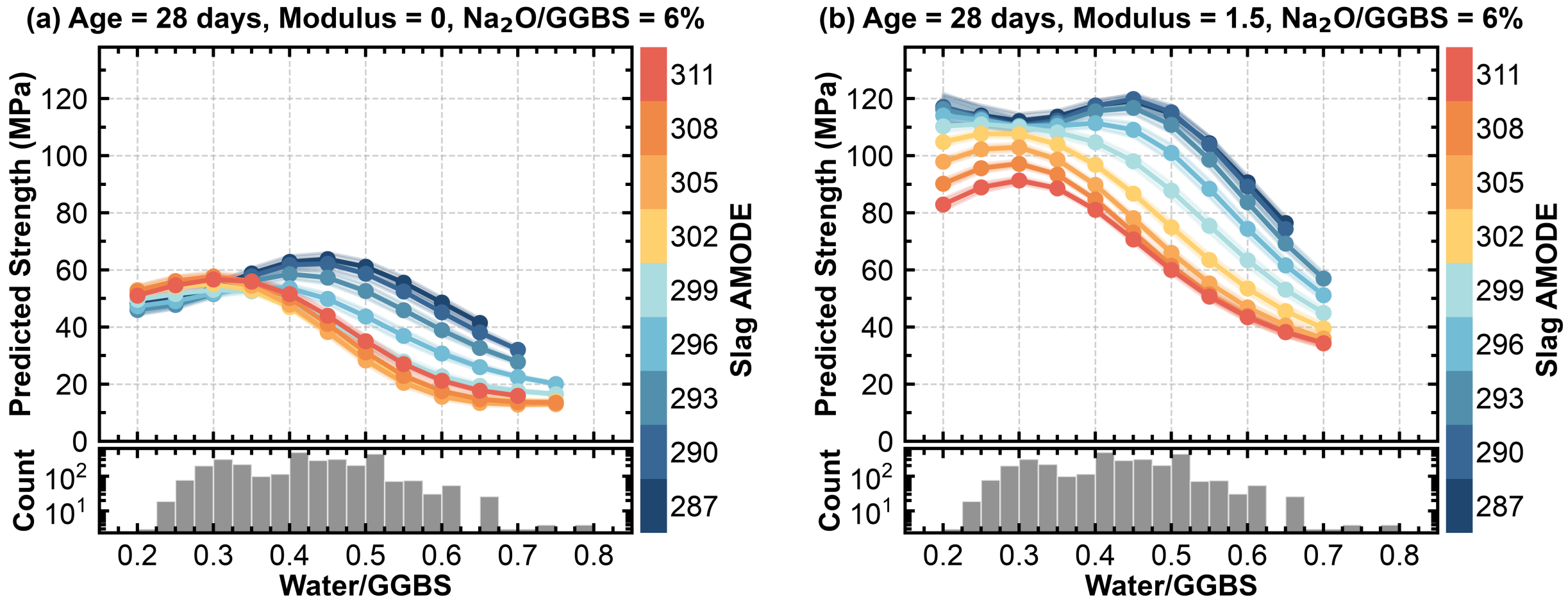

**Figure 11.** GPR-predicted 28-day strength as a function of water/GGBS ratios for (a) NaOH-activated and (b) $Na_2SiO_3$-activated AAS pastes, derived from design space exploration. The lower panel shows the histogram of water/GGBS ratios in the curated experimental dataset. The solid curve represents the mean prediction across 10 random states, while the shaded band indicates the corresponding standard deviation. Other fixed input values are listed in **Table S9**.

The sensitivity of predicted strength to water/GGBS ratio is more pronounced in $Na_2SiO_3$-activated systems (**Figure 11b**) than in NaOH-activated systems (**Figure 11a**). This observation is consistent with the isothermal conduction calorimetry results reported by Shi and Day [161], which showed substantial changes in induction period and reaction kinetics with varying water content in $Na_2SiO_3$-activated systems, but a weaker response in NaOH-activated systems. The overall decrease in strength at higher water/GGBS ratios is reproduced across models, while the low-ratio response is more model-dependent (**Section S5.4**).

### *3.5.3 Coupled Effects of $Na_2O$/GGBS and Ms*

Because $Na_2O$ dosage and Ms jointly define activator chemistry, their coupled effects cannot be fully captured through one-dimensional analyses. A two-dimensional design space exploration was therefore performed to examine how combined variations in $Na_2O$ dosage and Ms affect predicted compressive strength, with particular attention to the role of GGBS reactivity and curing age.

Results are shown for 30 mm paste specimens; comparisons with 40 and 50 mm specimens changed absolute predicted strengths but largely preserved the main mix-design trends (**Section S6.3**).

### *3.5.3.1 Reactivity-Dependent Shifts in the $Na_2O$/GGBS–Ms Design Space*

**Figure 12** presents the GPR-predicted 28-day compressive strengths across the $Na_2O$–Ms design spaces at a fixed water/GGBS ratio of 0.4 for three representative GGBSs spanning high (AMODE = 290), medium (AMODE = 299), and low (AMODE = 311) reactivity. High-reactivity slags exhibit broader high-strength regions in the $Na_2O$–Ms design space, indicating greater tolerance to variations in activator chemistry. In contrast, low-reactivity slags require more narrowly tuned $Na_2O$–Ms combinations to achieve comparable strength.

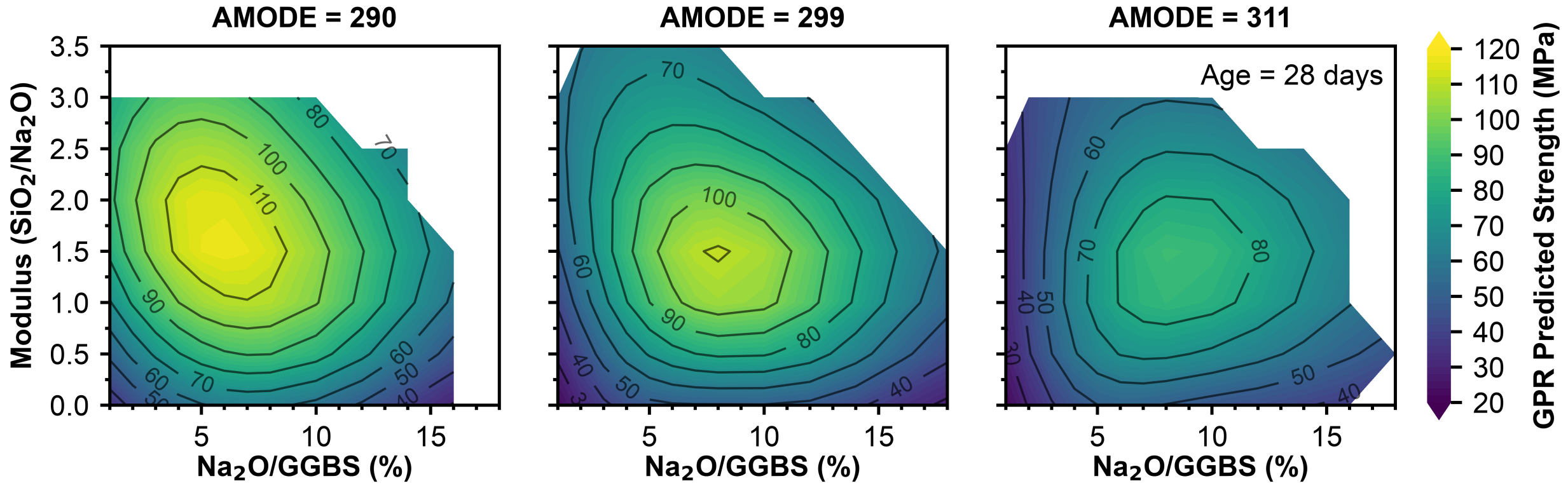

**Figure 12.** Two-dimensional mapping of GPR-predicted 28-day compressive strength as a function of $Na_2O$/GGBS and Ms for AAS pastes with varying GGBS reactivities: high (AMODE = 290), medium (AMODE = 299), and low (AMODE = 311). The water/GGBS ratio is fixed at 0.4. Other fixed input values are listed in **Table S9**. Strength values are mean predictions across 10 random states. Blank regions indicate combinations excluded by mass-balance screening or Mahalanobis distance filtering (**Section 2.5**).

The favorable activator region shifts with slag reactivity. For the high-reactivity slag, predicted peak strength occurs at $Na_2O \approx 6\%$ and Ms $\approx 1.5$, with near-peak strengths extending towards Ms $\approx 2.0$. Medium- and low-reactivity slags favor higher $Na_2O$ dosages (approximately 8%), with peak strengths also occurring near Ms ≈ 1.5. This shift reflects differences in alkalinity demand: highly reactive slags require less alkalinity to sustain dissolution and can tolerate higher silicate content, whereas less reactive slags require stronger alkaline activation to maintain sufficient dissolution kinetics. Similar reactivity-dependent patterns are observed for ANN (**Figure S38**). XGB and RF show broader, less localized high-strength regions (**Figures S39** and **S40**).

### *3.5.3.2 Age-dependent Shifts in the $Na_2O$/GGBS–Ms Design Space*

**Figure 13** shows the GPR-predicted compressive strength of a representative high-reactivity GGBS (AMODE = 290) across the $Na_2O$–Ms design space at different curing ages. The optimal range of activator chemistry shifts systematically towards lower $Na_2O$ dosage and higher Ms with increasing age. At 3 days, the favorable region is centered around $Na_2O$ ≈ 8% and Ms ≈ 1–1.5; at 28 days, it shifts towards $Na_2O$ ≈ 6% and Ms ≈ 1.5–2.0; and by 180 days, high strengths are achieved at $Na_2O$ ≈ 4–5% and Ms ≈ 1.5–2.0. These trends suggest that higher alkalinity (lower Ms and higher $Na_2O$) primarily enhances early-age strength through accelerated slag dissolution and rapid gel precipitation. However, such a rapid reaction can also promote more spatially heterogeneous gel formation and less efficient pore refinement [156,162], which may limit long-term strength development. In contrast, lower $Na_2O$ dosage and higher Ms reduce effective alkalinity and moderate slag dissolution kinetics [163], with the reaction progressively transitioning toward diffusion-controlled stages as reaction products accumulate [149], thereby enabling more gradual and sustained strength gain. Similar age-dependent patterns are reproduced by the ANN model (**Figure S41**), whereas XGB and RF resolve the optima less distinctly (**Figures S42** and **S43**). Comparable behavior is observed for medium-reactivity slags, while the age-dependent shift is less pronounced for low-reactivity slags (**Figures S44–S47**). Curing-age coverage statistics and GPR predictive uncertainty maps are provided in **Section S6.4**.

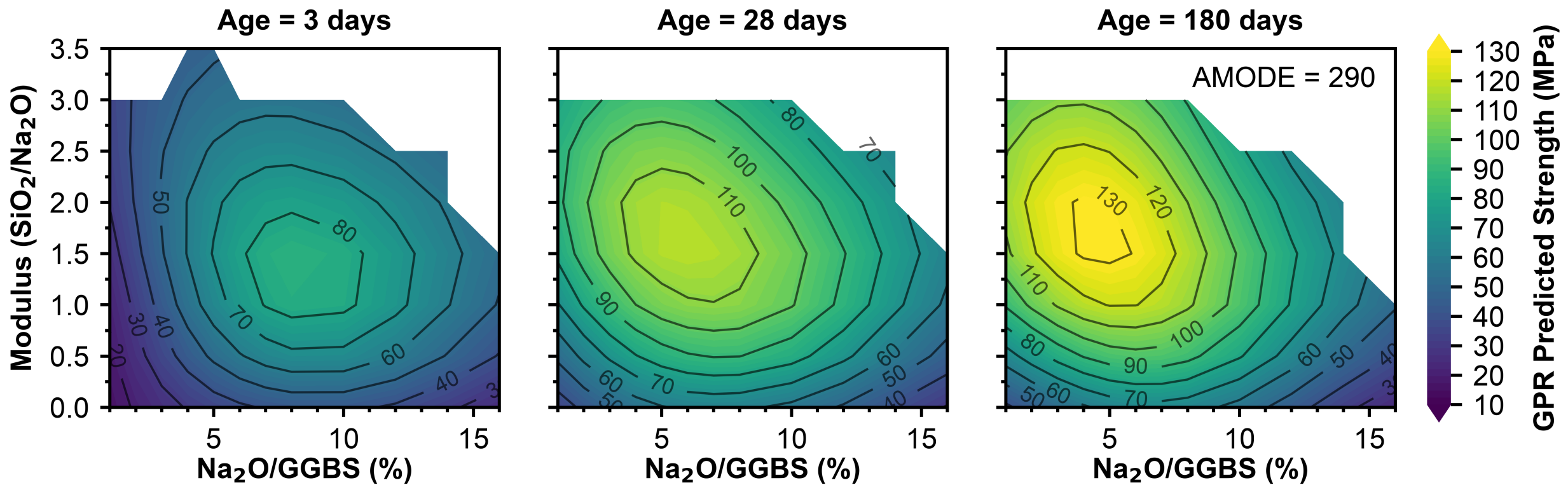

**Figure 13.** Two-dimensional mapping of GPR-predicted compressive strength at different curing ages (3, 28, and 180 days) as a function of $Na_2O$/GGBS and Ms for high-reactivity GGBS (AMODE = 290). The water/GGBS ratio is fixed at 0.4. Other fixed input values are listed in **Table S9**. Strength values are mean predictions across 10 random states. Blank regions indicate combinations excluded by mass-balance screening or Mahalanobis distance filtering (**Section 2.5**).

### *3.5.3.3 Strength–$CO_2$–Cost Trade-offs in the $Na_2O$/GGBS–Ms Design Space*

**Figure 14** illustrates the coupled environmental and economic implications of AAS mix designs across the $Na_2O$–Ms design space for slags of varying reactivity at a fixed water/GGBS ratio of 0.4. $CO_2$ emissions and cost are expressed relative to the same OPC paste baseline used in **Section 3.2**. The dashed contours denote GPR-predicted 28-day compressive strength levels of 50 MPa and 80 MPa, corresponding to the strength maps shown in **Figure 12**.

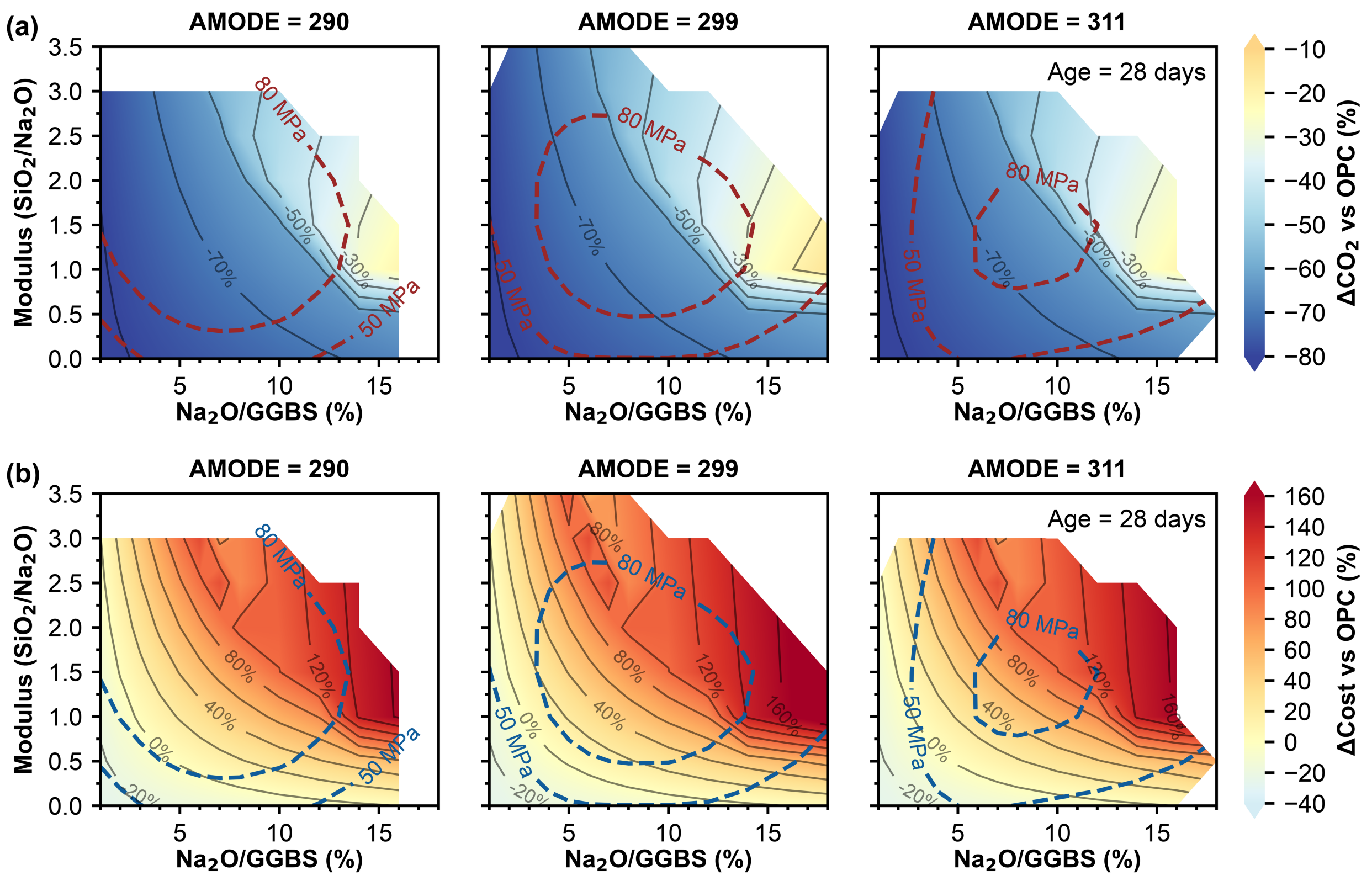

**Figure 14.** Two-dimensional mapping of the estimated (a) $CO_2$ emissions and (b) cost associated with varying $Na_2O$/GGBS and Ms combinations for AAS pastes with different GGBS reactivities. The water/GGBS ratio is fixed at 0.4. Other fixed input values are listed in **Table S9**. The calculations follow the same procedure described in **Section 2.6**, with both $CO_2$ emissions and cost expressed relative to the OPC paste benchmark derived from industrial data reported by Pfeiffer et al. [4]. The dashed contour lines represent the GPR-predicted 50 MPa and 80 MPa 28-day strength levels achieved at corresponding combinations of $Na_2O$/GGBS and Ms. Strength contours are based on mean predictions across 10 random states. Blank regions indicate combinations excluded by mass-balance screening or Mahalanobis distance filtering (**Section 2.5**).

Within the design space explored and the assumptions used here, AAS mixtures generally exhibit substantially lower estimated $CO_2$ emissions than the OPC reference (**Figure 14a**), highlighting the potential environmental advantage of clinker-free AAS systems. Although the estimated cost is generally higher than the OPC reference, **Figure 14b** reveals a distinct low-$Na_2O$, low-Ms region (bottom left corner) where cost approaches or falls below the OPC benchmark. However, this region does not coincide with optimal predicted strength, and the achievable strength depends largely on slag reactivity. This behavior reveals a trade-off among precursor characteristics, mechanical performance, environmental impact, and cost.

When specific performance targets are imposed, the trade-off becomes reactivity dependent. A moderate strength target (50 MPa) can be achieved within the low-$Na_2O$, low-Ms region, enabling approximately 70–80% $CO_2$ reductions relative to the OPC reference while maintaining lower cost, particularly for high-reactivity slags. For such slags, strengths approaching 80 MPa can also be attained at comparable cost while preserving similarly large $CO_2$ reductions. In contrast, achieving 80 MPa with medium- and low-reactivity slag requires shifting toward moderately higher $Na_2O$ and Ms values, increasing both cost and $CO_2$ emissions. Nevertheless, these slags can still achieve substantial $CO_2$ reductions relative to the OPC reference, although reaching 80 MPa requires a higher cost. The relative economic competitiveness of AAS further improves under a carbon pricing scenario, which broadens the portion of design space yielding cost parity or advantage depending on performance targets, as shown in **Section S7.2**.

Although demonstrated here for paste systems at a fixed water/GGBS ratio of 0.4, the same analysis can be readily extended to mortars and concretes, and broader feature spaces as data coverage permits.

# 4 BROADER IMPACT AND LIMITATIONS

## 4.1 Broader impact

This study demonstrates how knowledge-informed ML can synthesize heterogeneous, literature-derived AAS data and identify physically meaningful trends. By integrating precursor properties, mixture proportions, curing conditions, specimen geometry, compressive strength, embodied $CO_2$ emissions, and estimated cost within a unified workflow, the framework enables joint analysis of AAS strength development, emissions and cost across the chemically diverse domain represented by the curated dataset.

A central contribution is the explicit incorporation of GGBS reactivity through AMODE into predictive modeling. This is important because quantitatively linking slag chemistry, reactivity, and performance has long remained difficult in AAS research. By showing how slag reactivity modulates the effects of $Na_2O$ dosage, Ms, water/GGBS ratio, and curing age, the framework provides a systematic basis for evaluating source-dependent mixture behavior. This capability can help reduce reliance on empirical, slag-specific mixture tuning and guide more targeted experimental exploration. More broadly, the strategy of combining physically grounded reactivity descriptors with data-driven modeling may be transferable to other AAMs, blended cements, and cementitious systems where raw-material reactivity strongly influences performance, although the specific descriptor may need to reflect precursor mineralogy, amorphous content, and reaction mechanism.

The curated dataset is, to the authors' knowledge, the largest and most comprehensive AAS dataset reported to date. Its open availability provides a shared research resource for the AAM community, supporting benchmarking of new algorithms and descriptors; future expansion to additional activators, precursors, curing regimes, and performance metrics; and studies on uncertainty quantification and transfer learning. The feature-importance results also highlight the reporting requirements for reusable AAS data. Variables such as curing humidity, specimen geometry, GGBS fineness, activator composition, precursor chemistry, and mixture proportions affect model interpretation yet remain inconsistently reported across the literature.

Finally, the resulting design space maps provide industry-relevant guidance for screening low-carbon, cost-efficient mixtures that meet strength targets and selecting activator chemistry according to precursor reactivity. They can also help prioritize targeted experiments in underexplored regions of the AAS design space.

### 4.2 Limitations

Despite the breadth of the curated dataset and the overall robustness of the modeling framework, several limitations should be acknowledged. First, some potentially relevant variables were reported only in limited subsets of studies and therefore could not be systematically included in the dataset. These include particle size distribution, BET surface area, GGBS cooling history and mineralogy, aggregate gradation, and activator source. For variables retained in the dataset, some values were also unavailable; in particular, curing humidity and Blaine fineness values had to be approximated or imputed where not reported. Therefore, while the dataset provides a broad representation of the available AAS literature, its feature coverage remains constrained by the reporting practices of the original sources. Given the potential importance of these variables, more standardized and complete reporting would support future data-driven modeling and analyses.

Second, the curated dataset may contain uncertainties inherited from the source literature. For example, oxide compositions obtained by XRF can be affected by sample preparation and calibration uncertainty [164], while compressive strengths measured under standardized procedures are still subject to within- and between-laboratory variability [165]. These uncertainties may include random components, such as sample-to-sample or measurement variability, as well as systematic components associated with shared materials and experimental or reporting practices within the same study or laboratory. Because systematic uncertainties may appear in both the training and test sets under random train–test splits, the resulting model performance may overestimate predictive accuracy when transferring the model to entirely new studies, laboratories, or slag sources. Future work should assess these transferability limits using study-, slag-, and mixture-grouped validation and age-based extrapolation tests.

Third, the statistical structure of the dataset is uneven. Several input features exhibit non-uniform distributions, and some features are highly discrete (e.g., specimen aspect ratio and curing-condition descriptors), resulting in uneven local data density and potentially affecting model sensitivity and the stability of inferred response trends. While multiple model architectures and random train–test splits were benchmarked to enhance robustness, predictive reliability remains dependent on the density and coverage of the underlying data [166].

Fourth, surrogate model predictions are intrinsically constrained by the statistical support of the curated dataset. Design space exploration was therefore restricted using Mahalanobis distance filtering to reduce unsupported multivariate extrapolation. The retained regions provide a basis for examining conditional model trends, but predictions in sparsely populated regions or outside the observed domains should be interpreted with caution. Future work could incorporate additional domain constraints or mechanistic knowledge into model training [167], alongside targeted experimental validation, to improve model generalizability and predictive reliability.

Fifth, the AMODE descriptor, while physically grounded and effective for predominantly amorphous GGBS systems, is derived from bulk oxide composition and does not explicitly capture phase assemblage, crystalline fraction, cooling history or nanoscale structural heterogeneity. Although an adaptation of AMODE has been proposed for mixed crystalline–amorphous systems [77], this approach requires detailed mineralogical characterization that is typically unavailable in literature-derived datasets. The scope and limitations of both the original and modified AMODE formulations have been discussed in prior studies [79,80]. As more comprehensive mineralogical and structural data become available, adapting this descriptor for mixed-phase precursors offers a promising direction for future work.

Finally, this analysis focuses on compressive strength as the primary target, as with most existing ML studies, because it is the most widely reported performance metric in AAS literature. Other important metrics, including workability, setting time, shrinkage, and durability, were not considered and are targets for future extension of the framework. In addition, the $CO_2$ emission and cost factors used here were compiled from published sources and may not fully reflect current market conditions, regional variation, or differences in accounting boundaries. The estimated environmental benefits also depend on the adopted GGBS allocation approach. Alternative allocations of ironmaking burdens are not covered by the ±10% factor sensitivity analysis. Therefore, the environmental and economic assessments should be interpreted as comparative estimates. This study also focuses on GGBS activated with $Na_2SiO_3$ and NaOH. In practice, a wide range of precursors (e.g., fly ash, metakaolin) and activators (e.g., potassium-based or carbonate-based systems) are commonly explored. Expanding the framework to include other precursors, activators, and performance indicators would broaden its applicability and better support source-dependent AAM mix design.

# 5 CONCLUSIONS

This study establishes a reactivity-informed ML framework for analyzing and exploring AAS mix design within a broad literature-supported domain. A literature-derived dataset comprising over 3000 AAS compressive strength records with 24 input features was curated, covering mixture proportions, GGBS chemistry and fineness, precursor reactivity, curing conditions, and specimen geometry.

Building on this dataset, a suite of surrogate ML models was evaluated under progressively enriched input scenarios using random record-level splits. Incorporating curing conditions, specimen geometry, and GGBS fineness markedly improved predictive performance, while precursor chemistry provided additional predictive value. AMODE retained much of the predictive value of explicit oxide composition while providing a compact, physically grounded representation of GGBS reactivity for ML modeling. GPR was selected as the primary surrogate model based on its predictive performance, stability, and smooth response behavior.

SHAP and ALE analyses consistently identified curing age, activator chemistry, water/GGBS ratio, specimen geometry, and slag reactivity as important contributors to model predictions. The inferred trends included rapid early-age strength development followed by progressive saturation, non-monotonic effects of $Na_2O$ dosage and Ms, and lower predicted strength at higher water/GGBS ratio and larger specimen sizes. These data-derived trends are broadly consistent with established understanding of AAS chemistry and experimental observations reported in the literature, supporting their physical plausibility.

Design space exploration was conducted within a mass-balance-screened and Mahalanobis distance-filtered domain. Within this domain, the model identified systematic shifts in favorable $Na_2O$ dosage, Ms, and water/GGBS ranges with slag reactivity, activator type, and curing age. Higher-reactivity slags generally favored lower $Na_2O$ dosage and tolerated higher Ms, whereas less reactive slags required stronger alkaline activation to achieve comparable strength. Favorable activator ranges also differed between early and later curing ages. These model-inferred trends are largely consistent with reported AAS behavior; however, the precise locations and boundaries of the candidate favorable regions, and their transferability to new slag sources or experimental settings, require independent validation.

Finally, integrating embodied $CO_2$ emissions and cost into design space exploration revealed non-monotonic trade-offs among strength, environmental impact, and economic feasibility. The design space maps identified candidate regions combining high predicted strength with substantially lower $CO_2$ emissions than OPC-based references at similar costs. These favorable regions depend strongly on slag reactivity and activator demand. Together, the framework and curated dataset provide a foundation for source-dependent AAS mix design, with potential extension to other cementitious systems where raw-material variability, processing conditions, and sustainability constraints are strongly coupled.

## APPENDIX A. SUPPLEMENTARY MATERIAL

The supplementary material contains: **S1. Supplementary Methodological Details**, covering sensitivity analyses for Blaine fineness imputation, curing-condition assumptions, thermal-exposure representation, and data cleaning, together with hyperparameter tuning, AMODE-related analysis, model evaluation metrics, fixed inputs for design space exploration, and mixture conversion and mass-balance screening; **S2. Dataset Structure and Exploratory Relationships**, including feature–strength relationships and Pearson correlation analysis; **S3. Model Predictive Performance and Input-Representation Limits**, including empirical RMSE floors and complete predictive performance results; **S4. Extended Model Interpretation Analysis and Discussion**, including KernelSHAP sensitivity analysis, SHAP and ALE analyses across models, and discussions of GGBS fineness and oxide components; **S5. Extended One-Dimensional Design Space Exploration**, covering the effects of age, $Na_2O$/GGBS, Ms, and water/GGBS ratio; **S6. Extended Two-Dimensional Design Space Exploration**, covering optimal $Na_2O$/GGBS–Ms combinations across AMODE levels and curing ages, together with specimen-size sensitivity, curing-age coverage, and predictive uncertainty; and **S7. Extended Environmental and**

**Economic Implications**, including model-dependent sustainability analyses and carbon tax scenarios.

## DECLARATION OF GENERATIVE AI AND AI-ASSISTED TECHNOLOGIES

During the preparation of this work, the authors used ChatGPT to support language refinement and coding assistance. All AI-assisted outputs, including text and code, were critically reviewed, edited, and validated by the authors, who take full responsibility for the analysis, interpretation, and conclusions presented in the final manuscript.

## DECLARATION OF COMPETING INTEREST

All authors are named inventors on a pending patent application (No. 64/041,318) filed by Rice University on April 16, 2026. The authors declare no other competing interests.

## ACKNOWLEDGMENTS

This work was supported by the Department of Civil and Environmental Engineering at Rice University, the Gulf Research Program's Early-Career Research Fellowship, and the National Science Foundation under CAREER Award No. 2543840. Z.L.'s participation was partially supported by the Rice Academy Postdoctoral Fellowship.

## DATA AVAILABILITY

The dataset curated and analyzed in this study is publicly available on Zenodo at https://doi.org/10.5281/zenodo.20738377.

Supplementary Material for

# Reactivity-Informed Machine Learning for Performance Prediction and Design Space Exploration of Alkali-Activated Slag

Qiyao He[a,b,c], Zhanzhao Li[a,b,c], Kai Gong[a,b,c,*]

[a] Department of Civil and Environmental Engineering, Rice University, Houston, Texas 77005, United States

[b] Rice Advanced Materials Institute, Rice University, Houston, Texas 77005, United States

[c] Ken Kennedy Institute, Rice University, Houston, Texas 77005, United States

[*] Corresponding author. E-mail: kg51@rice.edu

**Contents**

## S1 Supplementary Methodological Details

### S1.1 Blaine Fineness Imputation and Sensitivity Analysis

Blaine fineness is reported for 1889 of the 3053 records in the curated dataset and is missing for the remaining 1164 records (38.1%). Missing values were assigned 4500 $cm^2/g$ to retain these records for analysis. This value serves as a practical placeholder rather than an estimate of the true fineness of each slag.

To assess sensitivity to this choice, three approaches were compared using Gaussian process regression (GPR) under Scenario 4: constant imputation at 4500 $cm^2/g$, multivariable imputation, and omission of the Blaine fineness feature. The models and feature scenarios are described in **Section 2.4** of the main manuscript. All approaches used the same training–test splits across ten random states. Multivariable imputation used Bayesian ridge regression within an iterative imputation procedure fitted only to the training data, using input features without compressive strength. The fitted imputer was applied to the test data, and observed fineness values were retained.

Constant and multivariable imputation produced nearly identical predictive performance, with mean test $R^2$ values of 0.932 for both approaches and mean test root-mean-square error (RMSE) values of 7.530 and 7.537 MPa, respectively. Omitting Blaine fineness modestly reduced predictive performance, yielding a mean test $R^2$ of 0.928 and RMSE of 7.746 MPa (**Table S1**). The principal SHapley Additive exPlanations (SHAP) importance patterns and accumulated local effects (ALE) trends remained similar across the three approaches (**Figures S1** and **S2**). These results support retaining constant imputation for the present analysis, while not establishing that fineness is unimportant or that missingness is unbiased.

**Table S1.** Gaussian Process Regression (GPR) predictive performance under Scenario 4 using different treatments of missing Blaine fineness. Values are means (standard deviations) across ten random states.

| Case | Train | | Test | | Extent of Overfitting | |
|---|---|---|---|---|---|---|
| | $R^2$ | RMSE (MPa) | $R^2$ | RMSE (MPa) | $\Delta R^2$ (%) | ΔRMSE (%) |
| Reference (constant imputation) | 0.965(0.001) | 5.481(0.046) | 0.932(0.007) | 7.530(0.380) | 3.387 | 37.396 |
| Multivariable imputation | 0.965(0.001) | 5.482(0.053) | 0.932(0.007) | 7.537(0.373) | 3.398 | 37.492 |
| Blaine omitted | 0.963(0.001) | 5.616(0.061) | 0.928(0.008) | 7.746(0.383) | 3.621 | 37.935 |

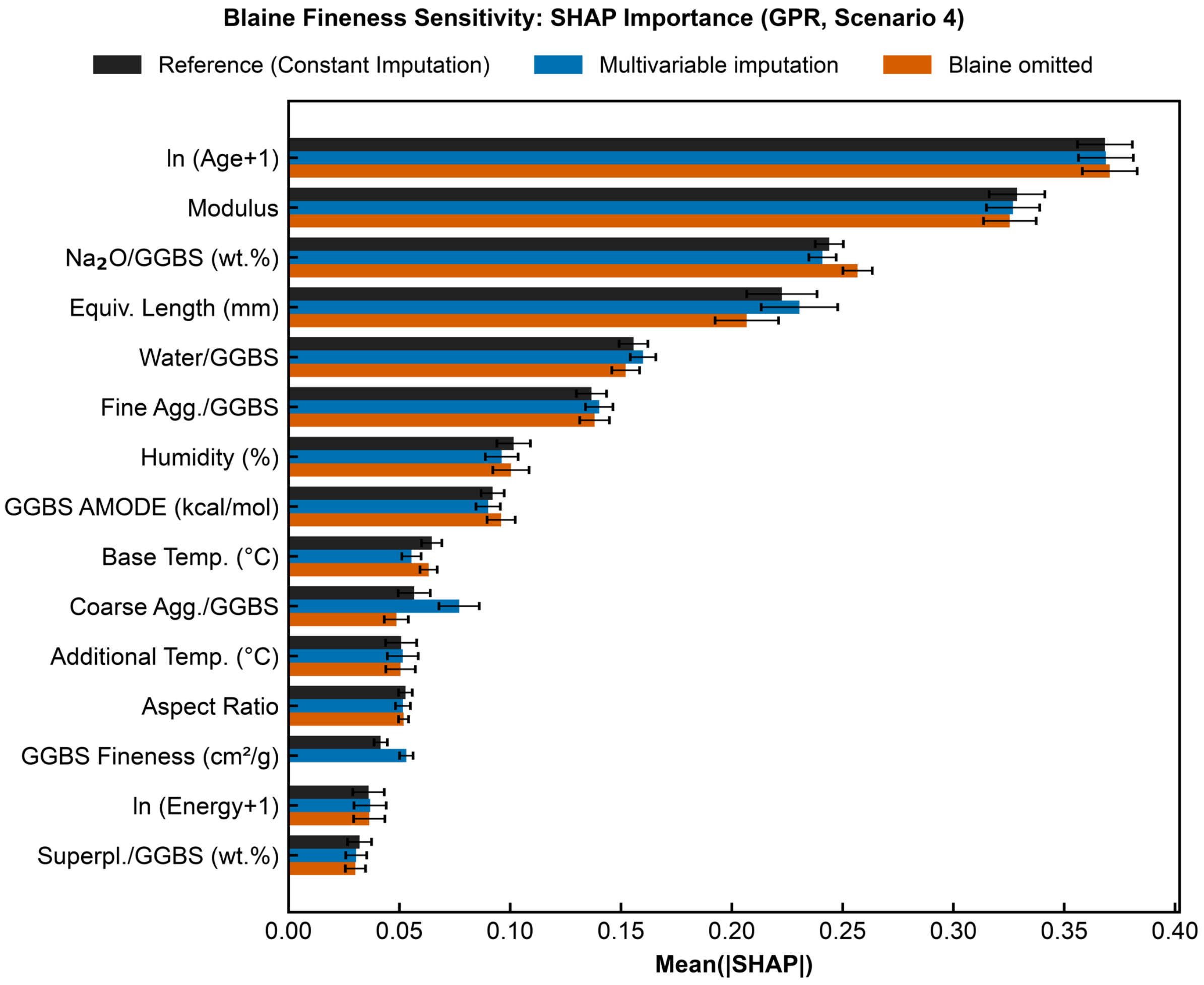

**Figure S1.** SHapley Additive exPlanations (SHAP) importance under three Blaine fineness treatments (Gaussian process regression, GPR, Scenario 4). Bars show mean absolute SHAP values averaged across ten random states; error bars indicate one standard deviation across states. Values are on the standardized strength scale.

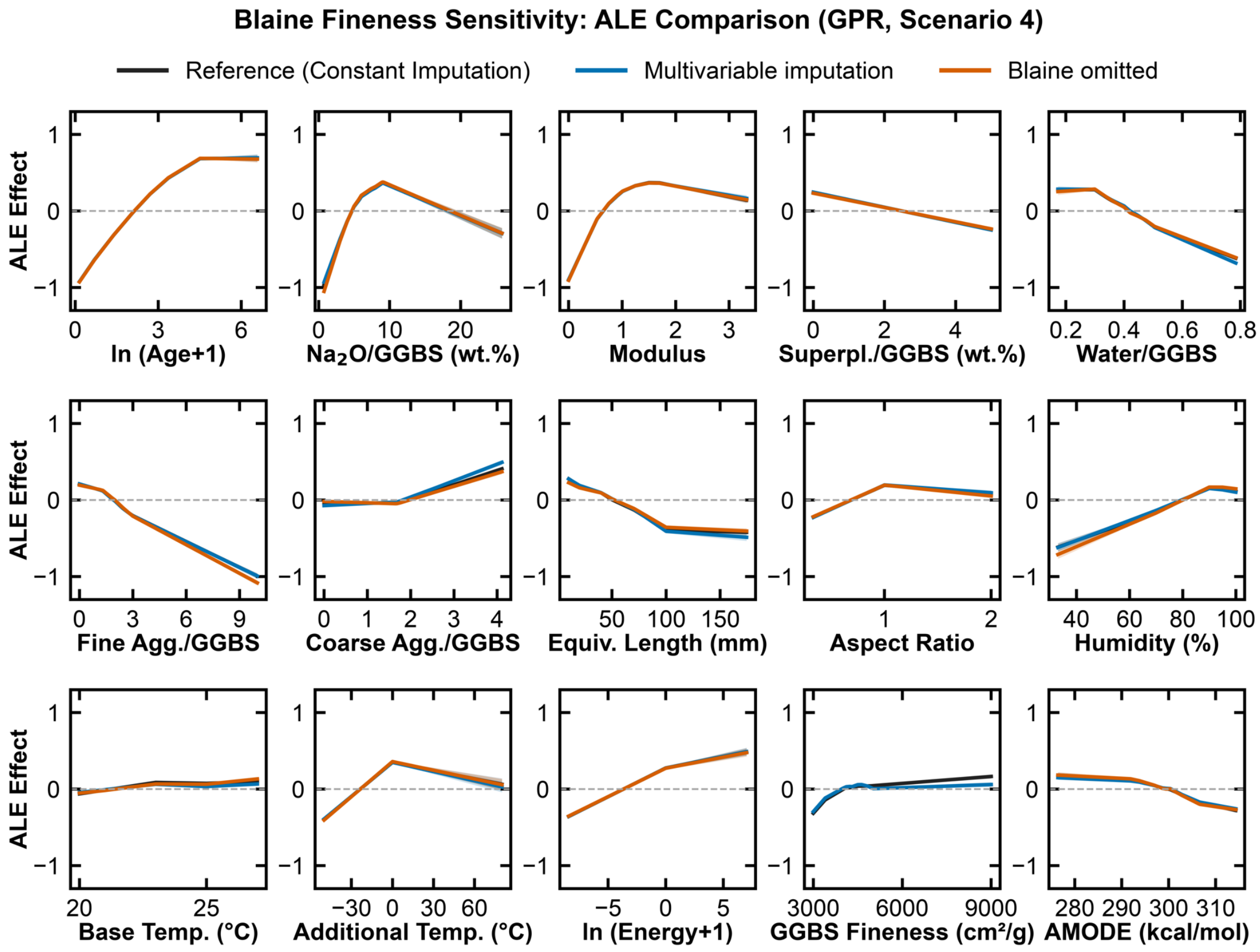

**Figure S2.** Accumulated local effects (ALE) comparison under three Blaine fineness treatments (GPR, Scenario 4). Lines and shaded bands show means and standard deviations across ten random states, on the standardized strength scale.

## S1.2 Curing-Condition Assumptions and Sensitivity Analysis

Sensitivity to relative humidity (RH) encoding and assumed values, unspecified room temperature, and thermal-exposure feature representation was assessed using GPR under Scenario 4. Each comparison used identical training–test splits across ten random states.

***Relative humidity***

The reference encoding assigned RH values of 90%, 95%, and 100% to sealed, moist-room, and water curing, respectively, when numerical RH was not reported. Two alternative encodings were examined: retaining RH while adding a binary indicator distinguishing inferred from explicitly reported values and omitting RH entirely. The indicator was assigned 1 for inferred values and 0 for reported values.

Adding the indicator provided only a small improvement in predictive performance, increasing mean test $R^2$ from 0.932 to 0.935 and reducing RMSE from 7.530 to 7.375 MPa. Omitting RH reduced mean test $R^2$ to 0.926 and increased RMSE to 7.867 MPa (**Table S2**). The principal SHAP importance patterns and ALE trends for the shared inputs remained broadly similar (**Figures S3** and **S4**). Given the small predictive improvement and the stability of the main trends, the reference encoding was retained for consistency and parsimony. The indicator distinguishes reporting status rather than an additional physical curing condition, and the assigned RH values remain approximations of the reported curing environments.

**Table S2.** GPR predictive performance under Scenario 4 using alternative relative humidity (RH) encodings. Values are means (standard deviations) across ten random states.

| Case | Train | | Test | | Extent of Overfitting | |
|---|---|---|---|---|---|---|
| | $R^2$ | RMSE (MPa) | $R^2$ | RMSE (MPa) | $\Delta R^2$ (%) | ΔRMSE (%) |
| Reference (original imputation) | 0.965(0.001) | 5.481(0.046) | 0.932(0.007) | 7.530(0.380) | 3.387 | 37.396 |
| RH omitted | 0.963(0.001) | 5.584(0.064) | 0.926(0.007) | 7.867(0.333) | 3.890 | 40.873 |
| RH + inferred indicator | 0.965(0.001) | 5.430(0.046) | 0.935(0.007) | 7.375(0.366) | 3.162 | 35.808 |

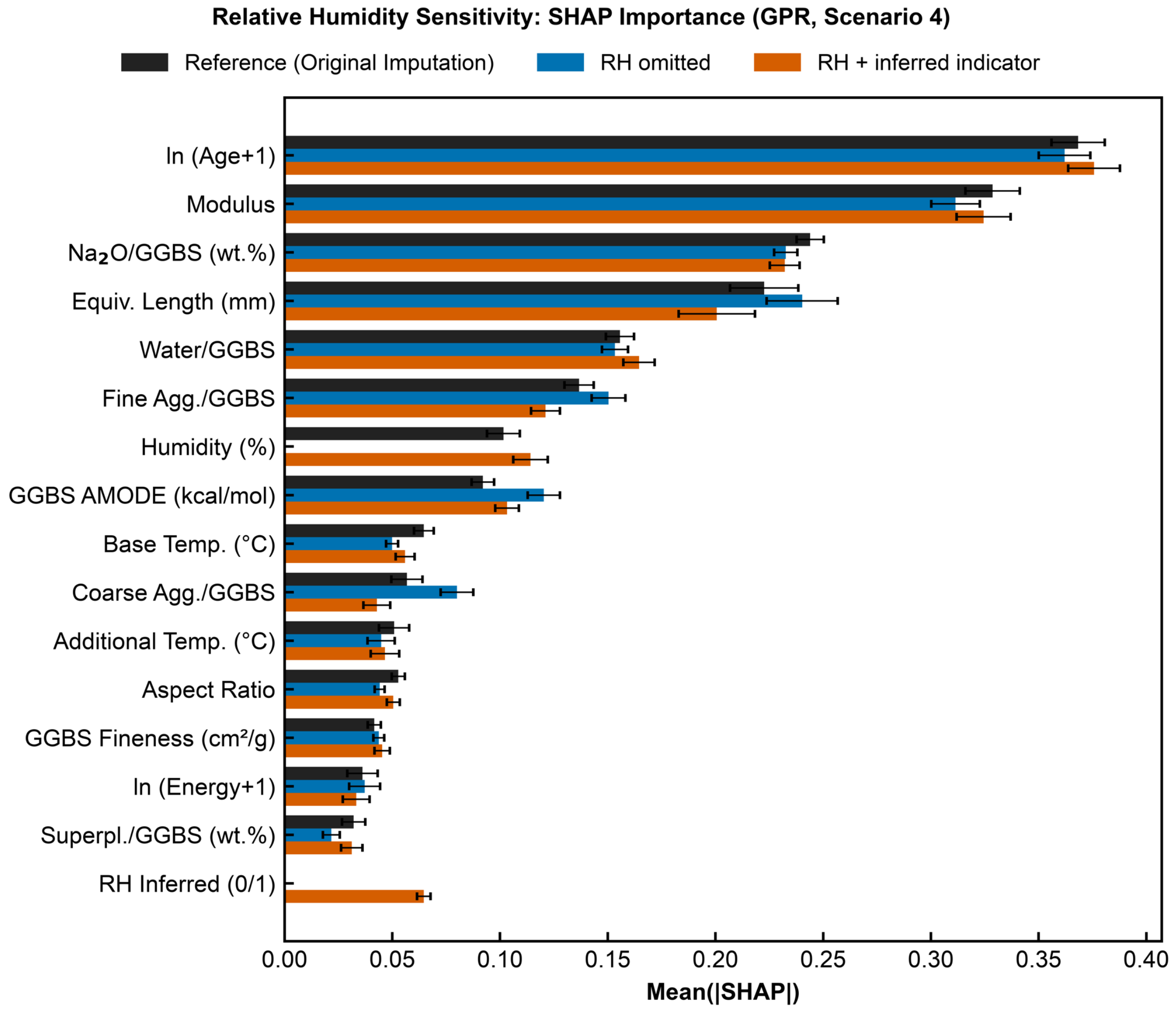

**Figure S3.** SHAP importance under alternative relative humidity (RH) encodings (GPR, Scenario 4). Bars show mean absolute SHAP values averaged across ten random states; error bars indicate one standard deviation across states. Values are on the standardized strength scale.

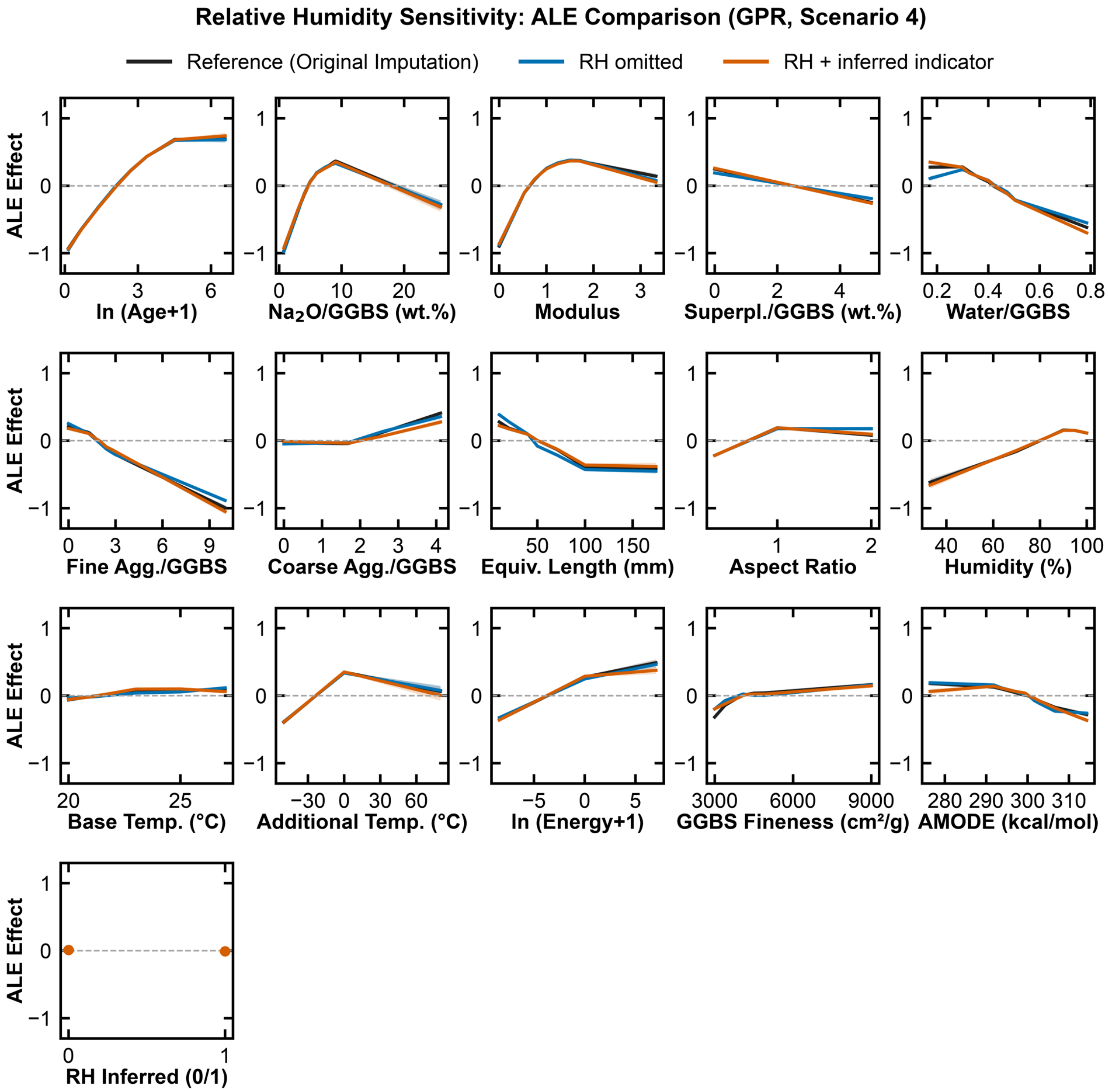

**Figure S4**. ALE comparison under alternative RH encodings (GPR, Scenario 4). Lines and shaded bands show means and standard deviations across ten random states, on the standardized strength scale. No RH effect is shown when the feature is omitted.

We additionally assessed sensitivity to the RH assumption used primarily for sealed curing by replacing the inferred reference value of 90% with 80%, 85%, and 95%, while retaining all other inputs and the same training–test splits. Mean test $R^2$ remained within 0.932–0.933, and RMSE ranged from 7.480 to 7.542 MPa (**Table S3**). The principal SHAP importance patterns and ALE trends remained broadly similar (**Figures S5** and **S6**), indicating limited sensitivity of predictive performance and the main model interpretations to the tested RH assumptions.

**Table S3.** GPR predictive performance under Scenario 4 using alternative RH assumptions for sealed curing, comparing 80%, 85%, and 95% with the 90% reference. Values are means (standard deviations) across ten random states.

| Case | Train | | Test | | Extent of Overfitting | |
|---|---|---|---|---|---|---|
| | $R^2$ | RMSE (MPa) | $R^2$ | RMSE (MPa) | $\Delta R^2$ (%) | ΔRMSE (%) |
| Reference (90%) | 0.965(0.001) | 5.481(0.046) | 0.932(0.007) | 7.530(0.380) | 3.387 | 37.396 |
| 80% | 0.965(0.001) | 5.474(0.046) | 0.933(0.007) | 7.480(0.370) | 3.300 | 36.638 |
| 85% | 0.965(0.001) | 5.482(0.046) | 0.933(0.007) | 7.504(0.378) | 3.336 | 36.891 |
| 95% | 0.965(0.000) | 5.493(0.043) | 0.932(0.007) | 7.542(0.359) | 3.392 | 37.312 |

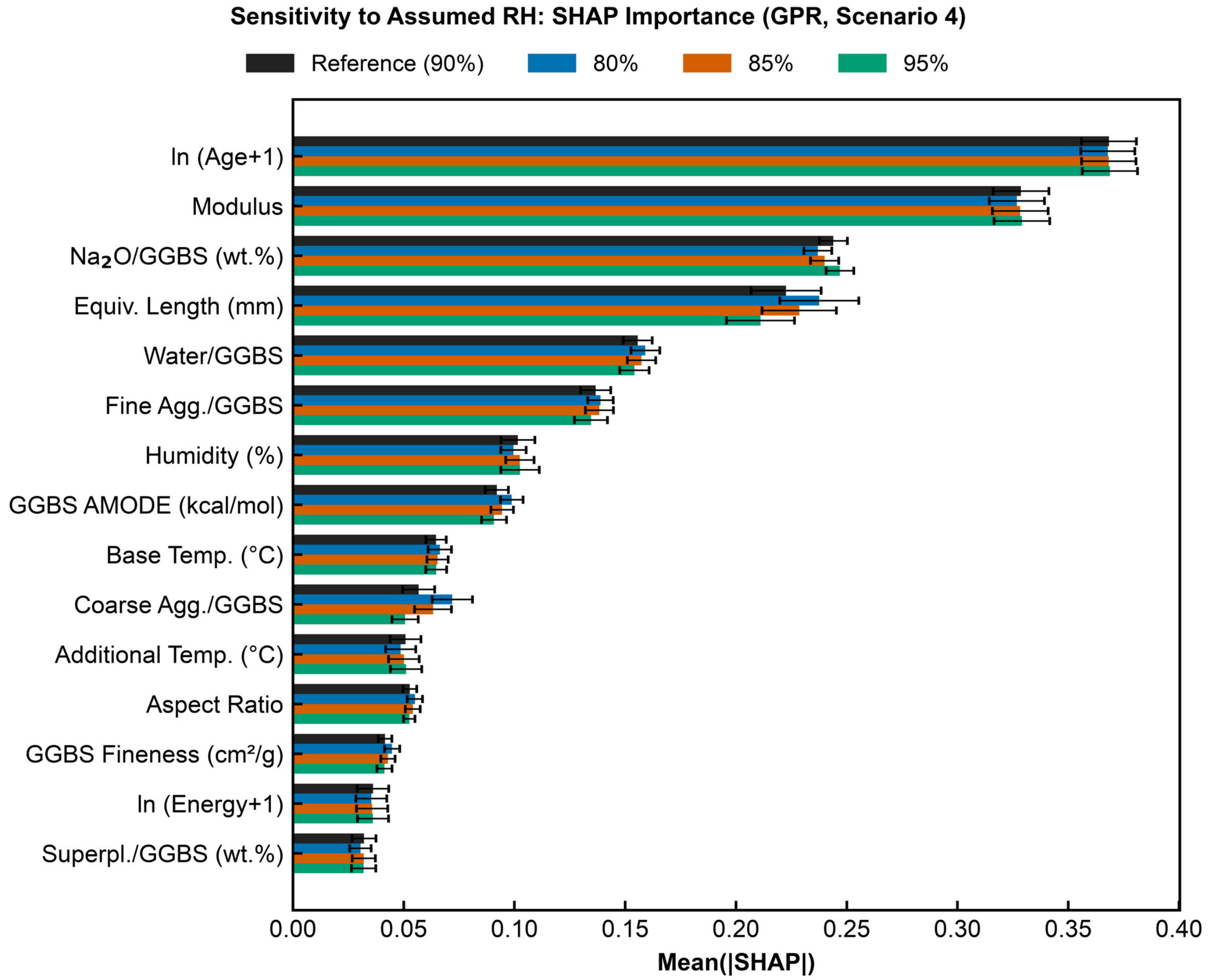

**Figure S5**. SHAP importance under alternative RH assumptions for sealed curing (GPR, Scenario 4). Bars show mean absolute SHAP values averaged across ten random states; error bars indicate one standard deviation across states. Values are on the standardized strength scale.

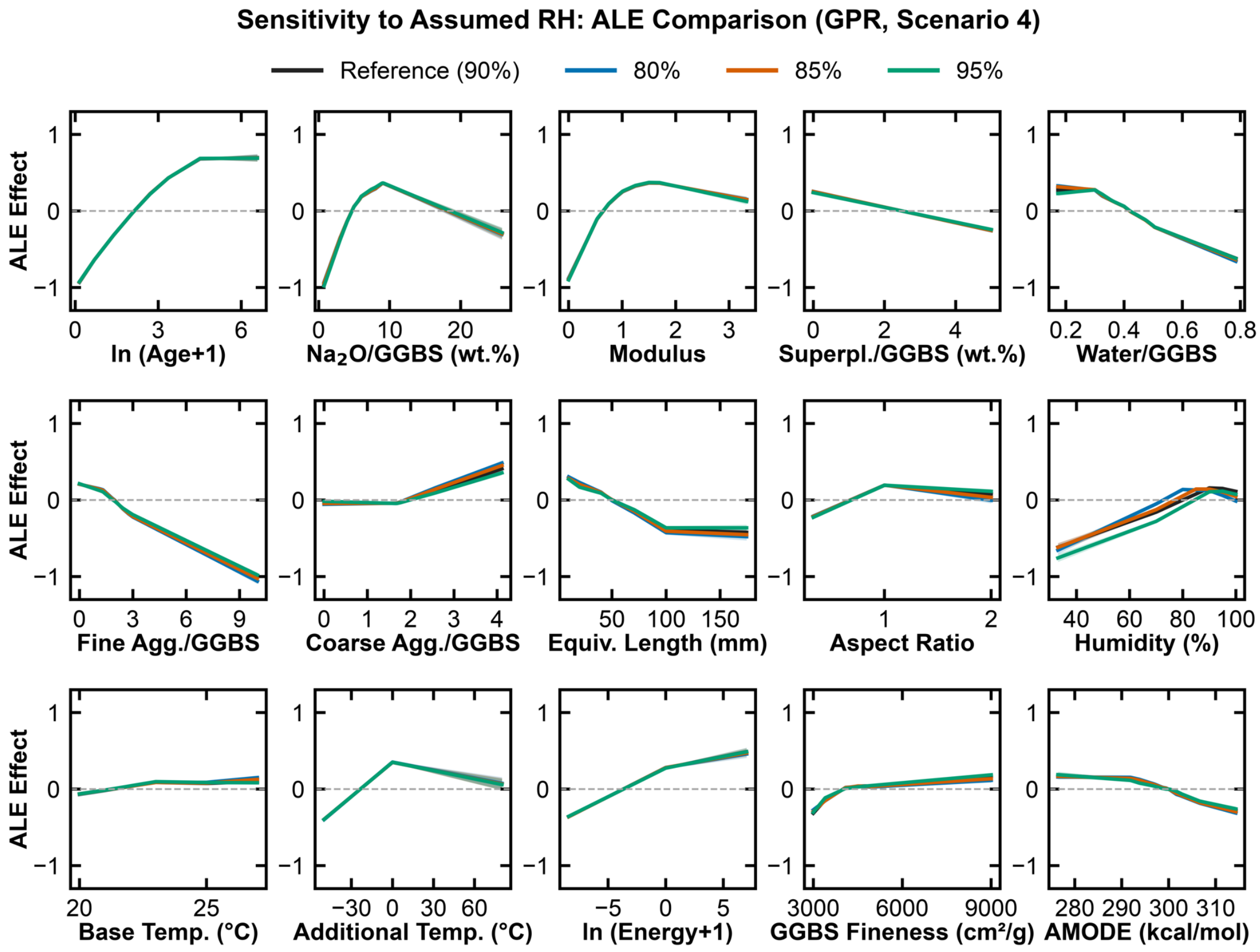

**Figure S6**. ALE comparison under alternative RH assumptions for sealed curing (GPR, Scenario 4). Lines and shaded bands show means and standard deviations across ten random states, on the standardized strength scale.

### *Base temperature*

For records with unspecified room temperature, alternative base temperatures of 18, 22, and 24 °C were compared with the reference assumption of 20 °C. Explicitly reported base temperatures were retained. For affected records with a nonambient curing stage, the absolute stage temperature was held fixed, and the temperature difference and derived thermal-exposure feature were recalculated consistently.

Predictive performance varied little across the tested assumptions, with mean test $R^2$ values of 0.932–0.935 and RMSE values of 7.393–7.530 MPa (**Table S4**). The principal SHAP importance patterns and ALE trends remained similar, although the base-temperature response itself varied with the assumed value (**Figures S7** and **S8**). These results indicate limited sensitivity to the tested room-temperature assumptions.

**Table S4.** GPR predictive performance under Scenario 4 using alternative assumptions for unspecified room temperature. Values are means (standard deviations) across ten random states.

| Case | Train | | Test | | Extent of Overfitting | |
|---|---|---|---|---|---|---|
| | $R^2$ | RMSE (MPa) | $R^2$ | RMSE (MPa) | $\Delta R^2$ (%) | ΔRMSE (%) |
| Reference (20 °C) | 0.965(0.001) | 5.481(0.046) | 0.932(0.007) | 7.530(0.380) | 3.387 | 37.396 |
| 18 °C | 0.965(0.001) | 5.427(0.047) | 0.933(0.007) | 7.470(0.358) | 3.342 | 37.641 |
| 22 °C | 0.965(0.001) | 5.433(0.046) | 0.934(0.007) | 7.430(0.375) | 3.261 | 36.768 |
| 24 °C | 0.965(0.001) | 5.440(0.046) | 0.935(0.007) | 7.393(0.346) | 3.182 | 35.907 |

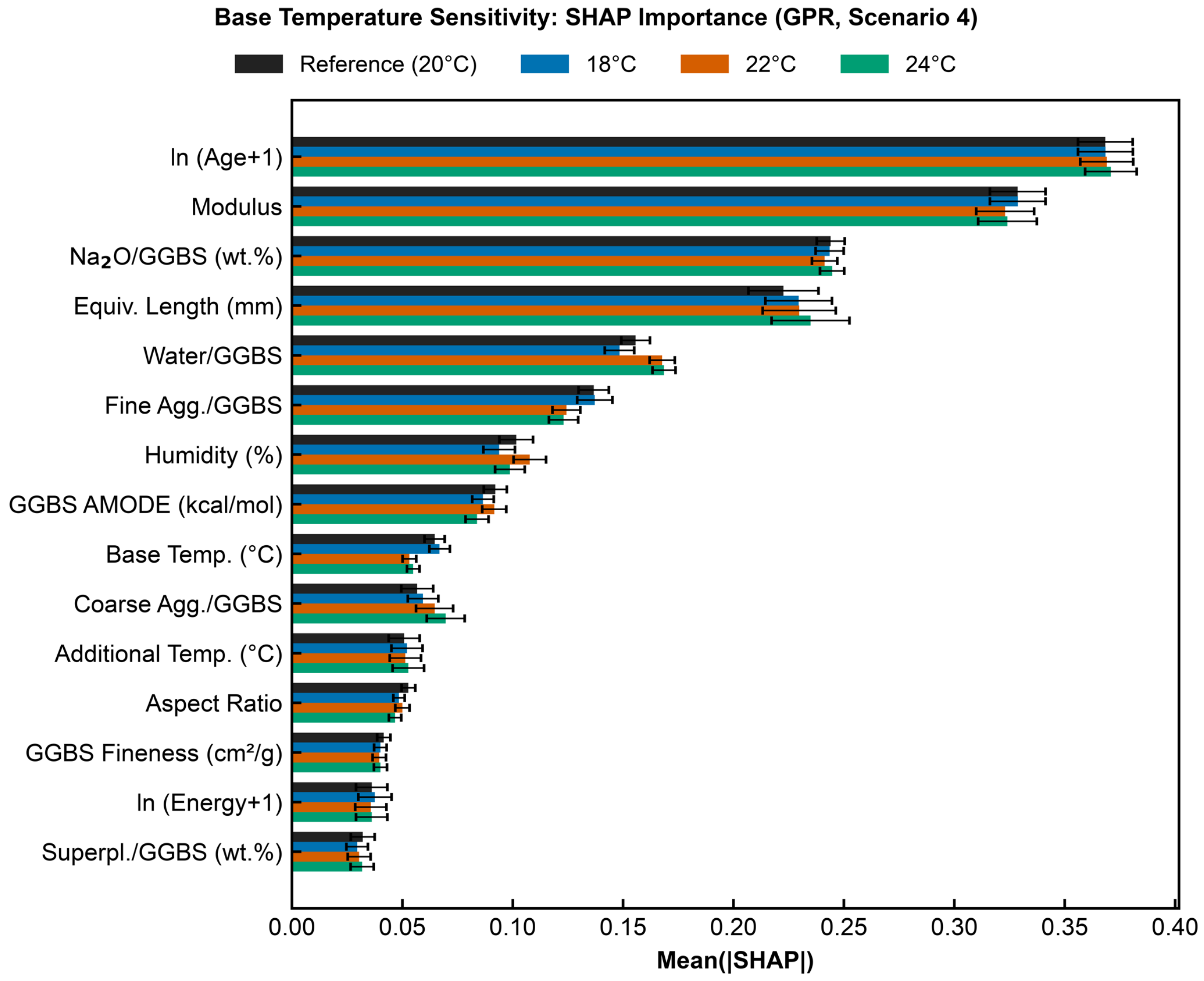

**Figure S7.** SHAP importance under alternative room-temperature assumptions (GPR, Scenario 4). Bars show mean absolute SHAP values averaged across ten random states; error bars indicate one standard deviation across states. Values are on the standardized strength scale.

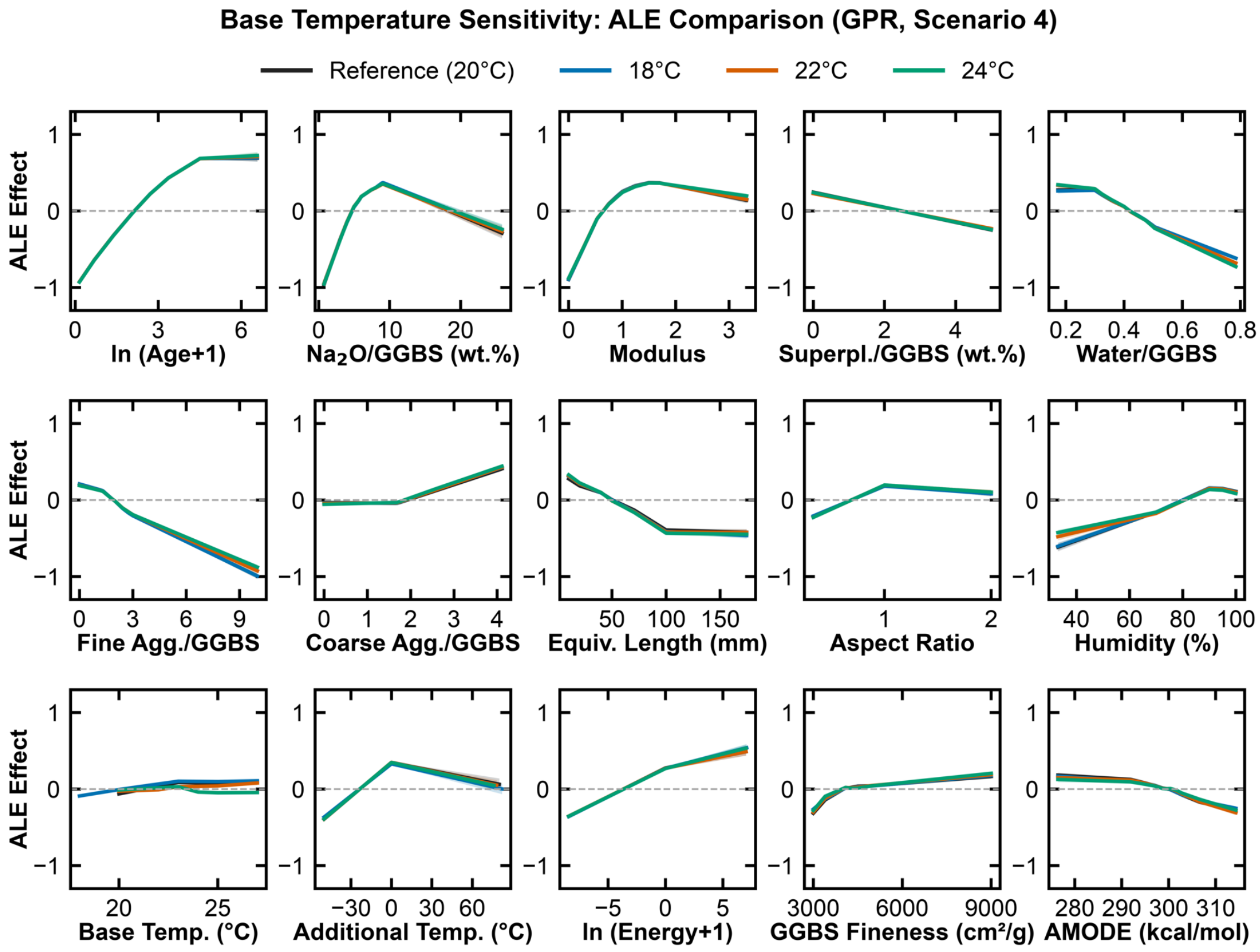

**Figure S8.** ALE comparison under alternative room-temperature assumptions (GPR, Scenario 4). Lines and shaded bands show means and standard deviations across ten random states, on the standardized strength scale.

### *Thermal-exposure feature*

To assess sensitivity to the temperature–time representation, the reference combination of Additional temperature and ln(Energy + 1) was compared with an alternative using Additional temperature and Additional time as separate inputs. All other features were retained, and both approaches used identical training–test splits across ten random states.

The two representations produced nearly identical predictive performance, with mean test $R^2$ values of 0.932 for both and RMSE values of 7.530 and 7.540 MPa for the reference and alternative, respectively (**Table S5**). The principal SHAP importance patterns and ALE trends for mixture proportions, curing age, and slag reactivity remained similar, although the ALE response for Additional temperature differed between representations (**Figures S9** and **S10**). These results indicate limited sensitivity of overall predictive performance and the main trends to the tested representation, while individual thermal-feature effects depend on how temperature and duration are encoded.

**Table S5.** GPR predictive performance under Scenario 4 using alternative temperature–time representations. Values are means (standard deviations) across ten random states.

| Case | Train $R^2$ | Train RMSE (MPa) | Test $R^2$ | Test RMSE (MPa) | Extent of Overfitting $\Delta R^2$ (%) | Extent of Overfitting ΔRMSE (%) |
|---|---|---|---|---|---|---|
| Reference (Additional Temp. + ln(Energy + 1)) | 0.965(0.001) | 5.481(0.046) | 0.932(0.007) | 7.530(0.380) | 3.387 | 37.396 |
| Additional Temp. + Additional Time | 0.965(0.001) | 5.474(0.047) | 0.932(0.008) | 7.540(0.403) | 3.416 | 37.734 |

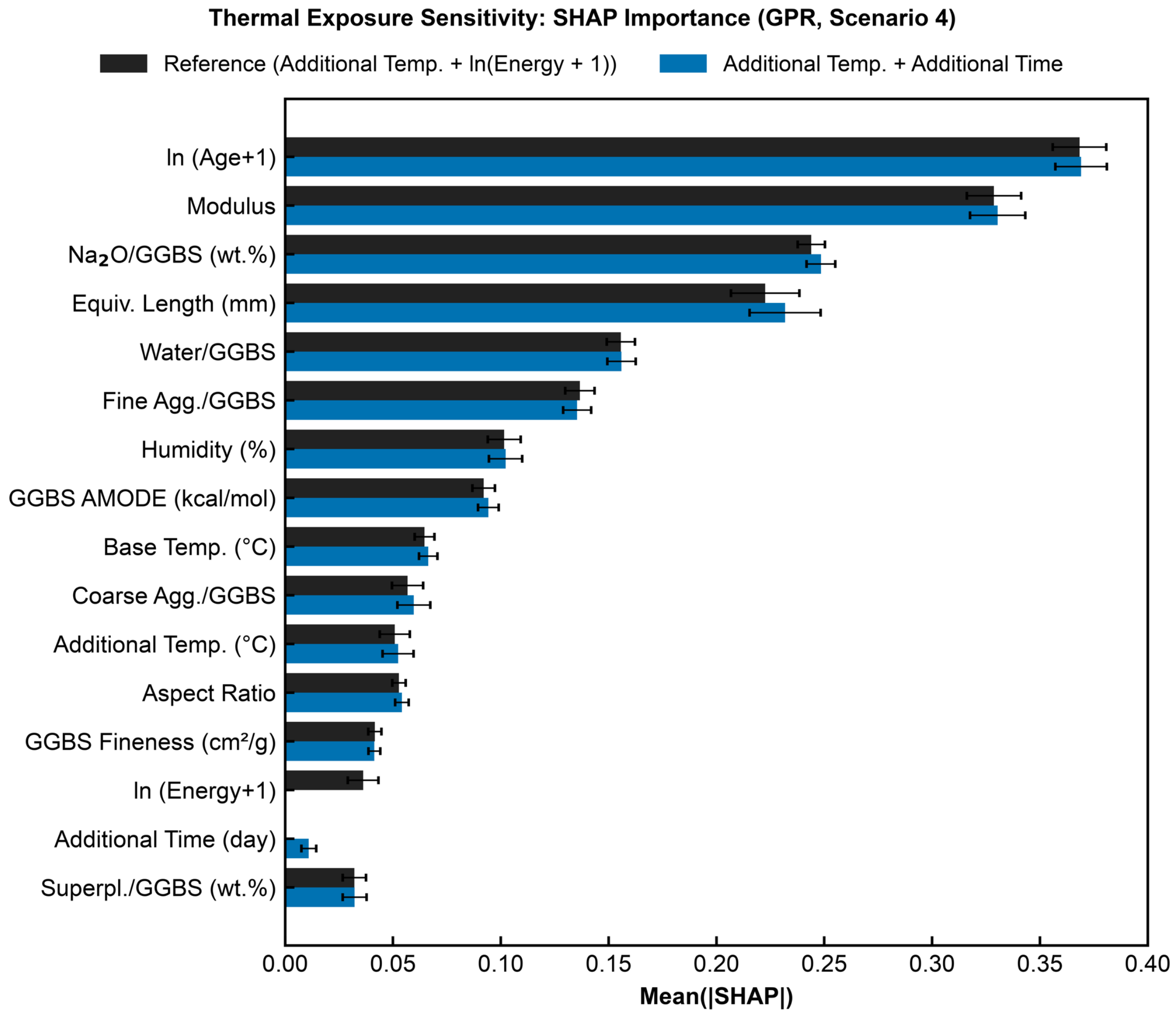

**Figure S9.** SHAP importance under alternative temperature–time representations (GPR, Scenario 4). Bars show mean absolute SHAP values averaged across ten random states; error bars indicate one standard deviation across states. Values are on the standardized strength scale.

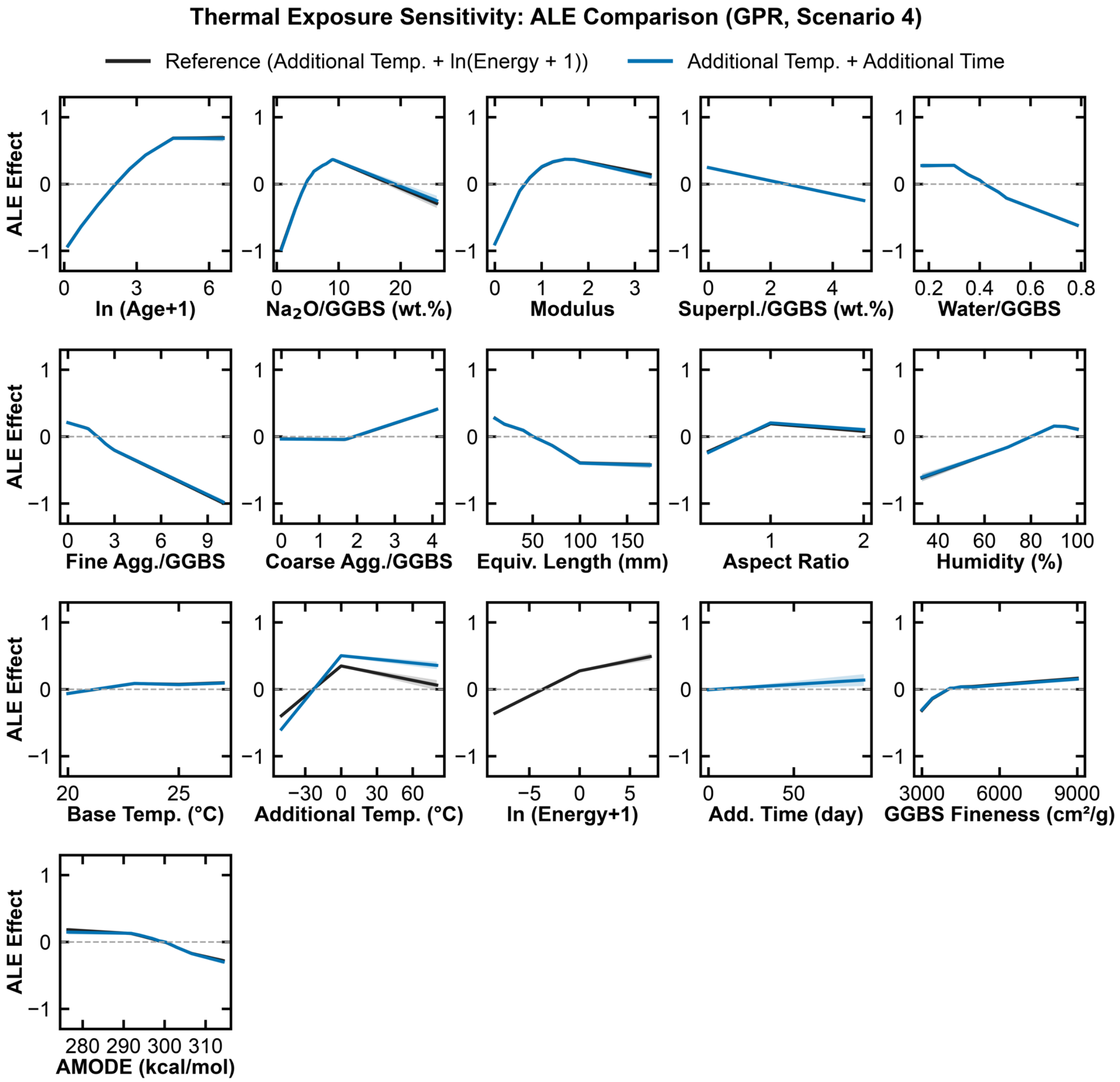

**Figure S10.** ALE comparison under alternative temperature–time representations (GPR, Scenario 4). Lines and shaded bands show means and standard deviations across ten random states, on the standardized strength scale. Effects are shown only for features included in each representation.

## S1.3 Hyperparameter Ranges and Tuning Strategies

The hyperparameter ranges and tuning strategies for the machine learning (ML) models are summarized in **Table S6**. For artificial neural network (ANN), extreme gradient boosting (XGB), and random forest (RF) models, hyperparameters were optimized using randomized search with 200 iterations, combined with stratified 10-fold cross-validation on the training set. The optimal hyperparameter configurations were identified as those with the highest average predictive performance across the validation folds.

In contrast, GPR does not employ randomized hyperparameter search. Instead, the GPR model was constructed using a predefined composite kernel, with physically and statistically motivated bounds placed on the kernel hyperparameters. This kernel structure was selected to balance model flexibility and numerical stability and was applied consistently across all feature scenarios.

**Table S6.** Hyperparameter ranges and tuning strategy for the machine learning (ML) models. Randomized search was applied to artificial neural network (ANN), extreme gradient boosting (XGB), and random forest (RF) (200 iterations each). GPR did not employ random search; its kernel configuration is explicitly defined.

| Model | Hyperparameter | Range | Iterations |
|---|---|---|---|
| GPR | Kernel | No random search applied. Kernel defined as: ConstantKernel (1.0, (0.1, 10.0)) * RationalQuadratic (length_scale = 5.0, length_scale_bounds = (1.0, 15.0), alpha=0.3, alpha_bounds = (0.3, 5.0)) + WhiteKernel (noise_level = 0.3, noise_level_bounds = (0.1, 3.0)) | N/A |
| ANN | "activation" | "relu" | 200 |
| | "hidden_layer_sizes" | (60, ), (80, ), (100, ), (120, ), (140, ), (100, 50), (120, 60), (150, 100), (200, 100) | |
| | "max_iter" | 300, 500, 600, 800, 1000 | |
| | "solver" | "adam" | |
| | "alpha" | 0.0001, 0.001, 0.01, 0.1 | |
| | "early_stopping" | "True" | |
| | "n_iter_no_change" | 20 | |
| | "learning_rate_init" | 0.1, 0.01, 0.001, 0.0001 | |
| | "validation_fraction" | 0.1 | |
| XGB | "n_estimators" | 300, 500, 800 | |
| | "learning_rate" | 0.01, 0.03, 0.05 | |
| | "max_depth" | 2, 3, 4, 5 | |
| | "subsample" | 0.6, 0.8, 1.0 | |
| | "colsample_bytree" | 0.6, 0.7, 0.8, 0.9 | |
| | "gamma" | 0, 0.1, 0.5, 1 | |

| | | |
|---|---|---|
| | "reg_alpha" | 0, 0.5, 1, 2, 3 |
| | "reg_lambda" | 1, 3, 5, 7, 10 |
| RF | "n_estimators" | 200, 400, 600, 800 |
| | "max_features" | 3, 4, 5, "sqrt" |
| | "min_samples_split" | 2, 4, 6, 8 |
| | "min_samples_leaf" | 2, 4, 6, 8 |
| | "max_depth" | None, 10, 20, 30, 40, 50 |

### S1.4 Data-Cleaning Exclusions and Sensitivity Analysis

**Table S7.** Points removed during the model-assisted, expert-reviewed data cleaning process. Only features directly relevant to the removal decisions are shown. Removal record IDs are table-specific and do not correspond to source-study identifiers or released dataset IDs.

| Removal record ID | $Na_2O$/GGBS (wt.%) | Modulus (Ms) | Water/ GGBS | Base Temp. (°C) | Additional Temp. (°C) | Additional Time (day) | Relative Humidity (%) | Age (day) | Compressive Strength (MPa) | Reason for removal |
|---|---|---|---|---|---|---|---|---|---|---|
| 1 | 1 | 1.5 | 0.38 | 23 | 0 | 0 | 95 | 3 | 1.56 | Reported error bar large relative to the near-zero mean strength; excluded because of limited quantitative precision. |
| 2 | 4 | 0 | 0.5 | 23 | 57 | 0.83 | 50–95 | 2 | 18.64 | Contains 4 wt.% $Ca(OH)_2$ and uses multi-stage curing RH (50 ± 5% for 2 h, then 95 ± 3% for 22 h), making activator and curing-condition features inconsistent with dataset definitions. |
| 3 | 4 | 0.67 | 0.5 | 23 | 57 | 0.83 | 50–95 | 2 | 32.54 | |
| 4 | 4 | 1.35 | 0.5 | 23 | 57 | 0.83 | 50–95 | 2 | 21.69 | |
| 5 | 4 | 2.02 | 0.5 | 23 | 57 | 0.83 | 50–95 | 2 | 36.78 | |
| 6 | 8 | 0 | 0.5 | 23 | 57 | 0.83 | 50–95 | 2 | 21.36 | |
| 7 | 8 | 1.35 | 0.5 | 23 | 57 | 0.83 | 50–95 | 2 | 64.41 | |
| 8 | 8 | 2.02 | 0.5 | 23 | 57 | 0.83 | 50–95 | 2 | 65.59 | |

To assess sensitivity to the exclusions in **Table S7**, GPR models under Scenario 4 were compared across ten random states (10, 20, …, 100). The reference used the cleaned dataset. In the restored case, all eight excluded records listed in **Table S7** were added to each training set. Test sets, raw SHAP background samples, ALE evaluation grids, and the 30 mm paste design-space grid were identical between cases.

Predictive performance changed little after restoration, with both cases yielding a mean test $R^2$ of 0.932 and a mean test RMSE of 7.530 MPa at the reported precision (**Table S8**). The ranking of mean absolute SHAP importance was unchanged, and the ALE curves closely overlapped (**Figures S11** and **S12**). Figure S13 illustrates the corresponding strength contours at an average metal oxide dissociation energy (AMODE) of 290 kcal/mol and water/GGBS = 0.4. The predicted optimal mix-design regions remained largely unchanged after restoration. Together, these results indicate limited sensitivity to the exclusions under the tested configuration.

**Table S8.** GPR predictive performance under Scenario 4 for the cleaned and **Table S7** restored cases. Values are means (standard deviations) across ten random states.

| Case | Train | | Test | | Extent of Overfitting | |
|---|---|---|---|---|---|---|
| | $R^2$ | RMSE (MPa) | $R^2$ | RMSE (MPa) | $\Delta R^2$ (%) | ΔRMSE (%) |
| Reference (cleaned) | 0.965(0.001) | 5.481(0.046) | 0.932(0.007) | 7.530(0.380) | 3.387 | 37.396 |
| Table S7 restored | 0.965(0.001) | 5.482(0.045) | 0.932(0.007) | 7.530(0.381) | 3.381 | 37.361 |

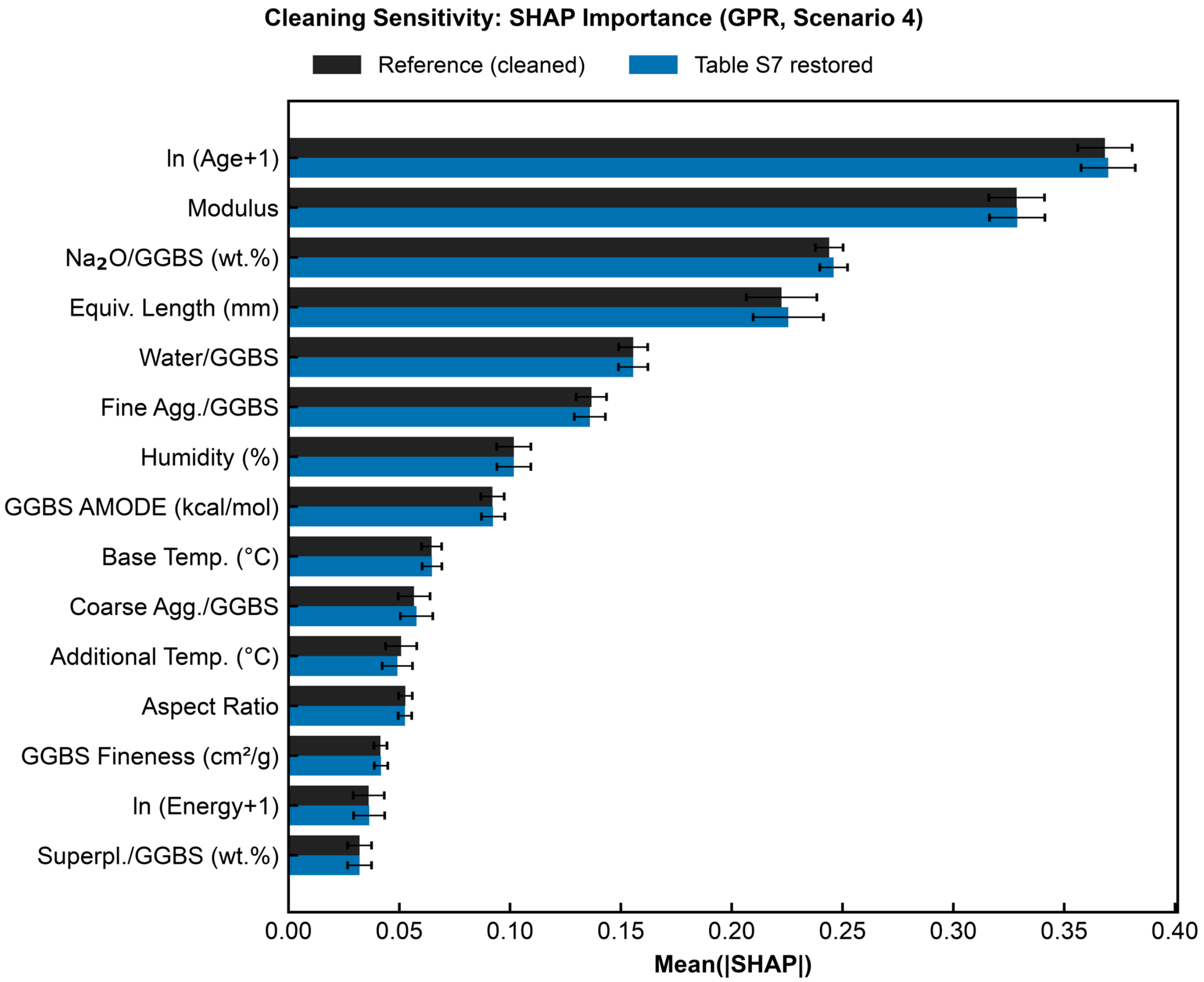

**Figure S11.** SHAP importance for the cleaned and **Table S7** restored cases (GPR, Scenario 4). Mean absolute SHAP values averaged across ten random states; error bars indicate one standard deviation across states. Values are on the standardized strength scale.

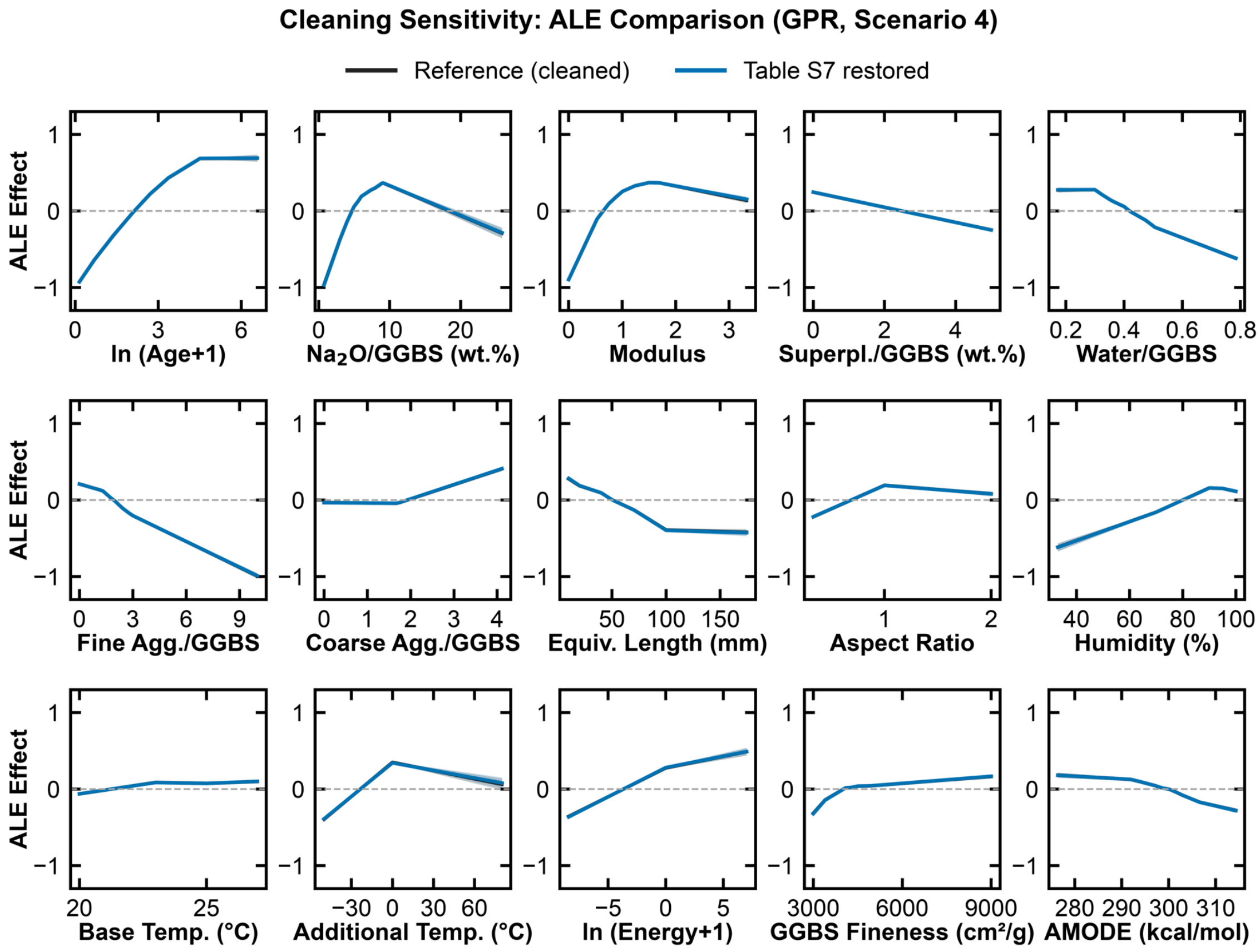

**Figure S12.** ALE comparison for the cleaned and **Table S7** restored cases (GPR, Scenario 4). Lines and shaded bands show means and standard deviations across ten random states, on the standardized strength scale.

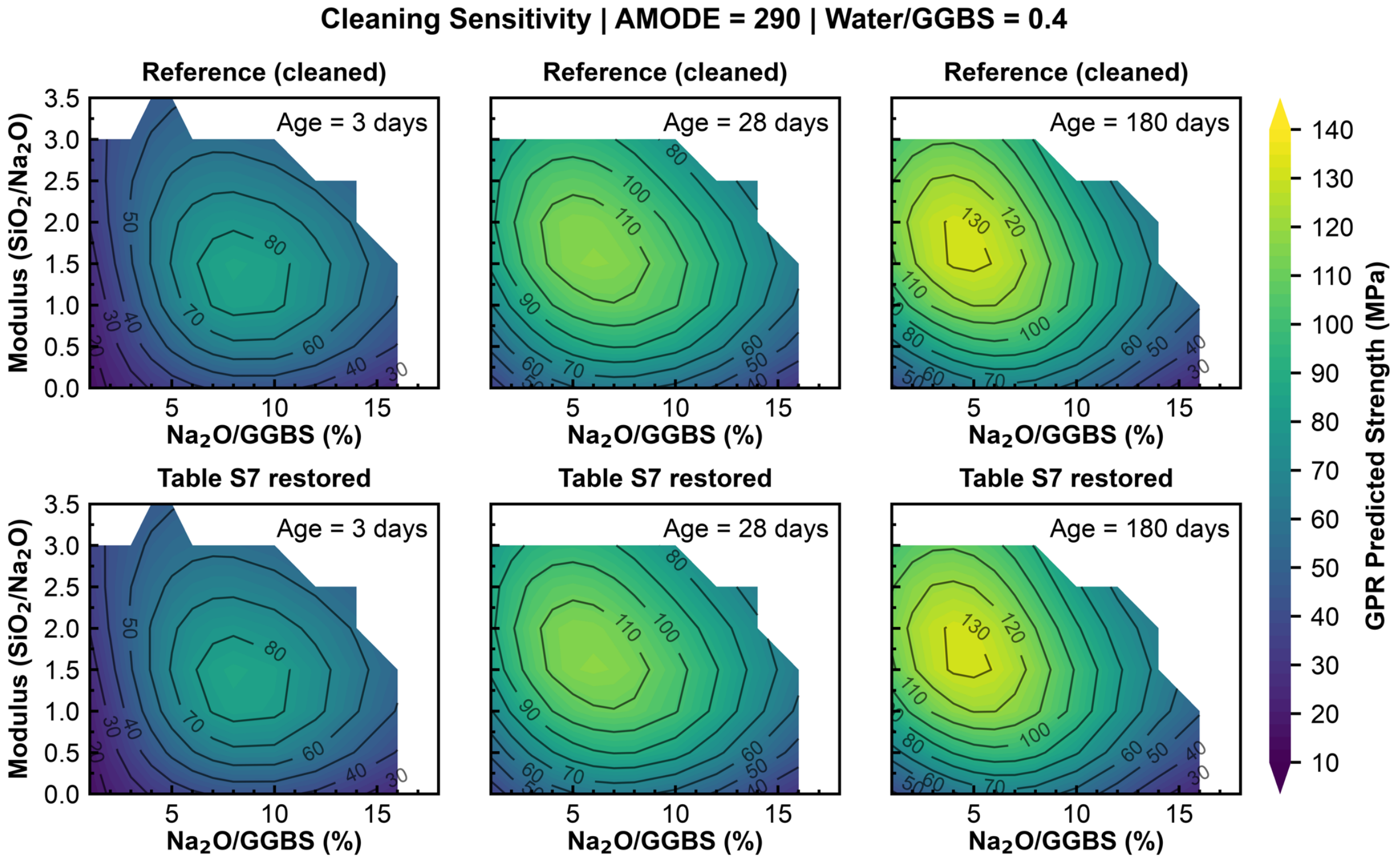

**Figure S13.** Mean GPR strength contours across ten random states for the cleaned (top) and **Table S7** restored (bottom) cases under Scenario 4. Results are shown for 30 mm paste specimens at an average metal oxide dissociation energy (AMODE) = 290 kcal/mol, water/GGBS = 0.4, and ages of 3, 28, and 180 days.

### S1.5 Correlation Between GGBS AMODE and Ca/Si Ratio

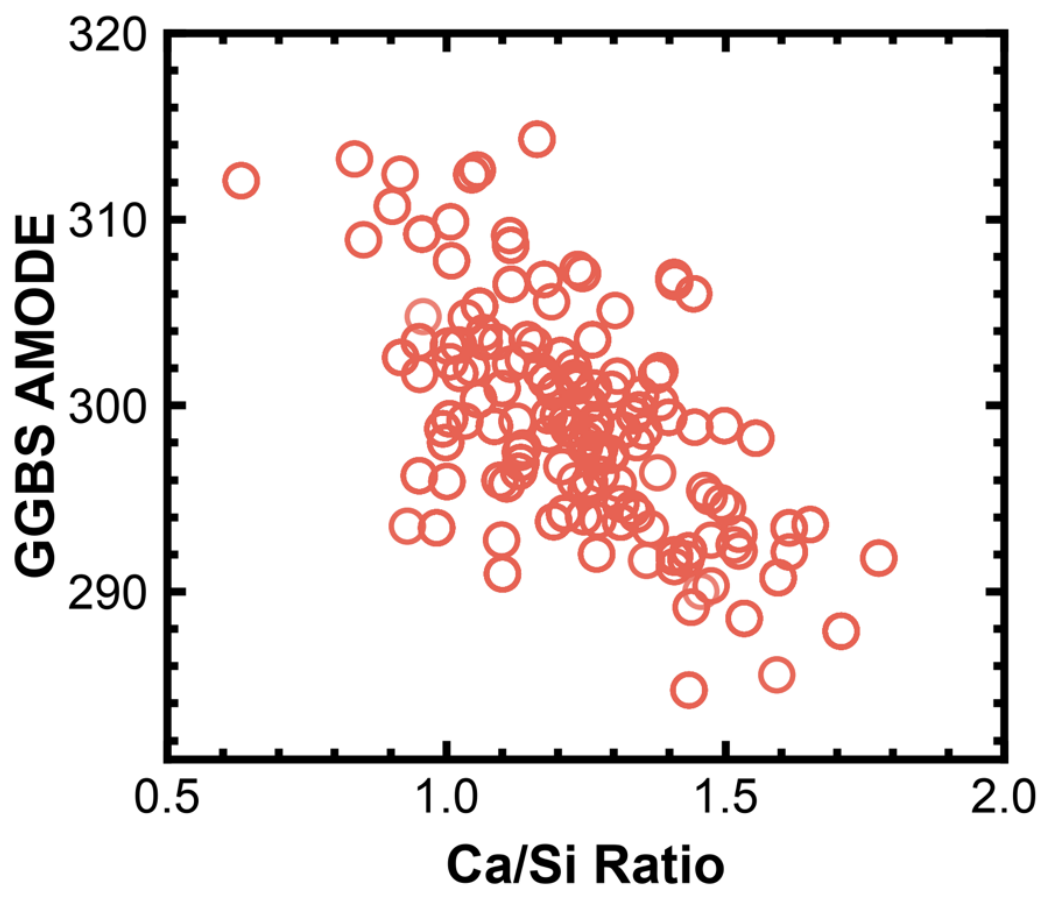

**Figure S14.** Correlation between the modified AMODE values and Ca/Si ratios of GGBSs in the dataset.

## S1.6 Calculation of Model Evaluation Metrics

Model performance was evaluated using the coefficient of determination (**$R^2$**) and the RMSE. Overfitting was quantified by the relative differences between training and test metrics ($\Delta R^2$ and ΔRMSE). $R^2$ and RMSE were calculated for each of the ten runs and are reported as means with standard deviations. $\Delta R^2$ and ΔRMSE were calculated from the unrounded mean training and test metrics across these runs.

### Coefficient of Determination (*$R^2$*)

$$R^2 = 1 - \frac{\sum_{i=1}^{n}(y_i - \hat{y}_i)^2}{\sum_{i=1}^{n}(y_i - \bar{y})^2} \tag{S1}$$

where $y_i$, $\hat{y}_i$, and $\bar{y}$ denote the observed, predicted, and mean observed values, respectively, and $n$ is the number of samples.

### Root-Mean-Square Error (RMSE)

$$\mathrm{RMSE} = \sqrt{\frac{1}{n}\sum_{i=1}^{n}(y_i - \hat{y}_i)^2} \tag{S2}$$

### Overfitting Metrics

$$\Delta R^2 = \left|\frac{R^2_{\mathrm{test}} - R^2_{\mathrm{train}}}{R^2_{\mathrm{train}}}\right| \times 100\% \tag{S3}$$

$$\Delta\mathrm{RMSE} = \left|\frac{\mathrm{RMSE}_{\mathrm{test}} - \mathrm{RMSE}_{\mathrm{train}}}{\mathrm{RMSE}_{\mathrm{train}}}\right| \times 100\% \tag{S4}$$

## S1.7 Constant Input Features Used in Design Space Exploration

**Table S9.** Constant input features for design space exploration (Scenario 4). The values here are values close to the dataset median (for pastes) or commonly used experimental conditions.

| Feature | Value |
|---|---|
| Fine aggregate/GGBS | 0 |
| Coarse aggregate/GGBS | 0 |
| Superplasticizer/GGBS (wt.%) | 0 |
| GGBS Blaine fineness ($cm^2/g$) | 4500 |
| Base temperature (°C) | 20 |
| Additional temperature (°C) | 0 |
| ln(Energy + 1) | 0 |
| Relative humidity (%) | 90 |
| Equivalent length (mm) | 30 |
| Aspect ratio | 1 |

## S1.8 Mixture Conversion and Mass-Balance Screening

All mixtures are expressed on a 100 g GGBS basis, and all ingredient masses are in g. Let $n$ denote the numerical value of $Na_2O$/GGBS expressed in wt.% (e.g., $n = 6$ for 6 wt.%), $w$ the total water/GGBS mass ratio, and Ms the target molar $SiO_2/Na_2O$ ratio.

***Silicate sources***

The $Na_2SiO_3$ solution is assumed to contain 37 wt.% solids with a $SiO_2/Na_2O$ mass ratio of 3.3. $Na_2SiO_3 \cdot 5H_2O$ replaces the $Na_2SiO_3$ solution when the combined water contribution from the $Na_2SiO_3$ and NaOH solutions exceeds the target total water content. Their component mass fractions are:

**Table S10.** Component mass fractions of the silicate sources used in mixture conversion.

| Silicate source | $f_{Na_2O}$ | $f_{SiO_2}$ | $f_{water}$ |
|---|---|---|---|
| $Na_2SiO_3$ solution | 0.37/4.3 | 0.37×3.3/4.3 | 0.63 |
| $Na_2SiO_3 \cdot 5H_2O$ | 62/212 | 60/212 | 90/212 |

The pentahydrate water fraction represents its crystallization water.

***Ingredient masses***

Let $m_{silicate}$, $m_{NaOH}$, $m_{additional\ SiO_2}$, and $m_{additional\ water}$ denote the masses of the selected silicate source, NaOH solution, additional $SiO_2$, and separately added water, respectively. Assuming a 50 wt.% NaOH solution and nominally pure additional $SiO_2$:

$$m_{silicate} = min\ (\frac{n}{f_{Na_2O}}, \frac{nMs(60/62)}{f_{SiO_2}})$$

$$m_{NaOH} = \frac{n - f_{Na_2O} m_{silicate}}{0.775 \times 0.50}$$

$$m_{additional\ SiO_2} = nMs(60/62) - f_{SiO_2} m_{silicate}$$

$$m_{additional\ water} = 100w - f_{water} m_{silicate} - 0.50 m_{NaOH}$$

Here, $60/62$ converts the molar $SiO_2/Na_2O$ ratio into a mass ratio. The factor $0.775 = 62/(2 \times 40)$ converts pure NaOH mass to $Na_2O$-equivalent mass. Total water includes water supplied by the selected silicate source and NaOH solution, together with separately added water.

***Source selection and screening***

First calculate the ingredient masses using the $Na_2SiO_3$ solution. Retain this formulation if:

$$0.63m_{solution} + 0.50m_{NaOH} \leq 100w$$

Otherwise, replace the solution entirely with $Na_2SiO_3 \cdot 5H_2O$ and recalculate all ingredient masses using the pentahydrate mass fractions. Retain this formulation only if:

$$\frac{90}{212}m_{pentahydrate} + 0.50m_{NaOH} \leq 100w$$

Reject formulations that remain infeasible. All ingredient masses must be non-negative:

$$m_{silicate}, m_{NaOH}, m_{additional\ SiO_2}, m_{additional\ water} \geq 0$$

Only formulations satisfying these conditions and the $Na_2O$-equivalent, $SiO_2$, and water balances proceed to Mahalanobis filtering.

***Illustrative examples***

**Table S11** illustrates three outcomes of source selection and mass-balance screening: retaining the $Na_2SiO_3$ solution, replacing it with $Na_2SiO_3 \cdot 5H_2O$, and rejecting an infeasible formulation.

**Table S11.** Illustrative examples of mixture conversion and mass-balance screening on a 100 g ground granulated blast-furnace slag (GGBS) basis. All ingredient masses are in grams.

| **Parameter / Ingredient** | **Example 1** | **Example 2** | **Example 3** |
|---|---|---|---|
| $Na_2O$/GGBS (wt.%) | 6 | 16 | 16 |
| Target Ms | 1.5 | 3.0 | 3.0 |
| Target water/GGBS | 0.40 | 0.40 | 0.20 |
| Target total water | 40.000 | 40.000 | 20.000 |
| Water supplied by the initial $Na_2SiO_3$ solution | 19.324 | 103.061 | 103.061 |
| **Ingredient masses after source selection** | | | |
| $Na_2SiO_3$ solution | 30.673 | 0 | 0 |
| $Na_2SiO_3 \cdot 5H_2O$ | 0 | 54.710 | 54.710 |
| NaOH solution | 8.673 | 0 | 0 |
| Additional $SiO_2$ | 0 | 30.968 | 30.968 |
| Water supplied by the selected silicate source | 19.324 | 23.226 | 23.226 |
| Water supplied by the NaOH solution | 4.336 | 0 | 0 |
| Required additional water | 16.340 | 16.774 | -3.226 |
| Screening outcome | Retain solution | Use pentahydrate | Reject |

Values are rounded to three decimal places for presentation; screening used unrounded values. Example 3 is rejected because the pentahydrate crystallization water exceeds the target total water content. Water supplied by the selected silicate source and NaOH solution is included in their respective product masses and is not a separate ingredient.

## S2 Dataset Structure and Exploratory Relationships

### S2.1 Relationship Between 28-Day Compressive Strength and Individual Features

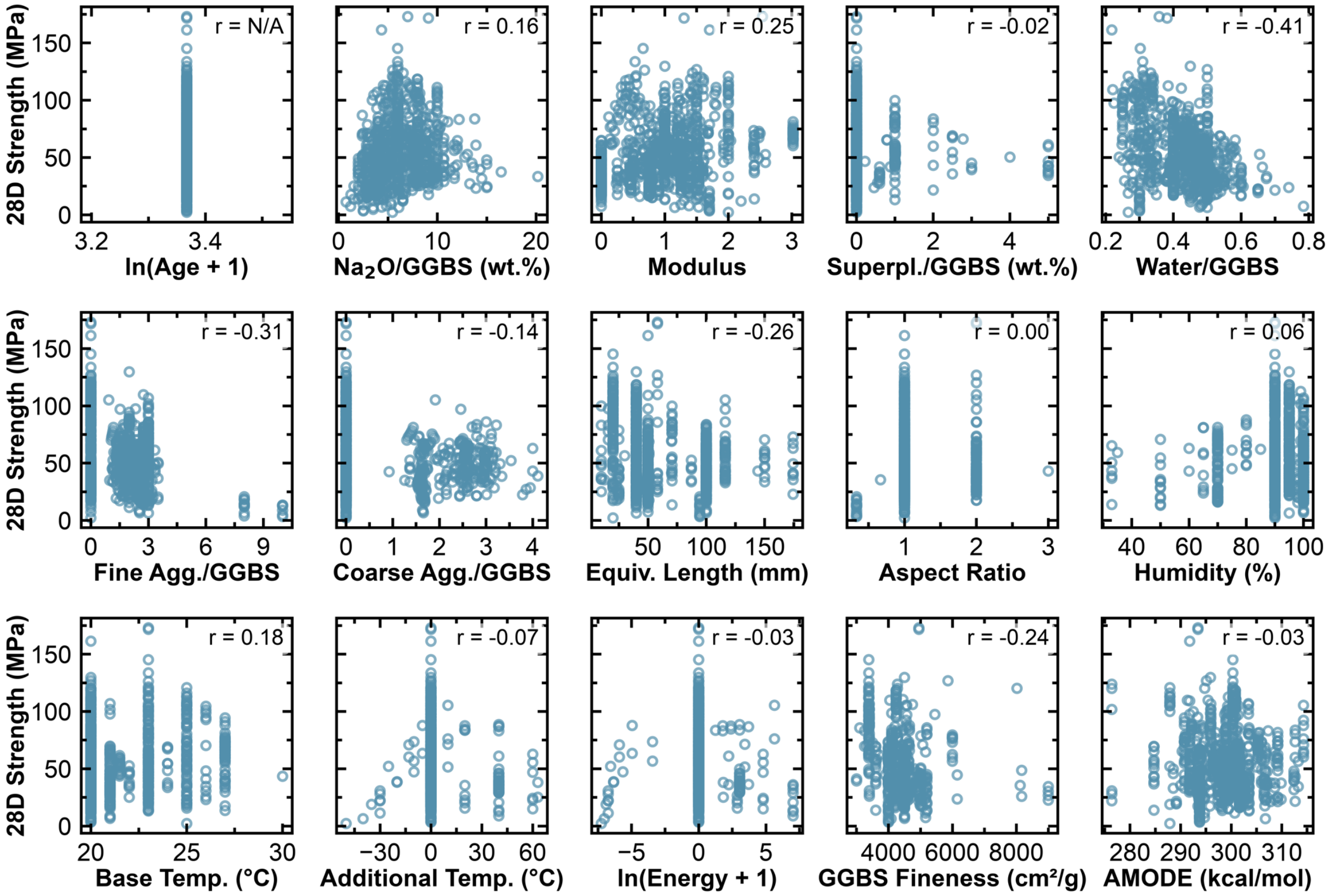

**Figure S15.** Scatter plots illustrating the relationships between 28-day compressive strength and individual input features in the curated dataset. Only mixtures with reported 28-day strength are included. The Pearson correlation coefficient (*r*) for each feature–strength pair is indicated in the upper-right corner of each subfigure. N/A indicates undefined correlation because curing age is constant.

## S2.2 Pearson Correlation Analysis of All Features in the Curated Dataset

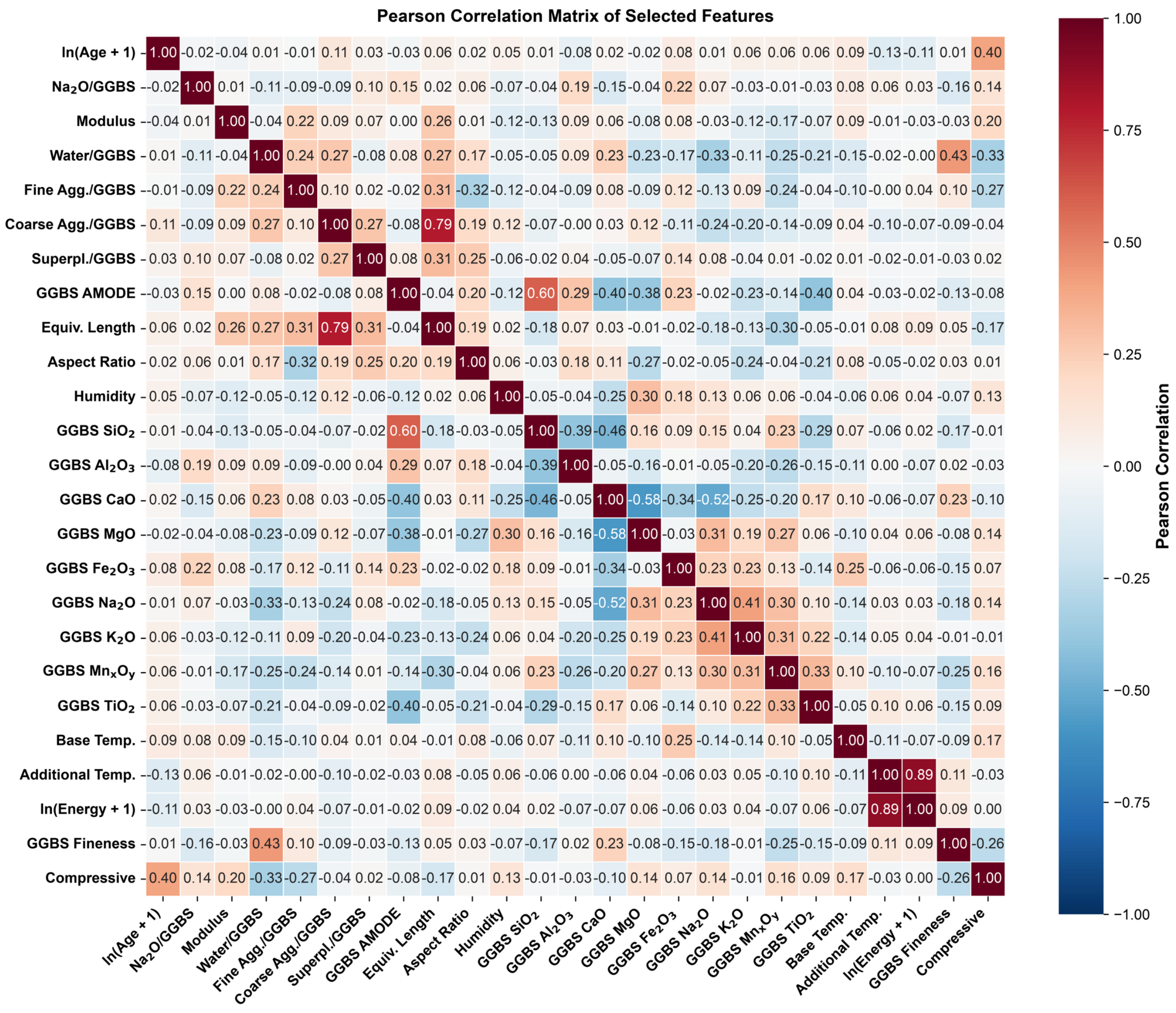

**Figure S16.** Pearson correlation heatmap for all input features and the compressive strength target in the curated dataset. Color denotes the Pearson correlation coefficient, where values close to ±1 indicate strong linear correlation, values around ±0.5 indicate moderate correlation, and values near 0 indicate weak or negligible correlation. Positive and negative correlations are shown by warm and cool colors, respectively.

## S3 Model Predictive Performance and Input-Representation Limits

### S3.1 Duplicate-Input Groups and Empirical RMSE Floors

Scenarios 1 and 2 use reduced input sets, so records with identical inputs may have different measured compressive strengths. To quantify this variation, duplicate-input groups were identified as groups containing at least two records with identical values for all scenario-specific inputs, without rounding. All 3053 records were included.

For a deterministic model, identical inputs must receive the same prediction. Within each group, the mean measured strength minimizes the sum of squared errors. The empirical RMSE floor was therefore calculated as:

$$RMSE_{floor} = \sqrt{\frac{1}{N}\sum_{g=1}^{G}\sum_{i\in g}(y_i - \bar{y}_g)^2}$$

where $N$ is the total number of records, $G$ is the number of distinct input groups, $y_i$ is the measured compressive strength of record $i$, and $\bar{y}_g$ is the mean measured strength within group $g$. Singleton groups contribute zero squared error.

Scenario 1 contained 275 duplicate-input groups involving 820 records (26.86%), whereas Scenario 2 contained 26 groups involving 78 records (2.55%). The corresponding empirical RMSE floors were 6.649 and 0.927 MPa (**Table S12**). The lower value in Scenario 2 indicates that its additional inputs distinguish records that are indistinguishable under Scenario 1. The remaining variation may reflect omitted variables, experimental variability, and other unrecorded factors. These empirical floors apply to the current dataset and input representations; they are not lower bounds on held-out test RMSE.

**Table S12.** Duplicate-input groups and empirical root-mean-square error (RMSE) floors for Scenarios 1 and 2.

| Measure | Scenario 1 | Scenario 2 |
|---|---|---|
| Number of features | 7 | 14 |
| Total records | 3053 | 3053 |
| Duplicate-input groups | 275 | 26 |
| Records in duplicate-input groups, n (%) | 820 (26.86%) | 78 (2.55%) |
| Duplicate-input groups with different strengths | 272 | 26 |
| Empirical RMSE floor (MPa) | 6.649 | 0.927 |

### S3.2 Complete Model Predictive Performances

**Table S13.** Summary of the models' predictive performances. Values are means (standard deviations) across 10 random states.

| Model | Scenario | Train | | Test | | Extent of Overfitting | |
|---|---|---|---|---|---|---|---|
| | | $R^2$ | RMSE (MPa) | $R^2$ | RMSE (MPa) | $\Delta R^2$ (%) | ΔRMSE (%) |
| GPR | 1 | 0.871(0.008) | 10.487(0.314) | 0.753(0.033) | 14.343(1.023) | 13.487 | 36.773 |
| | 2 | 0.959(0.001) | 5.897(0.077) | 0.922(0.009) | 8.091(0.442) | 3.926 | 37.223 |
| | 3 | 0.967(0.001) | 5.300(0.048) | 0.940(0.006) | 7.093(0.348) | 2.820 | 33.832 |
| | 4 | 0.965(0.001) | 5.481(0.046) | 0.932(0.007) | 7.530(0.380) | 3.387 | 37.396 |
| ANN | 1 | 0.849(0.043) | 11.273(1.461) | 0.747(0.033) | 14.533(0.951) | 11.998 | 28.911 |
| | 2 | 0.970(0.004) | 5.027(0.382) | 0.921(0.009) | 8.093(0.467) | 5.022 | 61.005 |
| | 3 | 0.977(0.003) | 4.451(0.327) | 0.939(0.008) | 7.140(0.444) | 3.868 | 60.397 |
| | 4 | 0.970(0.008) | 5.031(0.639) | 0.928(0.012) | 7.747(0.640) | 4.333 | 53.963 |
| XGB | 1 | 0.917(0.010) | 8.420(0.483) | 0.837(0.024) | 11.637(0.880) | 8.637 | 38.204 |
| | 2 | 0.971(0.006) | 4.948(0.536) | 0.928(0.010) | 7.733(0.506) | 4.401 | 56.280 |
| | 3 | 0.977(0.005) | 4.382(0.498) | 0.937(0.011) | 7.221(0.562) | 4.080 | 64.776 |
| | 4 | 0.975(0.005) | 4.628(0.496) | 0.935(0.010) | 7.379(0.498) | 4.099 | 59.454 |
| RF | 1 | 0.903(0.006) | 9.087(0.264) | 0.785(0.038) | 13.387(1.199) | 13.122 | 47.316 |
| | 2 | 0.958(0.001) | 5.983(0.091) | 0.888(0.013) | 9.685(0.573) | 7.347 | 61.868 |
| | 3 | 0.954(0.001) | 6.281(0.078) | 0.888(0.012) | 9.655(0.504) | 6.854 | 53.721 |
| | 4 | 0.961(0.001) | 5.784(0.081) | 0.895(0.012) | 9.345(0.527) | 6.804 | 61.559 |
| LR | 1 | 0.391(0.013) | 22.778(0.260) | 0.386(0.055) | 22.652(1.028) | 1.272 | 0.553 |
| | 2 | 0.467(0.012) | 21.302(0.258) | 0.467(0.050) | 21.114(1.023) | 0.177 | 0.884 |
| | 3 | 0.478(0.013) | 21.084(0.267) | 0.470(0.053) | 21.042(1.061) | 1.707 | 0.199 |
| | 4 | 0.472(0.013) | 21.210(0.268) | 0.469(0.053) | 21.061(1.063) | 0.607 | 0.703 |

## S4 Extended Model Interpretation Analysis and Discussion

### S4.1 Sensitivity Analysis of KernelSHAP Background Size

In KernelSHAP, background samples define the reference distribution used to approximate model predictions when input features are masked, thereby providing the baseline against which feature contributions are quantified. The choice of background size (BG) can influence the stability of SHAP attributions, while larger BG values incur substantially higher computational costs, particularly for GPR models. To determine an appropriate background size for all subsequent analyses, a sensitivity study was performed by varying BG = 50, 100, 150, and 200 while keeping the model architecture, SHAP sampling budget, and data-splitting protocol fixed. For each BG, SHAP values were computed on the held-out test set across ten random train–test splits, and global feature importance was quantified using the mean absolute SHAP value. As shown in **Figure S17**, all four feature scenarios exhibit very similar SHAP magnitudes and feature rankings across the tested BG values, indicating rapid convergence of global attributions. In particular, the negligible differences observed between BG = 150 and BG = 200 across all scenarios indicate that BG = 200 provides a conservative yet computationally tractable choice, and this background size is therefore adopted consistently in the main analysis.

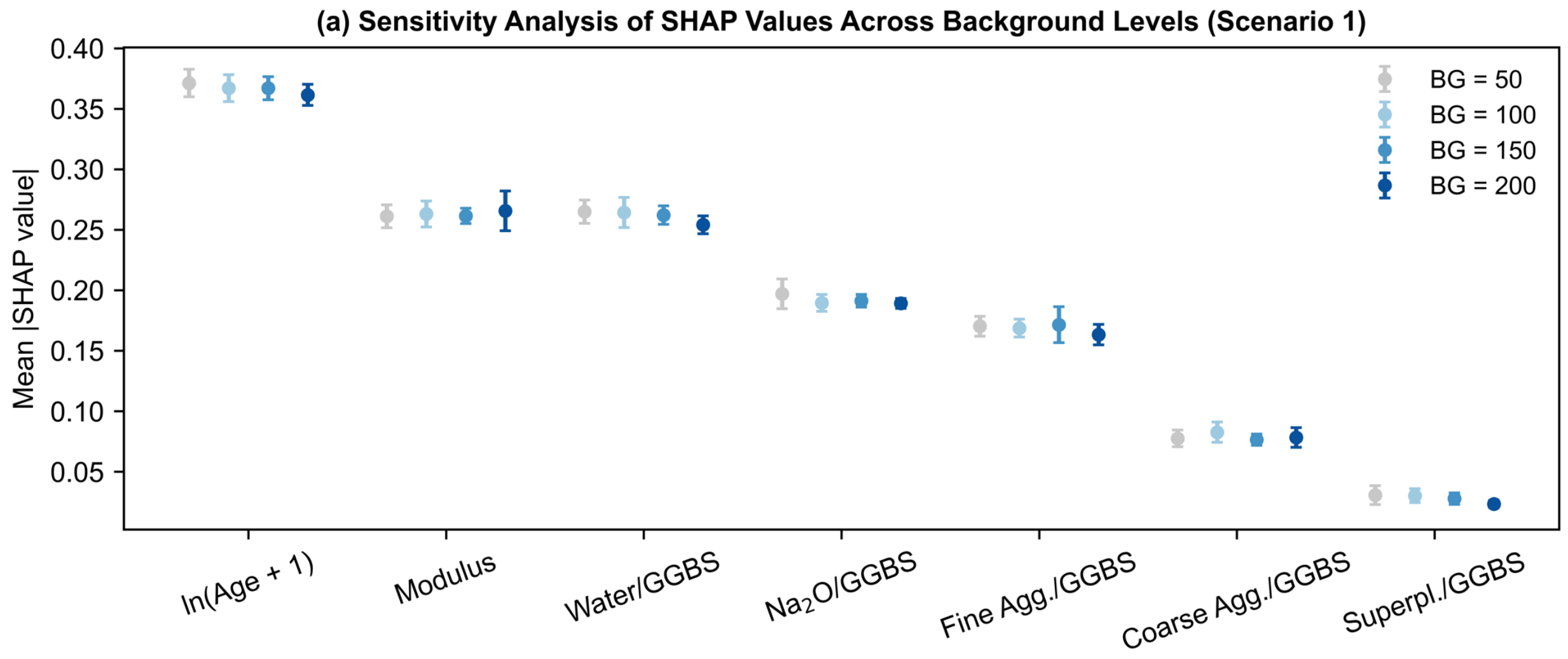

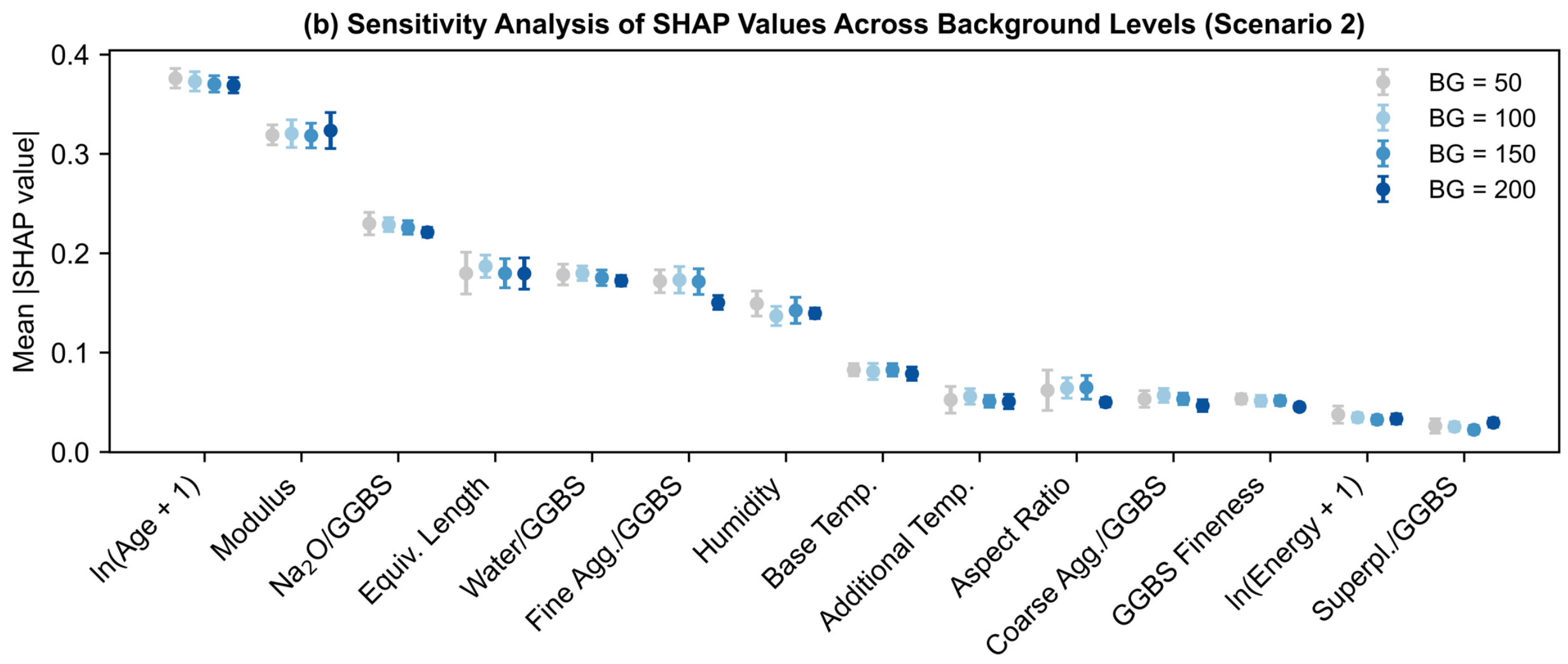
(b) Sensitivity Analysis of SHAP Values Across Background Levels (Scenario 2)
Mean |SHAP value|
BG = 50
BG = 100
BG = 150
BG = 200
ln(Age + 1)
Modulus
Na2O/GGBS
Equiv. Length
Water/GGBS
Fine Agg./GGBS
Humidity
Base Temp.
Additional Temp.
Aspect Ratio
Coarse Agg./GGBS
GGBS Fineness
ln(Energy + 1)
Superpl./GGBS

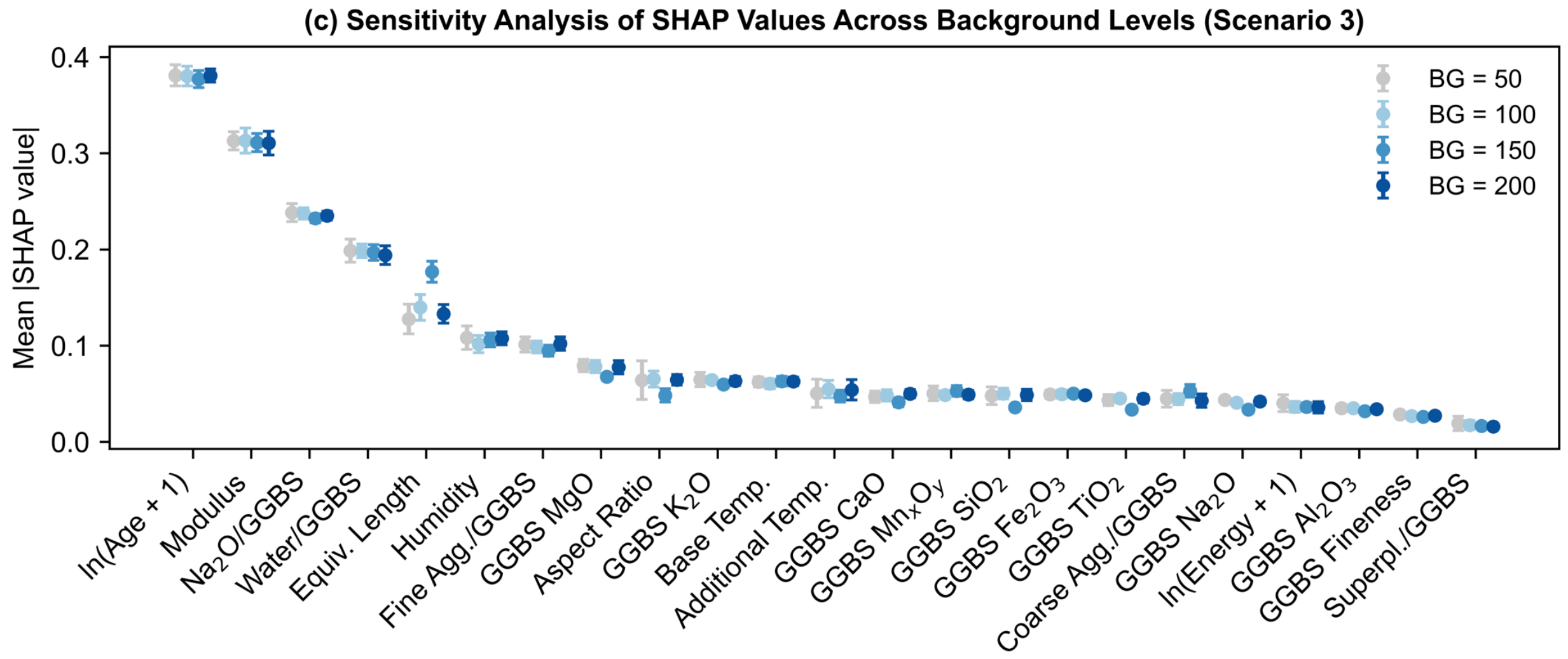
(c) Sensitivity Analysis of SHAP Values Across Background Levels (Scenario 3)
Mean |SHAP value|
BG = 50
BG = 100
BG = 150
BG = 200
ln(Age + 1)
Modulus
Na2O/GGBS
Water/GGBS
Equiv. Length
Humidity
Fine Agg./GGBS
GGBS MgO
Aspect Ratio
GGBS K2O
Base Temp.
Additional Temp.
GGBS CaO
GGBS MnxOy
GGBS SiO2
GGBS Fe2O3
GGBS TiO2
Coarse Agg./GGBS
GGBS Na2O
ln(Energy + 1)
GGBS Al2O3
GGBS Fineness
Superpl./GGBS

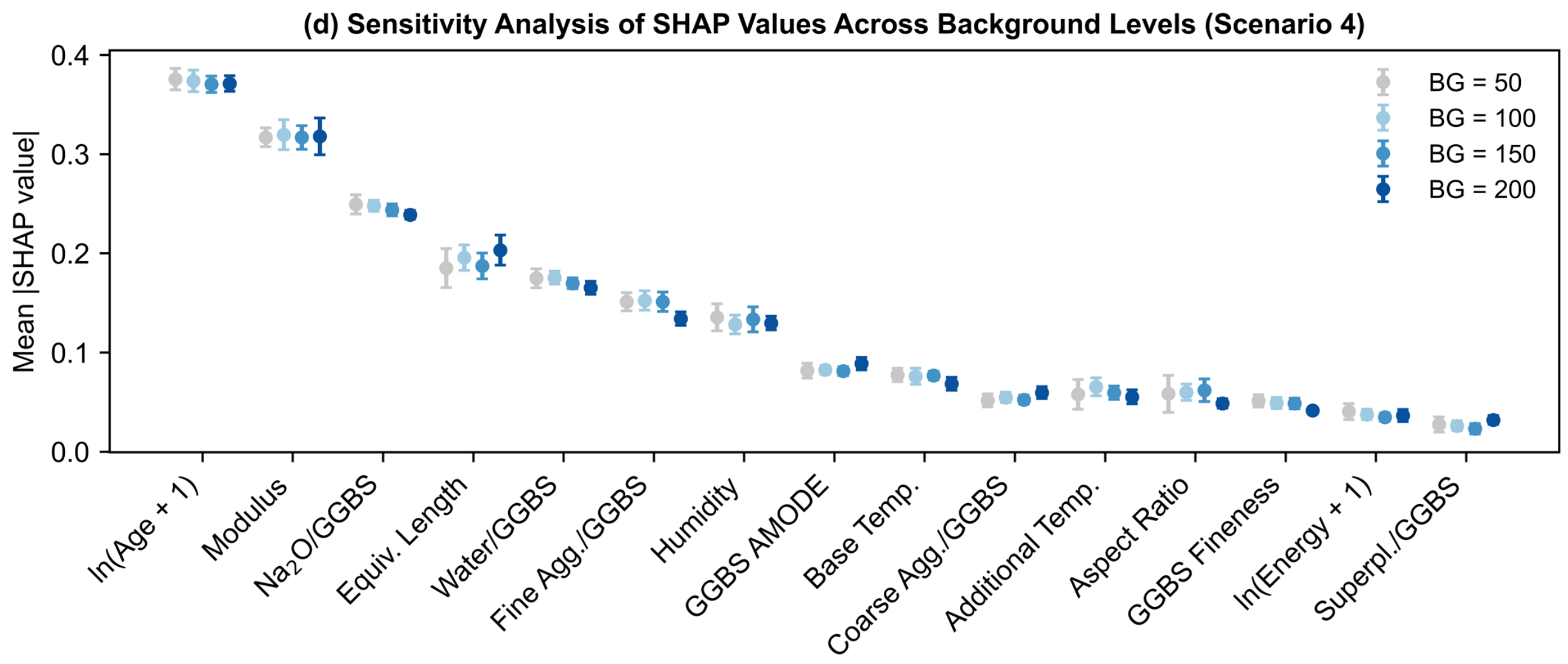

**Figure S17.** Sensitivity analysis of KernelSHAP feature attributions for the GPR model with respect to background size (BG) for: (a) Scenario 1, Baseline Mix Design and Age (7 features), (b) Scenario 2, Physical Property, Curing, and Geometry (14 features), (c) Scenario 3, Explicit Chemical Composition (23 features), and (d) Scenario 4, Physics-Informed Reactivity (15 features). For each BG, markers denote the mean absolute SHAP value averaged across ten random states, and error bars represent one standard deviation. The close agreement of SHAP magnitudes and feature rankings across increasing BG indicates convergence of global attributions by BG = 200.

## S4.2 SHAP Analysis of ANN, XGB, and RF Models

To examine the robustness of the interpretability results across different models, SHAP analysis was performed for ANN, XGB, and RF under all feature scenarios. SHAP summary plots for each model and scenario are presented in **Figures S18–S20**.

Across all models, the most influential features are largely consistent, with curing age (ln(Age + 1)) consistently identified as the dominant contributor to compressive strength. Activator chemistry ($Na_2O$/GGBS and silicate modulus, Ms = $SiO_2/Na_2O$), key mix-design ratios (water/GGBS and fine aggregate/GGBS ratios), equivalent length, and relative humidity also rank among the most important features, although their exact ordering varies slightly across models and scenarios. The directional effects of these features are consistent with those observed for the GPR model in the main text: curing age, $Na_2O$/GGBS, and Ms generally contribute positively to predicted strength, whereas water/GGBS ratio, equivalent length, and fine aggregate/GGBS ratio exhibit negative contributions.

For scenarios involving explicit GGBS chemical compositions (Scenario 3), the SHAP results for ANN, XGB, and RF exhibit trends that are broadly consistent with those of GPR. In particular, CaO and MgO tend to exhibit positive contributions, while $SiO_2$ shows a negative contribution across models. Minor differences in feature ranking and attribution magnitude are observed, especially for RF, reflecting model-specific sensitivity to correlated inputs. Overall, the SHAP results confirm that the primary drivers of strength identified in the main text are robust across model classes. In Scenario 4, the modified AMODE index shows consistent predictive relevance across models with negative contributions, consistent with the GPR results, confirming its robustness as a compact and physically meaningful descriptor of GGBS reactivity.

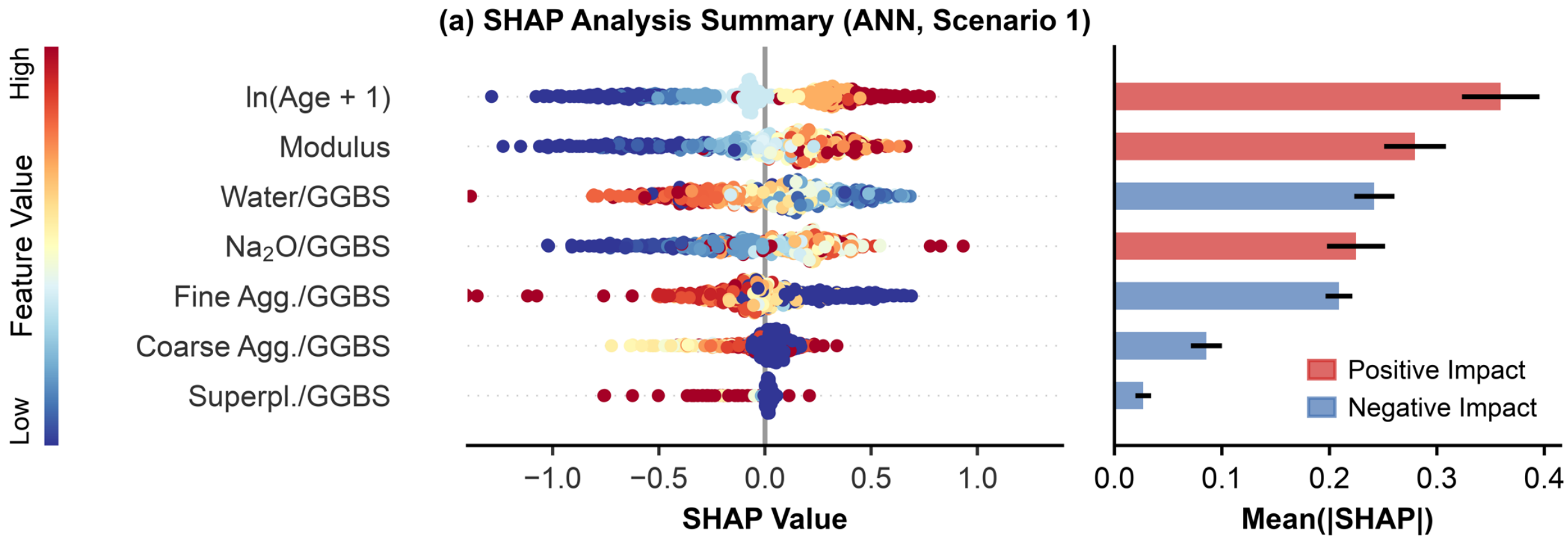

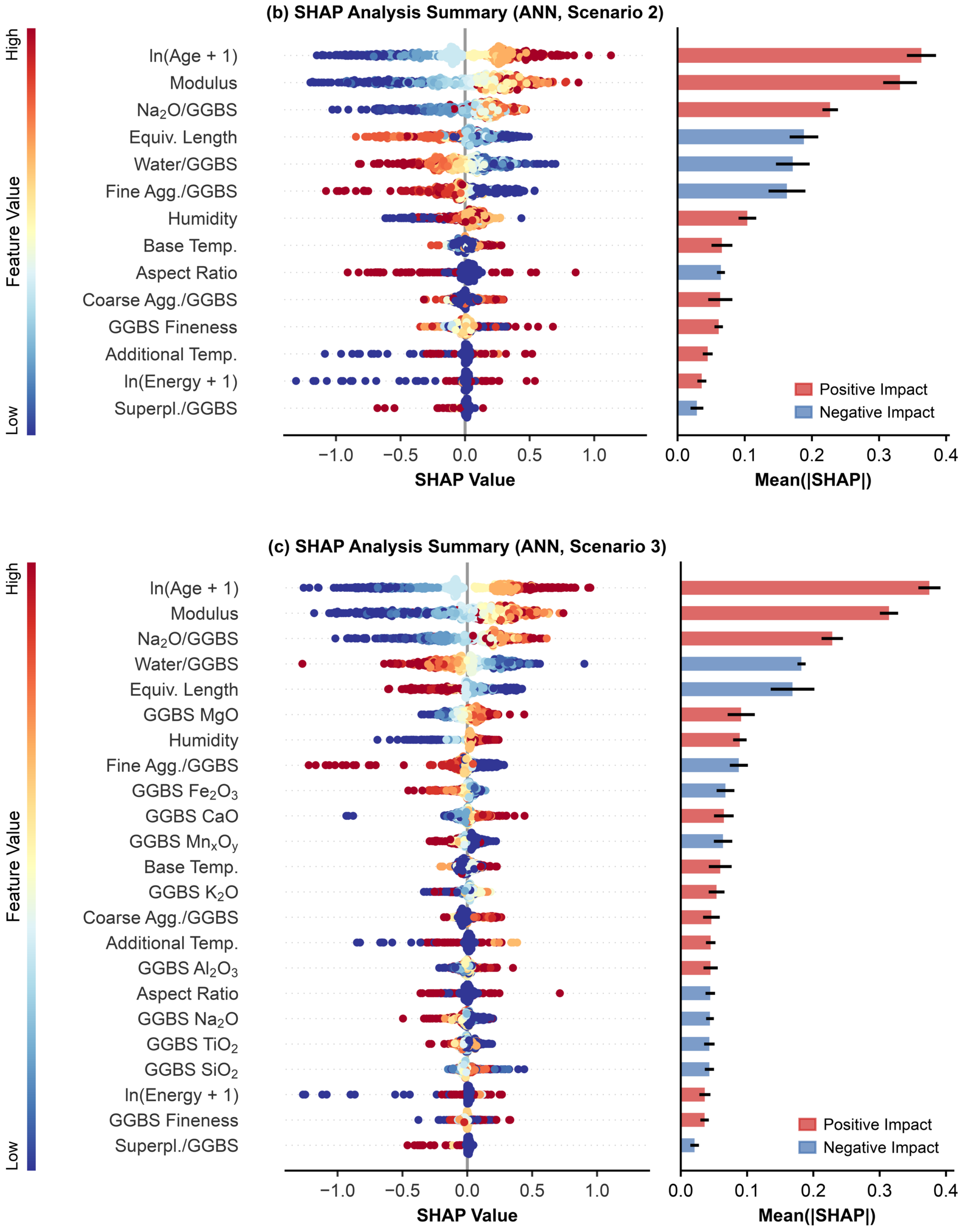

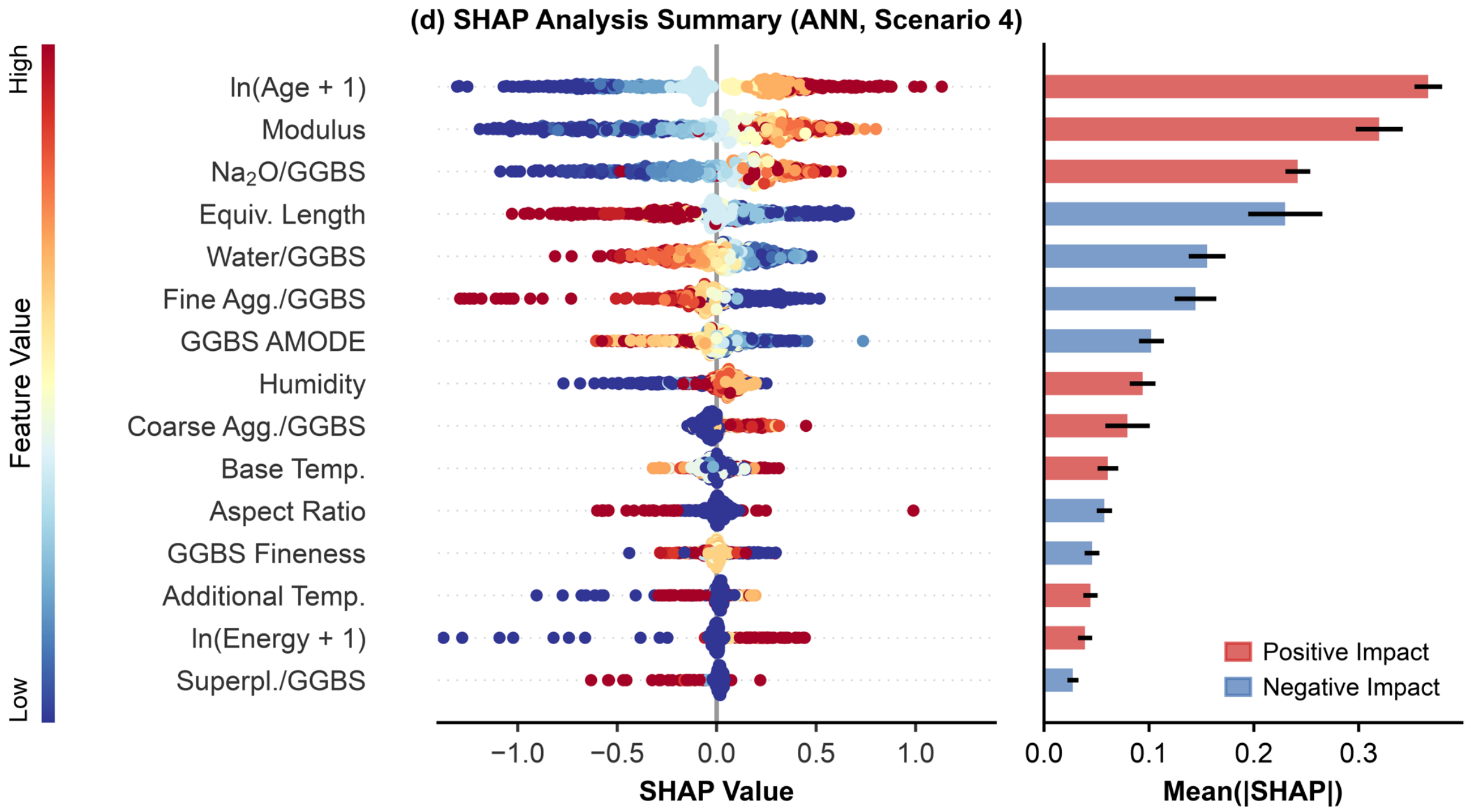

**Figure S18.** Summary of SHAP analysis for the artificial neural network (ANN) model under (a) Scenario 1, Baseline Mix Design and Age (7 features), (b) Scenario 2, Physical Property, Curing, and Geometry (14 features), (c) Scenario 3, Explicit Chemical Composition (23 features), and (d) Scenario 4, Physics-Informed Reactivity (15 features). All input features included in each model are shown. The left panels present representative SHAP summary plots computed from the test set, while the right panels display the corresponding mean absolute SHAP values averaged across 10 random states. Error bars denote one standard deviation across runs.

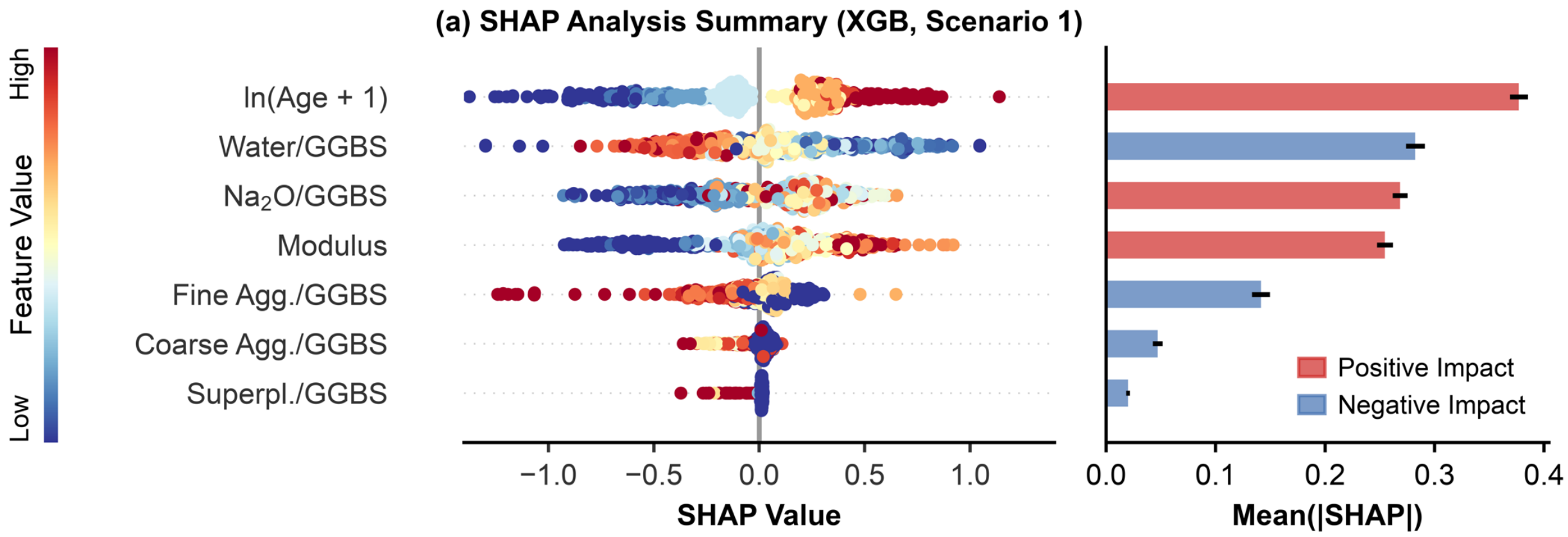

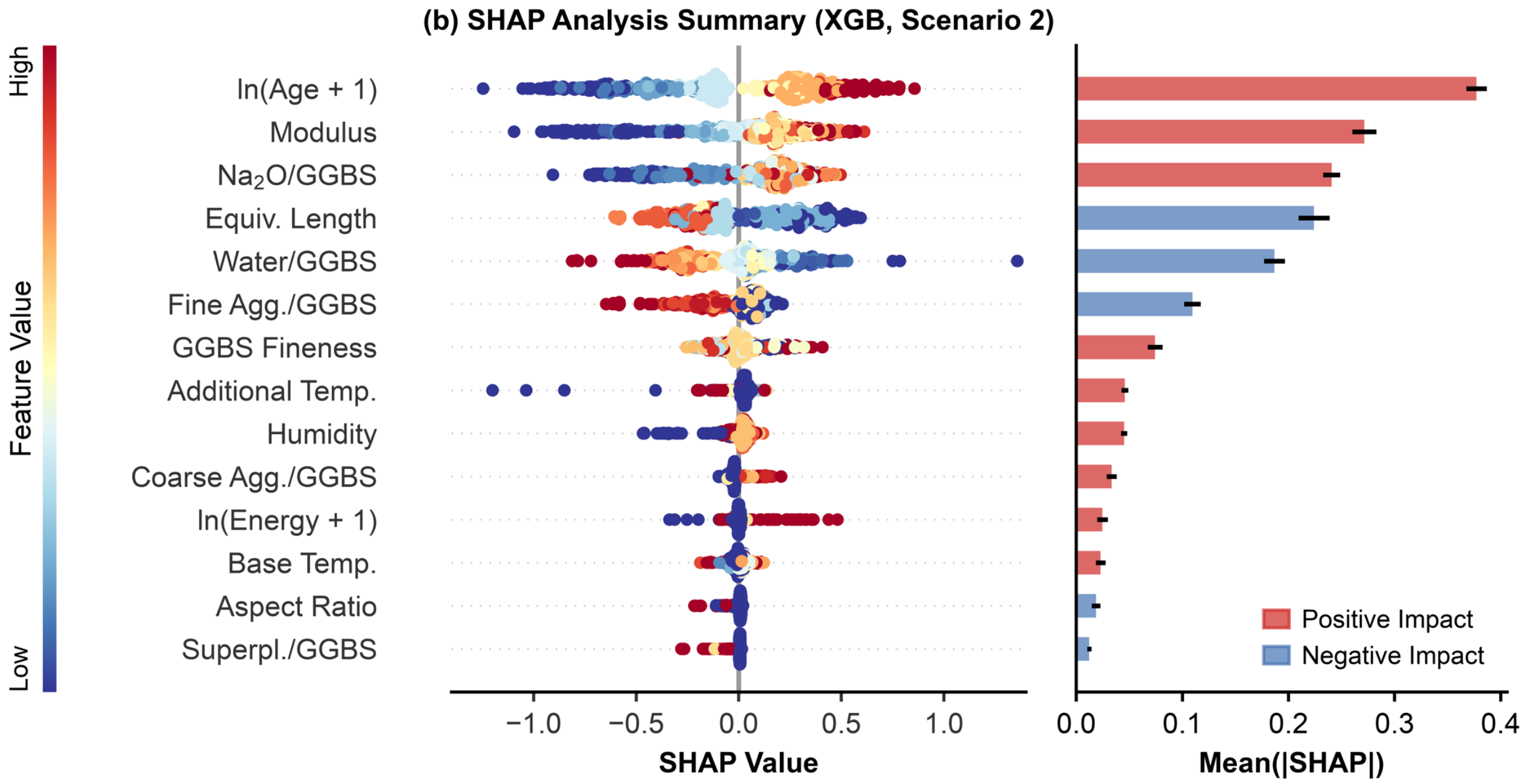

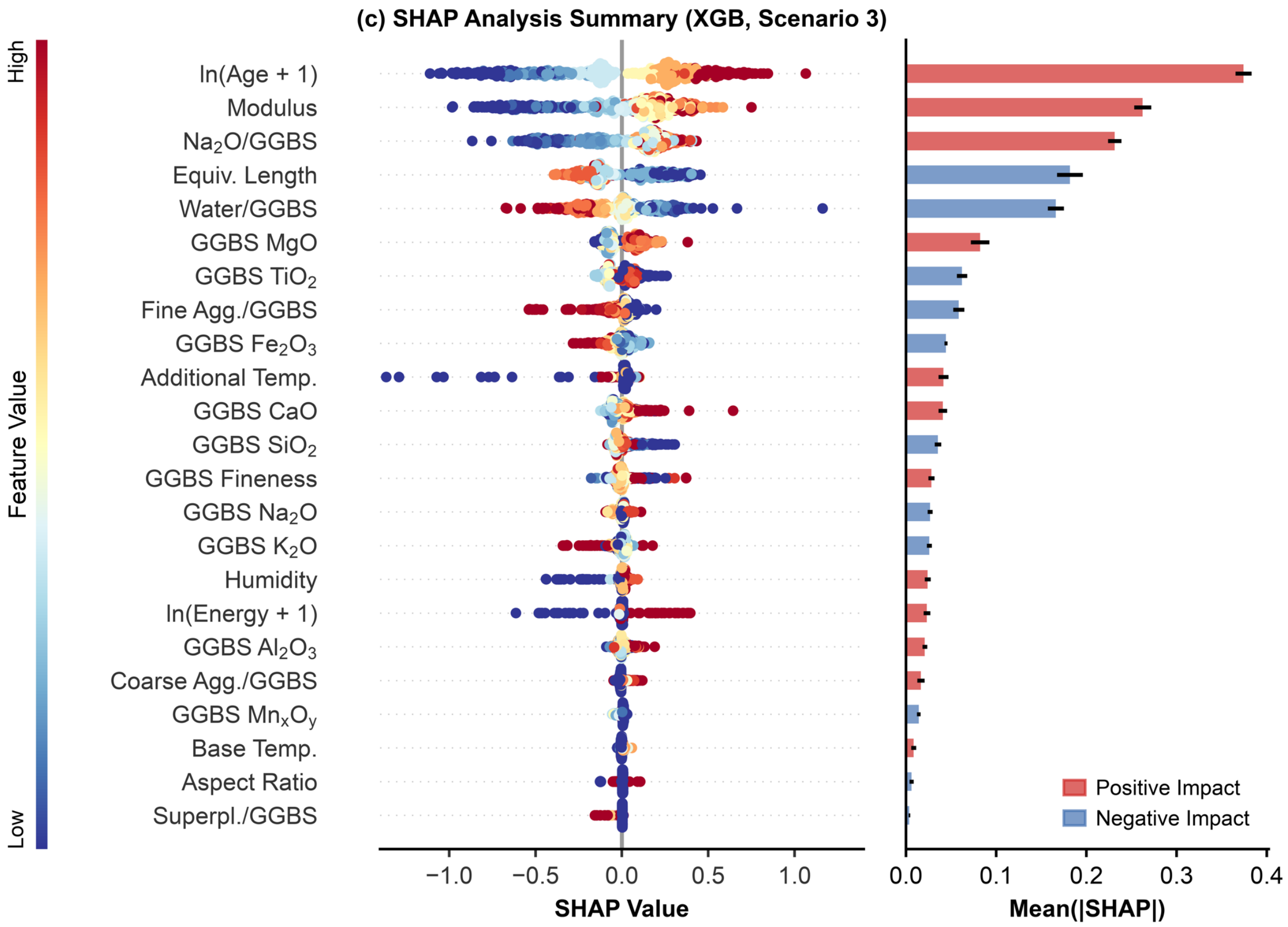

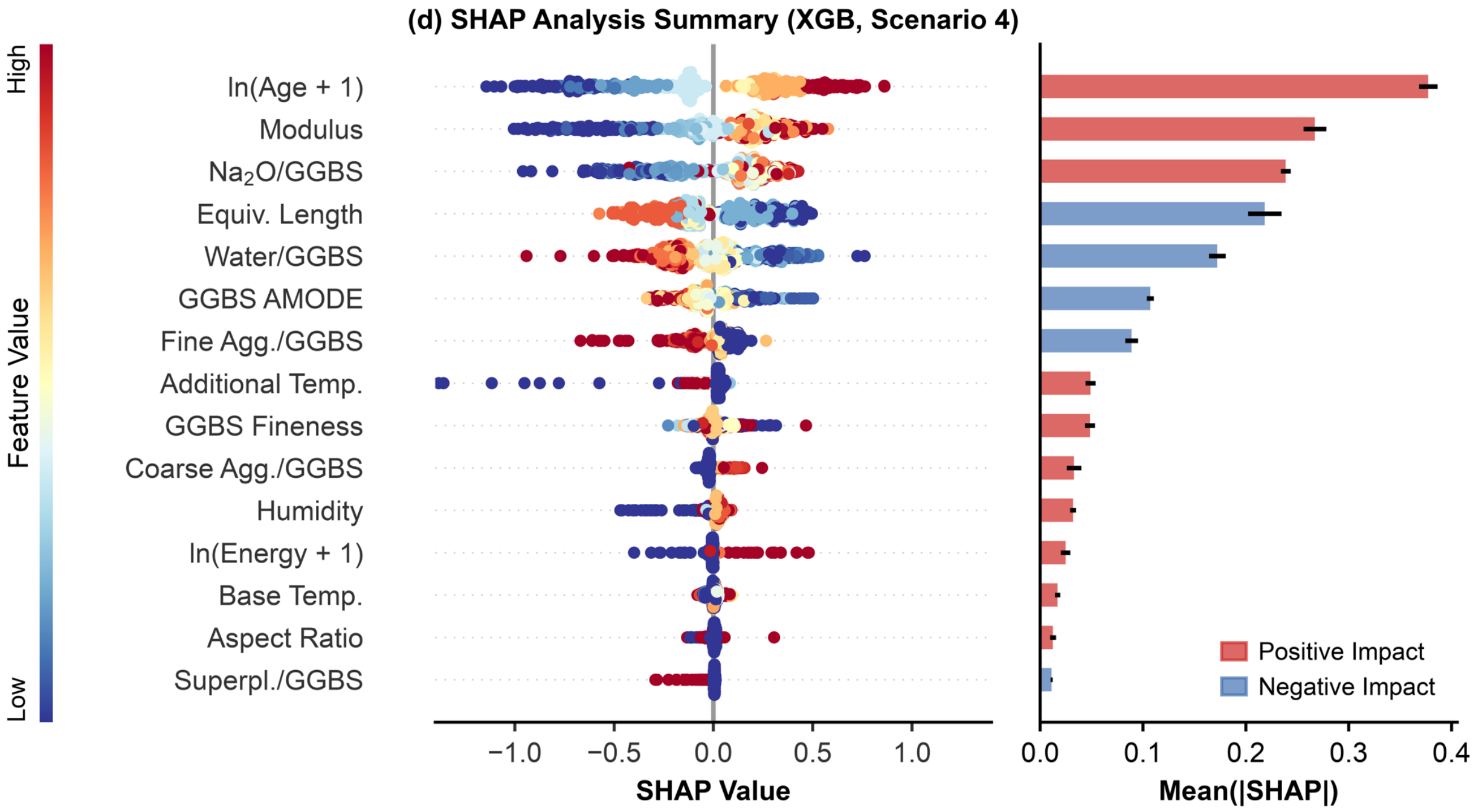

**Figure S19.** Summary of SHAP analysis for the extreme gradient boosting (XGB) model under (a) Scenario 1, Baseline Mix Design and Age (7 features), (b) Scenario 2, Physical Property, Curing, and Geometry (14 features), (c) Scenario 3, Explicit Chemical Composition (23 features), and (d) Scenario 4, Physics-Informed Reactivity (15 features). All input features included in each model are shown. The left panels present representative SHAP summary plots computed from the test set, while the right panels display the corresponding mean absolute SHAP values averaged across 10 random states. Error bars denote one standard deviation across runs.

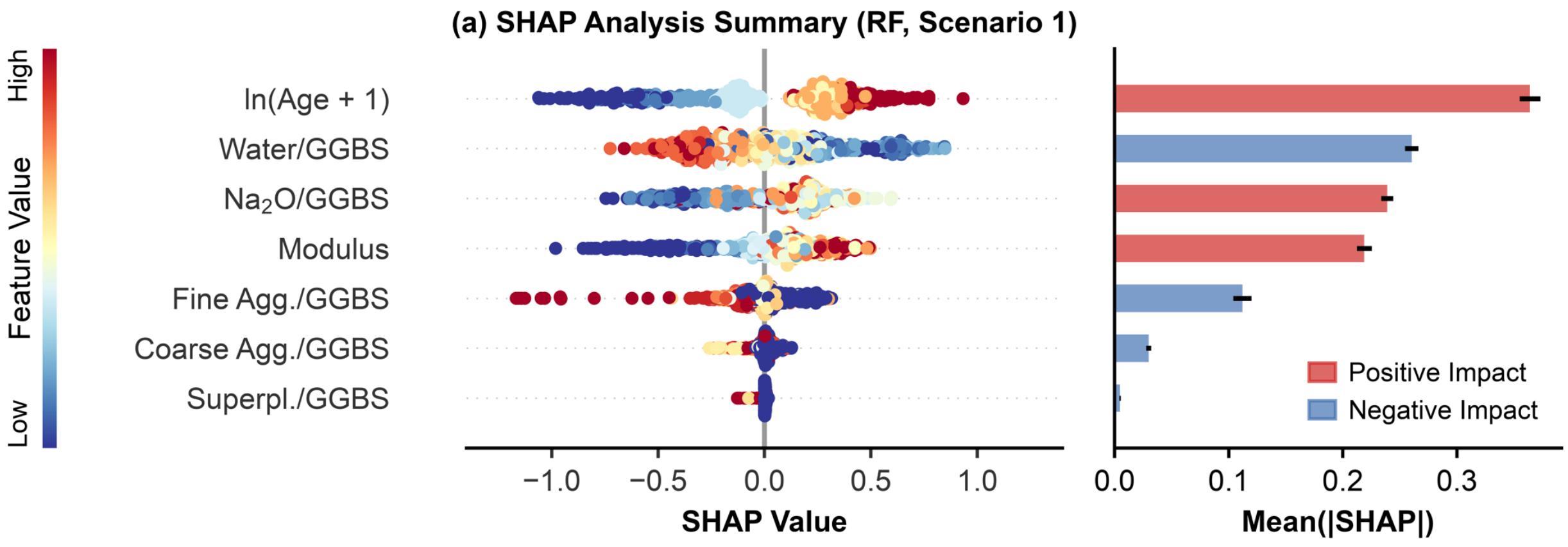

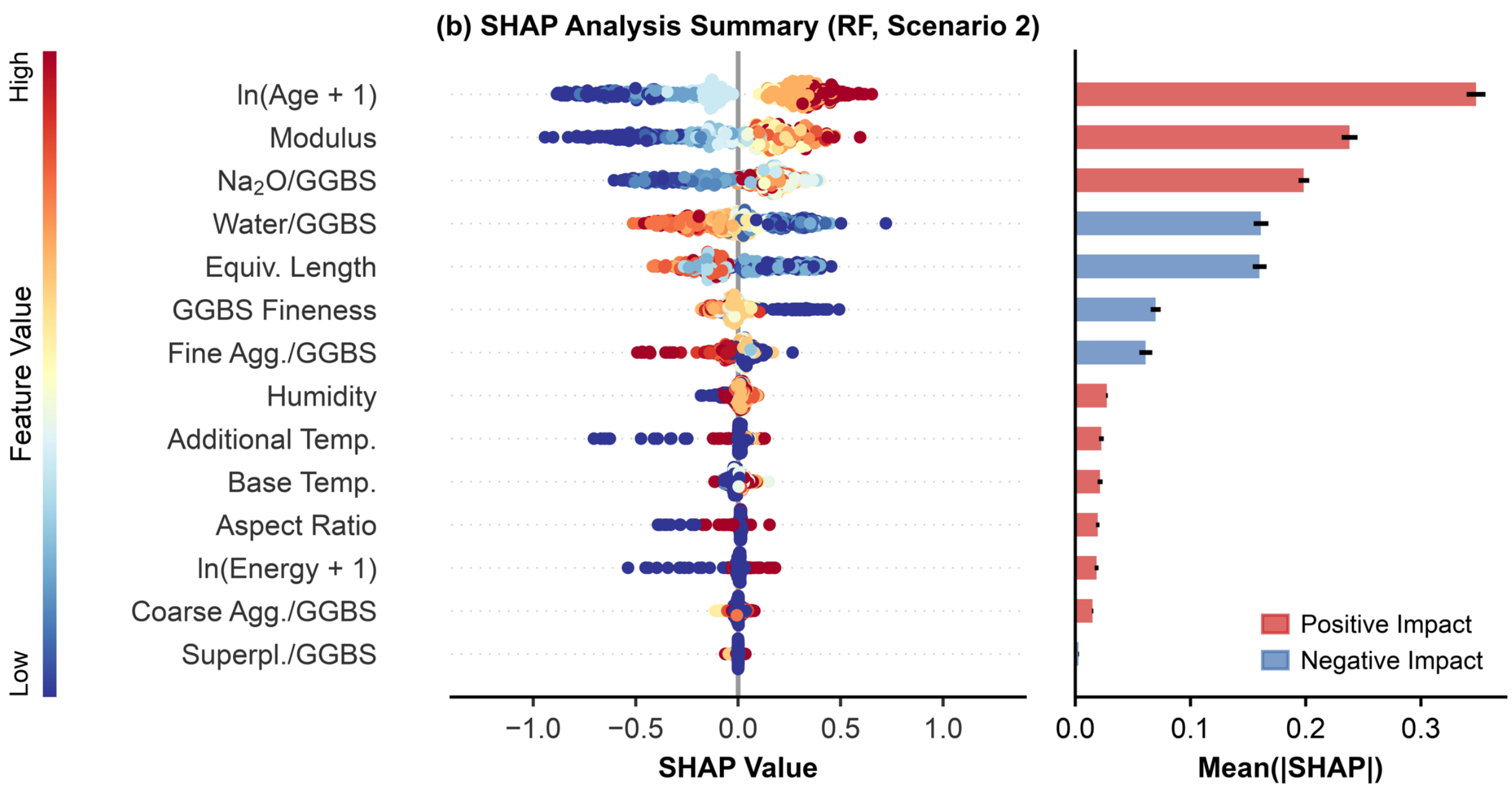

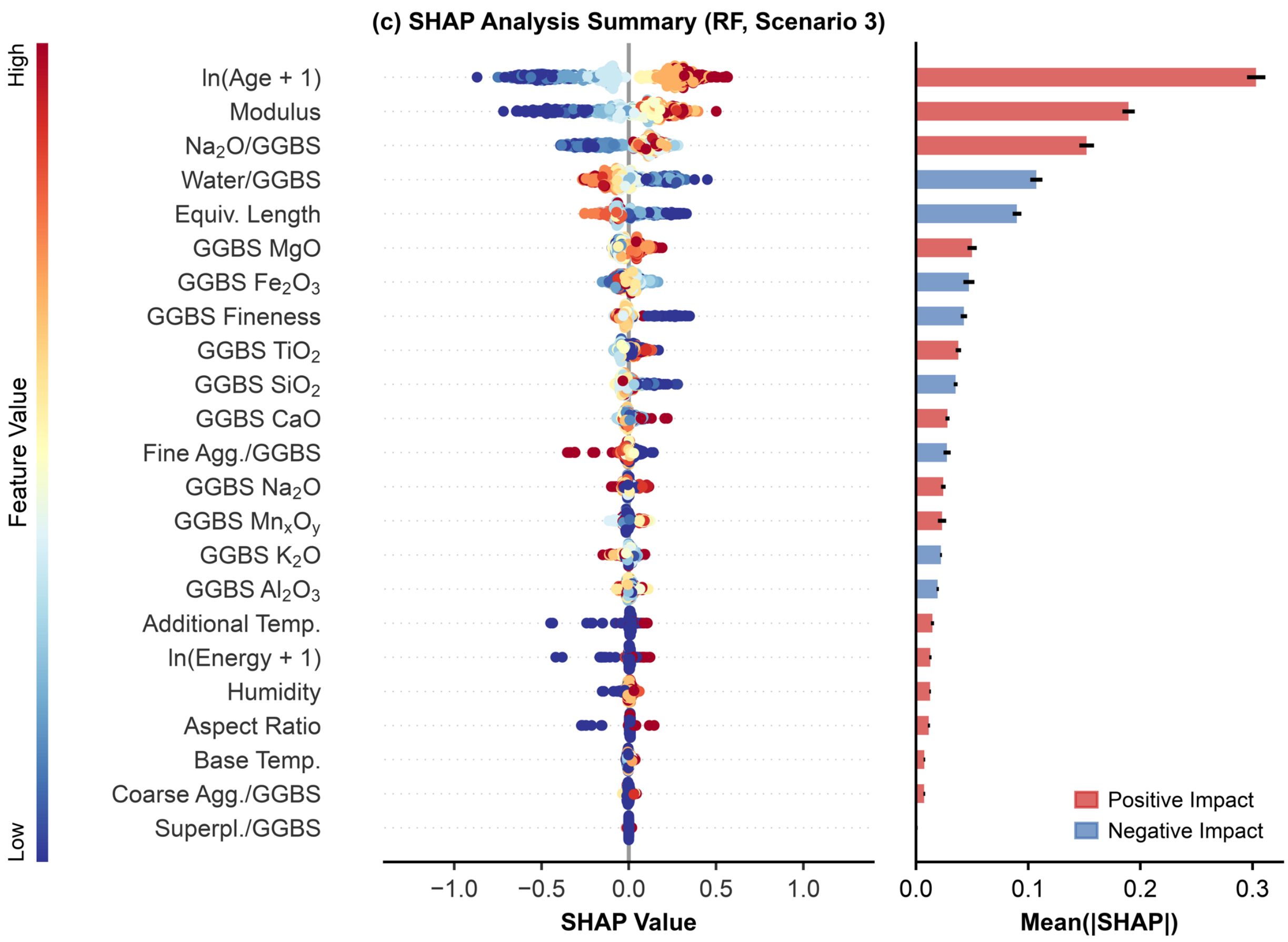

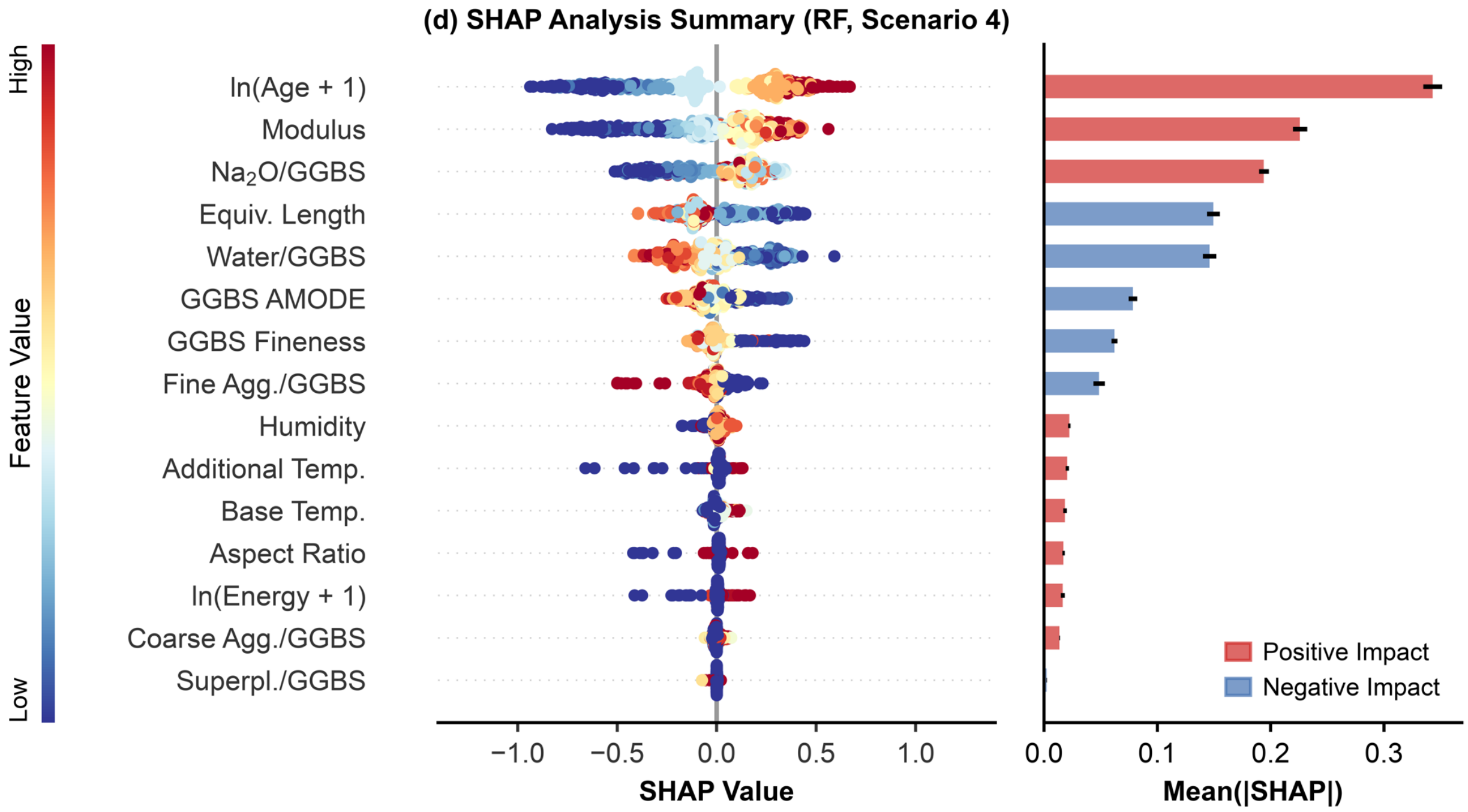

**Figure S20.** Summary of SHAP analysis for the random forest (RF) model under (a) Scenario 1, Baseline Mix Design and Age (7 features), (b) Scenario 2, Physical Property, Curing, and Geometry (14 features), (c) Scenario 3, Explicit Chemical Composition (23 features), and (d) Scenario 4, Physics-Informed Reactivity (15 features). All input features included in each model are shown. The left panels present representative SHAP summary plots computed from the test set, while the right panels display the corresponding mean absolute SHAP values averaged across 10 random states. Error bars denote one standard deviation across runs.

### S4.3 ALE Analysis Across Models

ALE analysis was conducted for the GPR, ANN, XGB, and RF models to further examine the directionality and potential nonlinearity of feature effects. ALE plots for all models and scenarios are shown in **Figures S21–S24**.

For dominant features, consistent qualitative trends are observed across models. Curing age shows an initial rapid increase in predicted strength, followed by gradual leveling off at later ages. Activator-related variables ($Na_2O$/GGBS and Ms) display nonlinear responses, with strength increasing over intermediate ranges and diminishing or plateauing at higher values. Water/GGBS and fine aggregate/GGBS ratios show predominantly negative associations with predicted strength. The modified AMODE generally shows a negative relationship with predicted strength, indicating higher strength at higher GGBS reactivity, although this trend is less pronounced for XGB and RF. These patterns are consistent with the SHAP-based importance rankings and with established experimental understanding.

By contrast, ALE responses for individual GGBS oxide contents are generally weak and relatively flat for the tree-based models (XGB and RF), with larger variability across models. This behavior likely reflects the narrow compositional ranges, feature correlations, and lack of systematic independent variation of individual oxides in the compiled literature dataset. As a result, oxide-level ALE trends are interpreted qualitatively and are discussed in greater detail in the follow-up section (**Section S4.5**).

Taken together, the ALE results complement the SHAP analysis by confirming the directionality and nonlinearity of the dominant features, while also highlighting the limited interpretability of weak and highly correlated inputs at the oxide level.

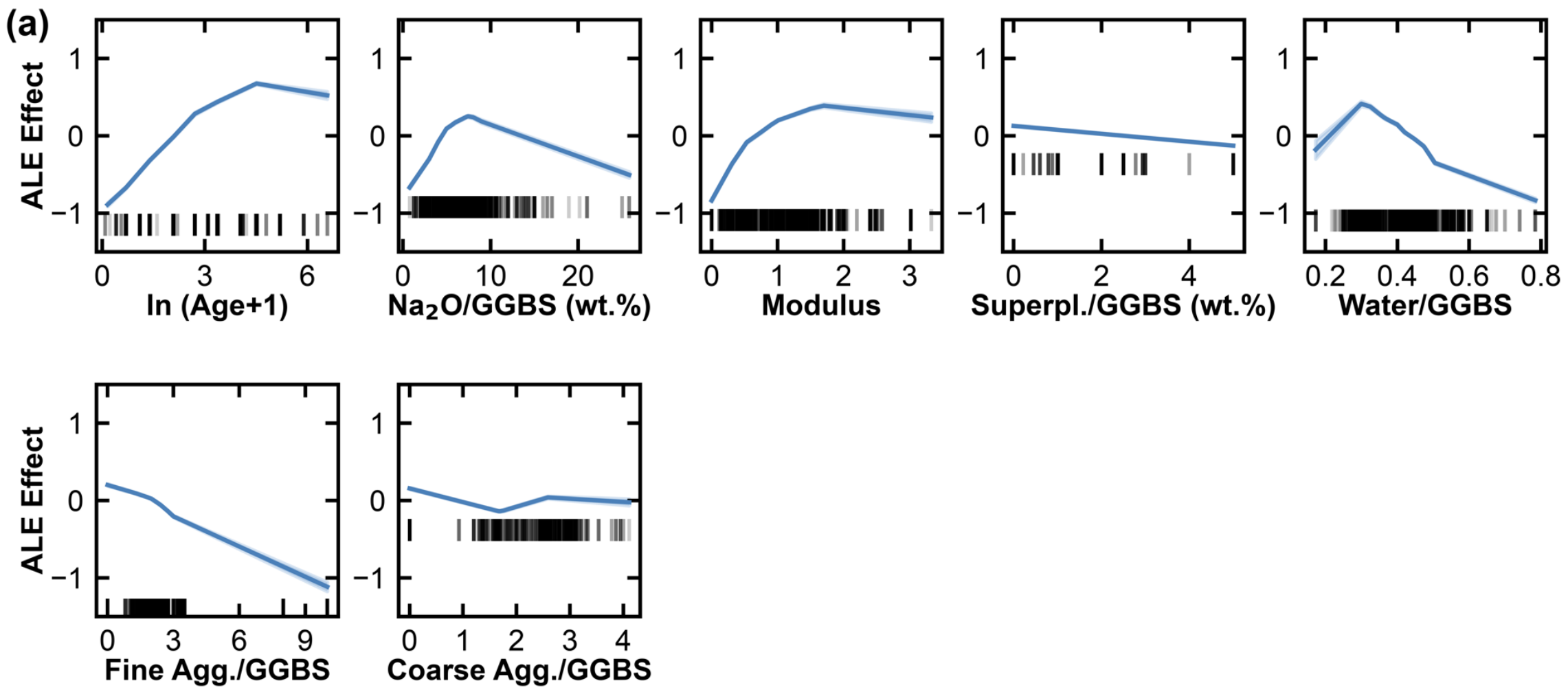

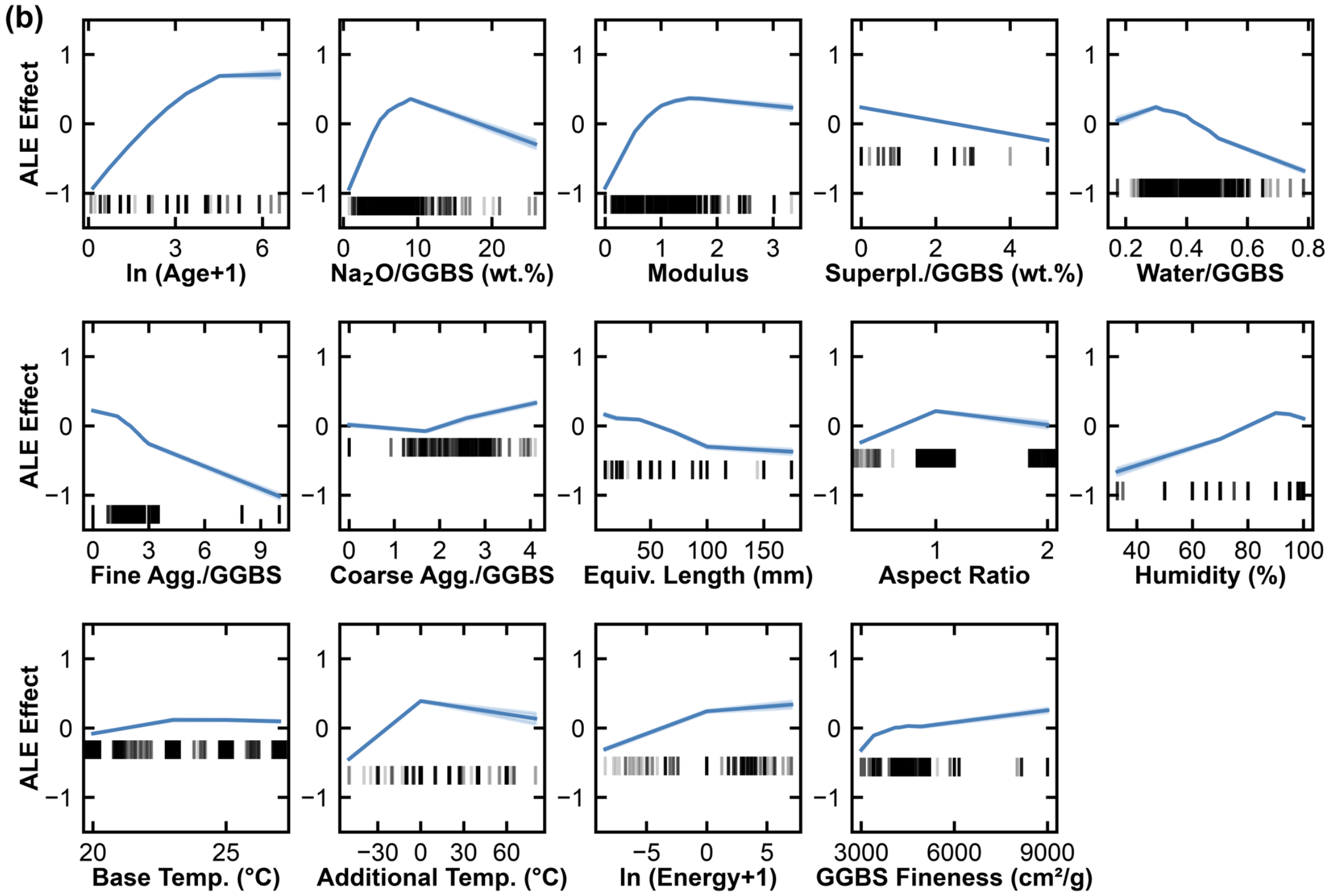

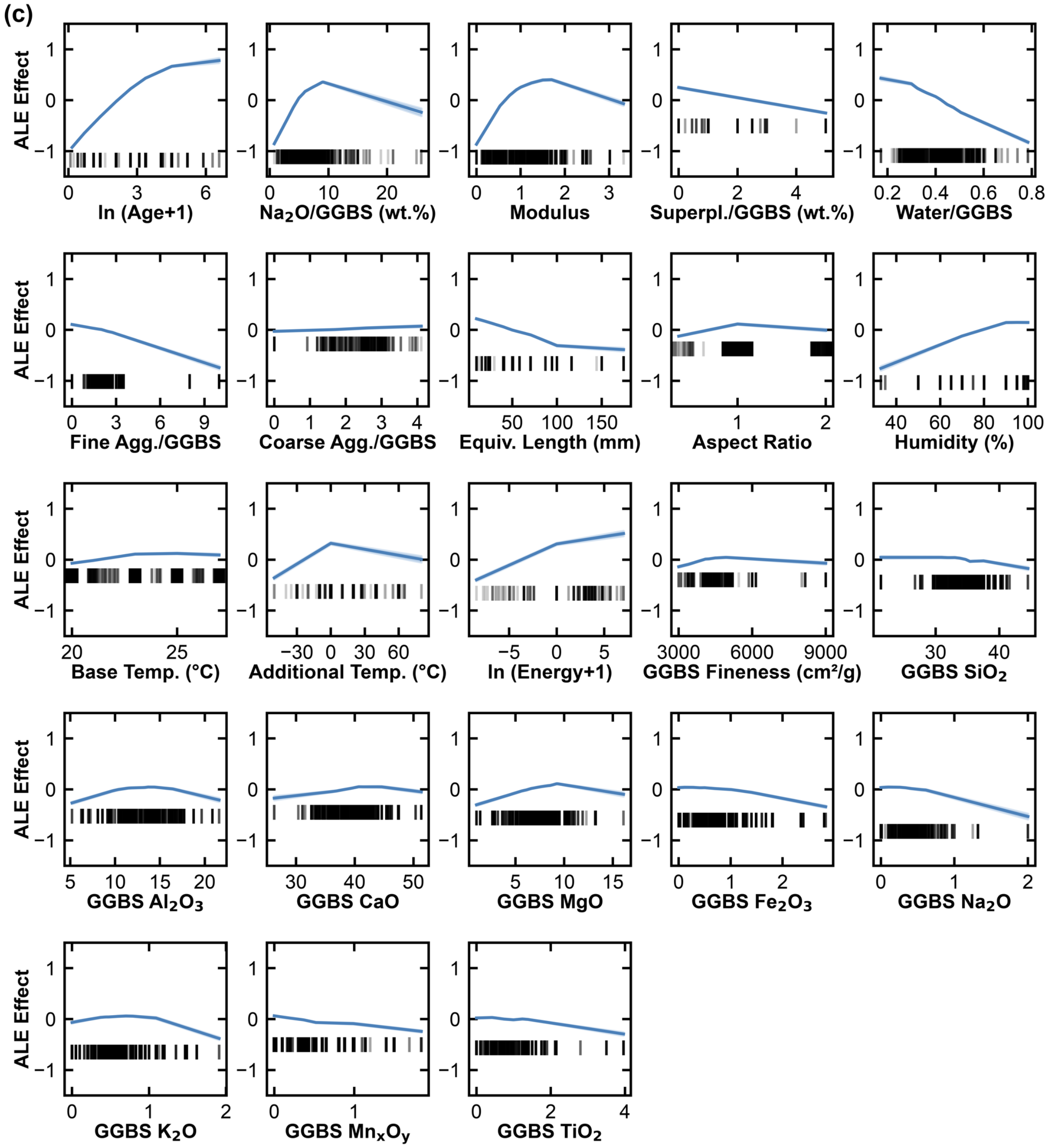

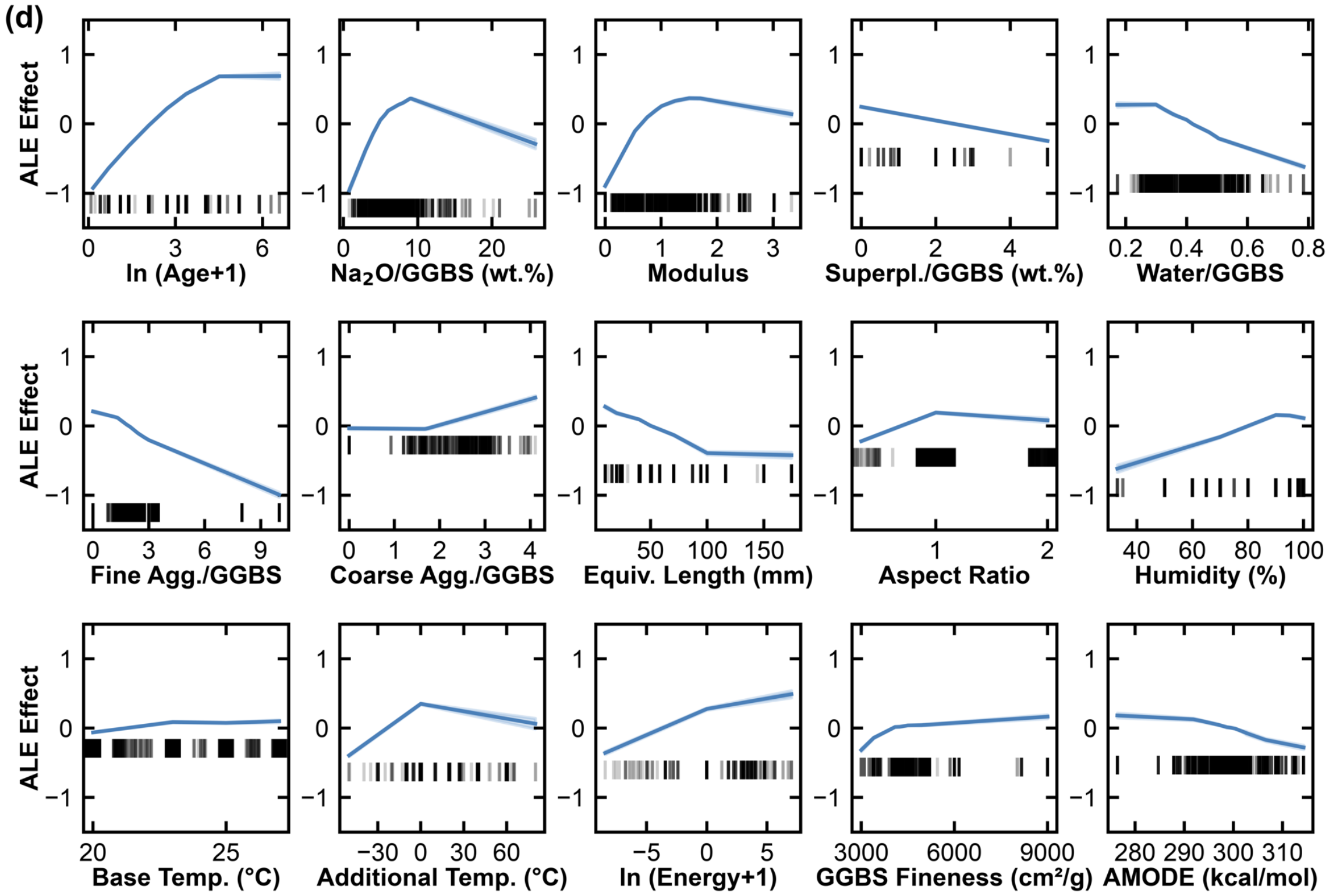

**Figure S21.** ALE analysis results of the GPR model under (a) Scenario 1, Baseline Mix Design and Age (7 features), (b) Scenario 2, Physical Property, Curing, and Geometry (14 features), (c) Scenario 3, Explicit Chemical Composition (23 features), and (d) Scenario 4, Physics-Informed Reactivity (15 features). All features used in each scenario are presented. Lines and shaded bands indicate means and ±1 standard deviation across ten random states.

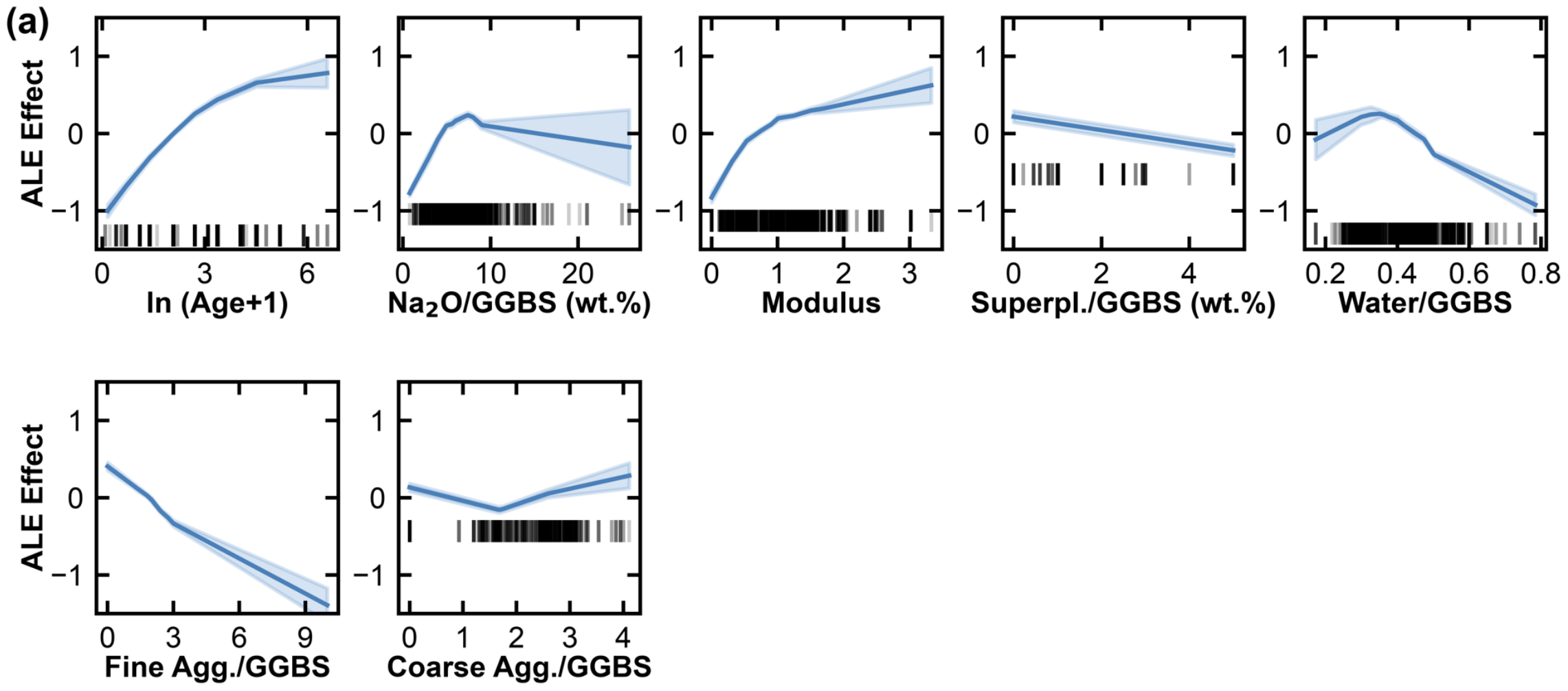

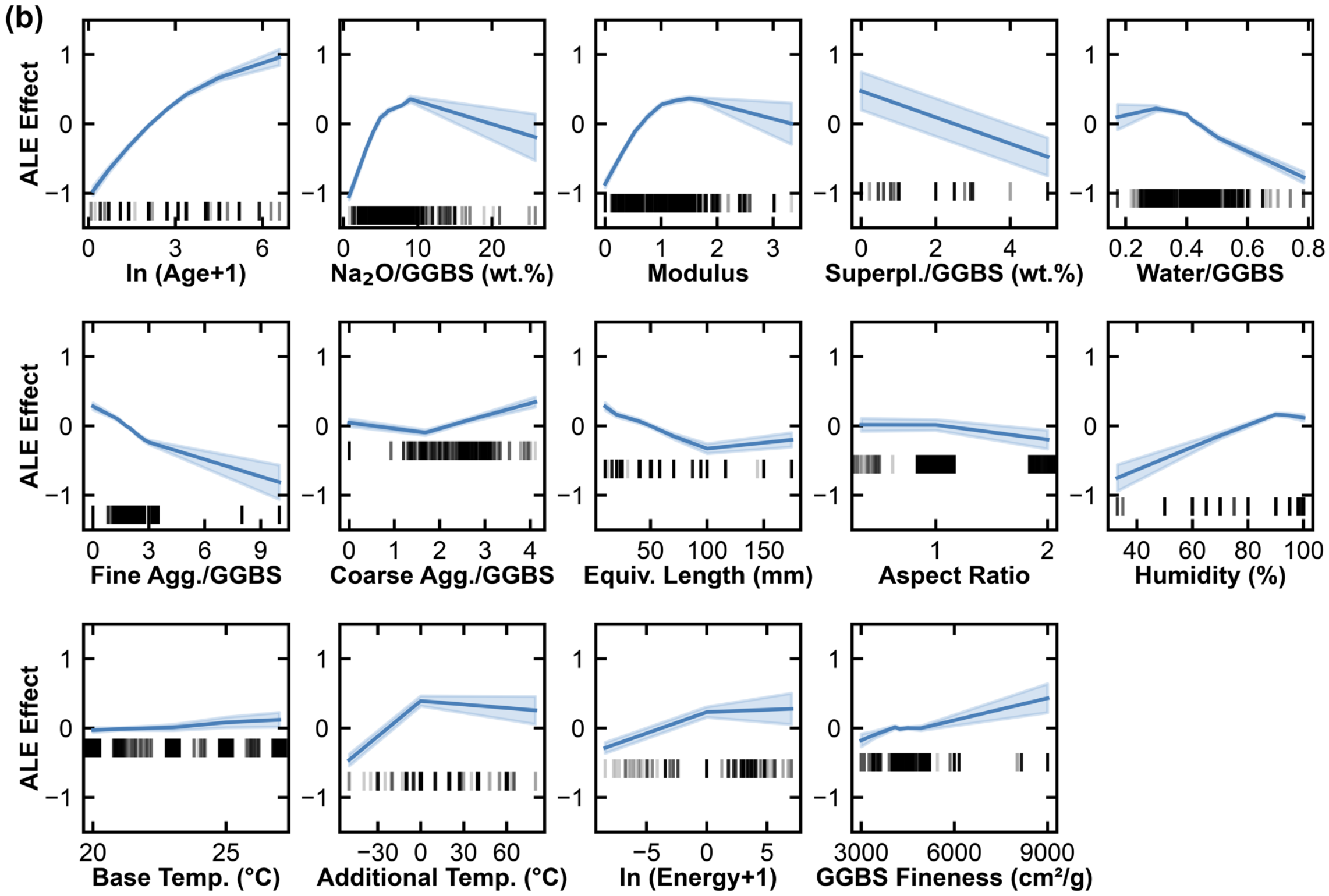

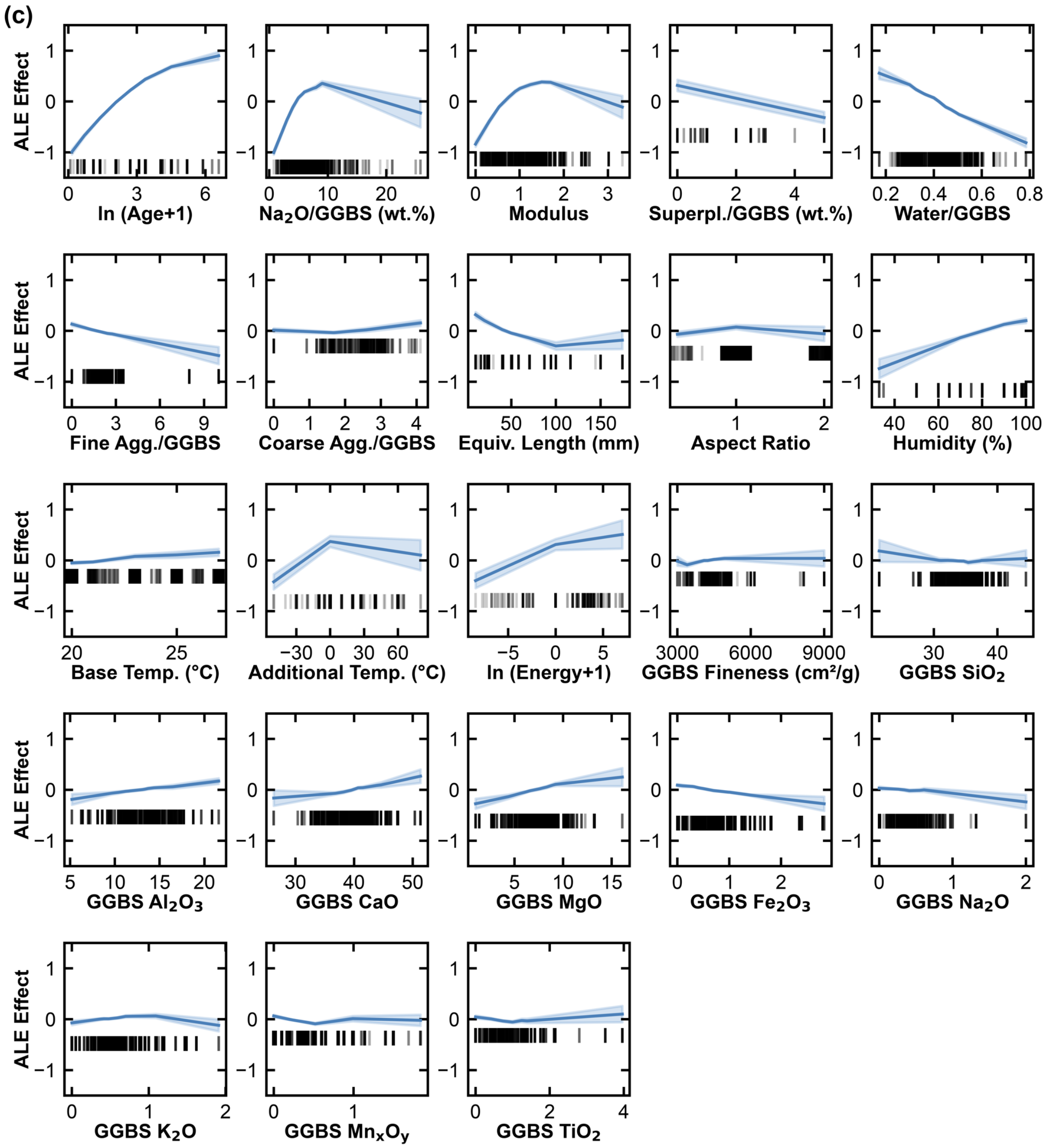

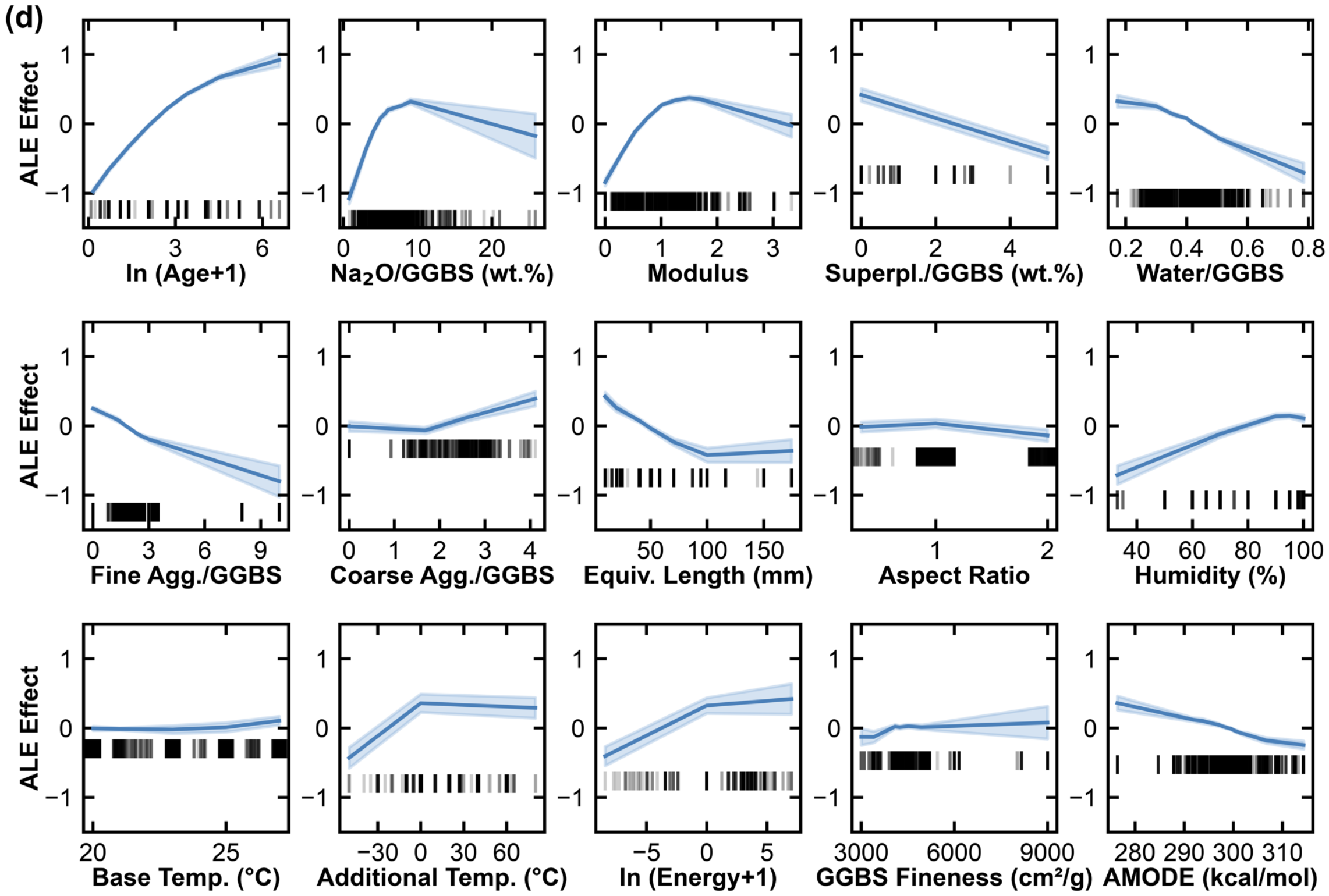

**Figure S22.** ALE analysis results of the ANN model under (a) Scenario 1, Baseline Mix Design and Age (7 features), (b) Scenario 2, Physical Property, Curing, and Geometry (14 features), (c) Scenario 3, Explicit Chemical Composition (23 features), and (d) Scenario 4, Physics-Informed Reactivity (15 features). All features used in each scenario are presented. Lines and shaded bands indicate means and ±1 standard deviation across ten random states.

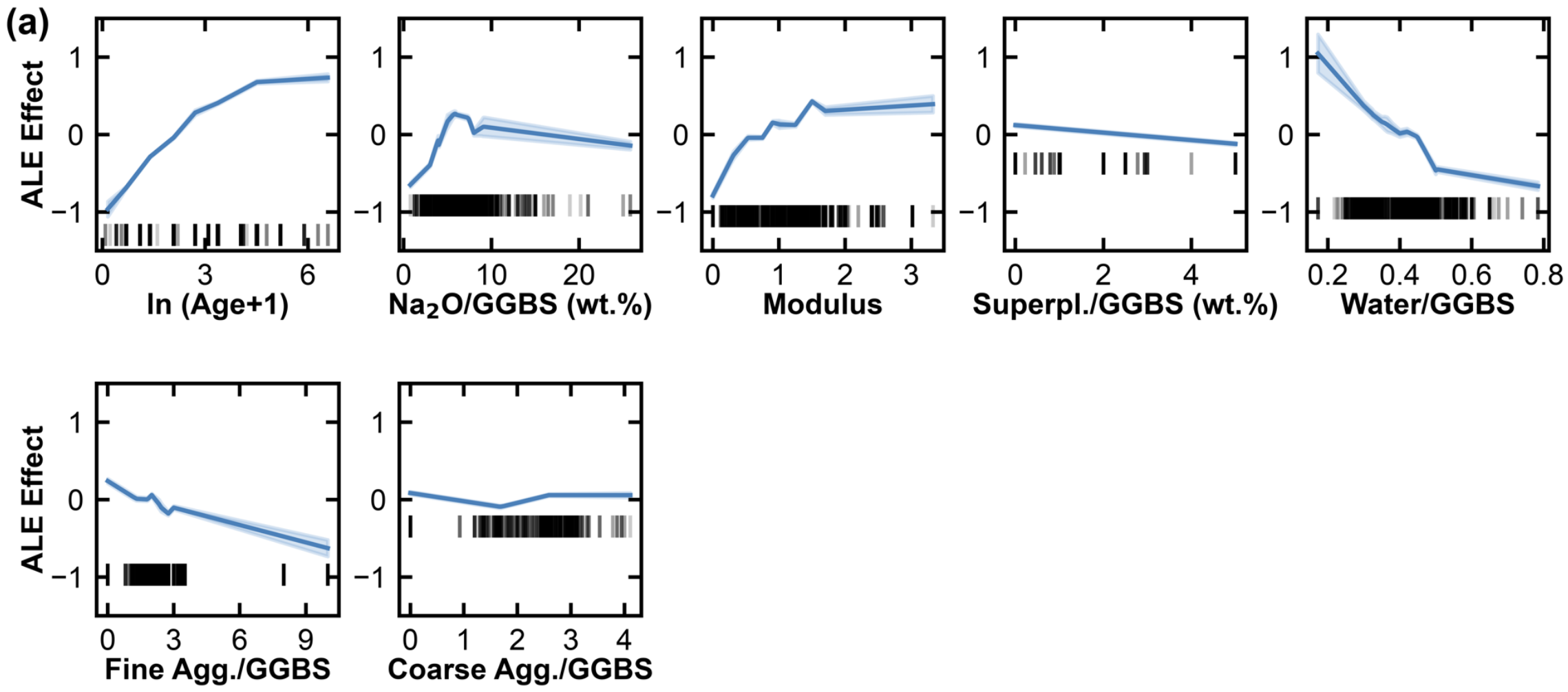

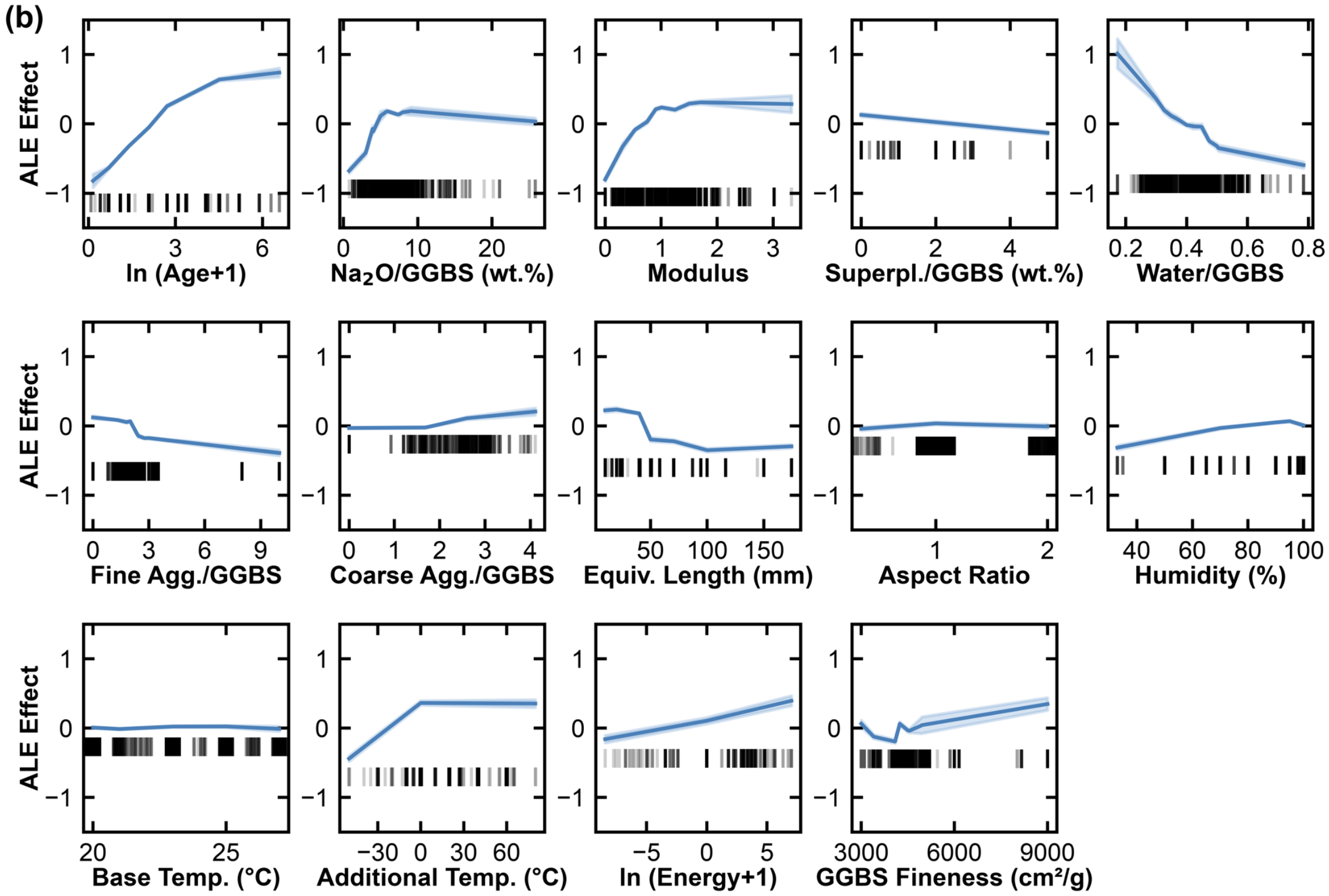

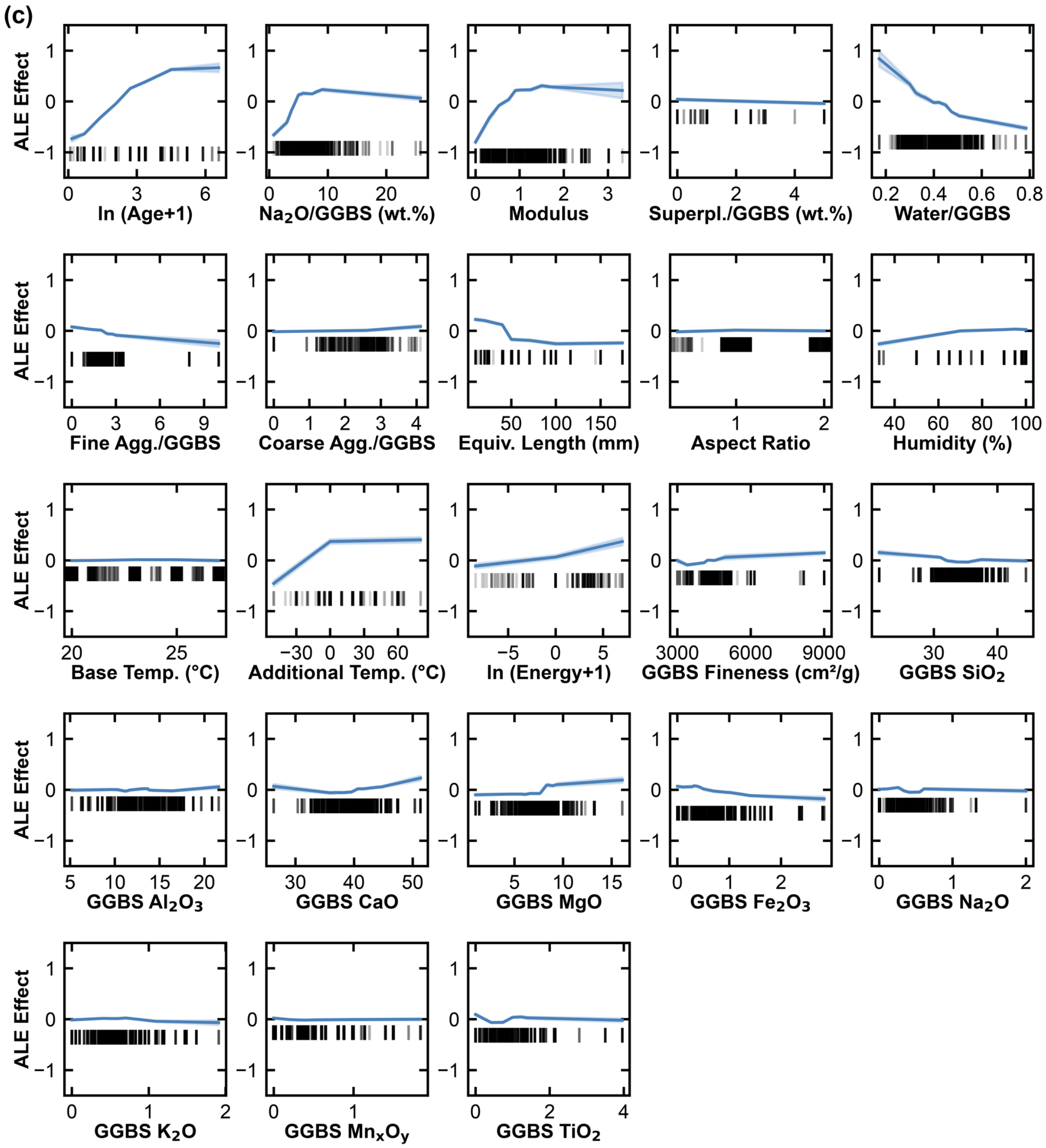

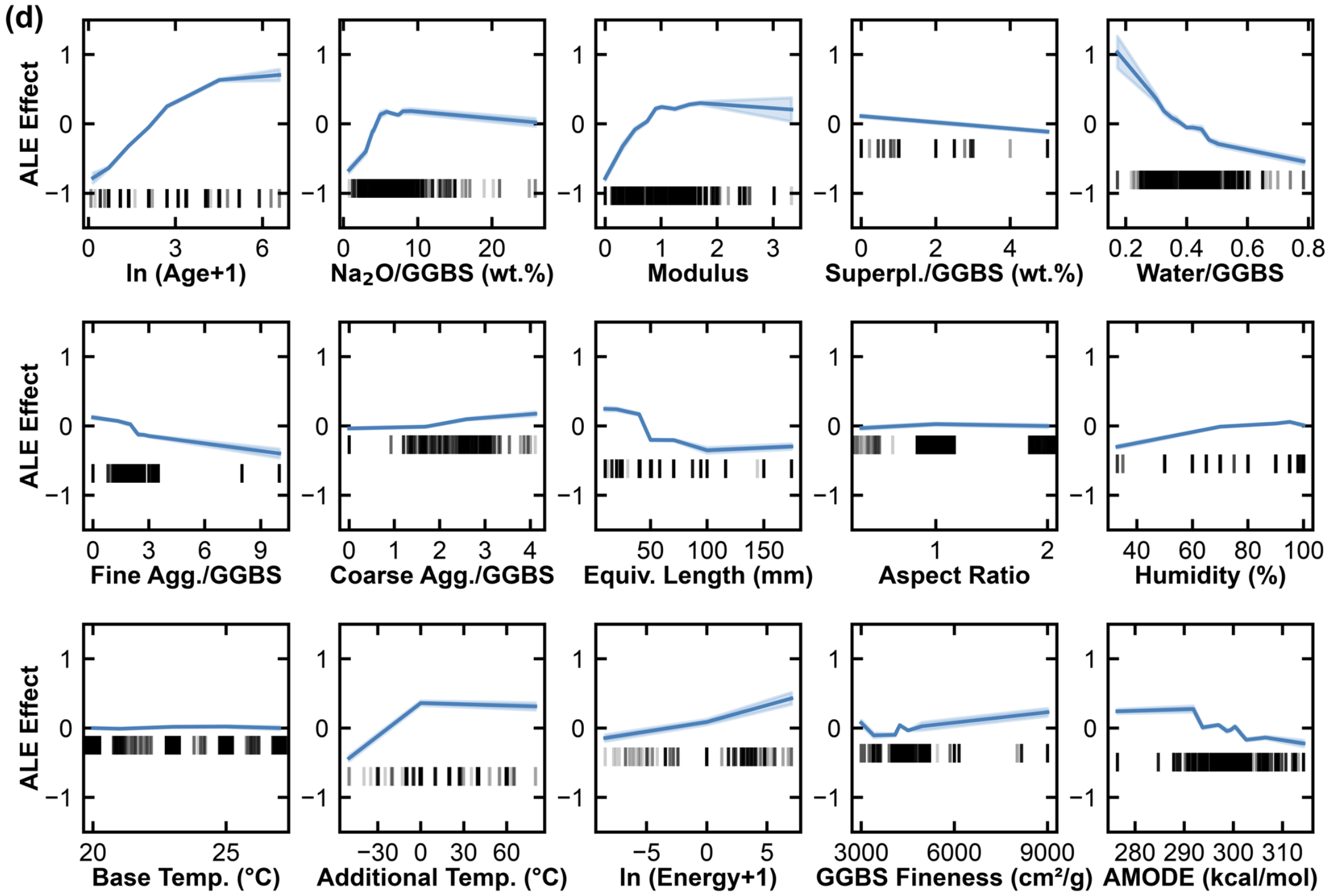

**Figure S23.** ALE analysis results of the XGB model under (a) Scenario 1, Baseline Mix Design and Age (7 features), (b) Scenario 2, Physical Property, Curing, and Geometry (14 features), (c) Scenario 3, Explicit Chemical Composition (23 features), and (d) Scenario 4, Physics-Informed Reactivity (15 features). All features used in each scenario are presented. Lines and shaded bands indicate means and ±1 standard deviation across ten random states.

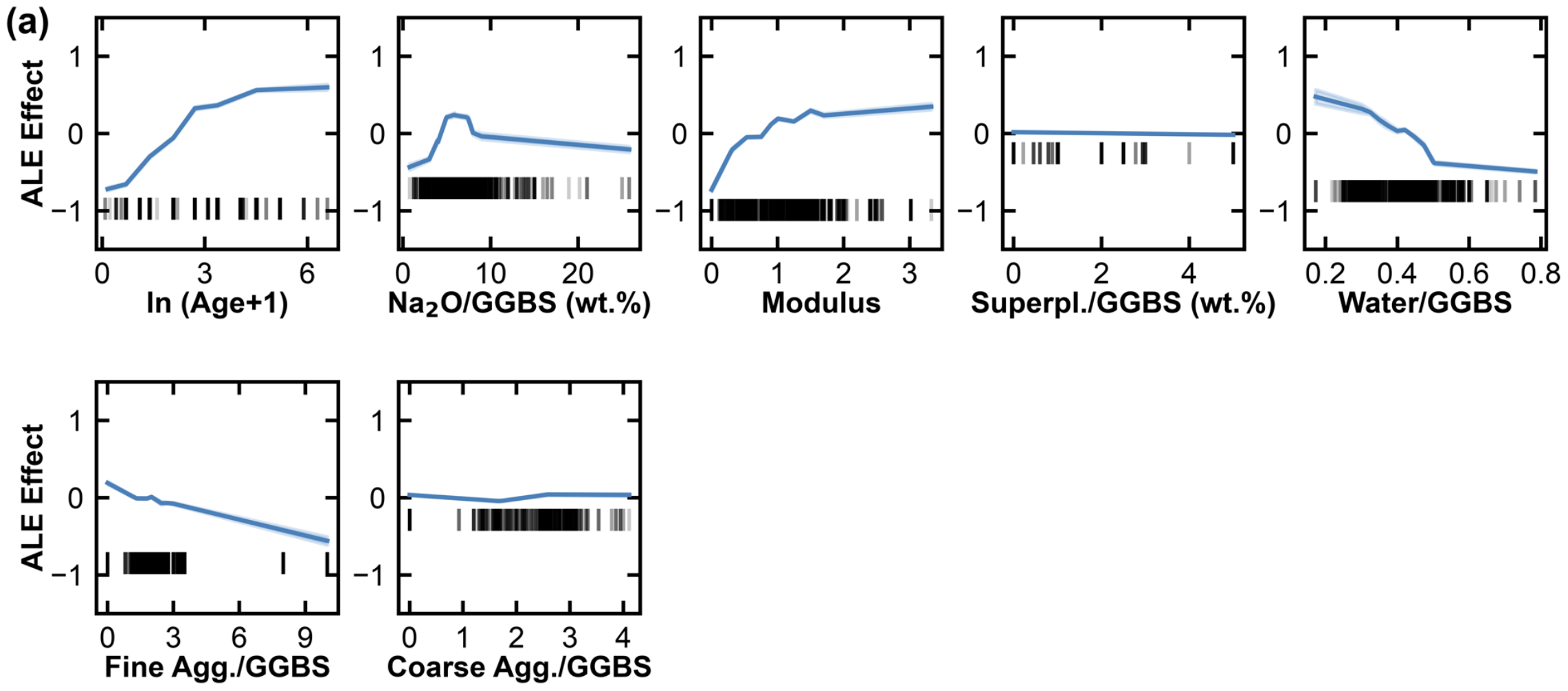

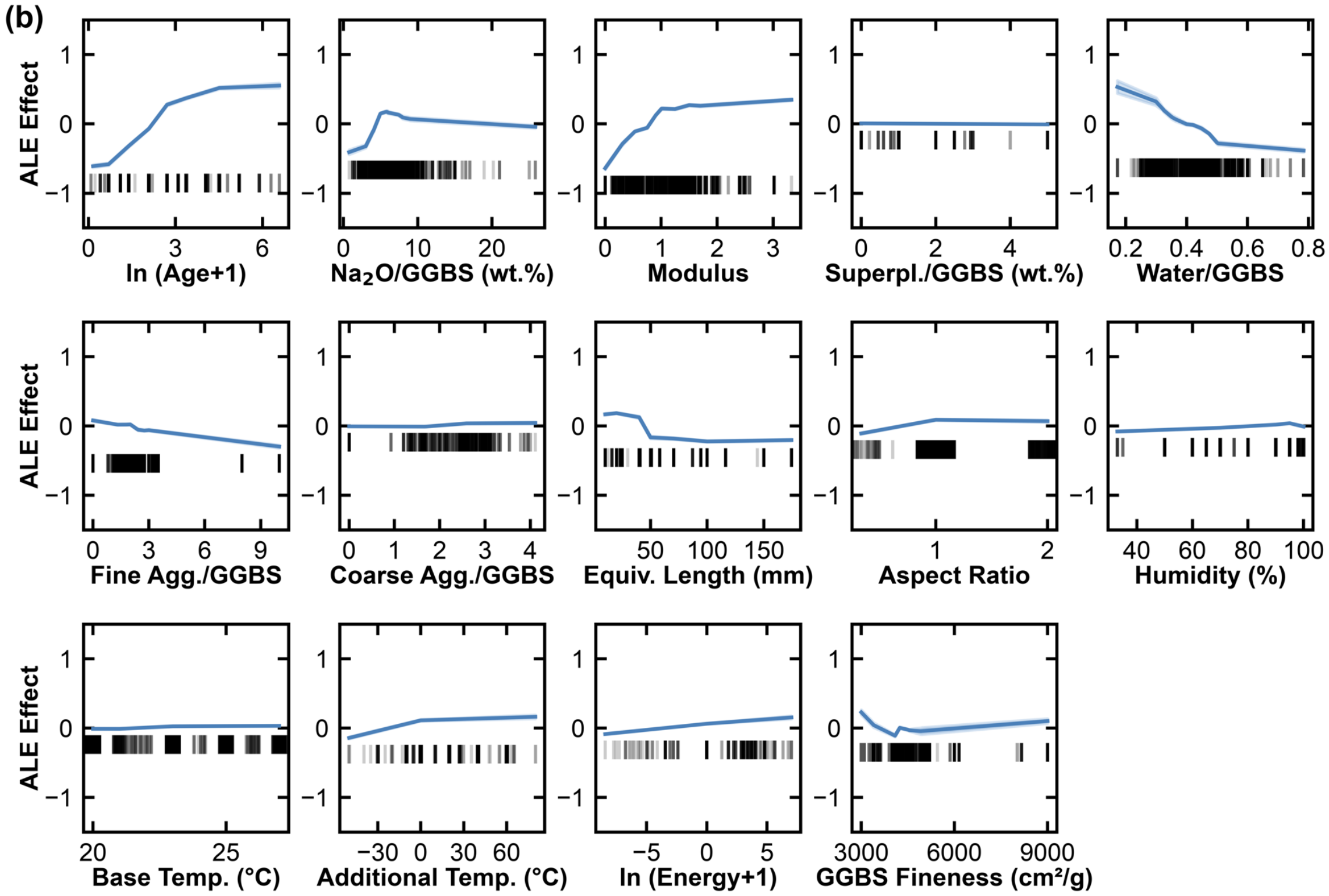

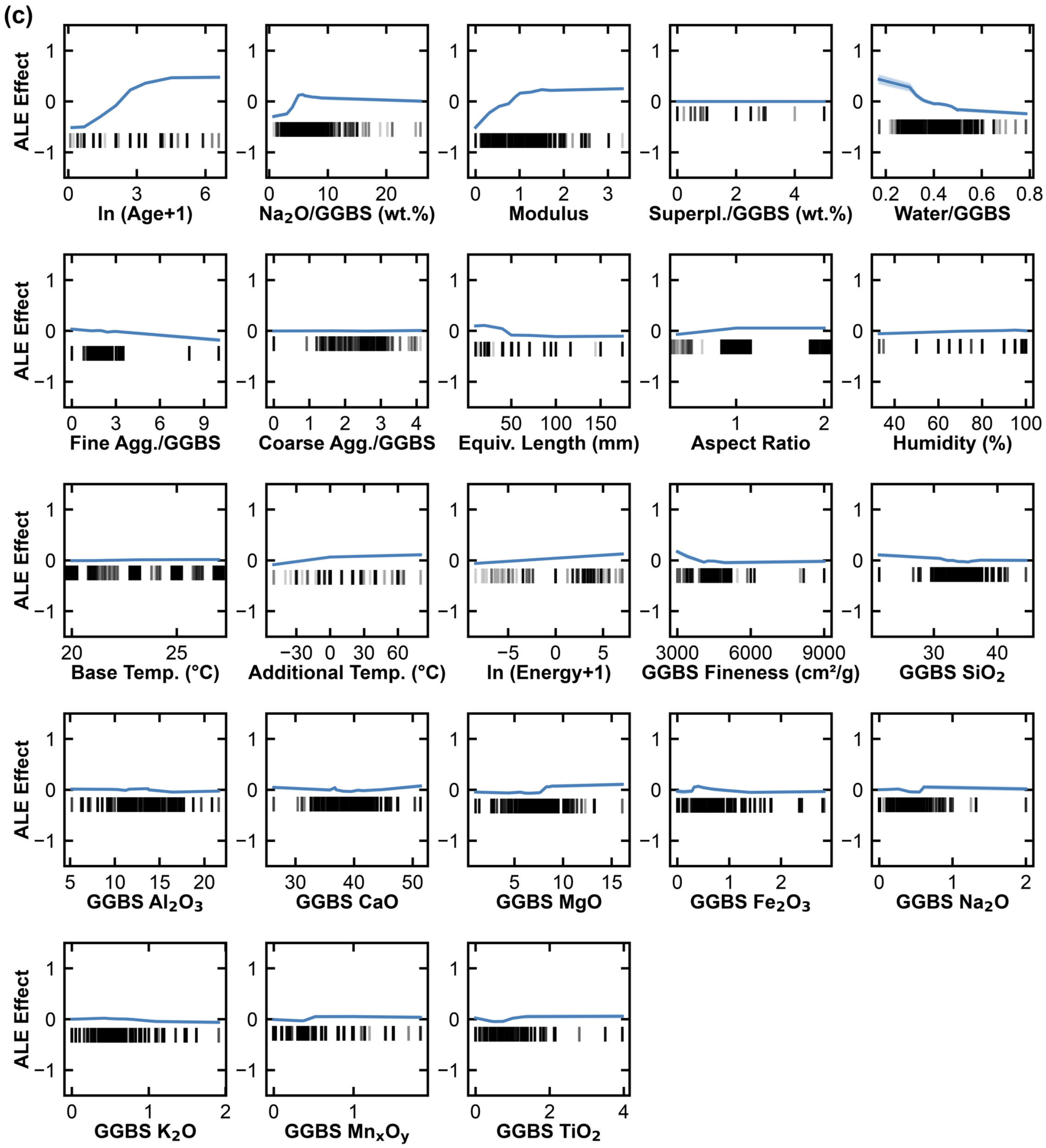

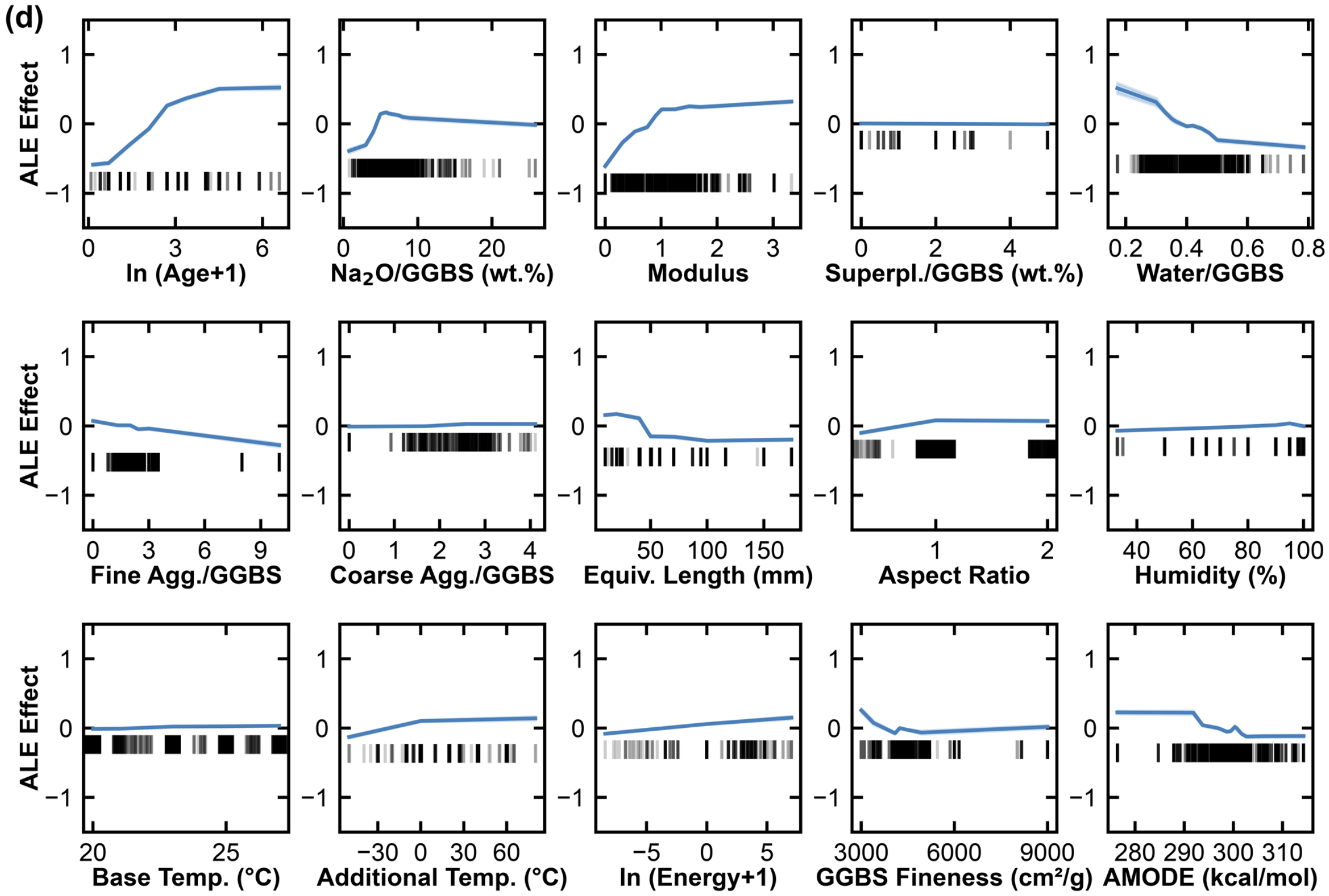

**Figure S24.** ALE analysis results of the RF model under (a) Scenario 1, Baseline Mix Design and Age (7 features), (b) Scenario 2, Physical Property, Curing, and Geometry (14 features), (c) Scenario 3, Explicit Chemical Composition (23 features), and (d) Scenario 4, Physics-Informed Reactivity (15 features). All features used in each scenario are presented. Lines and shaded bands indicate means and ±1 standard deviation across ten random states.

### S4.4 Follow-up Discussion on GGBS Blaine Fineness and Particle Size Characteristics

In Scenarios 2–4, the contribution of GGBS Blaine fineness is model-dependent, with an overall positive association in GPR but less consistent trends across the other models. The positive association observed in GPR aligns with the general expectation that increased fineness enhances reaction rate and extent, thereby promoting strength development, especially at early ages [1]. However, it ranks among the low-importance features overall.

This low importance may be partly attributed to the relatively narrow range (4000–6000 $cm^2/g$), within which variations in fineness do not strongly influence strength, as also reported in previous studies [2,3]. In addition, the comparability of Blaine fineness across studies and laboratories is inherently limited, and its correspondence with actual particle size characteristics is imperfect. Consistent with this limitation, mixtures for which particle size distribution data were available show no clear correlation between Blaine fineness and characteristic particle sizes (**Figure S25**), supporting prior findings that fineness effects on cement strength cannot be adequately described using Blaine fineness alone and require consideration of the full particle size distribution [4].

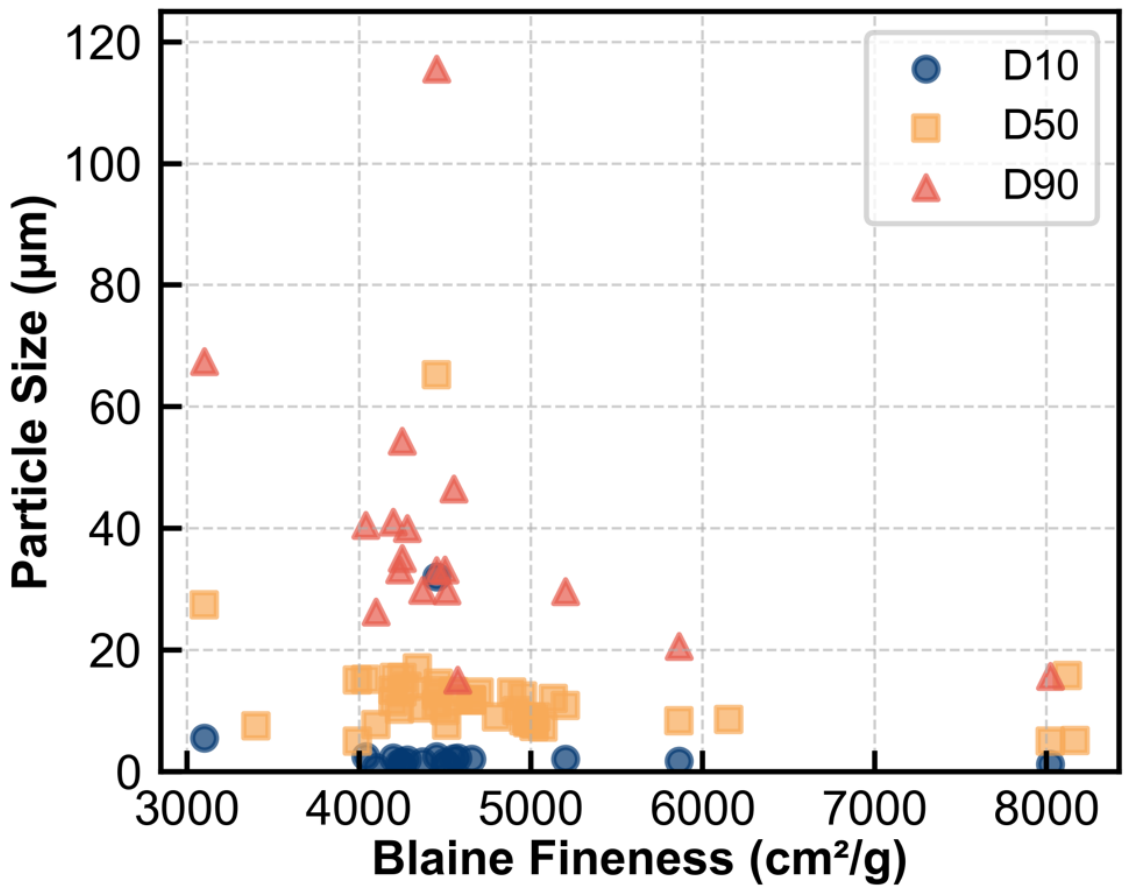

**Figure S25.** Scatter plot of GGBS Blaine fineness versus characteristic particle sizes (D10, D50, and D90), illustrating the limited correspondence between Blaine fineness and particle size distribution in the compiled dataset.

### S4.5 Follow-up Discussion on Effects of Individual GGBS Oxide Components

The impact of GGBS chemical composition on compressive strength is examined in Scenario 3, where slag chemistry is represented explicitly by nine oxide mass fractions. MgO has the highest mean absolute SHAP importance among the oxides across all four models, while the rankings of the remaining oxides vary between models. In GPR, $K_2O$, $Mn_xO_y$, and $Fe_2O_3$ follow MgO, whereas the other oxide contributions are smaller and relatively close in magnitude. These rankings reflect contributions to model predictions within the compiled dataset rather than the absolute abundance of each oxide.

MgO generally shows positive associations with SHAP values across models. CaO also shows positive associations in GPR, ANN, and XGB, but its association is weak in RF. These trends are broadly consistent with the role of network modifiers in promoting glass depolymerization and slag reactivity [5,6]. $SiO_2$, in contrast, generally shows negative associations, reflecting its function as the primary network former. This behavior underpins its placement in the denominator of many slag reactivity indices, as summarized by Blotevogel et al. [7]. $TiO_2$ and $Mn_xO_y$ show negative associations in GPR, ANN, and XGB but positive associations in RF. $Fe_2O_3$ also shows negative associations in GPR, ANN, and XGB, with a weak association in RF. Thus, the negative trends reported for these oxides in previous studies [6,8,9] are not uniformly reproduced across models.

The effects of intrinsic alkali oxides ($Na_2O$ and $K_2O$) are weak and model-dependent. $Na_2O$ shows negative associations with SHAP values in GPR, ANN, and XGB but a positive association in RF; $K_2O$ shows a positive association in ANN and negative associations in the other models. Their moderate mutual correlation (Pearson $r = 0.41$, **Figure S16**) and correlations with other compositional variables complicate the interpretation of their individual contributions.

Oxide-level ALE responses are generally more pronounced in GPR and ANN (**Figures S21** and **S22**) than in XGB and RF (**Figures S23** and **S24)**. $Al_2O_3$ has relatively low SHAP importance and shows a shallow bell-shaped ALE response in GPR, a generally increasing response in ANN, and weaker responses in the tree-based models. This model dependence is consistent with interpreting its contribution in the context of activator chemistry and curing age [2,7,10,11]. For MgO, the ANN ALE response generally increases with content, whereas GPR shows an initial increase followed by a decline at higher contents. These differences indicate that the detailed shape of the inferred MgO–strength relationship depends on the model.

Overall, individual oxide effects are secondary in magnitude, sensitive to feature correlations, and difficult to interpret in isolation. This observation highlights the limitations of oxide-level interpretability in literature-derived datasets and provides context for the use of integrated reactivity descriptors in the main analysis.

## S5 Extended One-Dimensional Design Space Exploration

**Sections S5.1–S5.4** present design space exploration results from the ANN, XGB, and RF models. Some general observations can be made in the comparison of different models:

ANN yielded predictions similar to those of GPR, indicating that both network-based and kernel-based approaches effectively captured the underlying pattern. The close agreement between two fundamentally different algorithms reinforces the robustness of the learned structure–property relationships. ANN predictions showed greater variability across random states than GPR in some regions of the explored design space.

In contrast, the two tree-based models (XGB and RF) produced results that were similar to each other but deviated from the GPR/ANN trends. Generally, they both exhibited smaller differences across GGBS reactivity indices, and RF systematically provided lower strength predictions. These discrepancies reflect the intrinsic piecewise-constant nature of tree-based models, which average responses within terminal leaves and thus struggle to represent smoothly varying relationships or capture subtle interaction effects. Moreover, tree ensembles lack the functional continuity and extrapolation capacity of kernel- or network-based models, which may affect their predictions in sparsely represented regions of the feature space [12]. The relatively lower predicted strengths observed for RF are consistent with its lower predictive accuracy reported in **Table S13**. Overall, the deviations of tree-based models from ANN/GPR trends are consistent with previous observations by Gong and Olivetti [13].

### S5.1 Strength Development with Age

**Figures S26–S28** present the predicted compressive strength with curing age from ANN, XGB, and RF under NaOH and $Na_2SiO_3$ activation.

All models successfully reproduced the characteristic "rapid initial increase followed by plateau" trend in strength development with curing age. They also consistently captured the larger influence of GGBS reactivity (AMODE) on $Na_2SiO_3$-activated systems compared to NaOH-activated ones. The model-specific deviations (tree-based vs. ANN/GPR) described in **Section S5** were observed in this parameter space.

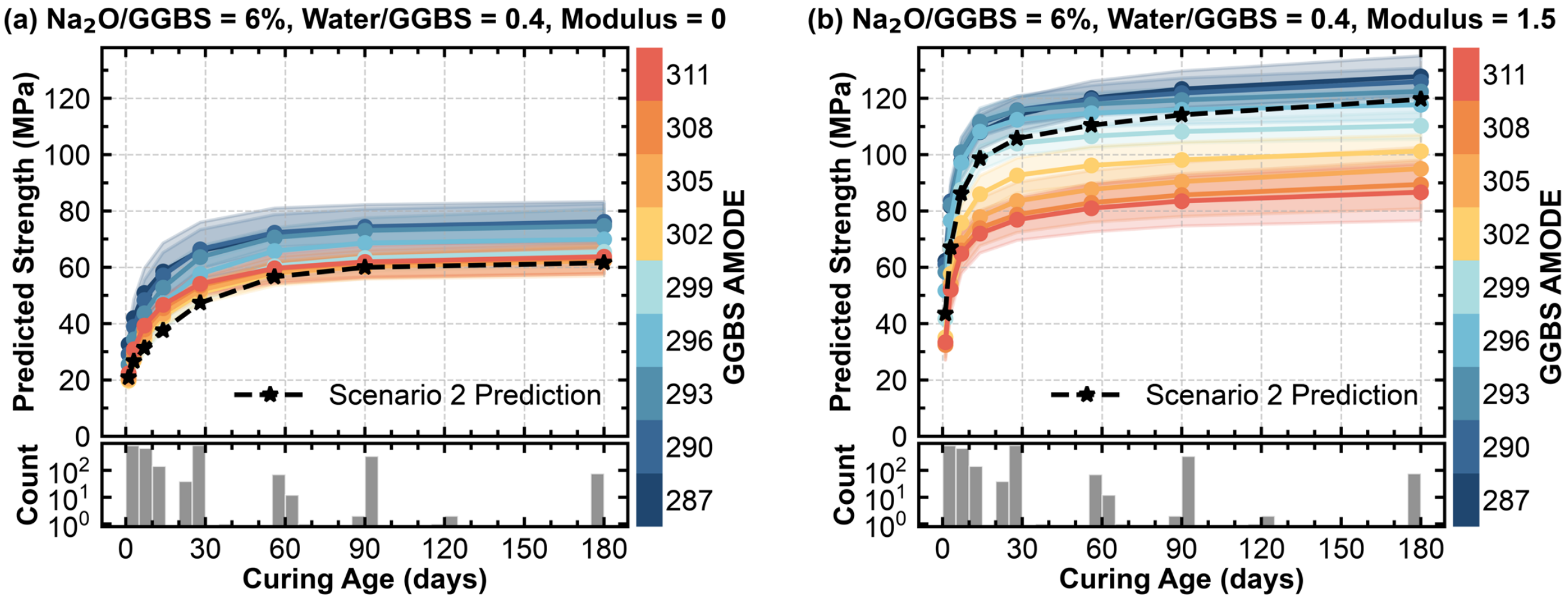

**Figure S26.** ANN-predicted strength development curves with increasing curing age for (a) NaOH-activated and (b) $Na_2SiO_3$-activated samples, based on the design space exploration. The lower subplot shows the histogram of curing age in the experimental dataset used for model training and validation. The solid curve represents the mean prediction across 10 random states, while the shaded band indicates the corresponding standard deviation.

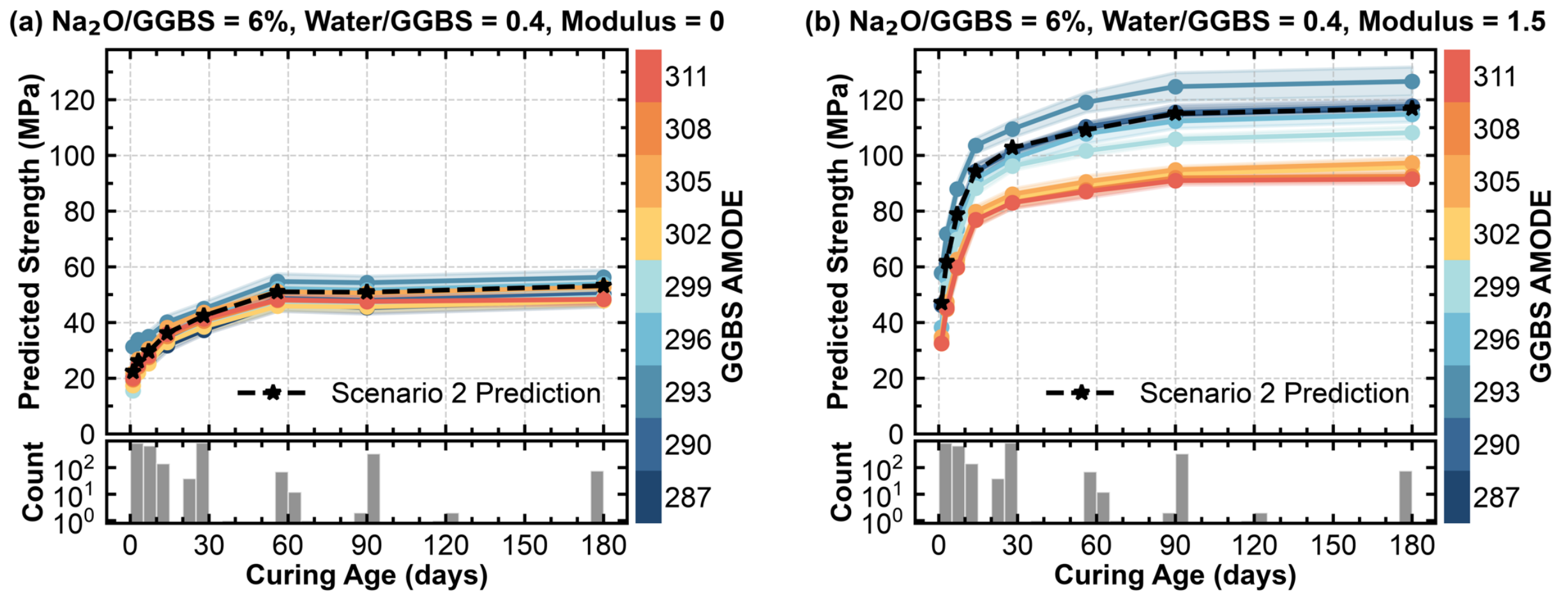

**Figure S27.** XGB-predicted strength development curves with increasing curing age for (a) NaOH-activated and (b) $Na_2SiO_3$-activated samples, based on the design space exploration. The lower subplot shows the histogram of curing age in the experimental dataset used for model training and validation. The solid curve represents the mean prediction across 10 random states, while the shaded band indicates the corresponding standard deviation.

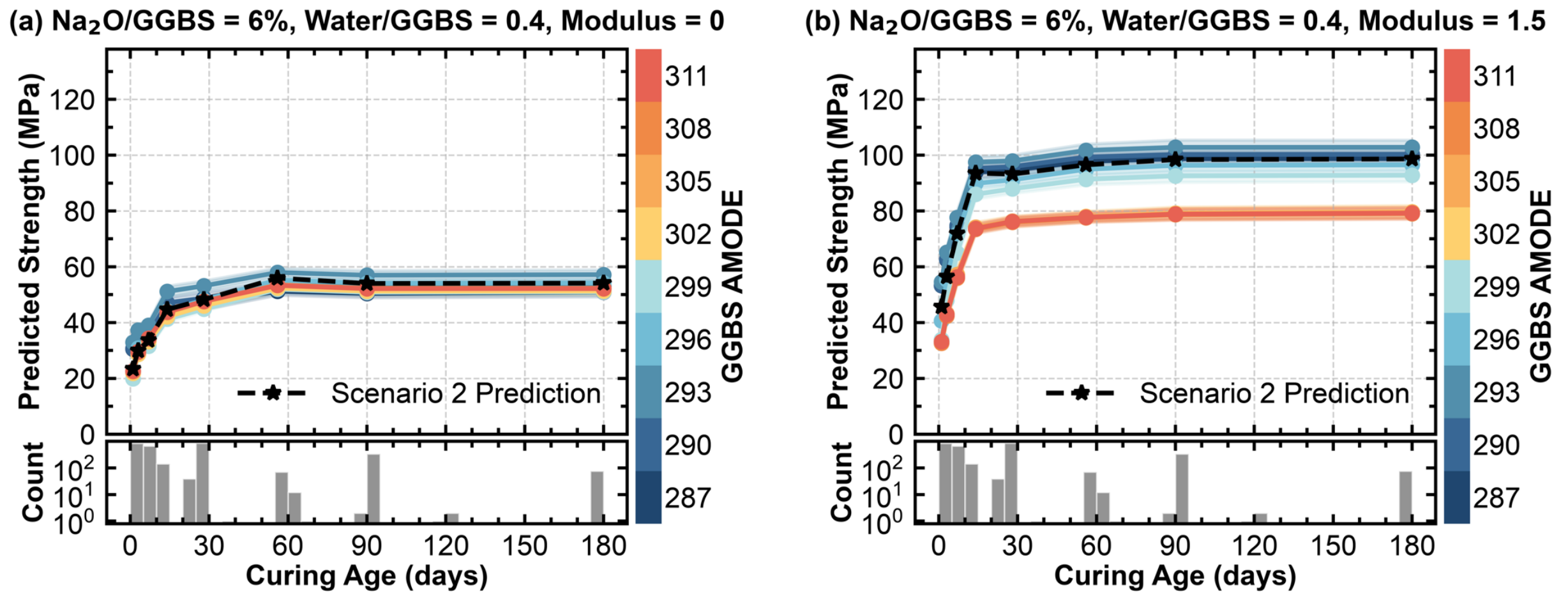

**Figure S28.** RF-predicted strength development curves with increasing curing age for (a) NaOH-activated and (b) $Na_2SiO_3$-activated samples, based on the design space exploration. The lower subplot shows the histogram of curing age in the experimental dataset used for model training and validation. The solid curve represents the mean prediction across 10 random states, while the shaded band indicates the corresponding standard deviation.

## S5.2 Effect of $Na_2O$/GGBS

**Figures S29–S31** present the 28-day compressive strength predictions at varying $Na_2O$/GGBS ratios from ANN, XGB, and RF under NaOH and $Na_2SiO_3$ activation.

All models consistently indicated an optimal $Na_2O$ dosage that maximizes compressive strength. Beyond the general model-specific deviations discussed in **Section S5**, several additional observations can be made.

First, GPR and ANN predict more pronounced strength declines beyond the optimal $Na_2O$ dosage, whereas XGB and RF generally show smaller declines and flatter responses at higher dosages. These flatter responses are consistent with the piecewise-constant structure of tree-based models and should be interpreted cautiously in sparsely sampled regions.

Second, shifts in the optimal $Na_2O$/GGBS with varying GGBS reactivity are less pronounced in the tree-based predictions than in GPR and ANN. This indicates weaker predicted sensitivity to slag reactivity in these comparisons.

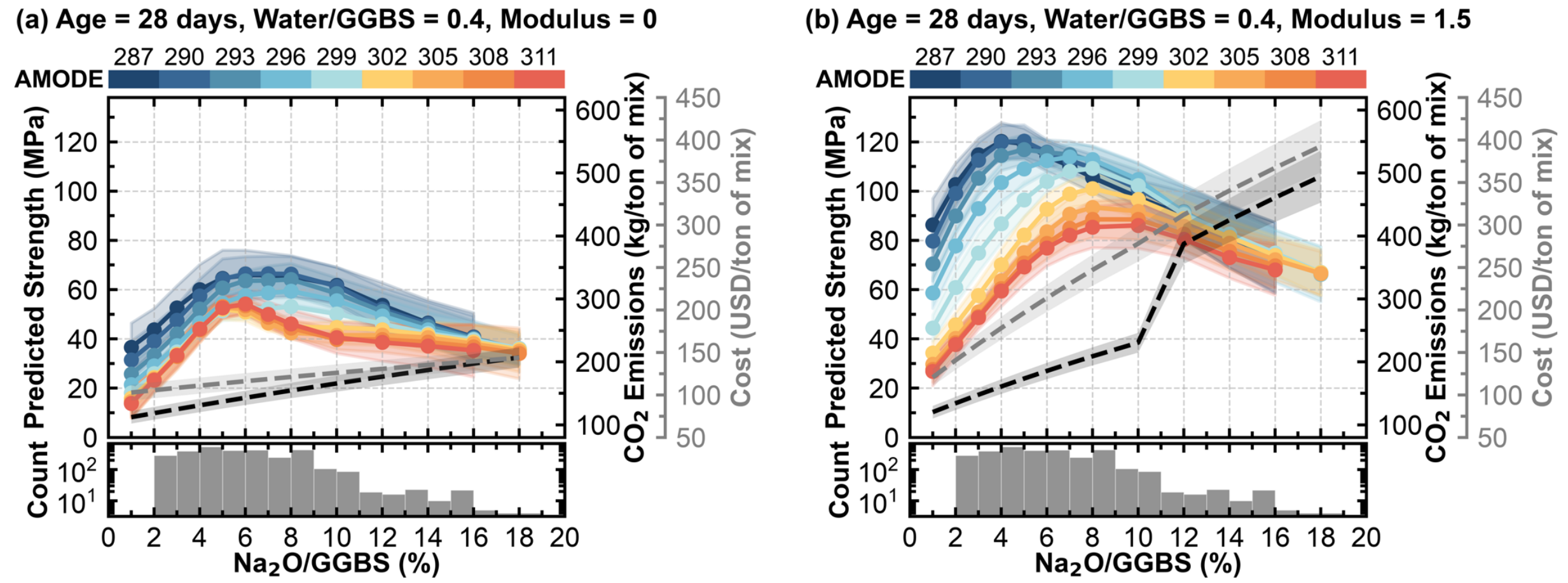

**Figure S29.** ANN-predicted 28-day strength at varying $Na_2O$/GGBS for (a) NaOH-activated pastes and (b) $Na_2SiO_3$-activated pastes, based on the design space exploration. The lower subplot shows the histogram of $Na_2O$/GGBS in the curated dataset. Solid curves and colored bands show mean predicted strengths and standard deviations across 10 random states. Black and grey dashed lines represent the associated $CO_2$ emissions and cost, respectively. Shading around the dashed lines indicates the 2.5th–97.5th percentile ranges from $CO_2$ and cost factor sensitivity analysis.

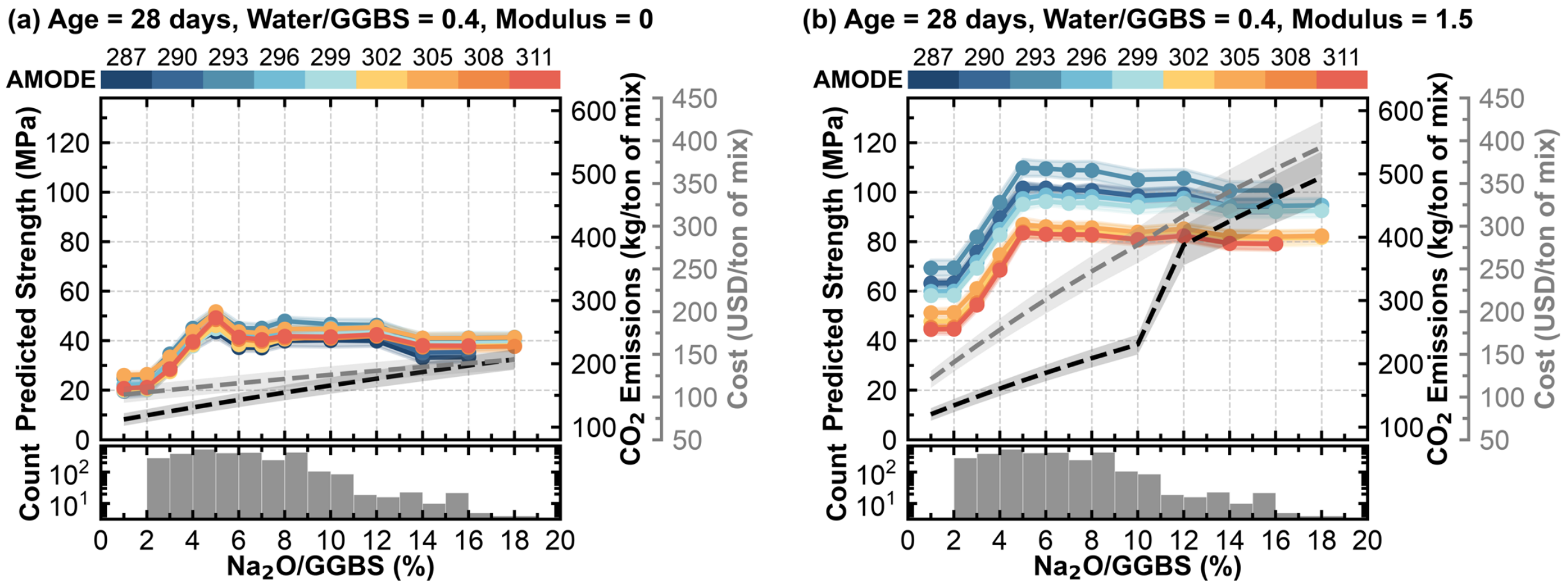

**Figure S30.** XGB-predicted 28-day strength at varying $Na_2O$/GGBS for (a) NaOH-activated pastes and (b) $Na_2SiO_3$-activated pastes, based on the design space exploration. The lower subplot shows the histogram of $Na_2O$/GGBS in the curated dataset. Solid curves and colored bands show mean predicted strengths and standard deviations across 10 random states. Black and grey dashed lines represent the associated $CO_2$ emissions and cost, respectively. Shading around the dashed lines indicates the 2.5th–97.5th percentile ranges from $CO_2$ and cost factor sensitivity analysis.

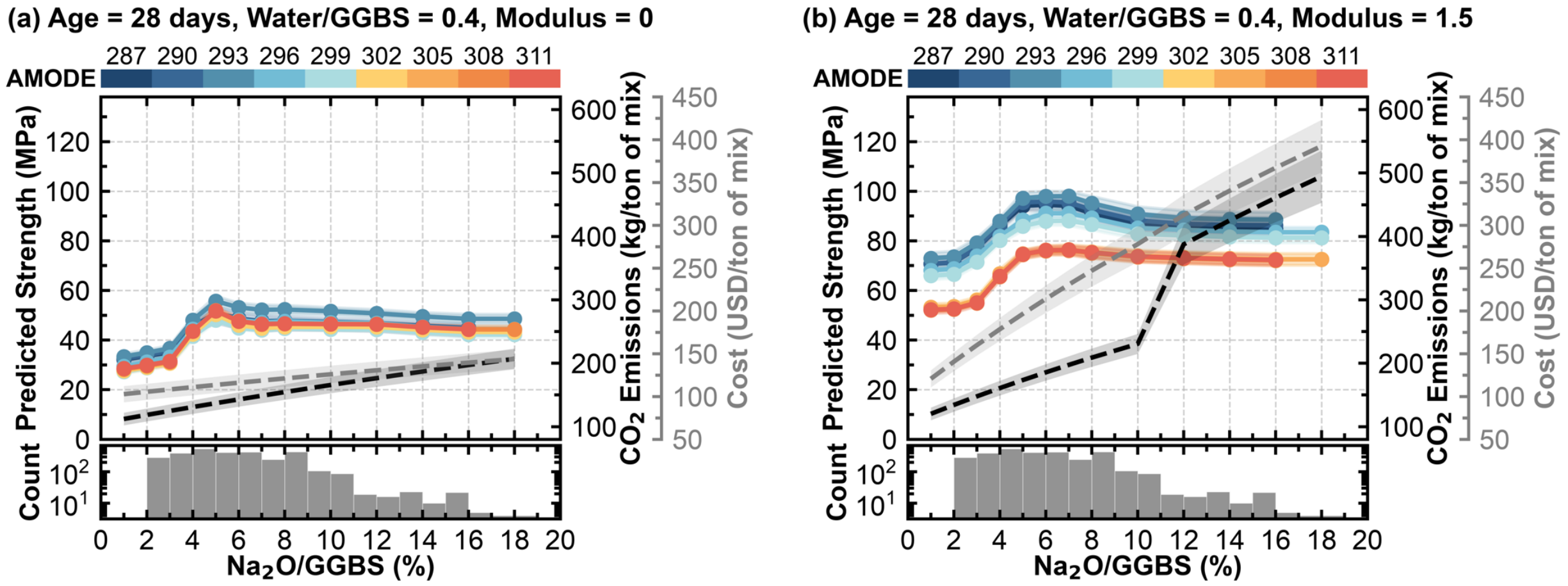

**Figure S31.** RF-predicted 28-day strength at varying $Na_2O$/GGBS for (a) NaOH-activated pastes and (b) $Na_2SiO_3$-activated pastes, based on the design space exploration. The lower subplot shows the histogram of $Na_2O$/GGBS in the curated dataset. Solid curves and colored bands show mean predicted strengths and standard deviations across 10 random states. Black and grey dashed lines represent the associated $CO_2$ emissions and cost, respectively. Shading around the dashed lines indicates the 2.5th–97.5th percentile ranges from $CO_2$ and cost factor sensitivity analysis.

## S5.3 Effect of Silicate Modulus

**Figures S32–S34** present the predicted variation of 28-day compressive strength with the activator Ms from ANN, XGB, and RF models.

Both the ANN and GPR models indicate a distinct optimal Ms, reflecting their ability to capture the nonlinear dependency of strength on activator chemistry. In contrast, XGB and RF generally show broader high-strength plateaus and less pronounced declines at higher Ms. The flattened trends resemble those observed in the $Na_2O$/GGBS exploration (**Section S5.2**) and should likewise be interpreted cautiously near the data boundaries.

Other general discrepancies between the tree-based and ANN/GPR predictions, as described in **Section S5**, can again be identified in this parameter space.

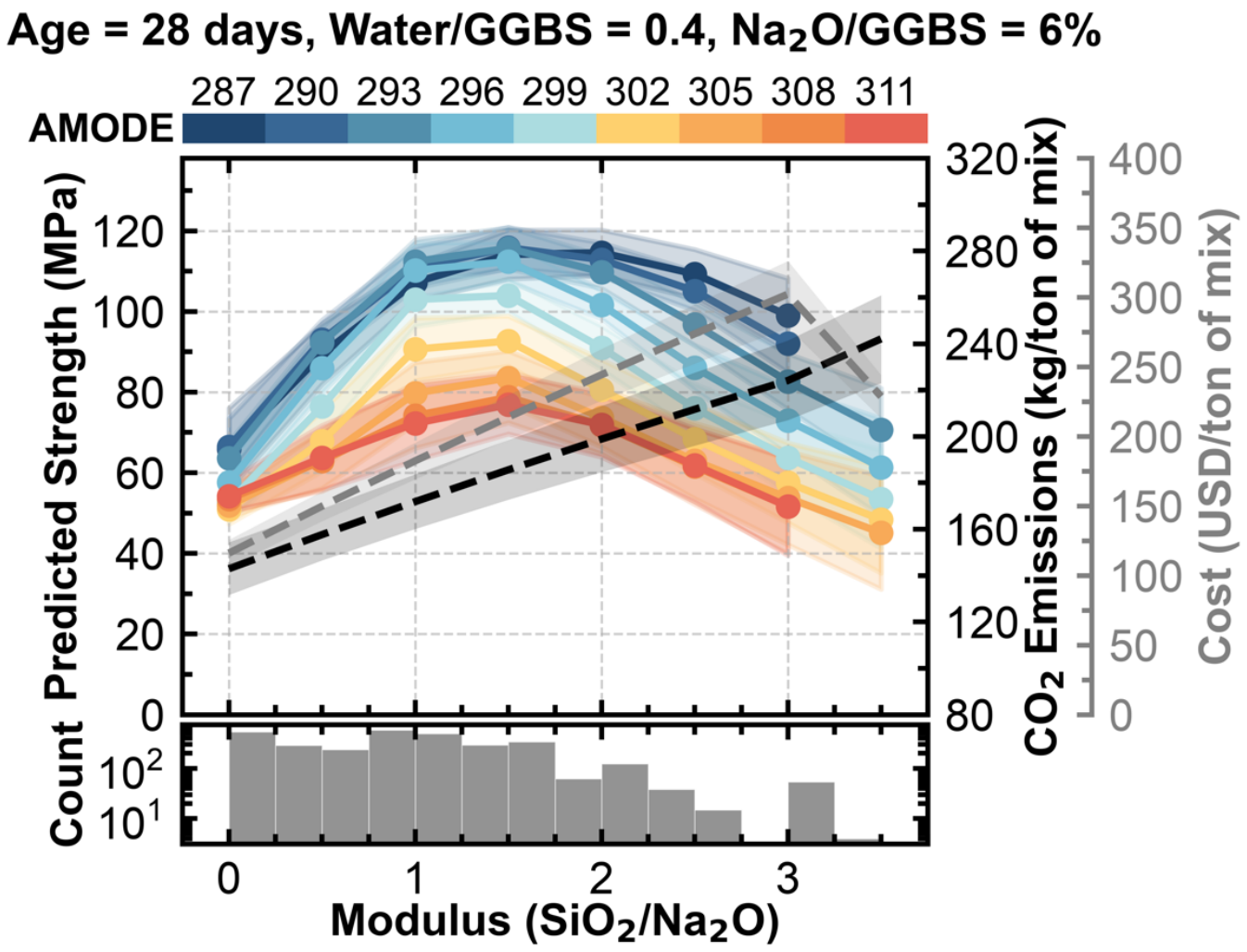

**Figure S32.** ANN-predicted 28-day strength at varying silicate modulus (Ms) for $Na_2SiO_3$-activated pastes, based on the design space exploration. The lower subplot shows the histogram of Ms in the curated dataset. Solid curves and colored bands show mean predicted strengths and standard deviations across 10 random states. Black and grey dashed lines represent the associated $CO_2$ emissions and cost, respectively. Shading around the dashed lines indicates the 2.5th–97.5th percentile ranges from $CO_2$ and cost factor sensitivity analysis.

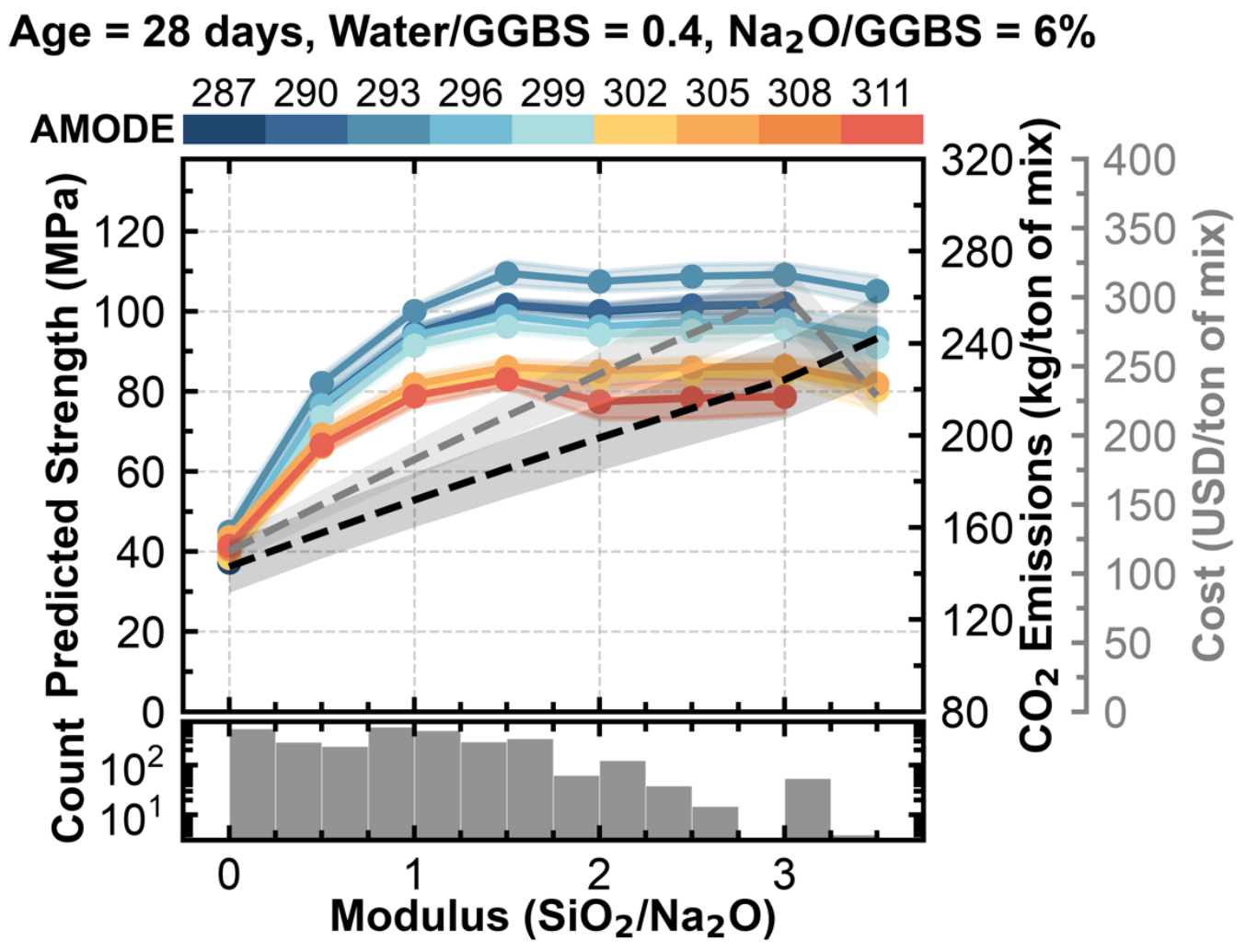

**Figure S33.** XGB-predicted 28-day strength at varying Ms for $Na_2SiO_3$-activated pastes, based on the design space exploration. The lower subplot shows the histogram of Ms in the curated dataset. Solid curves and colored bands show mean predicted strengths and standard deviations across 10 random states. Black and grey dashed lines represent the associated $CO_2$ emissions and cost, respectively. Shading around the dashed lines indicates the 2.5th–97.5th percentile ranges from $CO_2$ and cost factor sensitivity analysis.

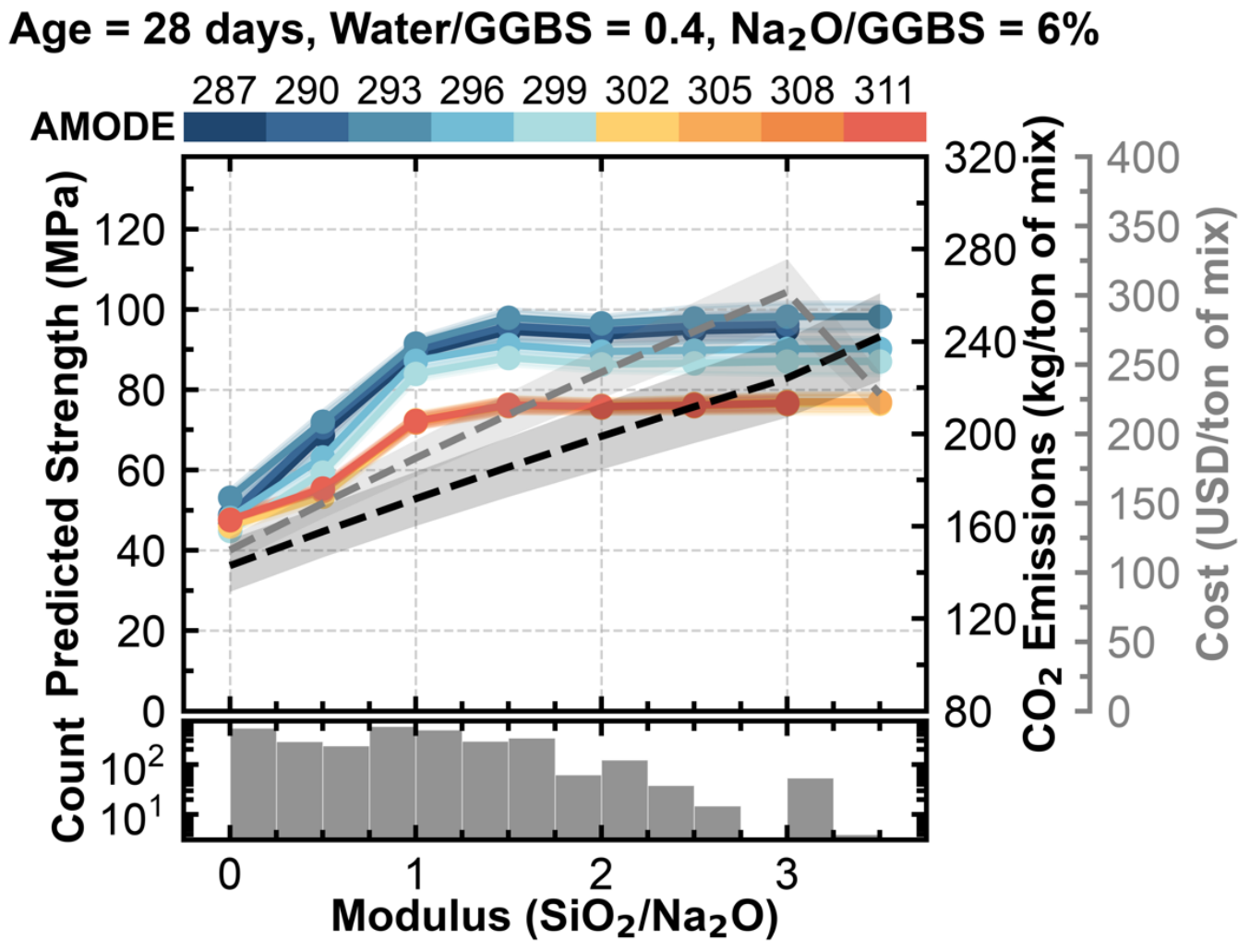

**Figure S34.** RF-predicted 28-day strength at varying Ms for $Na_2SiO_3$-activated pastes, based on the design space exploration. The lower subplot shows the histogram of Ms in the curated dataset. Solid curves and colored bands show mean predicted strengths and standard deviations across 10 random states. Black and grey dashed lines represent the associated $CO_2$ emissions and cost, respectively. Shading around the dashed lines indicates the 2.5th–97.5th percentile ranges from $CO_2$ and cost factor sensitivity analysis.

## S5.4 Effect of Water/GGBS Ratio

**Figures S35–S37** present the 28-day compressive strength predictions as a function of the water/GGBS ratio under NaOH- and $Na_2SiO_3$-activation from ANN, XGB, and RF models.

All four models generally predicted lower strength at higher water/GGBS ratios, although the location and prominence of the transition to declining strength varied with model, activator type, and slag reactivity.

In addition to the general model-specific differences summarized in **Section S5**, further discrepancies emerge near the boundaries of the design space. At very low water/GGBS ratios (<0.25), the tree-based models, especially XGB, predicted increasing strength with decreasing ratio. This behavior should be interpreted cautiously because strength in this range may also depend on workability, air entrainment, porosity, and mixing quality, which are not explicitly represented in the models. At higher ratios (>0.6), the tree-based predictions tended to flatten. These differences highlight model-dependent behavior near the design space boundaries.

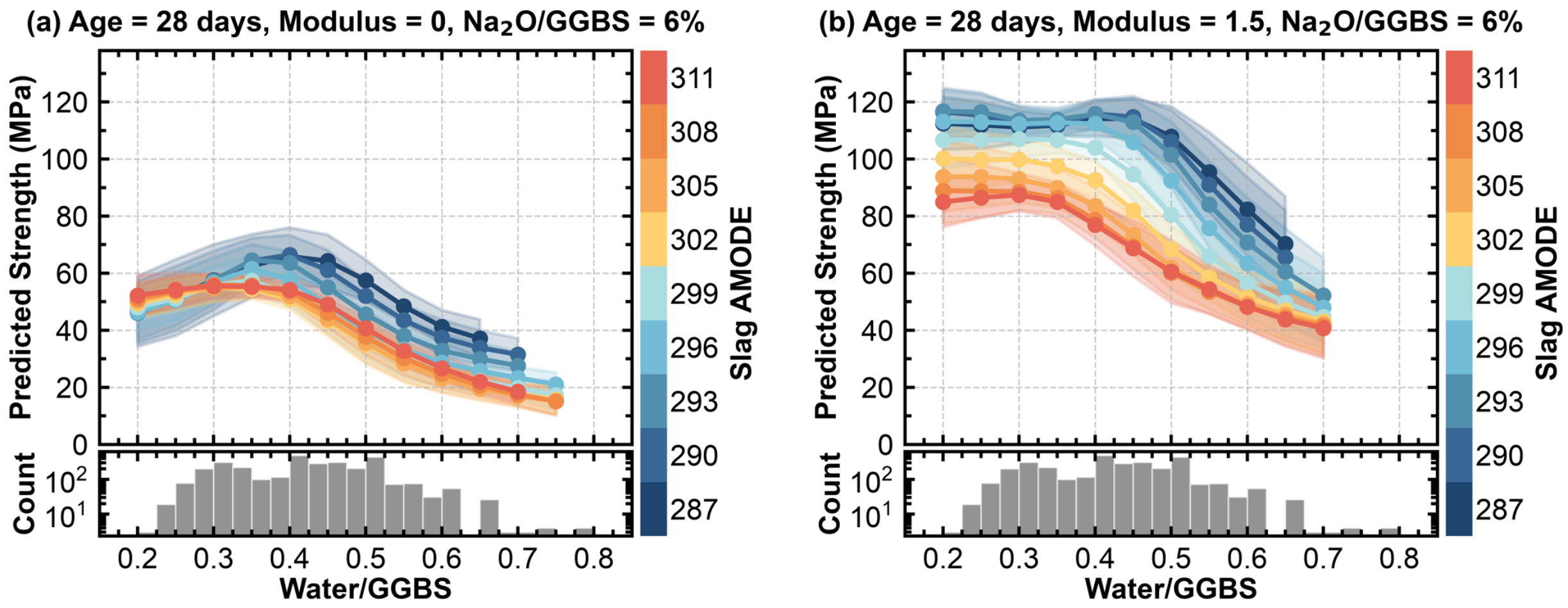

**Figure S35.** ANN-predicted 28-day strength at varying water/GGBS ratios for (a) NaOH-activated and (b) $Na_2SiO_3$-activated pastes, derived from the design space exploration. The lower panel shows the histogram of water/GGBS ratios in the curated experimental dataset. The solid curve represents the mean prediction across 10 random states, while the shaded band indicates the corresponding standard deviation.

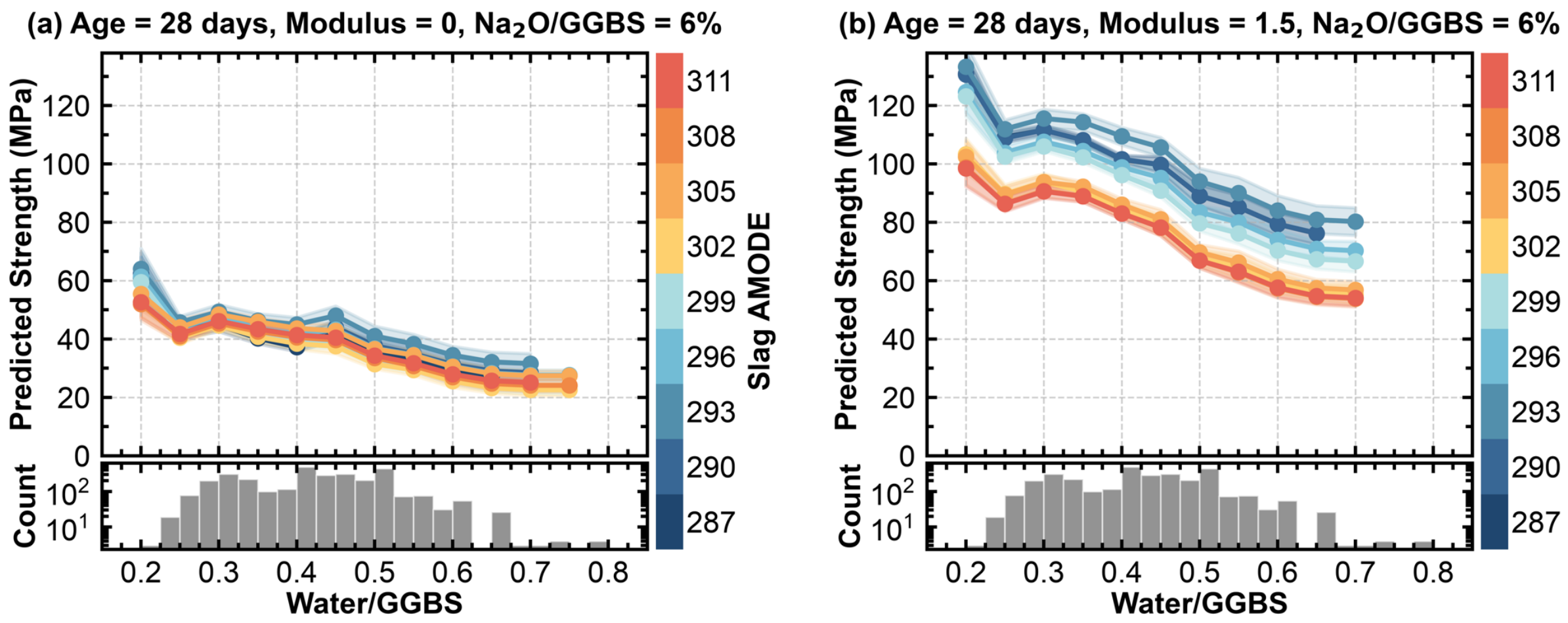

**Figure S36.** XGB-predicted 28-day strength at varying water/GGBS ratios for (a) NaOH-activated and (b) $Na_2SiO_3$-activated pastes, derived from the design space exploration. The lower panel shows the histogram of water/GGBS ratios in the curated experimental dataset. The solid curve represents the mean prediction across 10 random states, while the shaded band indicates the corresponding standard deviation.

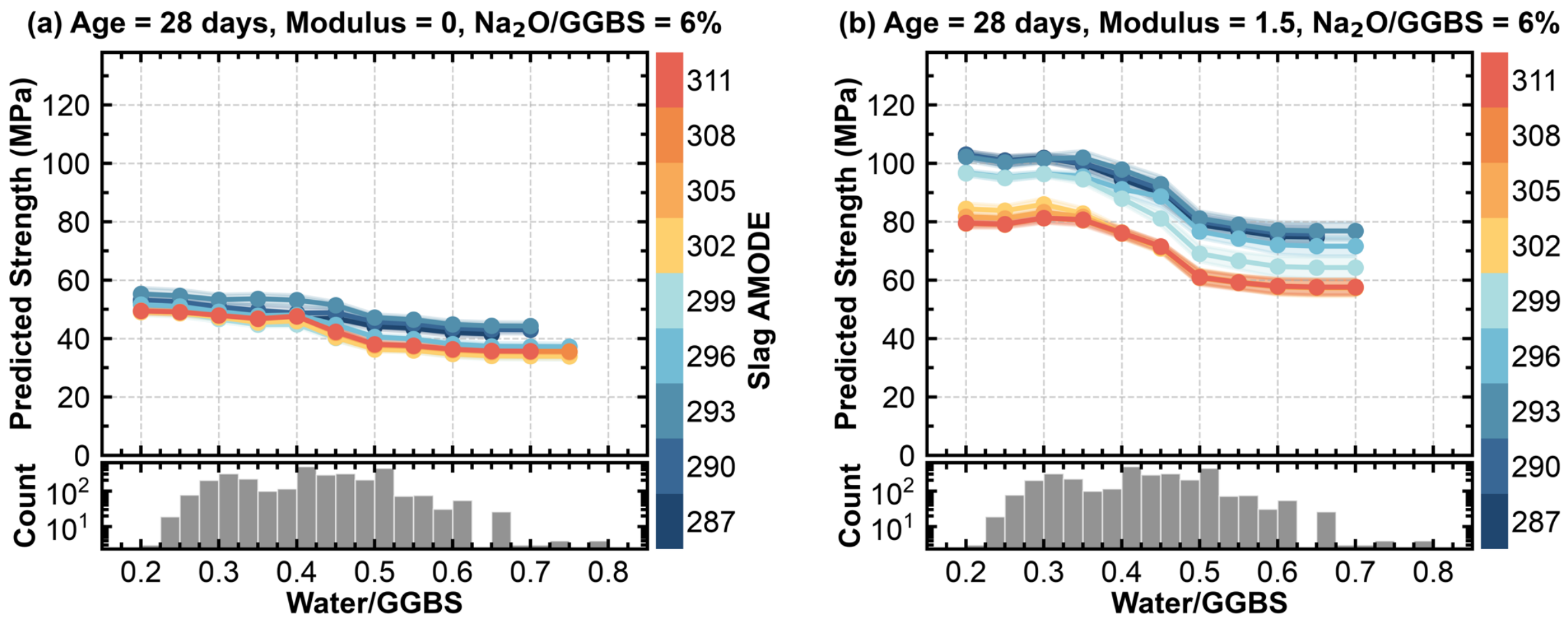

**Figure S37.** RF-predicted 28-day strength at varying water/GGBS ratios for (a) NaOH-activated and (b) $Na_2SiO_3$-activated pastes, derived from the design space exploration. The lower panel shows the histogram of water/GGBS ratios in the curated experimental dataset. The solid curve represents the mean prediction across 10 random states, while the shaded band indicates the corresponding standard deviation.

## S6 Extended Two-Dimensional Design Space Exploration

### S6.1 Optimal $Na_2O$/GGBS and Ms Combinations at Different AMODE Levels

**Figures S38–S40** present the predicted 28-day compressive strength across varying combinations of $Na_2O$ content and Ms for different GGBS reactivity levels from ANN, XGB, and RF models.

Similar to GPR, ANN predicts a shift towards lower $Na_2O$ dosages with increasing GGBS reactivity, while high-reactivity slags retain high predicted strengths over a broader range of Ms.

In contrast, XGB and RF produce broader, band-like high-strength regions with less clearly localized optima. These patterns are consistent with the piecewise-constant structure of tree-based models and should be interpreted cautiously near the design space boundaries.

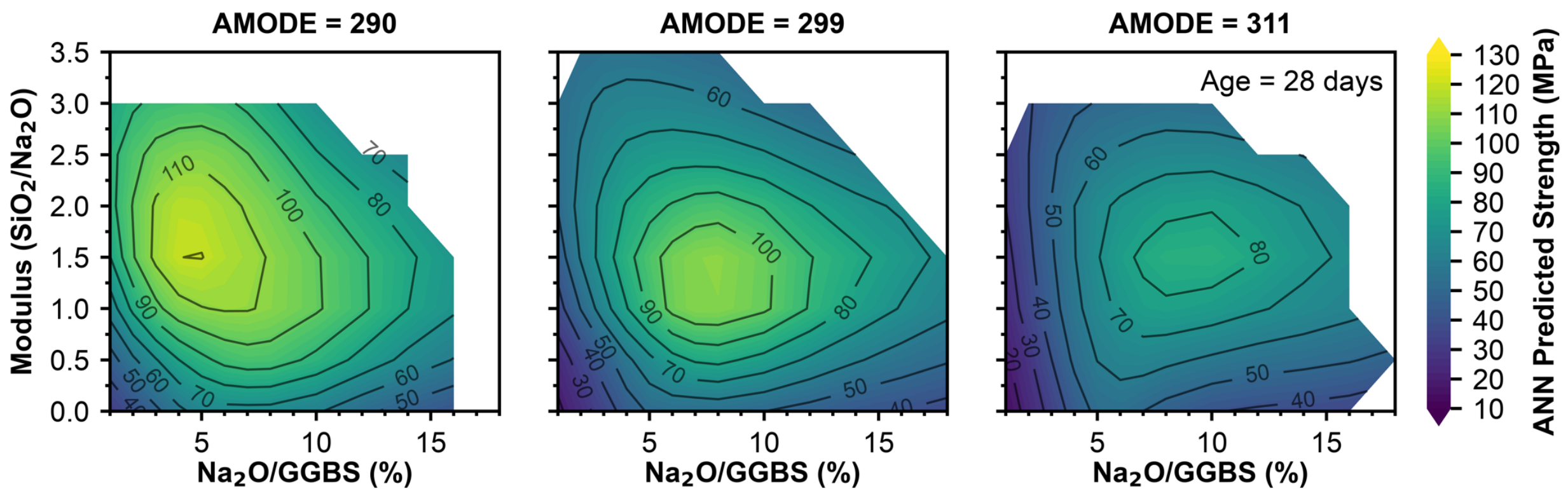

**Figure S38.** Two-dimensional mapping of ANN-predicted 28-day compressive strength as a function of $Na_2O$/GGBS and Ms for alkali-activated slag (AAS) pastes with varying GGBS reactivities: high (AMODE = 290), medium (AMODE = 299), and low (AMODE = 311). The water/GGBS ratio is fixed at 0.4. Strength values are mean predictions across 10 random states. Blank regions indicate combinations excluded by mass-balance screening or Mahalanobis distance filtering (**Section 2.5**).

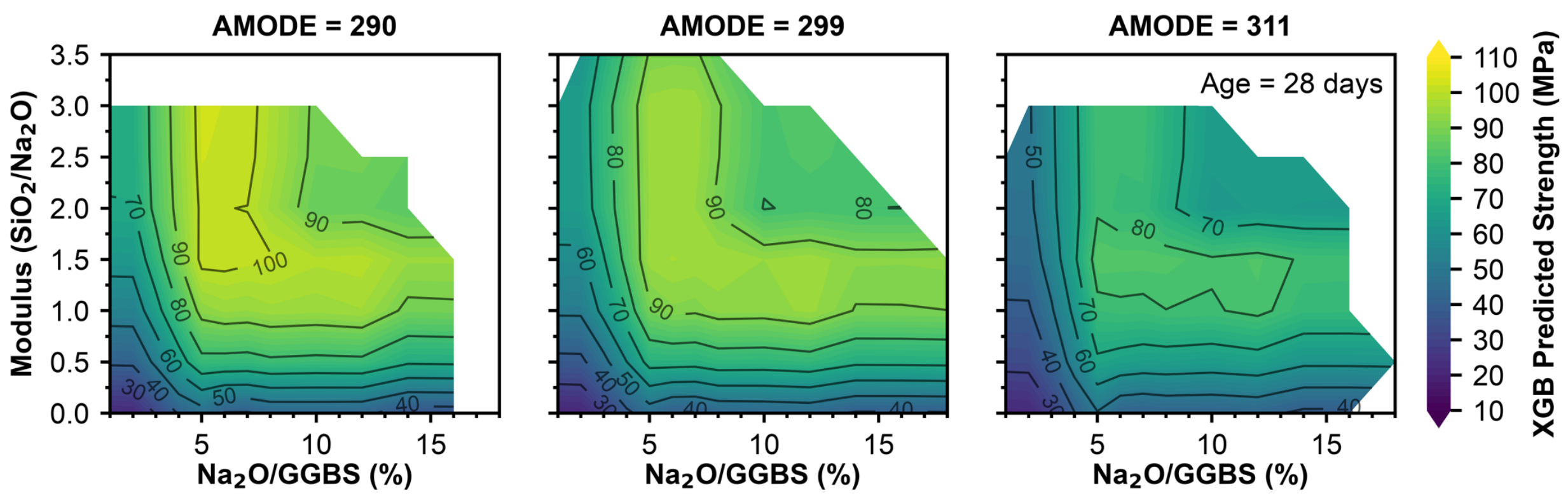

**Figure S39.** Two-dimensional mapping of XGB-predicted 28-day compressive strength as a function of $Na_2O$/GGBS and Ms for AAS pastes with varying GGBS reactivities: high (AMODE = 290), medium (AMODE = 299), and low (AMODE = 311). The water/GGBS ratio is fixed at 0.4. Strength values are mean predictions across 10 random states. Blank regions indicate combinations excluded by mass-balance screening or Mahalanobis distance filtering (**Section 2.5**).

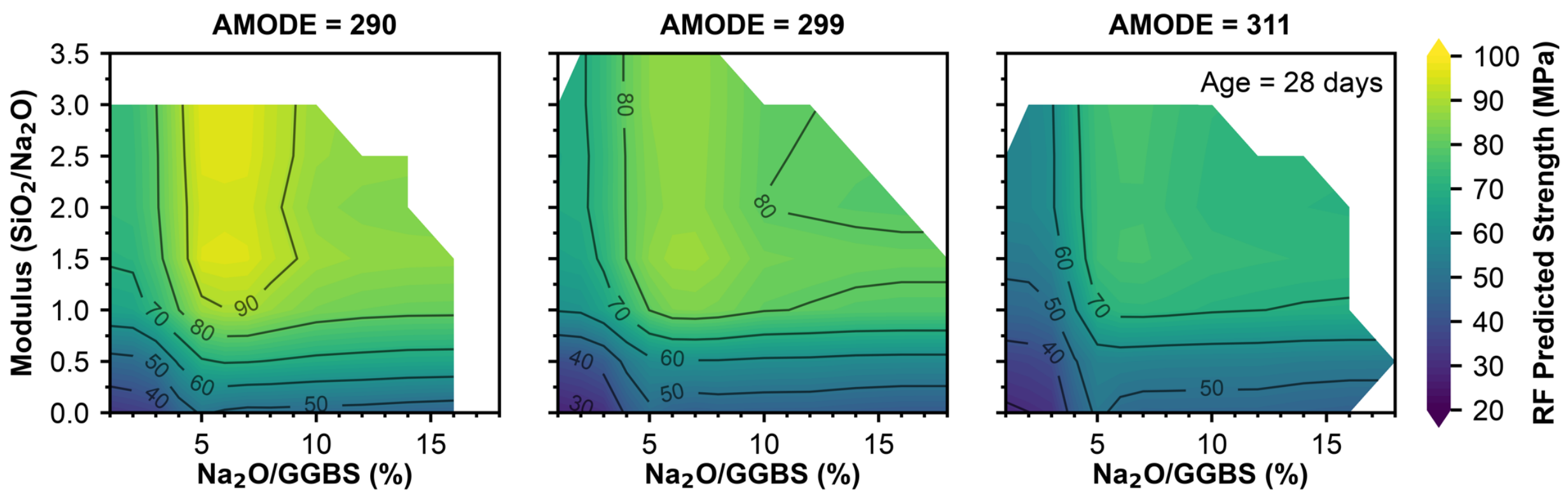

**Figure S40.** Two-dimensional mapping of RF-predicted 28-day compressive strength as a function of $Na_2O$/GGBS and Ms for AAS pastes with varying GGBS reactivities: high (AMODE = 290), medium (AMODE = 299), and low (AMODE = 311). The water/GGBS ratio is fixed at 0.4. Strength values are mean predictions across 10 random states. Blank regions indicate combinations excluded by mass-balance screening or Mahalanobis distance filtering (**Section 2.5**).

## S6.2 Optimal $Na_2O$/GGBS and Ms Combinations at Different Ages

### *S6.2.1 High-Reactivity GGBS*

**Figures S41–S43** present the predicted strengths with varying $Na_2O$/GGBS and Ms combinations of high-reactivity slag at different ages from ANN, XGB, and RF models.

Consistent with the observations in the previous section, ANN identified optimal $Na_2O$/GGBS and Ms combinations, along with a systematic shift toward lower alkalinity (i.e., lower $Na_2O$ dosage and higher Ms) as the curing age increased. XGB and RF also predict age-dependent shifts in the high-strength regions, particularly between 3 and 28 days, although these regions are broader and less localized than those predicted by GPR and ANN.

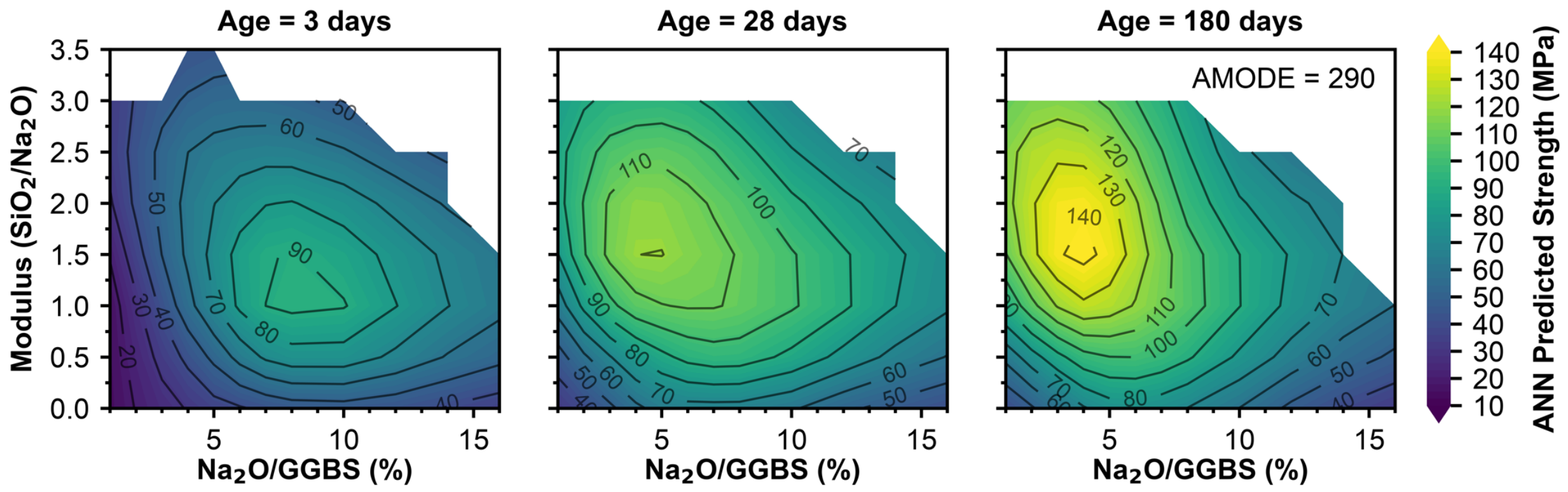

**Figure S41.** Two-dimensional mapping of ANN-predicted compressive strength at different curing ages (3, 28, and 180 days) as a function of $Na_2O$/GGBS and Ms for high-reactivity GGBS (AMODE = 290). The water/GGBS ratio is fixed at 0.4. Strength values are mean predictions across 10 random states. Blank regions indicate combinations excluded by mass-balance screening or Mahalanobis distance filtering (**Section 2.5**).

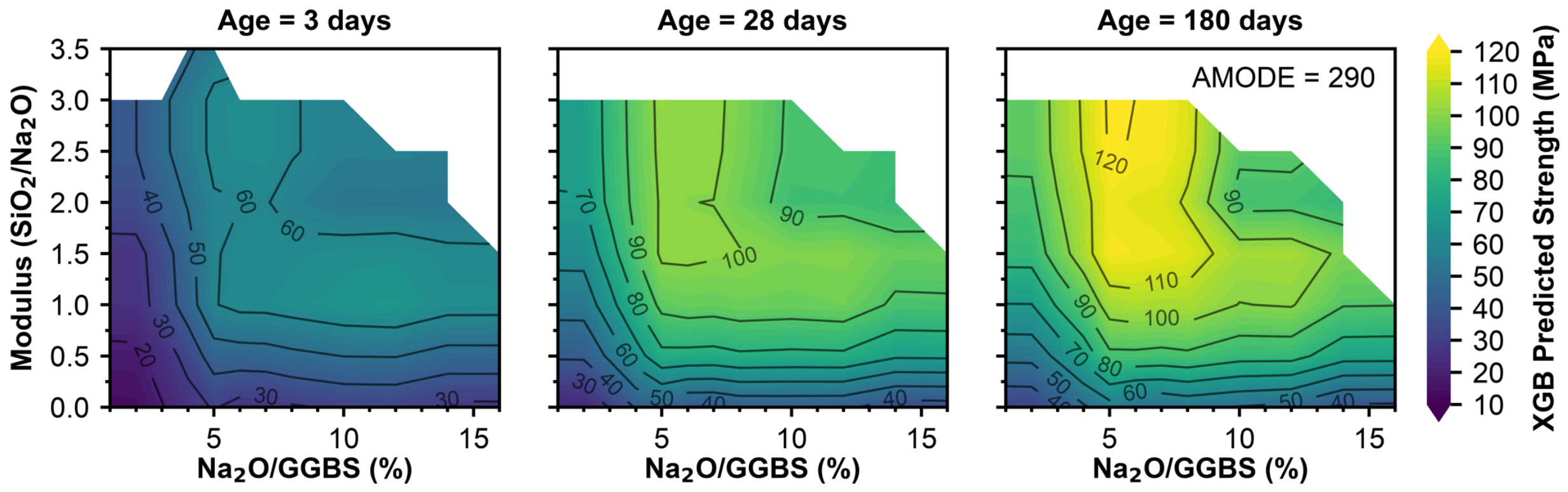

**Figure S42.** Two-dimensional mapping of XGB-predicted compressive strength at different curing ages (3, 28, and 180 days) as a function of $Na_2O$/GGBS and Ms for high-reactivity GGBS (AMODE = 290). The water/GGBS ratio is fixed at 0.4. Strength values are mean predictions across 10 random states. Blank regions indicate combinations excluded by mass-balance screening or Mahalanobis distance filtering (**Section 2.5**).

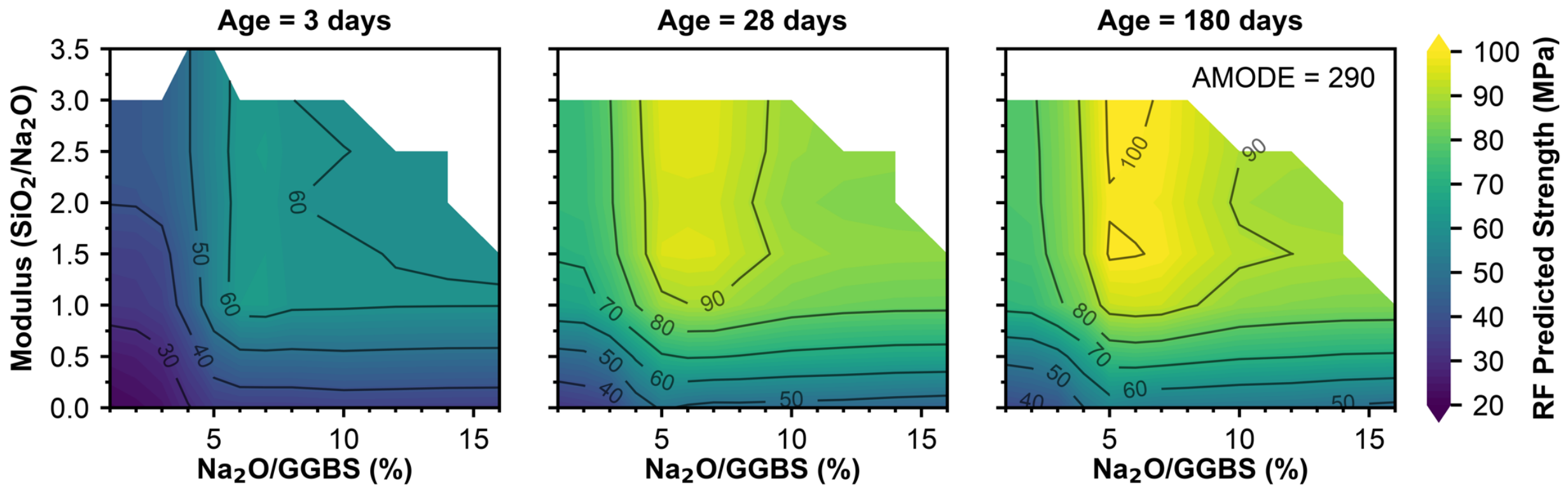

**Figure S43.** Two-dimensional mapping of RF-predicted compressive strength at different curing ages (3, 28, and 180 days) as a function of $Na_2O$/GGBS and Ms for high-reactivity GGBS (AMODE = 290). The water/GGBS ratio is fixed at 0.4. Strength values are mean predictions across 10 random states. Blank regions indicate combinations excluded by mass-balance screening or Mahalanobis distance filtering (**Section 2.5**).

*S6.2.2 Medium- and Low-Reactivity GGBS*

**Figures S44–S47** present the predicted compressive strength across varying $Na_2O$ and Ms combinations for medium- and low-reactivity GGBSs (AMODE = 299 and 311, respectively) from GPR, ANN, XGB, and RF models.

Both GPR and ANN captured a distinct shift in optimal activator composition toward lower alkalinity (primarily through lower $Na_2O$ dosages) with increasing curing age for the medium-reactivity GGBS (AMODE = 299), consistent with the pattern observed for high-reactivity slag in the previous subsection. For the low-reactivity GGBS (AMODE = 311), however, the shift is less pronounced, suggesting a weaker sensitivity of activator chemistry to curing age under slower reaction kinetics. XGB and RF also show age-dependent changes in the high-strength regions, although their predictions remain more band-like and less localized.

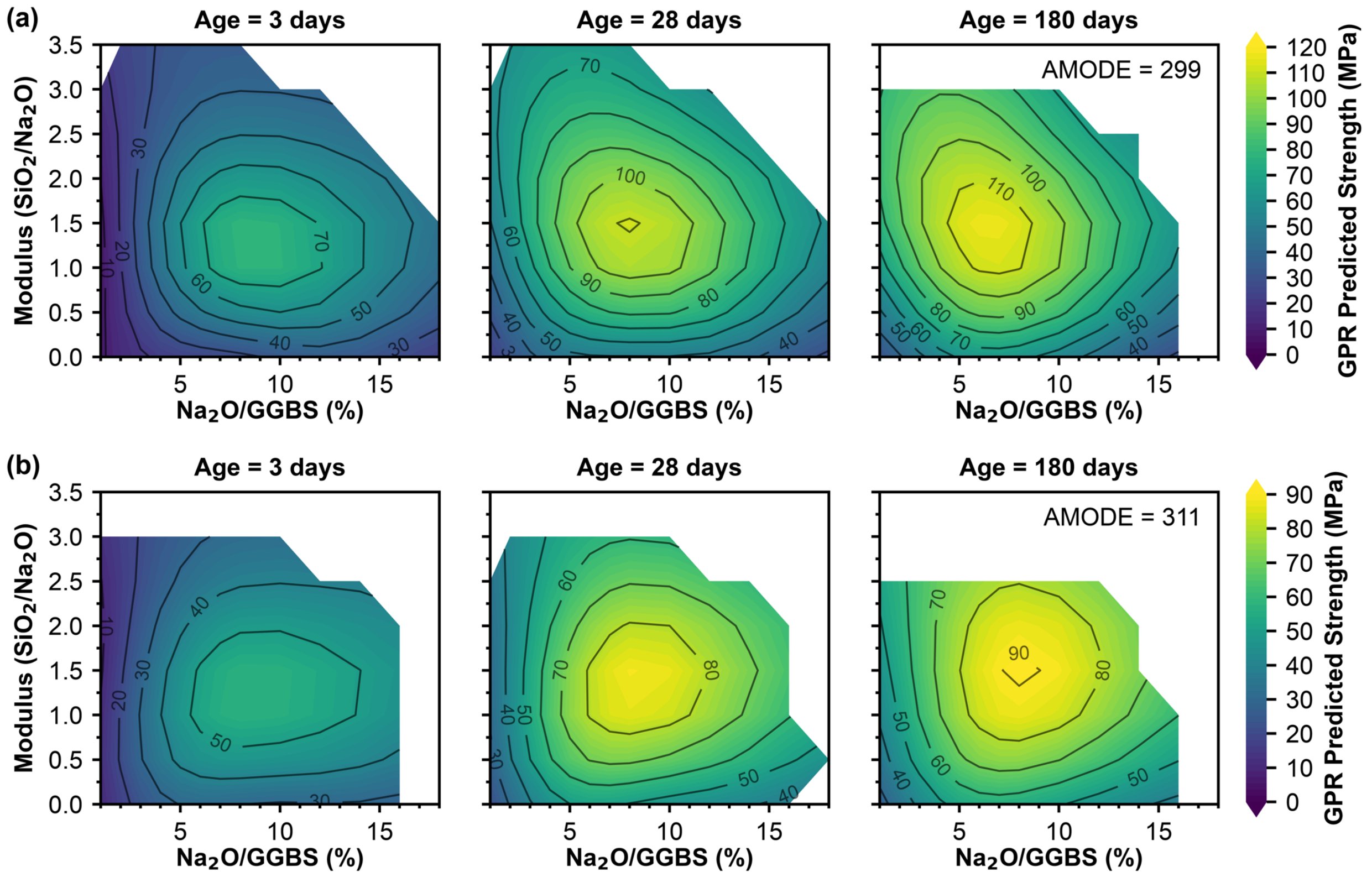

**Figure S44.** Two-dimensional mapping of GPR-predicted compressive strength at different curing ages (3, 28, and 180 days) across $Na_2O$/GGBS and Ms combinations for (a) medium- (AMODE=299) and (b) low-reactivity (AMODE=311) slags. The water/GGBS ratio is fixed at 0.4. Strength values are mean predictions across 10 random states. Blank regions indicate combinations excluded by mass-balance screening or Mahalanobis distance filtering (**Section 2.5**).

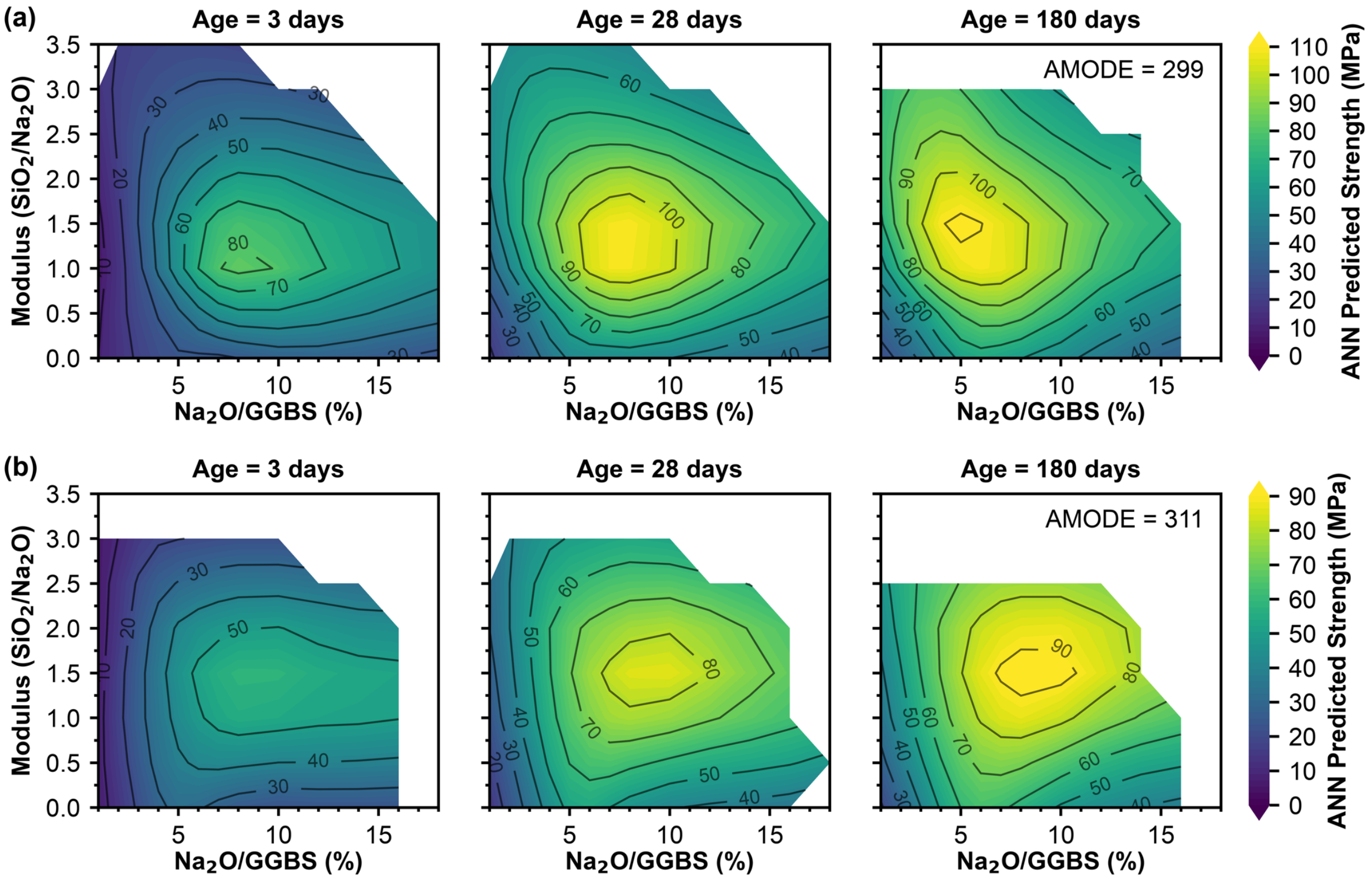

**Figure S45.** Two-dimensional mapping of ANN-predicted compressive strength at different curing ages (3, 28, and 180 days) across $Na_2O$/GGBS and Ms combinations for (a) medium- (AMODE=299) and (b) low-reactivity (AMODE=311) slags. The water/GGBS ratio is fixed at 0.4. Strength values are mean predictions across 10 random states. Blank regions indicate combinations excluded by mass-balance screening or Mahalanobis distance filtering (**Section 2.5**).

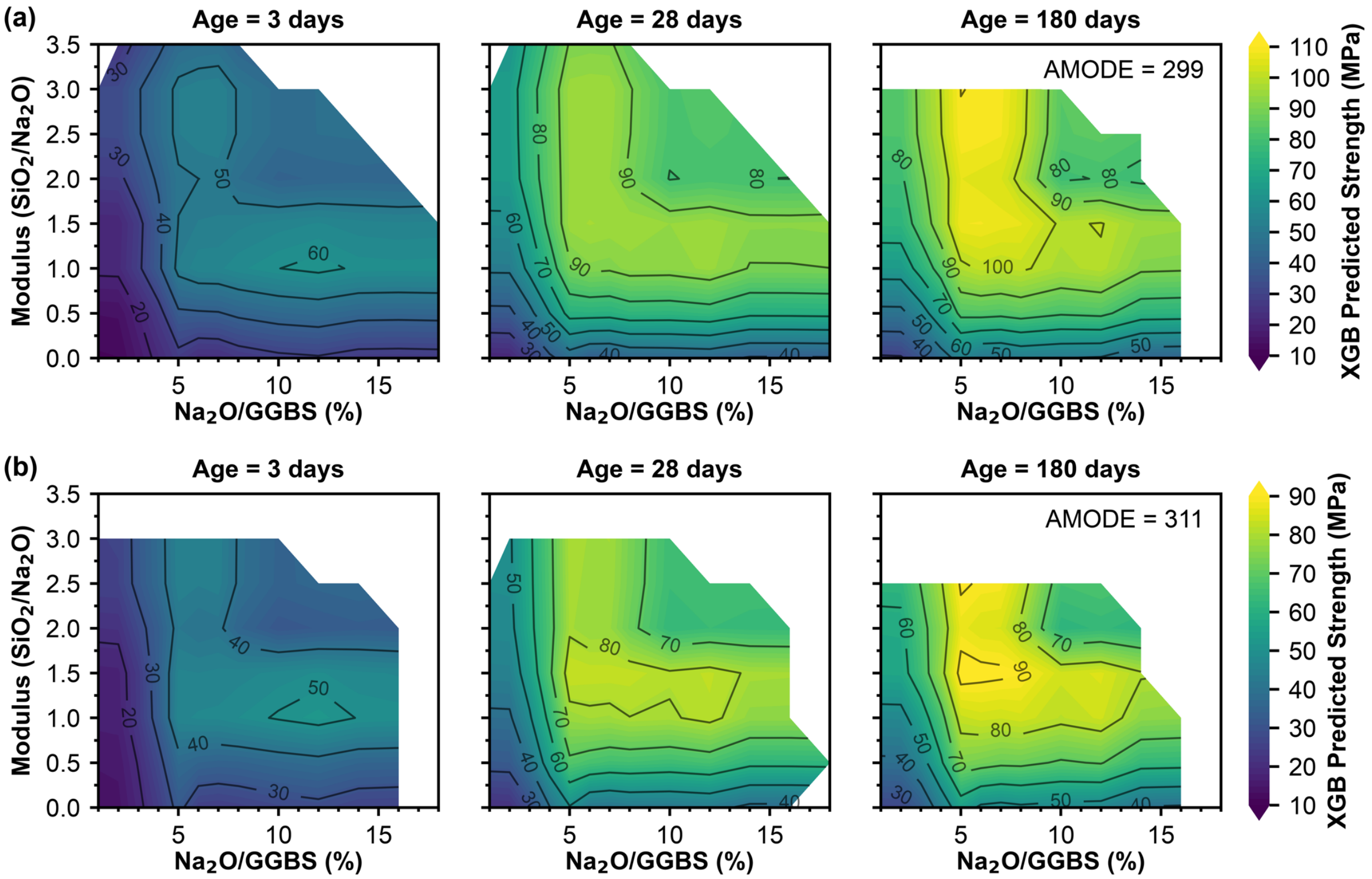

**Figure S46.** Two-dimensional mapping of XGB-predicted compressive strength at different curing ages (3, 28, and 180 days) across $Na_2O$/GGBS and Ms combinations for (a) medium- (AMODE=299) and (b) low-reactivity (AMODE=311) slags. The water/GGBS ratio is fixed at 0.4. Strength values are mean predictions across 10 random states. Blank regions indicate combinations excluded by mass-balance screening or Mahalanobis distance filtering (**Section 2.5**).

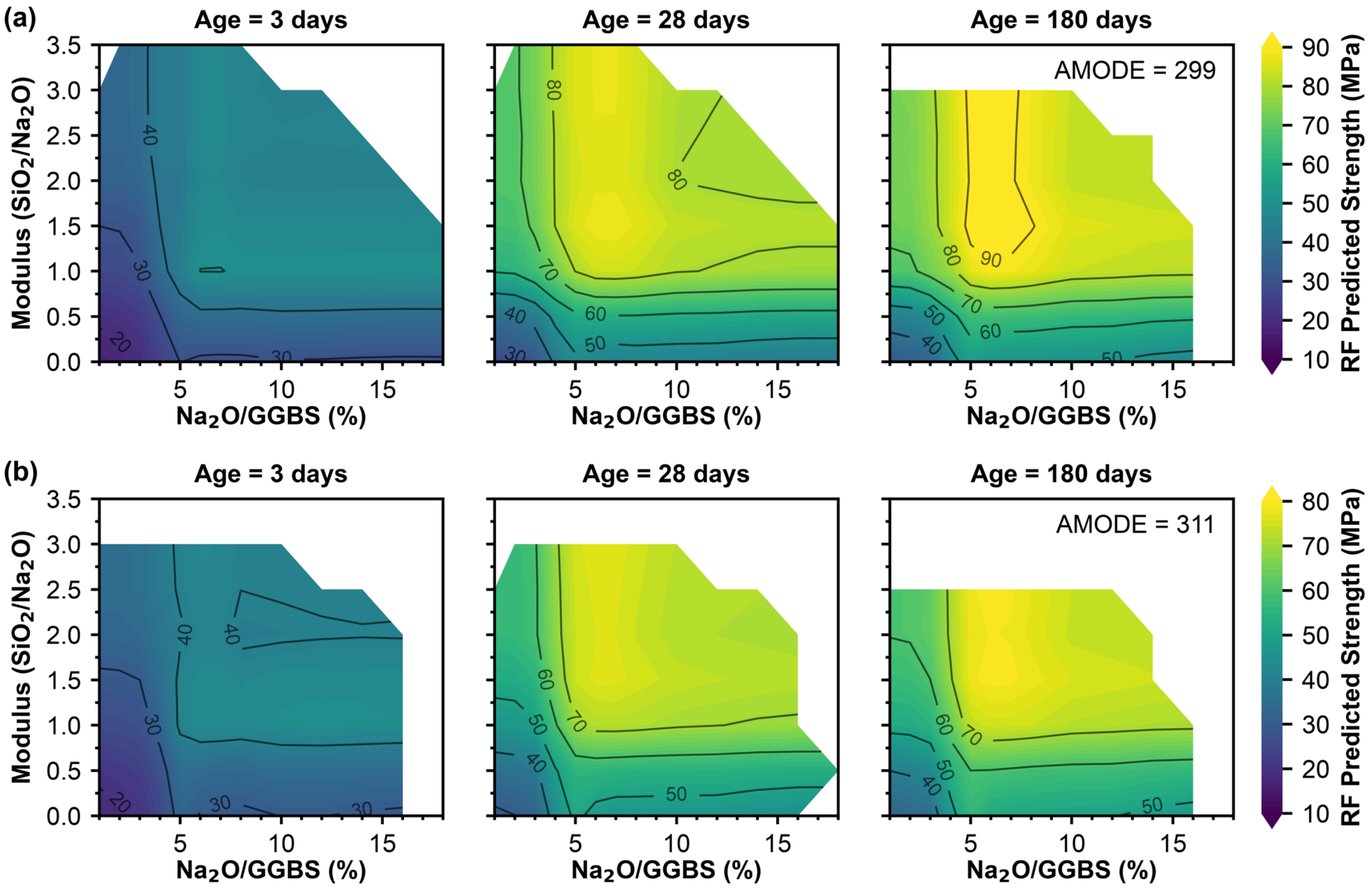

**Figure S47.** Two-dimensional mapping of RF-predicted compressive strength at different curing ages (3, 28, and 180 days) across $Na_2O$/GGBS and Ms combinations for (a) medium-(AMODE=299) and (b) low-reactivity (AMODE=311) slags. The water/GGBS ratio is fixed at 0.4. Strength values are mean predictions across 10 random states. Blank regions indicate combinations excluded by mass-balance screening or Mahalanobis distance filtering (see **Section 2.5**).

### S6.3 Sensitivity Analysis of Specimen Size

To assess whether the choice of 30 mm influenced the two-dimensional design space trends, GPR predictions under Scenario 4 were compared at specimen equivalent lengths of 30, 40, and 50 mm, with aspect ratio fixed at 1 and water/GGBS at 0.4. Comparisons covered curing ages of 3, 28, and 180 days at AMODE = 290 kcal/mol, and AMODE values of 290, 299, and 311 kcal/mol at 28 days. Within each figure, comparisons were restricted to common $Na_2O$–Ms combinations retained after screening.

Increasing specimen size generally lowered the absolute predicted strengths, while the main contour patterns and regions of high predicted strength remained broadly similar (**Figures S48** and **S49**). The shifts in favorable activator ranges with curing age and slag reactivity were also largely preserved. These results indicate that the principal two-dimensional mix-design trends were not specific to the 30 mm assumption within the tested size range.

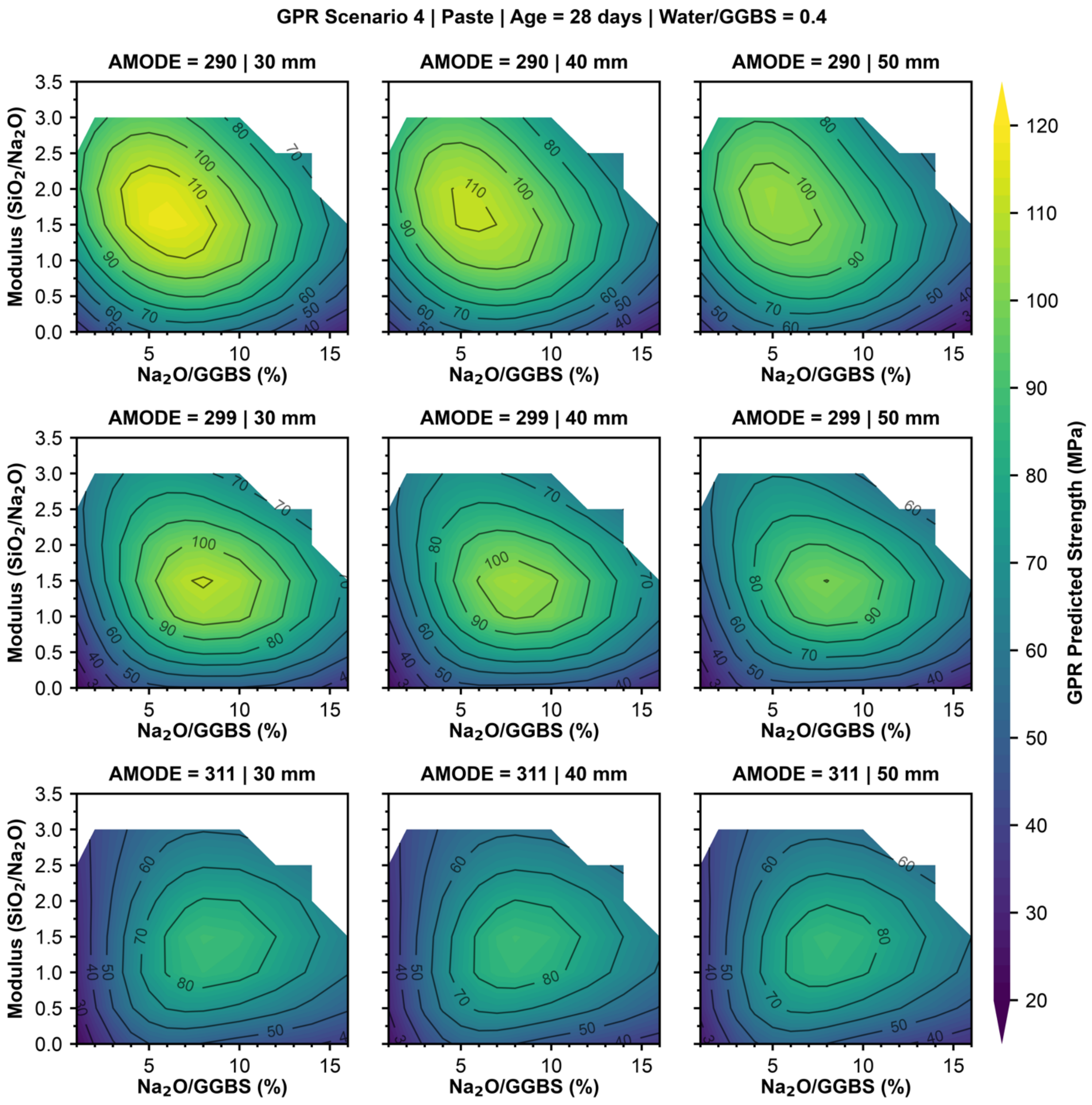

**Figure S48.** Sensitivity of GPR-predicted strength contours to specimen equivalent length at different slag reactivities (Scenario 4). Rows correspond to AMODE values of 290, 299, and 311 kcal/mol; columns to equivalent lengths of 30, 40, and 50 mm. Curing age = 28 days, water/GGBS = 0.4, and aspect ratio = 1.

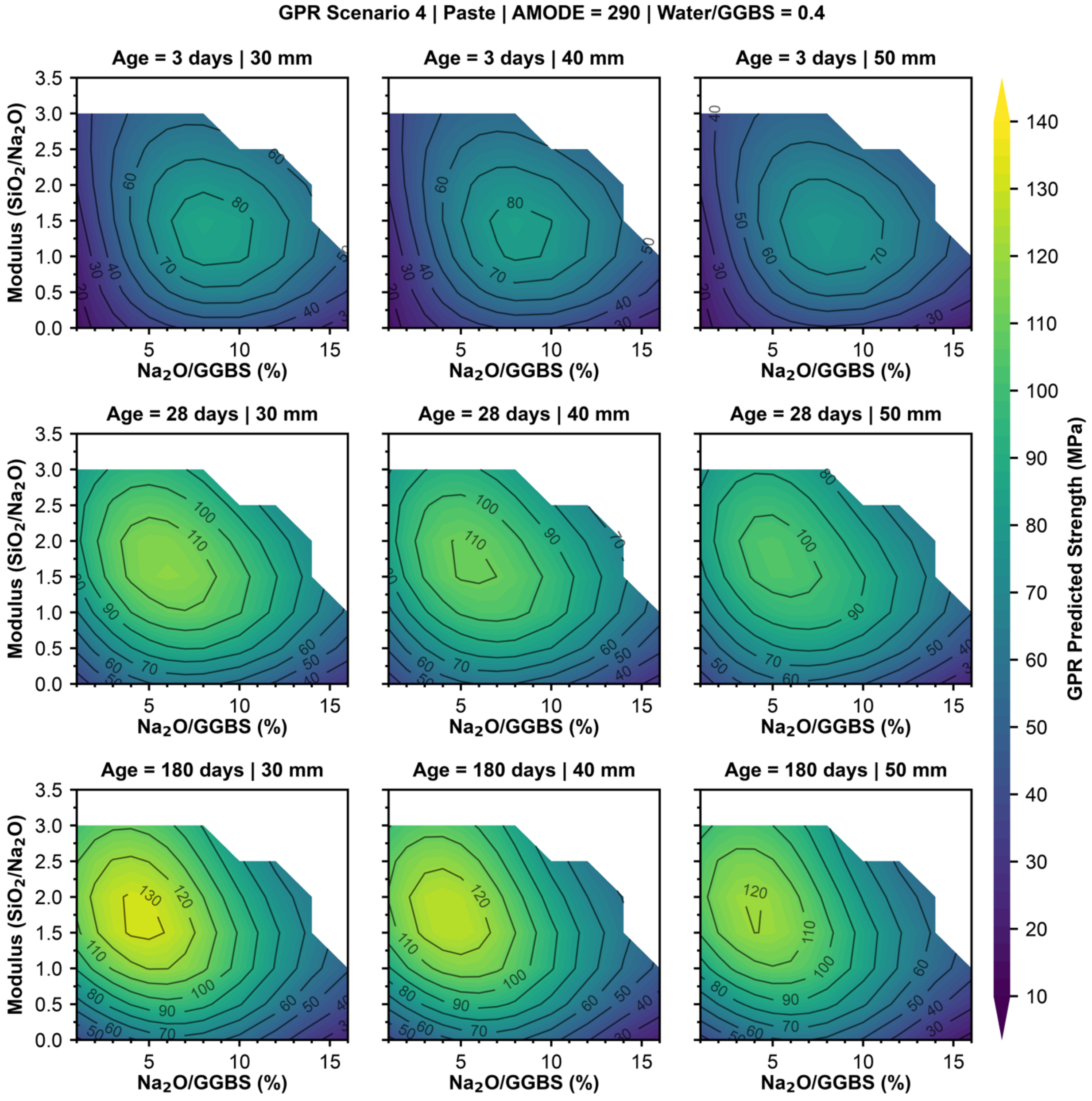

**Figure S49.** Sensitivity of GPR-predicted strength contours to specimen equivalent length at different curing ages (Scenario 4). Rows correspond to 3, 28, and 180 days; columns to equivalent lengths of 30, 40, and 50 mm. AMODE = 290 kcal/mol, water/GGBS = 0.4, and aspect ratio = 1.

### S6.4 Age Coverage and Predictive Uncertainty

**Table S14** summarizes the curing-age coverage and mixture-type distribution of the curated dataset. 2518 of the 3053 records (82.48%) were obtained at ages up to 28 days, while 499 records cover 56–180 days and originate from 58 studies. At 180 days specifically, the dataset contains 75 records from 18 studies, representing 70 distinct mixtures, including 35 paste records from 8 studies. Distinct mixtures were identified within each study using the recorded slag properties and mixture proportions, irrespective of curing age, curing conditions, or specimen geometry. These counts do not imply statistical independence.

**Table S14.** Curing-age coverage and mixture-type distribution of the curated dataset.

| Age, t (days) | Records | Mixture type | | | Source studies | Distinct mixtures |
|---|---|---|---|---|---|---|
| | | Paste | Mortar | Concrete | | |
| $t \leq 3$ | 836 (27.38%) | 345 (41.27%) | 361 (43.18%) | 130 (15.55%) | 88 | 563 |
| $3 < t < 28$ | 838 (27.45%) | 362 (43.20%) | 315 (37.59%) | 161 (19.21%) | 99 | 612 |
| $28 \leq t < 180$ | 1271 (41.63%) | 381 (29.98%) | 462 (36.35%) | 428 (33.67%) | 124 | 786 |
| $t \geq 180$ | 108 (3.54%) | 63 (58.33%) | 27 (25.00%) | 18 (16.67%) | 20 | 73 |
| Total | 3053 (100.00%) | 1151 (37.70%) | 1165 (38.16%) | 737 (24.14%) | 136 | 880 |

*Notes:* Record percentages refer to the full dataset; mixture-type percentages refer to each row. Studies and mixtures may span multiple age intervals; totals count unique entries.

**Figure S50** provides an additional diagnostic of GPR predictive uncertainty at 3, 28, and 180 days under Scenario 4. The models were trained on records spanning multiple curing ages, and each map uses the design domain retained after mass-balance screening and Mahalanobis distance filtering. Predictive standard deviation (SD) denotes the model-estimated SD averaged across ten fitted models, which differs from the variability of mean predictions across random states. Relative predictive SD is calculated by dividing the mean predictive SD by the mean strength prediction across the ten random states and expressing the result as a percentage.

Both absolute and relative uncertainty maps show generally lower uncertainty within the predicted favorable regions and higher uncertainty toward the boundaries of the retained design space. The maps also suggest that predictive uncertainty does not show a simple monotonic increase with curing age. These uncertainty patterns provide additional support for the internal consistency of the model-generated design maps but do not validate temporal extrapolation or the predicted age-dependent shifts.

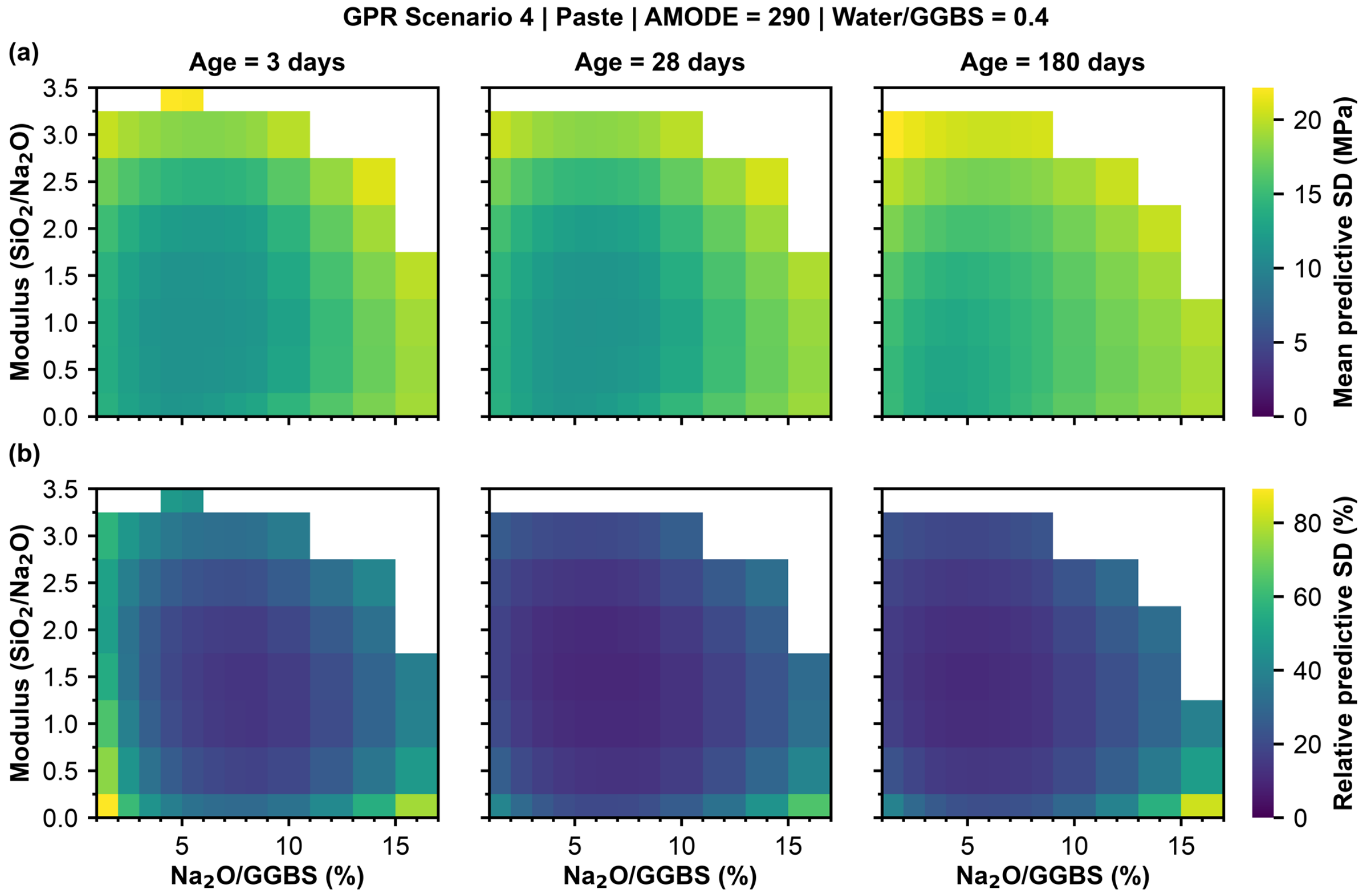

**Figure S50.** GPR predictive uncertainty at 3, 28, and 180 days (left to right) under Scenario 4: (a) predictive standard deviation (SD) averaged across ten random states; (b) relative predictive SD, calculated as the mean SD divided by the mean predicted strength and expressed as a percentage. Results are for paste with AMODE = 290 kcal/mol, water/GGBS = 0.4, specimen equivalent length = 30 mm, and aspect ratio = 1.

## S7 Extended Environmental and Economic Implications

### S7.1 Environmental and Economic Implications from ANN, XGB, and RF models

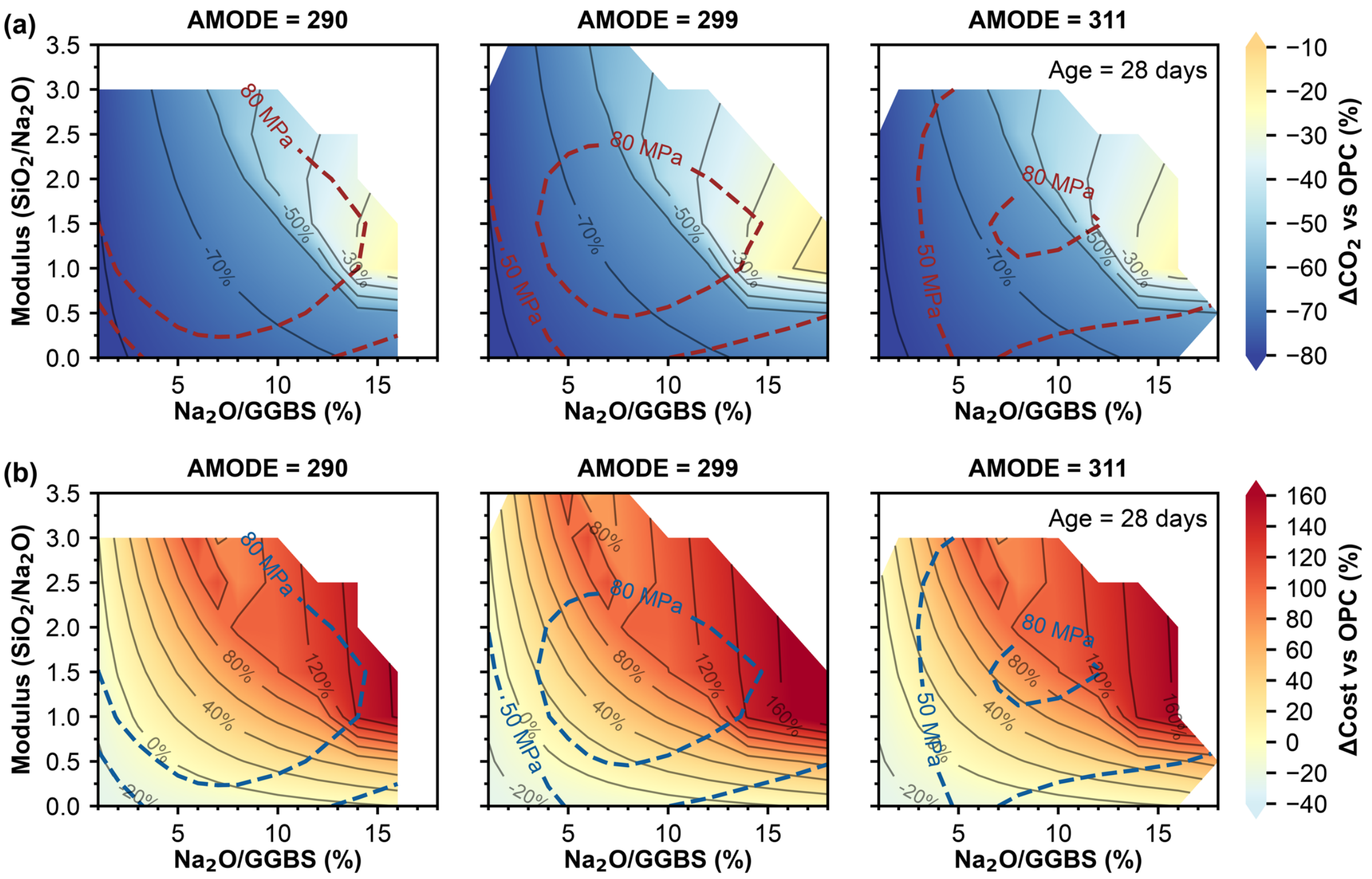

**Figure S51.** Two-dimensional mapping of the relative (a) $CO_2$ emissions and (b) cost associated with varying $Na_2O$/GGBS and Ms combinations for AAS pastes with different GGBS reactivities. The water/GGBS ratio is fixed at 0.4. The calculations follow the same procedure described in **Section 2.6**, with both $CO_2$ emissions and cost expressed relative to the ordinary Portland cement (OPC) paste benchmark derived from industrial data reported in [14]. The dashed contour lines represent the ANN-predicted 50 MPa and 80 MPa 28-day strength levels achieved at corresponding combinations of $Na_2O$/GGBS and Ms. Strength contours are based on mean predictions across 10 random states. Blank regions indicate combinations excluded by mass-balance screening or Mahalanobis distance filtering (**Section 2.5**).

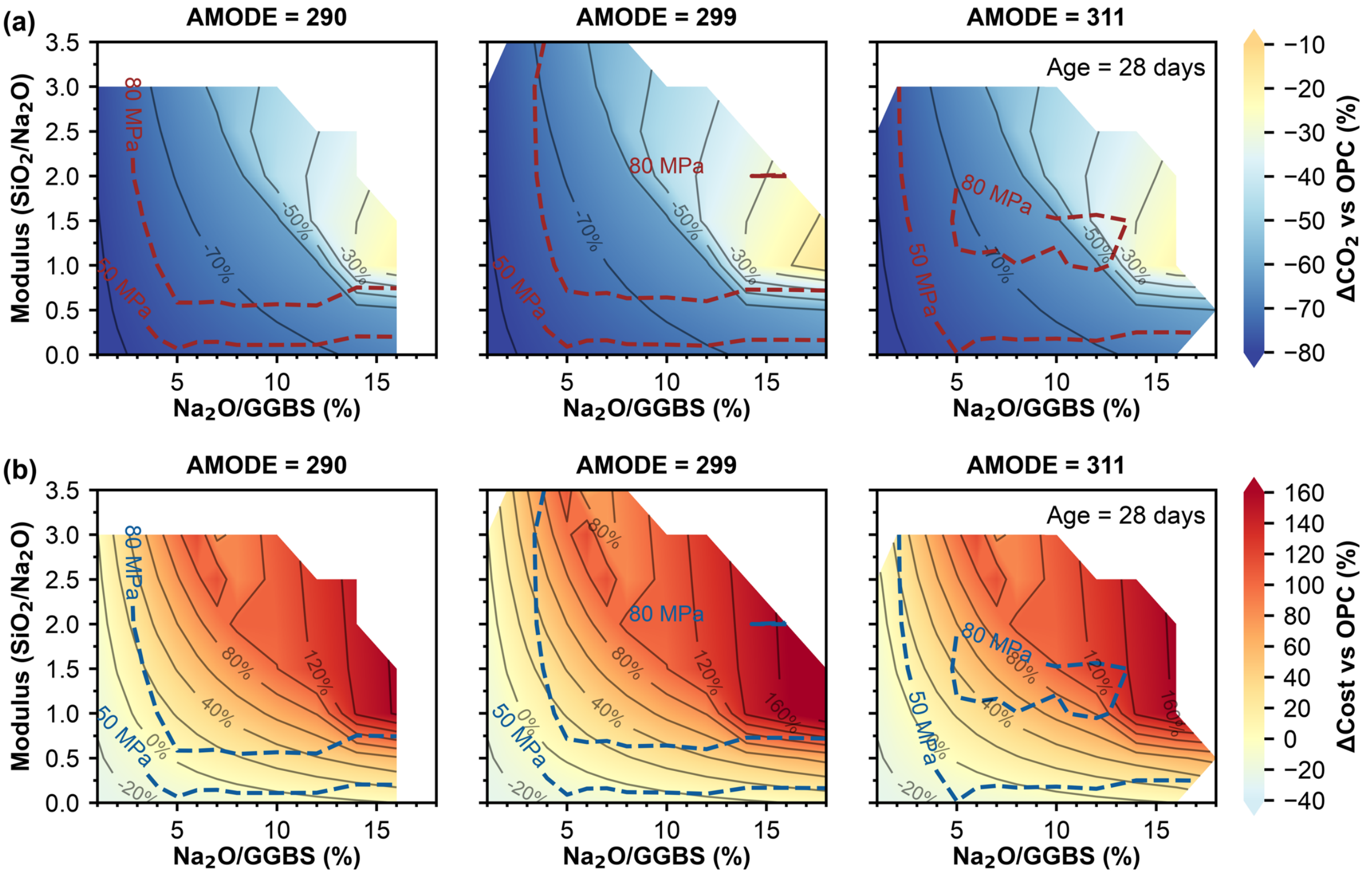

**Figure S52.** Two-dimensional mapping of the relative (a) $CO_2$ emissions and (b) cost associated with varying $Na_2O$/GGBS and Ms combinations for AAS pastes with different GGBS reactivities. The water/GGBS ratio is fixed at 0.4. The calculations follow the same procedure described in **Section 2.6**, with both $CO_2$ emissions and cost expressed relative to the OPC paste benchmark derived from industrial data reported in [14]. The dashed contour lines represent the XGB-predicted 28-day strength levels of 50 MPa and 80 MPa, achieved at corresponding combinations of $Na_2O$/GGBS and Ms. Strength contours are based on mean predictions across 10 random states. Blank regions indicate combinations excluded by mass-balance screening or Mahalanobis distance filtering (**Section 2.5**).

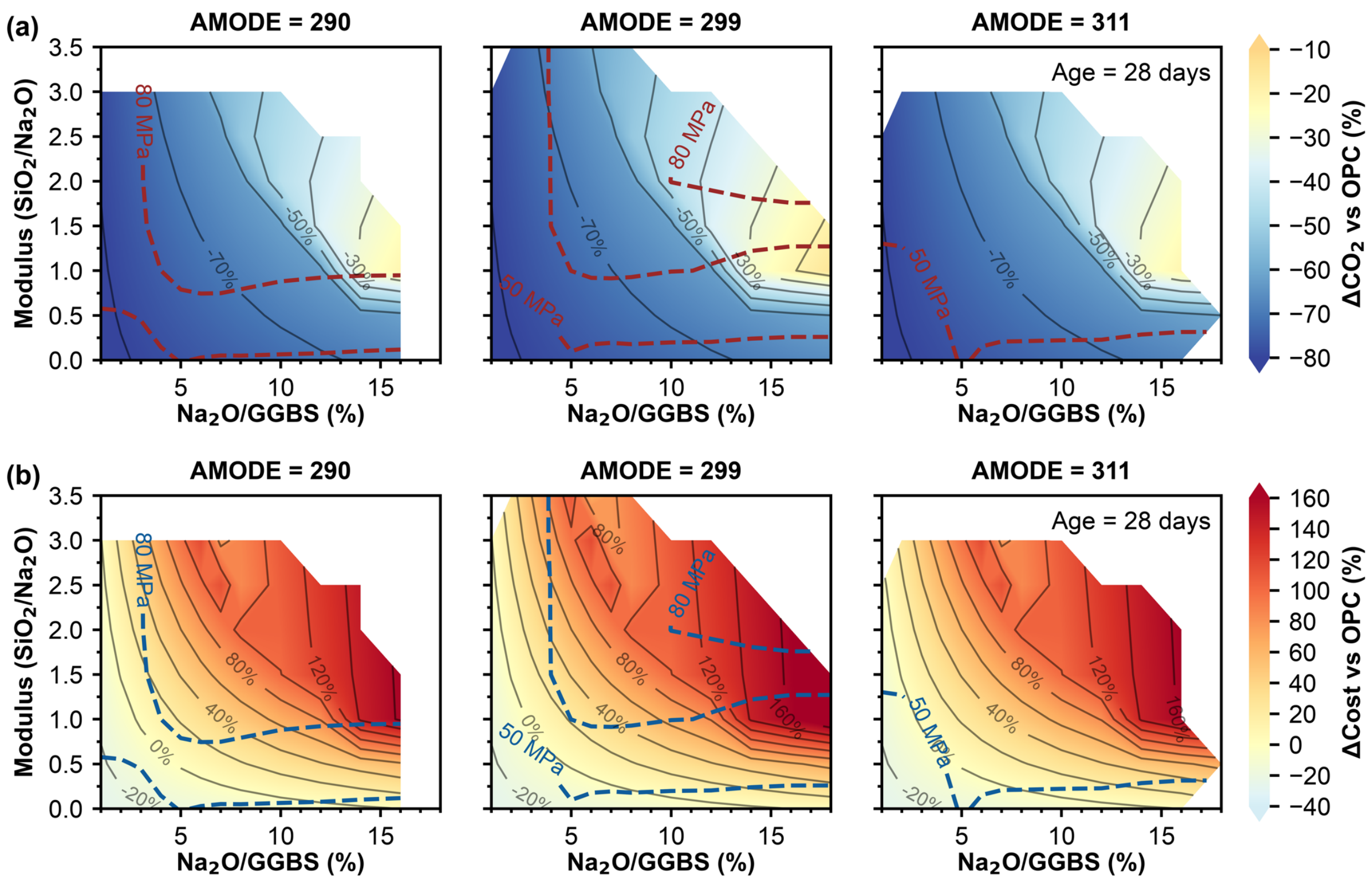

**Figure S53.** Two-dimensional mapping of the relative (a) $CO_2$ emissions and (b) cost associated with varying $Na_2O$/GGBS and Ms combinations for AAS pastes with different GGBS reactivities. The water/GGBS ratio is fixed at 0.4. The calculations follow the same procedure described in **Section 2.6**, with both $CO_2$ emissions and cost expressed relative to the OPC paste benchmark derived from industrial data reported in [14]. The dashed contour lines represent the RF-predicted 28-day strength levels of 50 MPa and 80 MPa, achieved at corresponding combinations of $Na_2O$/GGBS and Ms. Strength contours are based on mean predictions across 10 random states. Blank regions indicate combinations excluded by mass-balance screening or Mahalanobis distance filtering (**Section 2.5**).

## S7.2 Economic Impact Considering Carbon Tax Scenarios

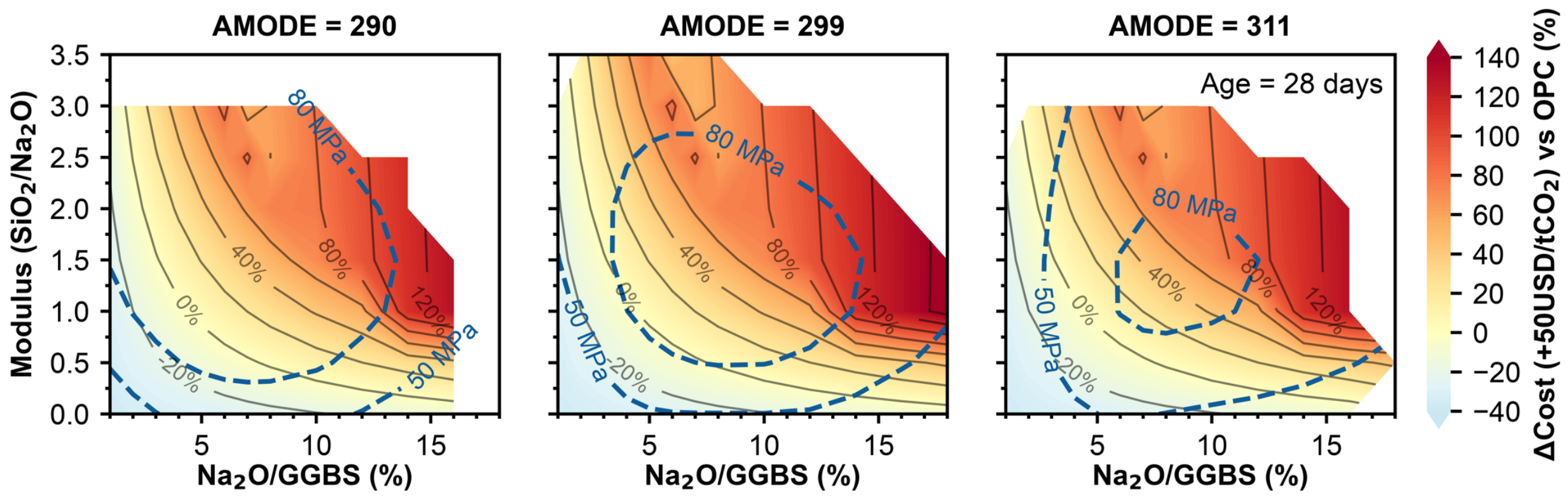

**Figure S54.** Two-dimensional mapping of cost after accounting for carbon tax, associated with varying $Na_2O$/GGBS and Ms combinations for AAS pastes with different GGBS reactivities. The water/GGBS ratio is fixed at 0.4. The carbon tax is 50 USD/ton $CO_2$, following the Rocky Mountain Institute (RMI)'s methodology [15]. The dashed contour lines represent the GPR-predicted 50 MPa and 80 MPa 28-day strength levels achieved at corresponding combinations of $Na_2O$/GGBS and Ms. Strength contours are based on mean predictions across 10 random states. Blank regions indicate combinations excluded by mass-balance screening or Mahalanobis distance filtering (**Section 2.5**).

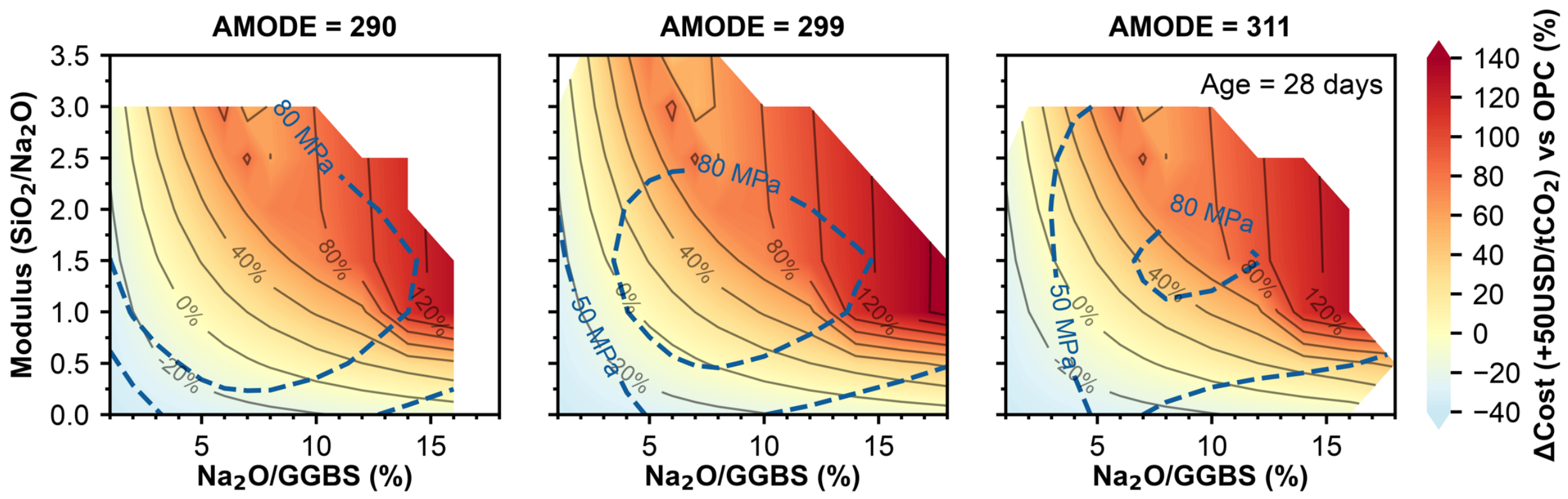

**Figure S55.** Two-dimensional mapping of cost after accounting for carbon tax, associated with varying $Na_2O$/GGBS and Ms combinations for AAS pastes with different GGBS reactivities. The water/GGBS ratio is fixed at 0.4. The carbon tax is 50 USD/ton $CO_2$, following the RMI's methodology [15]. The dashed contour lines represent the ANN-predicted 50 MPa and 80 MPa 28-day strength levels achieved at corresponding combinations of $Na_2O$/GGBS and Ms. Strength contours are based on mean predictions across 10 random states. Blank regions indicate combinations excluded by mass-balance screening or Mahalanobis distance filtering (**Section 2.5**).

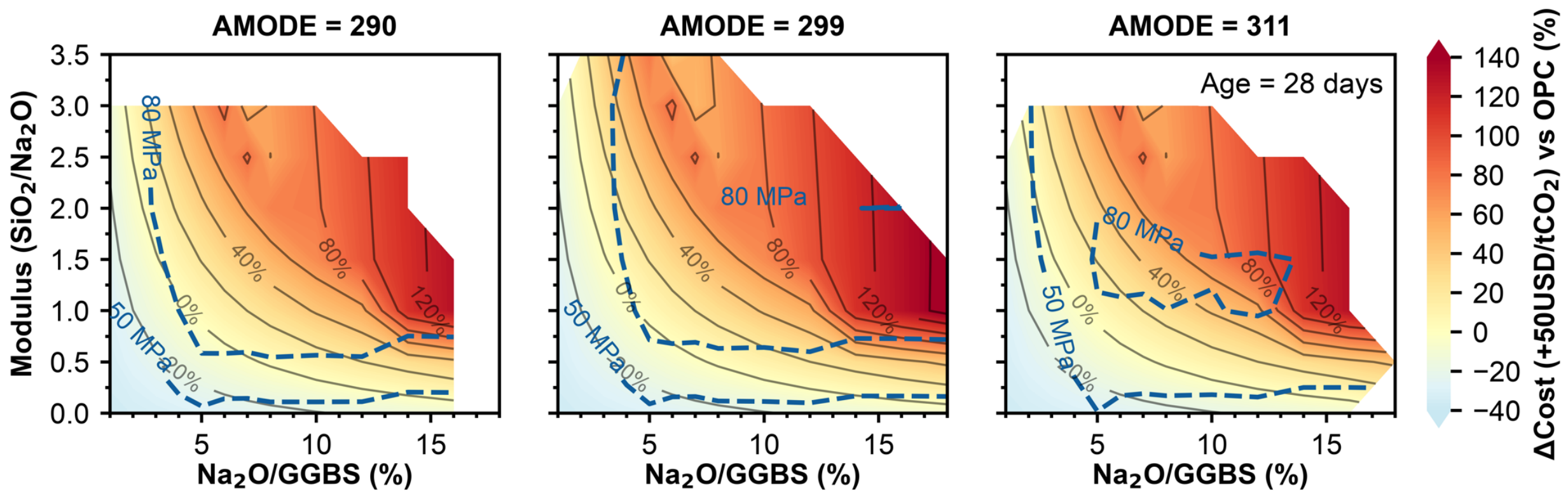

**Figure S56.** Two-dimensional mapping of cost after accounting for carbon tax, associated with varying $Na_2O$/GGBS and Ms combinations for AAS pastes with different GGBS reactivities. The water/GGBS ratio is fixed at 0.4. The carbon tax is 50 USD/ton $CO_2$, following the RMI's methodology [15]. The dashed contour lines represent the XGB-predicted 50 MPa and 80 MPa 28-day strength levels achieved at corresponding combinations of $Na_2O$/GGBS and Ms. Strength contours are based on mean predictions across 10 random states. Blank regions indicate combinations excluded by mass-balance screening or Mahalanobis distance filtering (**Section 2.5**).

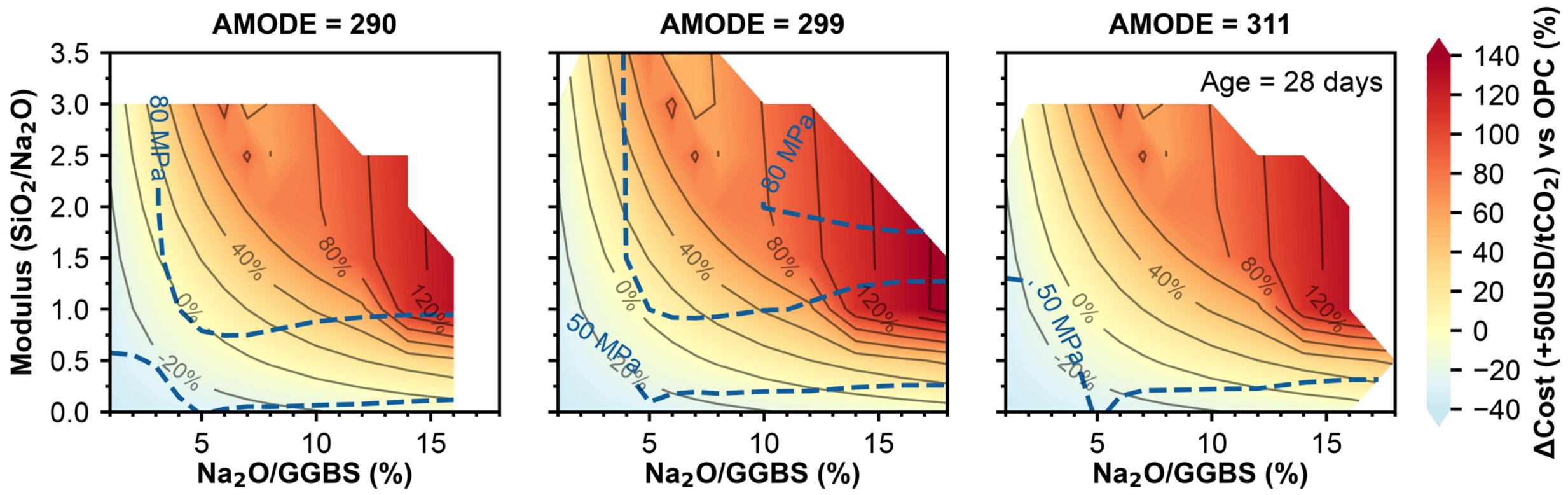

**Figure S57.** Two-dimensional mapping of cost after accounting for carbon tax, associated with varying $Na_2O$/GGBS and Ms combinations for AAS pastes with different GGBS reactivities. The water/GGBS ratio is fixed at 0.4. The carbon tax is 50 USD/ton $CO_2$, following the RMI's methodology [15]. The dashed contour lines represent the RF-predicted 50 MPa and 80 MPa 28-day strength levels achieved at corresponding combinations of $Na_2O$/GGBS and Ms. Strength contours are based on mean predictions across 10 random states. Blank regions indicate combinations excluded by mass-balance screening or Mahalanobis distance filtering (**Section 2.5**).